\documentclass[iop,appendixfloats]{emulateapj} 
\usepackage[T1]{fontenc}
\usepackage{amsmath}
\usepackage{amssymb}	
\usepackage{graphicx,aas_macros}
\usepackage{subfigure}
\usepackage{bm}
\usepackage{color}
\usepackage[dvipsnames]{xcolor}
\usepackage{setspace}

\usepackage[breaklinks,colorlinks,citecolor=blue]{hyperref}
\usepackage{footnotebackref}
\usepackage[all]{hypcap}

\def\stacksymbols #1#2#3#4{\def\theguybelow{#2}
        \def\verticalposition{\lower#3pt}
        \def\spacingwithinsymbol{\baselineskip0pt\lineskip#4pt}
        \mathrel{\mathpalette\intermediary#1}}
\def\intermediary #1#2{\verticalposition\vbox{\spacingwithinsymbol
        \everycr={}\tabskip0pt
        \halign{$\mathsurround0pt#1\hfil##\hfil$\crcr#2\crcr
                \theguybelow\crcr}}}

\def\lta{\stacksymbols{<}{\sim}{2.5}{.2}}
\def\gta{\stacksymbols{>}{\sim}{2.5}{.2}}

\def\kms{{\rm km\,s^{-1}}}
\def\kmsd{{\rm km^2\,s^{-2}}}
\def\Tx{$T_{\rm x}$}
\def\tx{T_{\rm x}}
\def\Lx{$L_{\rm x}$}
\def\lx{L_{\rm x}}
\def\R500{$R_{500}$}
\def\r500{R_{500}}
\def\Sige{$\sigma_{\rm e}$}
\def\sige{\sigma_{\rm e}}
\def\Re{$R_{\rm e}$}
\def\re{R_{\rm e}}
\def\Rx{$R_{\rm x}$}
\def\rx{R_{\rm x}}
\def\Rxg{$R_{\rm x,g}$}
\def\rxg{R_{\rm x,g}}
\def\Rxc{$R_{\rm x,c}$}
\def\rxc{R_{\rm x,c}}

\newcommand{\be}{\begin{equation}}
\newcommand{\ee}{\end{equation}}
\newcommand{\bea}{\begin{eqnarray}}
\newcommand{\eea}{\end{eqnarray}}

\newcommand{\msun}{M_{\odot}}

\newcommand{\mbh}{M_\bullet}
\newcommand{\Mbh}{$M_\bullet$}

\newcommand{\cc}{{\rm cm}^{-3}}
\newcommand{\es}{{\rm erg\,s^{-1}}}

\newcommand{\Chandra}{\textit{Chandra}}
\newcommand{\Athena}{\textit{Athena}}
\newcommand{\Lynx}{\textit{Lynx}}

\begin{document}

\shortauthors{m.\,gaspari et al.}
\title{ 
The X-ray Halo Scaling Relations of Supermassive Black Holes}
\shorttitle{the link between smbhs and hot x-ray halos of galaxies, groups, and clusters}
\author{M.~Gaspari$^{1,*,\dagger}$,
D.~Eckert$^2$, 
S.~Ettori$^{3,4}$,
P.~Tozzi$^5$, 
L.~Bassini$^{6,7}$, 
E.~Rasia$^{6,8}$, 
F.~Brighenti$^{9}$,\\
M.~Sun$^{10}$,
S.~Borgani$^{6,7,8,11}$,
S.~D.~Johnson$^{1, 12, \ddagger}$,
G.~R.~Tremblay$^{13}$, 
J.~M.~Stone$^1$, \\
P.~Temi$^{14}$, 
H.-Y.~K.~Yang$^{15,16}$,
F.~Tombesi$^{17,15,18,19}$, 
M.~Cappi$^3$
\vspace{+0.2cm}
}
\affil{
\scriptsize$^1\,$Department of Astrophysical Sciences, Princeton University, 4 Ivy Lane, Princeton, NJ 08544-1001, USA \\
\scriptsize$^2\,$Department of Astronomy, University of Geneva, ch. d'Ecogia 16, 1290 Versoix, Switzerland \\
\scriptsize$^3\,$INAF, Osservatorio di Astrofisica e Scienza dello Spazio, via Pietro Gobetti 93/3, 40129 Bologna, Italy \\
\scriptsize$^4\,$INFN, Sezione di Bologna, viale Berti Pichat 6/2, I-40127 Bologna, Italy \\
\scriptsize$^5\,$INAF, Astronomy Observatory of Florence, Largo Enrico Fermi 5, 50125, Firenze, Italy\\
\scriptsize$^6\,$INAF - Osservatorio Astronomico di Trieste, via Tiepolo 11, 34122, Trieste, Italy \\
\scriptsize$^7\,$Astronomy Unit, Department of Physics, University of Trieste, via Tiepolo 11, I-34131 Trieste, Italy \\
\scriptsize$^8\,$IFPU - Institute for Fundamental, Physics of the Universe, Via Beirut 2, 34014 Trieste, Italy \\ 
\scriptsize$^9\,$Dipartimento di Fisica e Astronomia, Università di Bologna, via Gobetti 93, 40127 Bologna, Italy \\
\scriptsize$^{10}\,$Physics Department, University of Alabama in Huntsville, Huntsville, AL 35899, USA \\
\scriptsize$^{11}\,$INFN - National Institute for Nuclear Physics, Via Valerio 2, I-34127 Trieste, Italy \\
\scriptsize$^{12}\,$The Observatories of the Carnegie Institution for Science, 813 Santa Barbara Street, Pasadena, CA 91101, USA \\
\scriptsize$^{13}\,$Center for Astrophysics $|$ Harvard \& Smithsonian, 60 Garden St., Cambridge, MA 02138, USA \\
\scriptsize$^{14}\,$Astrophysics Branch, NASA Ames Research Center, Moffett Field, CA 94035, USA \\
\scriptsize$^{15}\,$Department of Astronomy, University of Maryland, College Park, MD 20742, USA \\
\scriptsize$^{16}\,$Joint Space-Science Institute, College Park, MD 20742, USA \\ 
\scriptsize$^{17}\,$Department of Physics, University of Rome "Tor Vergata", Via della Ricerca Scientifica 1, 00133, Rome, Italy \\\scriptsize$^{18}\,$X-ray Astrophysics Laboratory, NASA/Goddard Space Flight Center, Greenbelt, MD, 20771, USA \\
\scriptsize$^{19}\,$INAF - Osservatorio Astronomico di Roma, via Frascati 33, 00044, Monte Porzio Catone (Roma), Italy
 }

\altaffiltext{\hspace{-0.15in} * }{E-mail: mgaspari@astro.princeton.edu}
\altaffiltext{\hspace{-0.15in} $\dagger$ }{\textit{Lyman Spitzer Jr.}~Fellow}
\altaffiltext{\hspace{-0.15in} $\ddagger$ }{\textit{Hubble} and \textit{Carnegie-Princeton} Fellow}

\begin{abstract}
\noindent
We carry out a comprehensive Bayesian correlation analysis between hot halos and {\it direct} masses of supermassive black holes (SMBHs), by retrieving the X-ray plasma properties (temperature, luminosity, density, pressure, masses) over galactic to cluster scales for 85 diverse systems. We find new key scalings, with the tightest relation being the $\mbh - \tx$, followed by $\mbh - \lx$. The tighter scatter (down to 0.2 dex) and stronger correlation coefficient of all the X-ray halo scalings compared with the optical counterparts (as the $\mbh-\sige$) suggest that plasma halos play a more central role than stars in tracing and growing SMBHs (especially those that are ultramassive). Moreover, $\mbh$ correlates better with the gas mass than dark matter mass. We show the important role of the environment, morphology, and relic galaxies/coronae, as well as the main departures from virialization/self-similarity via the optical/X-ray fundamental planes. We test the three major channels for SMBH growth: hot/Bondi-like models have inconsistent anti-correlation with X-ray halos and too low feeding; cosmological simulations find SMBH mergers as sub-dominant over most of the cosmic time and too rare to induce a central-limit-theorem effect; the scalings are consistent with chaotic cold accretion (CCA), the rain of matter condensing out of the turbulent X-ray halos that sustains a long-term self-regulated feedback loop. The new correlations are major observational constraints for models of SMBH feeding/feedback in galaxies, groups, and clusters (e.g., to test cosmological hydrodynamical simulations), and enable the study of SMBHs not only through X-rays, but also via the Sunyaev-Zel'dovich effect (Compton parameter), lensing (total masses), and cosmology (gas fractions). 
\vspace{+0.23cm}
\end{abstract}


\section{Introduction} \label{s:intro}
\setcounter{footnote}{0}
Supermassive black holes (SMBHs) are found at the center of most -- if not all -- galaxies (e.g., \citealt{Kormendy:2013} for a review).  High-resolution observations of stellar and cold gas kinematics in the central regions of nearby galaxies have enabled dynamical measurements of central SMBH masses in over a hundred objects \citep[e.g.,][]{Magorrian:1998,Ferrarese:2000,Gebhardt:2000,Gultekin:2009,Beifiori:2012,Saglia:2016,vBosch:2016}. The measured masses of the SMBHs are correlated with the luminosity ($L_K$) and effective velocity dispersion (\Sige) of the host galaxy, suggesting a co-evolution between the SMBH and the properties of their host environments.
These findings further imply an interplay between the feeding/feedback mechanisms of the SMBH and its host galaxy \citep{Silk:1998}. 
During the active galactic nucleus (AGN) phase,
outflows and jets from the central SMBH are thought to play a fundamental role in establishing the multiphase environment in their host halo, quenching cooling flows/star formation, and shaping the galaxy luminosity function
\citep[e.g.,][]{Fabian:2012,Tombesi:2013,Tombesi:2015,McNamara:2007,McNamara:2012,King:2015,Fiore:2017}.
For these reasons, AGN feedback has become a crucial ingredient in modern galaxy formation models \citep[e.g.,][]{Sijacki:2007,Borgani:2008,Booth:2009,Gaspari:2011a,Gaspari:2011b,Yang:2016a,Tremmel:2017}. 

While SMBH feedback is central to galaxy evolution, the mechanism through which the observed correlations between BH mass $\mbh$ and galaxy (or halo) properties are established is still debated. In the simple, idealized gravitational scenario, BH seeds are thought to grow rapidly at high redshift, with the scaling relations arising from the bottom-up structure formation process in which large structures are formed through the merging of smaller structures under the action of gravity (leading to virialization and self-similarity; e.g., \citealt{Kravtsov:2012}).
In this scenario, the central SMBHs of the merging systems settle to the bottom of the potential well of the newly formed halo and eventually merge, inducing hierarchical scaling relations between $\mbh$ and galaxy properties (e.g., \citealt{Peng:2007,Jahnke:2011}).

However, in recent years, measurements of BH masses in the most massive local galaxies (ultramassive black holes -- UMBHs) have challenged the hierarchical formation scenario \citep[e.g.,][]{McConnell:2011,Hlavacek-Larrondo:2012,Hlavacek-Larrondo:2015,McConnell:2013,Thomas:2016}. Some studies reported dynamical masses in excess of $10^{10}M_\odot$, i.e. about an order of magnitude greater than expected from the $M_\bullet-\sige$ and $M_\bullet-L_K$ relations. 
A prominent example of such an outlier is M87 (NGC\,4486), for which the BH mass of $6.5\times10^9M_\odot$ lies an order of magnitude above that expected from the $M_\bullet-\sige$ relation \citep{Gebhardt:2011}. Spectacular observations by the Event Horizon Telescope (EHT) have recently confirmed the extreme mass of this object \citep{EHT:2019I,EHT:2019VI}. Recent works have suggested that the environment and location of such UMBHs at the bottom of the potential well of galaxy clusters and groups, where the most massive galaxies are formed (known as brightest cluster/group galaxies -- BCGs/BGGs), could be responsible for the observed deviations \citep{Gaspari:2017_uni,Bogdan:2018,Bassini:2019}. 

Beyond the stellar component, an important ingredient for SMBH feeding is the surrounding X-ray emitting plasma halo.
At scales beyond the effective galactic radius, the majority of baryons are found in the form of a diffuse ($n_{\rm e}<0.1\;\cc$) and hot ($\tx>0.1$\;keV) plasma, often referred to as the circumgalactic (CGM), intragroup (IGrM), or intracluster medium (ICM -- e.g., \citealt{Sarazin:1986,Mathews:2003,Kravtsov:2012,Sun:2012,Gonzalez:2013,Eckert:2016}).
In the central regions of relaxed, cool-core (CC) systems, the plasma densities are such that the cooling time of the hot ICM/IGrM becomes much smaller than the Hubble time. Thus, a fraction of the hot gas\footnote{For consistency with the literature, we refer interchangeably to the diffuse plasma component by using the `gas' nomenclature.} cools and condenses in the central galaxy, forming extended warm filaments detected in H$\alpha$ and cold molecular clouds that fuel star formation (e.g., \citealt{Fabian:2002,Peterson:2006,Combes:2007,McDonald:2010,McDonald:2011a,McDonald:2018,Gaspari:2015_xspec,Molendi:2016,Temi:2018,Tremblay:2015,Tremblay:2018,Nagai:2019}). A portion of the cooling gas ignites the central AGN, which triggers the SMBH response via outflowing material that regulates the cooling flow of the macro-scale gaseous halo \citep[e.g.,][]{Birzan:2004,Rafferty:2006,McNamara:2007,Gaspari:2017_uni}. Such SMBHs follow an intermittent duty cycle
\citep{Birzan:2012}, as evidenced by the common presence of radio-emitting AGN, especially in massive galaxies 
\citep{Burns:1990,Mittal:2009,Bharadwaj:2014,Main:2017}.

Over the past decade, extensive investigations have been carried out in order to understand the mechanism through which AGN inject energy into the surrounding medium and how the condensed filaments/clouds form out of the hot halos \citep{Gaspari:2009,Gaspari:2011a,Gaspari:2011b,Gaspari:2017_cca,Pizzolato:2010,McCourt:2012,Sharma:2012,Li:2014,Prasad:2015,Prasad:2017,Voit:2015_nat,Voit:2017,Valentini:2015,Soker:2016,Yang:2019}. A novel paradigm has emerged in which the AGN feedback cycle operates through chaotic cold accretion (CCA; \citealt{Gaspari:2013_cca,Gaspari:2015_cca,Gaspari:2017_cca}), where turbulent eddies induced by AGN outflows (and cosmic flows; \citealt{Lau:2017}) are responsible for the condensation of multiphase gas out of the hot halos via nonlinear thermal instability.
The condensed gas then rapidly cools and rains toward the central SMBH. Within $r<100$ pc, the clouds start to collide inelastically and get efficiently funneled inward within a few tens of Schwarzschild radii, where an accretion torus rapidly pushes the gas through the BH horizon via magneto-rotational instability (MRI; e.g., \citealt{Balbus:2003,Sadowski:2017}).
A growing body of studies suggests that, in spite of the mild average Eddington ratios\footnote{The Eddington ratio is defined as follows: $\dot M_{\rm Edd} \equiv L_{\rm Edd}/(0.1\,c^2)\simeq 22.8\,(\mbh/10^9\,\msun)\,\msun\,{\rm yr^{-1}}$, where $L_{\rm Edd}=10^{47}(\mbh/10^9\,\msun)\;\es$ is the Eddington luminosity.}, the mass accreted through CCA over long timescales can account for a substantial fraction of the SMBH masses 
(e.g., \citealt{Gaspari:2013_cca,Gaspari:2015_cca,Voit:2015_gE,Tremblay:2016,Tremblay:2018,Prasad:2017}).
Alternative models treat BH accretion purely from the single-phase, hot gas perspective, following the seminal work by \citet{Bondi:1952} and related variants (e.g., \citealt{Narayan:2011}), predicting unintermittent accretion rates inversely tied to the plasma entropy. Further models, such as hierarchical major/minor mergers and high-redshift quasars, are tackled in \S\ref{s:disc}, in particular by means of cosmological simulations.

This work is part of the \textit{BlackHoleWeather} program (PI: M.\,Gaspari), which aims at understanding the link between the central SMBH and its surrounding halo, both from the theoretical and observational points of view. 
Historically, this paper was initiated five years ago, inspired by the thorough review by \citet{Kormendy:2013}.
We make use of precise dynamical (direct) SMBH mass measurements collected from the literature \citep{Kormendy:2013,McConnell:2013,vBosch:2016} over a wide range of systems -- including central galaxies and satellites, early- and late-type galaxies (ETGs, LTGs) -- and correlate them with the properties of the surrounding hot X-ray atmosphere (X-ray luminosity, temperature, gas mass, pressure/thermal energy, and entropy).  
A tight correlation between $\mbh-\tx$ and $\mbh-\lx$ is indeed expected based on first-principle arguments initially proposed by \citet{Gaspari:2017_uni}.
We focus here only on hot X-ray plasma halos and related SMBHs, leaving halos falling below this band to future work (e.g., UV and related intermediate mass BHs -- IMBHs).
We further compare the X-ray scalings with the optical counterparts via both univariate and multivariate correlations.
We discover new correlations between the various hot gas properties and SMBH mass, which help us to test the main models of macro-scale BH feeding, i.e., hot Bondi-like accretion, CCA, and hierarchical mergers.
With the advent of gravitational-wave astronomy (LISA), direct SMBH imaging (EHT), and next-generation X-ray instruments with superb angular resolution and sensitivity (\Athena, XRISM, and the proposed \Lynx\ and AXIS), it is vital to understand how SMBHs form and grow.

This work is structured as follows.
In \S\ref{s:data}, we present the retrieved data sample (85 galaxies) from a thorough literature search.
In \S\ref{s:res}, we describe the main results in terms of a robust Bayesian statistical analysis of all the X-ray and optical properties (via univariate and multivariate correlations; Table \ref{tabsum}).
In \S\ref{s:disc}, we probe the main models of BH feeding and discuss key astrophysical insights arising from the presented correlations. 
In \S\ref{s:conc}, we summarize the major results of the study and provide concluding remarks.
As used in most literature studies,
we adopt throughout the work a flat concordance cosmology with $H_0 = 70\;{\rm km\,s^{-1}\,Mpc^{-1}}$ ($h=0.7$) and $(\Omega_{\rm m},\Omega_{\Lambda}) = (0.3, 0.7)$. The Hubble time is $t_{\rm H}\equiv1/H_0\simeq13.9$\,Gyr, which well approximates the age of the universe.


\section{Data Analysis} \label{s:data}

\subsection{Data sample and fundamental properties} \label{s:sample}
The main objective of this study is to measure the observed correlations between direct SMBH masses and both the stellar and plasma halo properties. To achieve this goal, we performed
a thorough search of the past two decades of the related observational literature
aimed at assembling the fundamental observables in both the optical and X-ray band for a large sample of (85) galaxies.
The selection is straightforward: we inspected any SMBH with a direct/dynamical BH mass measurement (\citealt{vBosch:2016}) and looked for an available X-ray detection, in terms of galactic, group, and cluster emission from diffuse hot plasma. 
We tested combinations of these datasets, with comparable results.
The potential role of selection effects is discussed in \S\ref{s:cav}.
The retrieved, homogenized fundamental variables are listed in Appendix~\ref{a:tables}, including the detailed references and notes for each galaxy in the sample. In the next two subsections, we describe their main optical and X-ray features.

\subsubsection{Optical stellar observables and BH masses}\label{s:ovar}
Table~\ref{tabop} lists all the optical properties and BH masses of the galaxies in our sample.
The vast majority of the BH masses come from \citet{vBosch:2016}, who compiled high-quality, dynamical measurements, mostly from \citet{Gultekin:2009}, \citet{Sani:2011}, \citet{Beifiori:2012}, \citet{McConnell:2013}, \citet{Kormendy:2013}, \citet{Rusli:2013}, and \citet{Saglia:2016}. 
Direct methods imply resolving the stellar or (ionized) gas kinematics shaped by the BH influence region $G\mbh/\sigma_v^2\sim1$\,-100\,pc (for a few galaxies, water masers or reverberation mapping are other feasible methods; see \citealt{Kormendy:2013} for a technical review).
Such scales require observations with arcsec/sub-arcsec resolution (the majority of which have been enabled by HST), thus limiting direct BH detections to the local universe (distance $D< 150$\,Mpc or redshift $z<0.04$).\footnote{The evolution factor is negligible, given the low-redshift sample: $E(z)=\sqrt{\Omega_{\rm m}(1+z)^3  +\Omega_\Lambda}<1.02$.}
One case (M87) includes the first direct imaging of the SMBH horizon available via EHT (\citealt{EHT:2019I}).
In this study we focus on X-ray halos and related SMBHs, leaving BHs associated with gaseous halos emitting below the X-ray band to future investigations (i.e., IMBHs with $\mbh \lta 3\times10^7\,\msun$).
Further, we do not include SMBH masses with major upper limits (e.g., NGC 4382, UGC 9799, NGC 3945) or which are substantially uncertain in the literature (e.g., Cygnus\,A, NGC\,1275). 
The direct BHs with reliable X-ray data are listed, in ascending order, in column (vi) of Table~\ref{tabop}, for a total robust sample of 85 BHs, spanning a wide range of masses $\mbh\sim4 \times10^7$\,-\,$2\times10^{10}\,\msun$.
We remark that it is crucial to adopt \textit{direct} BH mass measurements, instead of converting a posteriori from the $\mbh-\sige$ and $\mbh - L_K$ relations, or the AGN fundamental plane (\citealt{Merloni:2003_FP}), which can lead to biased, non-independent correlations with unreliable conversion uncertainty $\gta1$\,dex (\citealt{Fujita:2004,Mittal:2009,Main:2017,Phipps:2019}).

Unless noted in Table~\ref{tabop}, the stellar velocity dispersion, effective radius, and total luminosity are from the collection by \citet{vBosch:2016}, who further expanded the optical investigations by \citet{Cappellari:2013}, \citet{Kormendy:2013}, \citet{McConnell:2013}, and \citet{Saglia:2016}.
All the collected properties are rescaled as per our adopted distances $D$ (column (v) in Tab.~\ref{tabop}; e.g., $\mbh \propto D$ and $L_K\propto D^2$).
The measurement of the (effective) stellar velocity dispersion $\sige$ is typically carried out via long-slit or integral-field-unit (IFU) spectroscopy, by measuring the optical emission-line broadening of the spectrum integrated within the effective half-light radius $\re$ (or by the luminosity-weighted -- LW --  average of its radial profile; \citealt{McConnell:2013}). 
Note that $\sige$ is a one-dimensional (1D), line-of-sight velocity dispersion.
An excellent proxy for the the total (bulge plus eventual disk) stellar luminosity is the near-infrared (NIR) luminosity in the $K_{\rm s}$ band (rest-frame 2.0\,-\,2.3\;$\mu$m)\footnote{We drop the `s' (short) subscript throughout the manuscript, using only the $K$ band nomenclature.}, 
given its very low sensitivity to dust extinction (and star formation efficiencies).
\citet{vBosch:2016} carried out a detailed photometric growth-curve analysis based on a non-parametric determination of the galaxy $L_K$ and half-light radii. This Monte-Carlo method fits each galaxy several hundred times with S\'ersic profiles in which the outermost index is incrementally varied ($0.5<n<4$) until convergence is reached, thus leading to the listed total luminosity and effective radius $\re$ (column (viii) and (x) in Tab.~\ref{tabop}; the latter related to the major axis of the isophote containing 50\% of the emitted NIR light). 
In our final sample, the range of stellar luminosities spans $L_K\sim10^{10}$\,-\,$10^{12}\, L_\odot$, with effective radii between 1\,-\,100\;kpc. 
Total luminosities can be further converted to stellar masses by using the mean stellar mass-to-light ratio (\citealt{Kormendy:2013}):
\begin{equation}\label{e:ML}
\log M_\ast/L_K = 0.29 \log (\sigma_{\rm e}/{\rm km\,s^{-1}}) - 0.64 \pm 0.09. 
\end{equation}
About 1/3 of our galaxies have a significant disk component, which is reflected in a bulge-to-total luminosity ratio $B/T<1$ (column (xi) in Tab.~\ref{tabop}, from the photometric decomposition, mostly by \citealt{Beifiori:2012,Kormendy:2013,Saglia:2016}). The collected $B/T$, together with the above mass-to-light ratio, also allows us to compute the bulge mass $M_{\rm bulge}$, in case they are not directly available (e.g., \citealt{Beifiori:2012}).

Besides the commonly adopted galaxy name (usually from NGC), Table~\ref{tabop} lists the PGC identification number (HyperLEDA), which is useful to track the position of the galaxy within the group/cluster halo (\S\ref{s:xvar}) and to identify the number of members $N_{\rm m}$ in the large-scale, gravitationally bound cosmic structure. As shown in column (iv), $N_{\rm m}$ covers values of 1\,-\,2 (isolated galaxies), 2\,-\,8 (fields), 8\,-\,50 (groups), and 50\,-\,1000 (clusters of galaxies). 
Because our search is carried out blindly in terms of the optical galactic properties, we have inherited a diverse mix of Hubble morphological types (\citealt{Beifiori:2012,Kormendy:2013,Saglia:2016}), spanning from strong ETGs (E0-E4, including massive cD galaxies), to intermediate lenticulars S0s, and non-barred/barred spiral LTGs (see column (iii) in Tab.~\ref{tabop}). 
Moreover, it is evident by visual inspection that late types/early types correlate with both low/high $B/T$ and $N_{\rm m}$ values (poor/rich environments), as well as low/high BH masses.
In all the subsequent correlation plots, LTGs, S0s, and ETGs are marked with cyan, green, and blue circles, respectively.
However, unlike a few previous works, we do not attempt to divide a posteriori LTGs and ETGs in the Bayesian analysis  in order to seek a smaller scatter, as we want to retain a sample that is as unbiased as possible.

\subsubsection{X-ray plasma observables}\label{s:xvar}
Table~\ref{tabx} lists all the fundamental X-ray and environmental properties of the 85 galaxies, groups, and clusters in our sample, as well as related references and single-object notes. As introduced in \S\ref{s:ovar}, we carefully inspected the literature, choosing representative works with the deeper high-resolution \Chandra\ and wide-field ROSAT/XMM-\textit{Newton} datasets of hot halos (e.g., for the galactic scale, \citealt{OSullivan:2003,Diehl:2008b,Nagino:2009,Kim:2015,Su:2015,Goulding:2016}, to name a few).

The first fundamental plasma observable is the X-ray luminosity, which is constrained from the X-ray photon flux
(through either the count rate estimated in CCD images or the normalization of the energy spectra),
$L_{\rm x}= f_{\rm x}\,4\pi D^2$. 
Wherever necessary, given the cosmological distance dependence, 
luminosities (and radii)
are rescaled via our adopted $D$ (or alternatively via our $h=0.7$)\footnote{Luminosities scale as $\propto h^{-2}$ and radii as $\propto h^{-1}$.}.
Several of the above studies constrain \Lx\ within the (\Chandra) X-ray broad band with rest-frame energy range ${\rm 0.3\,\text{-}\,7\;keV}$, i.e., above the UV regime and including both soft and hard X-rays), making it our reference band too. 
For data points in a different X-ray energy range (e.g., in the soft\footnote{It is interesting to note that, by using the soft X-ray band, one can better separate the diffuse gas component from that associated with contaminating AGN/shocks (which dominate in hard X-rays; e.g., \citealt{LaMassa:2012}).} 0.5\,-\,2\;keV band or pseudo-bolometric 0.05\,-\,50\;keV)
we apply the appropriate correction by using PIMMS\footnote{\url{http://cxc.harvard.edu/toolkit/pimms.jsp}.} tools. 
These corrections range between 5\,-\,30\%.
Further, the considered studies aim to remove the contaminations due to 
foreground/background AGN, low-mass X-ray binaries (LMXBs) and fainter sources as active binaries/cataclysmic variables (ABs/CVs; see \citealt{Goulding:2016}), as well as correcting for Galactic neutral hydrogen absorption ($\log N_{\rm H}/{\rm cm^{-2}}\sim20$\,-\,21).
As for the optical morphological types, 
our final sample contains both gas-poor and gas-rich galaxies and groups/clusters (though only a handful of very massive clusters),
spanning a wide range of (unabsorbed) luminosities $\lx\sim10^{38}$\,-\,$10^{45}$\;erg\,s$^{-1}$.

The second key observable of hot halos is the X-ray temperature, inferred from the detected energy spectrum (e.g., via ACIS-S or RGS instruments). Modern spectral codes with atomic lines libraries (including photoionization and recombination rates) are employed to achieve an accurate fitting, the majority using \texttt{XSPEC} with 1-$T$ (or seldom 2-$T$) \texttt{APEC} models.
A typical -- though not unique -- procedure among the collected studies of ETGs is as follows (e.g., \citealt{Kim:2015}). 
After removal of the X-ray point sources (including the central AGN, e.g., via \texttt{CIAO} \texttt{wavedetect}), the spectra are fitted with a multi-component model, including a thermal plasma \texttt{APEC} (diffuse plasma),
hard X-ray power laws (residual AGN and AB/CV), 
and thermal 7\,keV Bremsstrahlung (unresolved LMXBs). 
The assumed abundances (ranging between 0.3\,-\,1\,$Z_\odot$) are usually a source of significant uncertainty. 
The temperature (and emission measure) retrieved in varying annuli are then deprojected into a three-dimensional (3D) profile (e.g., via \texttt{XSPEC projct}). We remark that our focus is on the X-ray component related to the diffuse thermal gas.
As tested by \citet{Goulding:2016}, using alternative plasma models (e.g., \texttt{MEKAL} or variations in \texttt{AtomDB}) leads to \Tx\ variations of up to 10\%; 
hence, on top of statistical errors, we conservatively add (in quadrature%
\footnote{We checked that adding such errors in a linear way has a negligible impact on the correlation results, since systematic errors are $\sim$\,$2\times$ larger than statistical errors. Further, given that all these errors are relatively small (in log space), using the sole statistical error induces only minor variations in the fit, with posterior parameters remaining comparable within the 1-$\sigma$ fitting uncertainty.}) 
a 10\% systematic uncertainty to allow for a more homogeneous comparison. For analogous reason, we add a systematic error on \Lx\ by propagating the distance errors (10\,-\,20\%). 
We note whenever archival errors are given in linear space, we transform and symmetrize them in logarithmic space.
The final range of hot halo temperatures for our sample covers the entire X-ray regime, spanning $\tx\sim0.2$\,-\,8\;keV.

The diffuse hot plasma can fill different regions of the potential well, including the galactic scale, the core and outskirt regions of the macro group/cluster halo.
We thus use three main extraction regions as proxies for three characteristic X-ray radii within the potential well.
The radius \Rxg\ is that describing the galactic/CGM potential. 
We thus searched for studies with \Lx\ and \Tx\ covering the region within $\approx$\,$2\re$ ($\sim$\,0.03\,$R_{500}$; columns (iii-v) in Tab.~\ref{tabx}),
as X-ray halos are typically more diffuse than the stellar component. 
Beyond such radius, the background noise becomes significant for several of the isolated galaxies, and thus this radius defines a characteristic size within which most of the galactic X-ray halo is contained.
Using the CGM region also helps to avoid inner, residual AGN contaminations. 
The second and third scales are related to $\rxc\approx0.15\,R_{\rm 500}$ and $1\,R_{500}$%
\footnote{$R_{500}$ is the radius that confines an average (total) matter density $500\times$ the critical cosmic density $\rho_{\rm c}$, such that $M_{\rm tot, 500} = (4/3)\pi\,500\rho_{\rm c} R_{500}^3$, where $\rho_{\rm c}=3 H^2/(8\pi G)$.}, 
i.e., the core and outskirt radius of the group/cluster halo, respectively (columns (viii) and (x)).
The latter is also a good proxy for the virial radius, $R_{500}\simeq R_{\rm vir}/1.7$, and is directly given by the X-ray temperature, $R_{500} \simeq {\rm 860\;kpc}\,(T_{\rm x, 500}/{\rm 3\,keV})^{1/2}$ (0.03\,dex scatter; \citealt{Sun:2009a}).\footnote{
Errors on \Rxg\ and \Rxc\ are given by the systematic error on distance plus 
a random error with 0.2 dex RMS; error on \R500 is propagated from the \citet{Sun:2009a} scaling.
We note that non-BCGs/BGGs without a macro halo have core radius  $ = 0.15\,R_{\rm 500}$.} 
To recap, our final collected archival sample has the following three mean (with RMS) extraction radii: 
\begin{flalign}\label{e:Rxg}
\log \rxg &= (0.22 \pm 0.33) + \log \re \nonumber\\ 
              &= (-1.59 \pm 0.30) + \log R_{500} \nonumber \\
	      &= (1.00  \pm 0.35) + \log {\rm kpc}, &&
\end{flalign}
\vspace{-0.6cm}
\begin{flalign}\label{e:Rxc}
\log \rxc &=  (1.00 \pm 0.28)+ \log \re \nonumber \\ 
              &=(-0.82 \pm 0.13) + \log R_{500} \nonumber \\
              &= (1.70 \pm 0.23) + \log {\rm kpc}, &&
\end{flalign}
\vspace{-0.6cm}
\begin{flalign}\label{e:R500}
\log R_{500} = (2.55 \pm 0.16) + \log {\rm kpc}. &&
\end{flalign}
The three extraction radii cover a healthy geometric progression of one-order-of-magnitude increments in spatial scale,  $\sim$\,$10^{-2}$\,-\,$10^{-1}$\,-\,$10^0$\,$R_{500}$.
Note that in a literature search we can not control a sharp threshold, but only select approximate regions (Tab.~\ref{tabx}).
However, a sharp line in the sand is unphysical, since hot atmospheres are continuous in space.
Having a dispersion on extraction radii also corroborates the robustness of any retrieved low intrinsic scatter.

To assess whether the galaxy is central, satellite, or isolated, we use the \citet{Tully:2015} PGC1 catalog (and correlated $N_{\rm m}$ in Tab.~\ref{tabop}; in uncertain cases, we also inspected the X-ray halo peak). Whenever the considered system matches the brightest (PGC1) central galaxy of the cluster or group, we list it as BCG or BGG in column (ii) of Tab.~\ref{tabx};
we then search in the cluster/group literature for X-ray data of the core luminosity/temperature ($L_{\rm x,c}$/$T_{\rm x,c}$; columns (vi-vii)) and global $L_{\rm x,500}$ (column (ix); e.g., \citealt{Reiprich:2002,Osmond:2004,Hudson:2010,Panagoulia:2014,OSullivan:2017,Babyk:2018}). 
Given the declining emissivity profile toward the outskirts, $T_{\rm x,c}$ is comparable to $T_{\rm x,500}$ within typical uncertainty (\citealt{Vikhlinin:2009}). 
In the opposite regime, galaxies that are satellites (moving at hundreds $\kms$ relative to the macro weather), isolated ($N_{\rm }\le 2$) or brightest in a poor field (BFGs; both having rapidly dropping gas density beyond $2\re$) can only feed from the local hot halo, hence we use as `macro' X-ray properties (e.g., $L_{\rm x,500}$) the CGM observables (this also avoids uncertain extrapolations).

The large majority of the listed $\tx$/$\lx$ are retrieved via single-aperture X-ray spectroscopy/photometry. 
Whenever such values are not tabulated by the authors, we integrate the given luminosity/density profile (Eq.~\ref{e:Lx}) or compute the LW temperature from its deprojected profile within our median \Rxg\ or \Rxc\ (avoiding extrapolations beyond constrained data points). The 14 systems retrieved via this method are listed in the Notes of Tab.~\ref{tabx} (e.g., \citealt{Nagino:2009}).

Finally, as for the optical properties, we do not include galaxies with unconstrained or unavailable X-ray data on extended hot halos (e.g., NGC\,4751, NGC\,7457, NGC\,4486A). Given the significant pile-up effects, we exclude objects that are heavily contaminated or show purely an X-ray AGN point source (often hosting nuclear fast outflows/winds), as several of the available Markarian galaxies (e.g., \citealt{Tombesi:2013}). We also exclude a few systems with X-ray emission completely swamped by the large-scale FRII-jet lobes or bubbles (Cygnus\,A, 3C66B/UGC\,1841, NGC\,193), thus preventing a reliable determination of the diffuse hot halo.
Nevertheless, as discussed in \S\ref{s:cav}, we decided to include the vast majority of X-ray systems with robust hot halo constraints, regardless of the dynamical or evolutionary stage, thus considering quiescent, fossil, feedback- and merger-heated systems, as well as major outliers (e.g., NGC\,1600). 

\capstartfalse
\begin{center}
\begin{deluxetable*}{l c c c} 
\tablecaption{Summary table of the main analyzed correlations between direct SMBH masses, X-ray and optical properties for our sample of 85 galaxies, groups, and clusters (Tab.~\ref{tabop}\,-\,\ref{tabx}).}
\tablehead{
{Bayesian univariate fitting $Y = (\overline{\alpha} \pm \sigma_\alpha)  +  (\overline{\beta} \pm \sigma_{\beta}) X$} 
& \colhead{$\overline{\epsilon} \pm \sigma_\epsilon$}
& \colhead{$ \overline{\rm corr} \pm \sigma_{\rm c}$}
}
\startdata
X-ray/plasma correlations:& & \\
\vspace{-0.22cm} & & \\
$\log ({\mbh}/{\msun}) = (9.39{\scriptstyle \pm0.05}) \,\  + (2.70{\scriptstyle \pm0.17})\,\log ({T_{\rm x,g}}/{\rm keV})$ & $0.21{\scriptstyle \pm0.03}$  & $0.94 {\scriptstyle \pm 0.02}$  \\
$\log ({\mbh}/{\msun}) = (9.18{\scriptstyle \pm0.05}) \,\  + (2.14{\scriptstyle \pm0.13})\,\log ({T_{\rm x,c}}/{\rm keV})$ & $0.25{\scriptstyle \pm0.02}$  & $0.93 {\scriptstyle \pm 0.02}$  \\
$\log ({\mbh}/{\msun}) = (10.63{\scriptstyle \pm0.15})  + (0.51{\scriptstyle \pm0.04})\,\log({L_{\rm x,g}}/{10^{44}\,\es})$ & $0.30{\scriptstyle \pm0.02}$ & $0.87 {\scriptstyle \pm 0.02}$\\
$\log ({\mbh}/{\msun}) = (10.18{\scriptstyle \pm0.11})  + (0.42{\scriptstyle \pm0.03})\,\log({L_{\rm x,c}}/{10^{44}\,\es})$ & $0.29{\scriptstyle \pm0.02}$ & $0.88 {\scriptstyle \pm 0.03}$ \\
$\log ({\mbh}/{\msun}) = (10.00{\scriptstyle \pm0.11}) + (0.38{\scriptstyle \pm0.03})\,\log({L_{\rm x,500}}/{10^{44}\,\es})$&$0.31{\scriptstyle \pm0.02}$ &$0.86 {\scriptstyle \pm 0.03}$\\
$\log ({\mbh}/{\msun}) = (10.60{\scriptstyle \pm0.42})  + (0.78{\scriptstyle \pm0.17})\,\log({n_{\rm e,g}}/{\cc})$ & $0.56{\scriptstyle \pm0.04}$ & $0.54 {\scriptstyle \pm 0.10}$\\
$\log ({\mbh}/{\msun}) = (12.75{\scriptstyle \pm0.41})  + (1.13{\scriptstyle \pm0.11})\,\log({n_{\rm e,c}}/{{\cc}})$ & $0.39{\scriptstyle \pm0.03}$ & $0.80 {\scriptstyle \pm 0.05}$ \\
$\log ({\mbh}/{\msun}) = (13.53{\scriptstyle \pm0.49})  + (1.02{\scriptstyle \pm0.10})\,\log({n_{\rm e,500}}/{{\cc}})$&$0.40{\scriptstyle \pm0.03}$ &$0.79 {\scriptstyle \pm 0.05}$\\
$\log ({\mbh}/{\msun}) = (8.29{\scriptstyle \pm0.07})\,\   + (0.75{\scriptstyle \pm0.09})\,\log({P_{\rm x,g}}/{10^{-3}\,\rm keV\,\cc})$ & $0.45{\scriptstyle \pm0.03}$ & $0.73 {\scriptstyle \pm 0.06}$\\
$\log ({\mbh}/{\msun}) = (9.15{\scriptstyle \pm0.05})\,\  + (0.80{\scriptstyle \pm0.06})\,\log({P_{\rm x,c}}/{10^{-3}\,\rm keV\,\cc})$ & $0.33{\scriptstyle \pm0.02}$ &$0.87 {\scriptstyle \pm 0.03}$ \\
$\log ({\mbh}/{\msun}) = (9.96{\scriptstyle \pm0.10})\,\  + (0.75{\scriptstyle \pm0.05})\,\log({P_{\rm x,500}}/{10^{-3}\,\rm keV\,\cc})$&$0.33{\scriptstyle \pm0.02}$&$0.87 {\scriptstyle \pm 0.03}$\\
$\log ({\mbh}/{\msun}) = (3.88{\scriptstyle \pm0.42})\,\   + (0.57{\scriptstyle \pm0.05})\,\log({M_{\rm gas,g}}/{\msun})$ & $0.32{\scriptstyle \pm0.03}$ & $0.87 {\scriptstyle \pm 0.04}$\\
$\log ({\mbh}/{\msun}) = (2.46{\scriptstyle \pm0.46})\,\   + (0.64{\scriptstyle \pm0.05})\,\log({M_{\rm gas,c}}/{\msun})$ & $0.26{\scriptstyle \pm0.03}$ &$0.92 {\scriptstyle \pm 0.03}$ \\
$\log ({\mbh}/{\msun}) = (1.61{\scriptstyle \pm0.49})\,\   + (0.64{\scriptstyle \pm0.04})\,\log({M_{\rm gas,500}}/{\msun})$&$0.30{\scriptstyle \pm0.02}$&$0.90 {\scriptstyle \pm 0.03}$\\
$\log ({\mbh}/{\msun}) = (10.16{\scriptstyle \pm0.11})  + (0.46{\scriptstyle \pm0.03})\,\log({Y_{\rm x,g}}/{\rm 10^{60}\,erg})$ & $0.32{\scriptstyle \pm0.02}$ & $0.88 {\scriptstyle \pm 0.03}$\\
$\log ({\mbh}/{\msun}) = (9.54{\scriptstyle \pm0.07})\,\   + (0.45{\scriptstyle \pm0.03})\,\log({Y_{\rm x,c}}/{\rm 10^{60}\,erg})$ & $0.30{\scriptstyle \pm0.02}$ &$0.89 {\scriptstyle \pm 0.03}$ \\
$\log ({\mbh}/{\msun}) = (8.98{\scriptstyle \pm0.04})\,\    + (0.49{\scriptstyle \pm0.03})\,\log({Y_{\rm x,500}}/{\rm 10^{60}\,erg})$&$0.29{\scriptstyle \pm0.02}$&$0.90 {\scriptstyle \pm 0.03}$\\
$\log ({\mbh}/{\msun}) = (\text{-}6.03{\scriptstyle \pm1.45})\,  + (1.28{\scriptstyle \pm0.13})\,\log({M_{\rm tot,g}}/{\msun})$ & $0.36{\scriptstyle \pm0.03}$ & $0.83 {\scriptstyle \pm 0.05}$\\
$\log ({\mbh}/{\msun}) = (\text{-}5.78{\scriptstyle \pm1.18})\,  + (1.17{\scriptstyle \pm0.10})\,\log({M_{\rm tot,c}}/{\msun})$ & $0.35{\scriptstyle \pm0.03}$ &$0.86 {\scriptstyle \pm 0.04}$ \\
$\log ({\mbh}/{\msun}) = (\text{-}9.56{\scriptstyle \pm1.16})\, + (1.39{\scriptstyle \pm0.09})\,\log({M_{\rm tot,500}}/{\msun})$&$0.25{\scriptstyle \pm0.03}$&$0.91 {\scriptstyle \pm 0.02}$\\
$\log ({\mbh}/{\msun}) = (11.67{\scriptstyle \pm0.31})  + (0.95{\scriptstyle \pm0.10})\,\log({f_{\rm gas,g}})$ & $0.35{\scriptstyle \pm0.04}$ & $0.85 {\scriptstyle \pm 0.05}$\\
$\log ({\mbh}/{\msun}) = (11.79{\scriptstyle \pm0.26})  + (1.18{\scriptstyle \pm0.10})\,\log({f_{\rm gas,c}})$ & $0.25{\scriptstyle \pm0.04}$ &$0.92 {\scriptstyle \pm 0.04}$ \\
$\log ({\mbh}/{\msun}) = (11.02{\scriptstyle \pm0.24})  + (1.11{\scriptstyle \pm0.11})\,\log({f_{\rm gas,500}})$&$0.38{\scriptstyle \pm0.03}$&$0.81 {\scriptstyle \pm 0.05}$\\
$\log ({L_{\rm x,g}}/{10^{44}\,\es}) \ \;\; = (\text{-}2.64{\scriptstyle \pm0.10})  + (4.23{\scriptstyle \pm0.29})\,\log({T_{\rm x,g}}/{\rm keV})$ & $0.49{\scriptstyle \pm0.04}$ & $0.90 {\scriptstyle \pm 0.03}$\\
$\log ({L_{\rm x,c}}/{10^{44}\,\es}) \ \ \;= (\text{-}2.53{\scriptstyle \pm0.08})  + (4.42{\scriptstyle \pm0.23})\,\log({T_{\rm x,c}}/{\rm keV})$ & $0.49{\scriptstyle \pm0.04}$ & $0.93 {\scriptstyle \pm 0.02}$\\
$\log ({L_{\rm x,500}}/{10^{44}\,\es}) = (\text{-}2.34{\scriptstyle \pm0.09})  + (4.71{\scriptstyle \pm0.26})\,\log({T_{\rm x,c}}/{\rm keV})$ & $0.57{\scriptstyle \pm0.05}$ & $0.92 {\scriptstyle \pm 0.02}$\\
$\log ({\mbh}/{\msun}) = (7.69{\scriptstyle \pm0.44})  + (\text{-}0.55{\scriptstyle \pm0.24})\,\log({K_{\rm x,g}^{-3/2}}/{\rm keV^{-3/2}\,cm^{-3}})$ & $0.62{\scriptstyle \pm0.04}$ & $\text{-}0.31 {\scriptstyle \pm 0.14}$\\
$\log ({\mbh}/{\msun}) = (7.31{\scriptstyle \pm1.13})  + (\text{-}0.46{\scriptstyle \pm0.38})\,\log({K_{\rm x,c}^{-3/2}}/{\rm keV^{-3/2}\,cm^{-3}})$ & $0.64{\scriptstyle \pm0.04}$ & $\text{-}0.19 {\scriptstyle \pm 0.15}$\\
$\log ({\mbh}/{\msun}) = (8.64{\scriptstyle \pm1.25})  + (\text{-}0.01{\scriptstyle \pm0.30})\,\log({K_{\rm x,500}^{-3/2}}/{\rm keV^{-3/2}\,cm^{-3}})$ & $0.65{\scriptstyle \pm0.04}$ & $0.00 {\scriptstyle \pm 0.13}$\\
$\log ({\mbh}/{\msun}) = (\text{-}0.34{\scriptstyle \pm0.89})  + (1.10{\scriptstyle \pm0.11})\,\log({M_{\rm \bullet,cca,g}}/{\msun}) $ & $0.39{\scriptstyle \pm0.03}$ & $0.80 {\scriptstyle \pm 0.05}$\\
$\log ({\mbh}/{\msun}) = (0.45{\scriptstyle \pm0.71})  \ + (0.98{\scriptstyle \pm0.08})\,\log({M_{\rm \bullet,cca,c}}/{\msun}) $ & $0.30{\scriptstyle \pm0.03}$ & $0.90 {\scriptstyle \pm 0.04}$\\
$\log ({\mbh}/{\msun}) = (2.07{\scriptstyle \pm0.59})  \ + (0.77{\scriptstyle \pm0.07})\,\log({M_{\rm \bullet,cca,500}}/{\msun}) $ & $0.38{\scriptstyle \pm0.03}$ & $0.82{\scriptstyle \pm 0.05}$\\
$\log ({\mbh}/{\msun}) = (8.29{\scriptstyle \pm0.07})  \ + (0.98{\scriptstyle \pm0.10})\,\log({R_{\rm cond}}/{\rm kpc}) $ & $0.32{\scriptstyle \pm0.04}$ & $0.87{\scriptstyle \pm 0.05}$\\
\vspace{-0.14cm} & & \\
\hline
\vspace{-0.14cm} & & \\
Optical/stellar correlations: & & \\
\vspace{-0.24cm} & & \\
$\log ({\mbh}/{\msun}) = (\text{-}1.53{\scriptstyle \pm0.87}) + (4.36{\scriptstyle \pm0.37})\,\log ({\sige}/{\rm \kms})$ & $0.36{\scriptstyle \pm0.02}$  & $0.84 {\scriptstyle \pm 0.04}$  \\
$\log ({\mbh}/{\msun}) = (8.37{\scriptstyle \pm0.07}) \   + (1.11{\scriptstyle \pm0.14})\,\log({L_K}/{\rm 10^{11}\,L_\odot})$ & $0.46{\scriptstyle \pm0.03}$ & $0.71{\scriptstyle \pm0.07}$ \\
$\log ({\mbh}/{\msun}) = (8.56{\scriptstyle \pm0.05}) \   + (0.90{\scriptstyle \pm0.09})\,\log({M_{\rm bulge}}/{10^{11} \msun})$&$0.40{\scriptstyle \pm0.03}$&$0.79{\scriptstyle \pm0.05}$\\
\vspace{+0.1cm} & & \\
\hline
\hline
\vspace{-0.17cm} & & \\
          
{Bayesian multivariate fitting $Y = (\overline{\alpha} \pm \sigma_\alpha) + (\overline{\beta}_{1} \pm \sigma_{\beta_{1}})\,X_1 + (\overline{\beta}_{2} \pm \sigma_{\beta_{2}})\,X_2$ \quad\quad} 
& \colhead{$\overline{\epsilon} \pm \sigma_\epsilon$}
& \colhead{$ \overline{\rm pcorr}_1 \pm \sigma_{\rm p_1}$}
& \colhead{$ \overline{\rm pcorr}_2 \pm \sigma_{\rm p_2}$} \\
\vspace{-0.17cm} & & \\
\hline
\vspace{-0.14cm} & & \\
X-ray/plasma correlations:& & \\
\vspace{-0.22cm} & & \\
$\log ({T_{\rm x,g}}/{\rm keV}) = (0.58{\scriptstyle \pm0.14}) \;+ (0.21{\scriptstyle \pm0.02})\,\log ({L_{\rm x,g}}/{\rm 10^{44}\,\es})\; + (\text{-}0.07{\scriptstyle \pm0.07})\,\log ({\rxg}/{\rm kpc})$  & 
$0.11{\scriptstyle \pm0.01}$  & $0.88 {\scriptstyle \pm 0.06}$ & $\text{-}0.19 {\scriptstyle \pm 0.22}$  \\
$\log ({T_{\rm x,c}}/{\rm keV}) = (\text{-}0.15{\scriptstyle \pm0.32}) + (0.16{\scriptstyle \pm0.02})\,\log ({L_{\rm x,c}}/{\rm 10^{44}\,\es})\ + (0.28{\scriptstyle \pm0.14})\,\log ({\rxc}/{\rm kpc})$  & 
$0.10{\scriptstyle \pm0.01}$  & $0.78 {\scriptstyle \pm 0.07}$ & $0.33 {\scriptstyle \pm 0.17}$  \\
$\log ({T_{\rm x,g}}/{\rm keV}) = (1.85{\scriptstyle \pm0.25}) \;+ (0.21{\scriptstyle \pm0.02})\,\log ({I_{\rm x,g}}/{\rm \es\,cm^{-2}}) + (0.39{\scriptstyle \pm0.05})\,\log ({\rxg}/{\rm kpc})$  & 
$0.10{\scriptstyle \pm0.01}$  & $0.89 {\scriptstyle \pm 0.05}$ & $0.85 {\scriptstyle \pm 0.07}$  \\
$\log ({\mbh}/{\msun}) = (9.42{\scriptstyle \pm0.91}) + (0.37{\scriptstyle \pm0.06})\,\log ({L_{\rm x,c}}/{\rm 10^{44}\,\es}) + (0.34{\scriptstyle \pm0.42})\,\log ({\rxc}/{\rm kpc})$  & 
$0.32{\scriptstyle \pm0.02}$  & $0.71 {\scriptstyle \pm 0.09}$ & $0.15 {\scriptstyle \pm 0.18}$  \\
$\log ({\mbh}/{\msun}) = (9.43{\scriptstyle \pm0.19}) + (1.64{\scriptstyle \pm0.37})\,\log ({T_{\rm x,c}}/{\rm keV}) + (0.10{\scriptstyle \pm0.08})\,\log ({L_{\rm x,c}}/{\rm 10^{44}\,\es})$  & 
$0.24{\scriptstyle \pm0.02}$  & $0.66 {\scriptstyle \pm 0.14}$ & $0.24 {\scriptstyle \pm 0.20}$  \\
$\log ({\mbh}/{\msun}) = (9.17{\scriptstyle \pm0.85}) + (2.13{\scriptstyle \pm0.33})\,\log ({T_{\rm x,c}}/{\rm keV}) + (0.01{\scriptstyle \pm0.45})\,\log (\rxc/{\rm kpc})$  & 
$0.26{\scriptstyle \pm0.02}$  & $0.79 {\scriptstyle \pm 0.10}$ & $0.00 {\scriptstyle \pm 0.21}$  \\
\vspace{-0.14cm} & & \\
\hline
\vspace{-0.14cm} & & \\
Optical/stellar correlations: & & \\
\vspace{-0.24cm} & & \\
$\log ({\sige^2}/{\kmsd}) = (5.22{\scriptstyle \pm0.11}) + (1.43{\scriptstyle \pm0.17})\,\log ({L_K}/{\rm 10^{11}\,L_\odot}) + (\text{-}1.40{\scriptstyle \pm0.22})\,\log ({\re}/{\rm kpc})$  & 
$0.10{\scriptstyle \pm0.02}$  & $0.92 {\scriptstyle \pm 0.05}$ & $\text{-}0.88 {\scriptstyle \pm 0.08}$  \\
$\log ({\mbh}/{\msun}) = (9.55{\scriptstyle \pm0.35})\  + (2.99{\scriptstyle \pm0.58})\,\log ({L_K}/{\rm 10^{11}\,L_\odot}) + (\text{-}2.56{\scriptstyle \pm0.75})\,\log ({\re}/{\rm kpc})$  & 
$0.39{\scriptstyle \pm0.04}$  & $0.77 {\scriptstyle \pm 0.11}$ & $\text{-}0.64 {\scriptstyle \pm 0.16}$  \\
$\log ({\mbh}/{\msun}) = (0.54{\scriptstyle \pm1.11}) \ + (1.71{\scriptstyle \pm0.24})\,\log ({\sige^2}/{\rm \kmsd}) \;+ (0.42{\scriptstyle \pm0.15})\,\log ({L_K}/{\rm 10^{11}\,L_\odot})$  & 
$0.34{\scriptstyle \pm0.02}$  & $0.75 {\scriptstyle \pm 0.10}$ & $0.44 {\scriptstyle \pm 0.18}$  \\
$\log ({\mbh}/{\msun}) = (\text{-}0.77{\scriptstyle \pm0.88}) + (1.96{\scriptstyle \pm0.20})\,\log ({\sige^2}/{\rm \kmsd}) \; + (0.43{\scriptstyle \pm0.16})\,\log ({\re}/{\rm kpc})$  & 
$0.34{\scriptstyle \pm0.03}$  & $0.85 {\scriptstyle \pm 0.07}$ & $0.44 {\scriptstyle \pm 0.18}$  \\
\vspace{-0.14cm} & & \\
\hline
\vspace{-0.14cm} & & \\
Mixed X-ray and optical correlations: & & \\
\vspace{-0.24cm} & & \\
$\log ({\mbh}/{\msun}) = (10.23{\scriptstyle \pm0.24}) + (0.43{\scriptstyle \pm0.05})\log ({L_{\rm x,c}}/{\rm 10^{44}\,\es}) + (\text{-}0.06{\scriptstyle \pm0.19})\log ({L_K}/{\rm 10^{11}\,L_\odot})$  & 
$0.32{\scriptstyle \pm0.02}$  & $0.80 {\scriptstyle \pm 0.09}$ & $\text{-}0.06 {\scriptstyle \pm 0.21}$  \\
$\log ({\mbh}/{\msun}) = (5.67{\scriptstyle \pm1.14}) + (1.65{\scriptstyle \pm0.21})\,\log ({T_{\rm x,c}}/{\rm keV}) + (0.73{\scriptstyle \pm0.23})\,\log ({\sige^2}/{\kmsd})$  & 
$0.23{\scriptstyle \pm0.02}$  & $0.83 {\scriptstyle \pm 0.07}$ & $0.49 {\scriptstyle \pm 0.15}$  \\
$\log ({\mbh}/{\msun}) = (4.67{\scriptstyle \pm0.59}) + (2.19{\scriptstyle \pm0.39})\,\log ({\rxc}/{\rm kpc}) + (0.27{\scriptstyle \pm0.23})\,\log ({\re}/{\rm kpc})$  & 
$0.42{\scriptstyle \pm0.03}$  & $0.70 {\scriptstyle \pm 0.10}$ & $0.19 {\scriptstyle \pm 0.16}$  \\

\vspace{-0.3cm}
\enddata
\tablenotetext{}{
\small 
\textbf{\textit{    Notes.}} 
Additional complementary univariate and multivariate correlations can be found in Appendix~\ref{a:uextra}.
}
\label{tabsum}
\end{deluxetable*}
\end{center}
\capstarttrue

\vspace{-0.85cm}
\subsection{Data fitting: Bayesian estimator} \label{s:corr}
One of the major advancements in statistical astronomy of the last decade has been the leverage of Bayesian inference methods, which substantially depart from classical methods (such as the simple least-squares estimator). 
As we are here concerned purely with linear fitting (in $\log_{10}$ space\footnote{Throughout the manuscript, we drop the `10' subscript and use the formalism $\log_{10}\equiv \log$.}), we adopt the widely tested and robust formalism proposed by \citet{Kelly:2007}, which is coded into the (IDL\footnote{
IDL {\tt linmix}:~\url{https://idlastro.gsfc.nasa.gov/ftp/pro/math/linmix_err.pro};
IDL {\tt mlinmix}:~\url{https://idlastro.gsfc.nasa.gov/ftp/pro/math/mlinmix_err.pro}.} 
or Python\footnote{
Python {\tt linmix}:~\url{https://github.com/jmeyers314/linmix}.
}) procedures {\tt linmix} and {\tt mlinmix} -- for univariate and multivariate fitting, respectively.
A key reason to use the Bayesian formalism for linear regression is that the intrinsic scatter ($\epsilon$) is treated as a free parameter, together with the normalization/intercept ($\alpha$) and slope ($\beta$). At the same time, {\tt linmix}/{\tt mlinmix} accounts for measurement errors in both the dependent and independent variable/s. Compared with previous statistical methods, as the BCES estimator (Bivariate Correlated Errors and intrinsic Scatter; \citealt{Akritas:1996}), {\tt linmix} is more robust and unbiased even for small samples and large uncertainties (e.g., \citealt{Sereno:2016}). 

Formally, the Bayesian {\tt linmix} method has the objective to find the regression coefficients of the form (for univariate fitting)
\begin{equation}\label{e:uni}
Y = \alpha + \beta\,X + \epsilon,
\end{equation}
with the measured values $x = X \pm \sigma_{x}$ and $y = Y \pm \sigma_y$; while 
for our multivariate model
\begin{equation}\label{e:multi}
Y = \alpha + \beta_1\,X_1 + \beta_2\,X_2 + \epsilon,
\end{equation}
with the measured values $x_1 = X_1 \pm \sigma_{x_1}$, $x_2 = X_2 \pm \sigma_{x_2}$, and $y = Y \pm \sigma_y$. The covariance terms in measurement errors are typically negligible (e.g., \citealt{Saglia:2016}). 
In this work, all the carried-out regressions are linear in logarithmic space (e.g., $y=\log M_\bullet \pm \sigma_{\log M_\bullet}$ and $x=\log T_{\rm x} \pm \sigma_{\log T_{\rm x}}$). 

Procedurally, 
the {\tt linmix} algorithm first approximates the independent variable distribution as a mixture of Gaussian distribution functions (three\footnote{We tested a larger number of initial Gaussians, finding no significant differences in the correlation results.} is typically sufficient). 
The posterior probability distributions are then efficiently constrained
through a Markov Chain Monte Carlo (MCMC) method known as Gibbs sampler\footnote{A minimum/maximum number of MCMC iterations set as 5000/100000 (with four chains) is sufficient in most cases to reach convergence.}. 
We quote as best-fit parameters (in the top-left inset of each correlation plot; see Fig.~\ref{MbhTx}) the averages of these distributions with 1-$\sigma$ errors given by the related standard deviation. For the univariate fitting we use
\begin{flalign}\label{e:linmix}
&\alpha     = \overline{\alpha} \pm \sigma_\alpha \hspace{+1.37cm} {\rm [intercept]},  \\
&\beta       = \overline{\beta} \pm \sigma_{\beta} \hspace{+1.4cm} {\rm [slope\ on}\ X], \nonumber \\ 
&\epsilon   = \overline{\epsilon} \pm \sigma_\epsilon \hspace{+1.6cm}  {\rm [intrinsic\ scatter]}, \nonumber \\
&{\rm corr} = \overline{\rm corr} \pm \sigma_{\rm c} \hspace{+0.3cm} [{\rm correlation\ coefficient}], \nonumber &&
\end{flalign}
while for the multivariate fitting we use
\begin{flalign}\label{e:mlinmix}
&\alpha     = \overline{\alpha} \pm \sigma_\alpha \hspace{+1.37cm} {\rm [intercept]},  \\
&\beta_{1,2} = \overline{\beta}_{1,2} \pm \sigma_{\beta_{1,2}}  \hspace{+0.32cm} {\rm [slope\ on}\ X_1\ {\rm and}\ X_2{\rm ],} \nonumber\\
&\epsilon   = \overline{\epsilon} \pm \sigma_\epsilon \hspace{+1.6cm}  {\rm [intrinsic\ scatter]}, \nonumber \\
&{\rm pcorr_{1,2}} = \overline{\rm pcorr}_{1,2} \pm \sigma_{\rm p_{1,2}}  \hspace{+0.2cm} [{\rm partial\ correlation\ coeff.}], \nonumber &&
\end{flalign}
where it is important to note that in the multi-dimensional fitting the meaningful correlation coefficient is the partial (conditional) pcorr, i.e., we want to understand the correlation between $Y$ and one of the independent variables given the second control variable.

Unlike the classic Pearson correlation analysis, the Bayesian inference gives us precise errors (and distributions) on the correlation coefficient, bounded between $[-1,+1]$, which we can use to compare in a clear way the significance of multiple correlations.
We quantify the strength of a positive correlation as follows: `very strong' (${\rm corr}> 0.85$), `strong' ($0.7<{\rm corr}< 0.85$), `mild' ($0.5<{\rm corr}< 0.7$), `weak' ($0.3<{\rm corr}< 0.5$), and `absent' ($0.0<{\rm corr}< 0.3$). Anti-correlations have simply the sign (negative) reversed. 
We remark the importance of providing uncertainties for all the parameters.
In \S\ref{s:res}, we will dissect three kinds of major correlations, the univariate fitting between two fundamental X-ray/optical variables (Eq.~\ref{e:uni}), the univariate fitting between composite variables
(again via Eq.~\ref{e:uni}, which has minimal number of free parameters), and the multivariate fitting between the fundamental X-ray/optical observables (Eq.~\ref{e:multi}).\\

\section{Results} \label{s:res}
We start the presentation of the results with the correlations of the fundamental X-ray/optical variables (\S\ref{s:uni}), namely temperature/velocity dispersion and luminosities/masses.
We then continue with the univariate correlations of the derived variables (such as gas density and pressure) and conclude the analysis with the higher-dimensional correlations (\S\ref{s:multi}).
A synoptic table with all the analyzed correlations is given in Table~\ref{tabsum}, which may be directly used in other studies.
The reader can also find in the top-left inset of each correlation plot all the posterior regression coefficients (Eq.~\ref{e:uni}\,-\,\ref{e:multi}). 
Needless to say, correlation does not necessarily imply causation. On the other hand, the combination of a tight intrinsic scatter, large correlation coefficient, and non-zero slope, all with small statistical errors, accumulate evidence that some properties are more central than others in shaping the growth of SMBHs (and vice versa).\\

\subsection{Univariate correlations: fundamental variables} \label{s:uni}

\begin{figure*}[!ht]
\subfigure{\includegraphics[width=1.02\columnwidth]{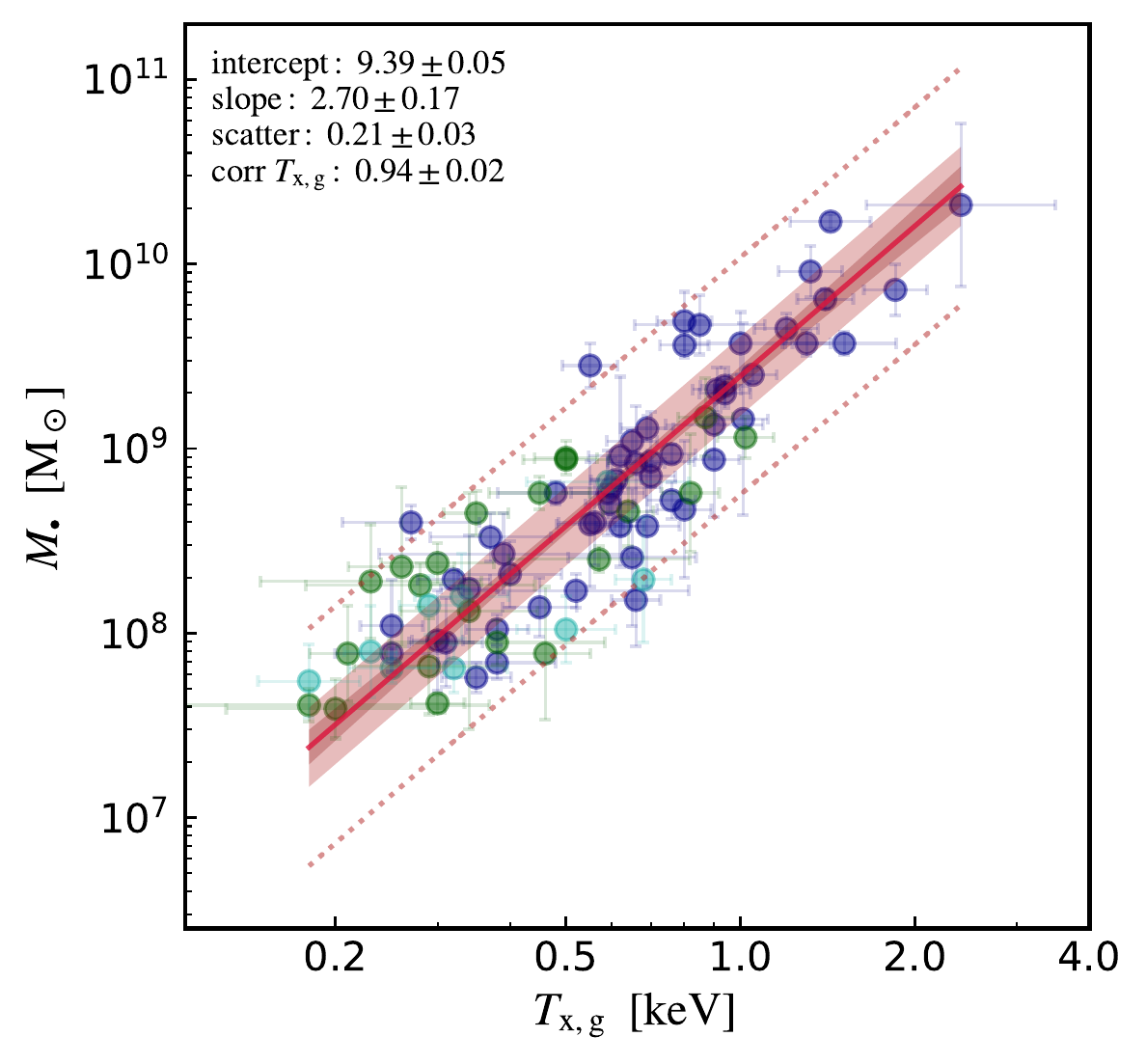}}
\subfigure{\includegraphics[width=1.0\columnwidth]{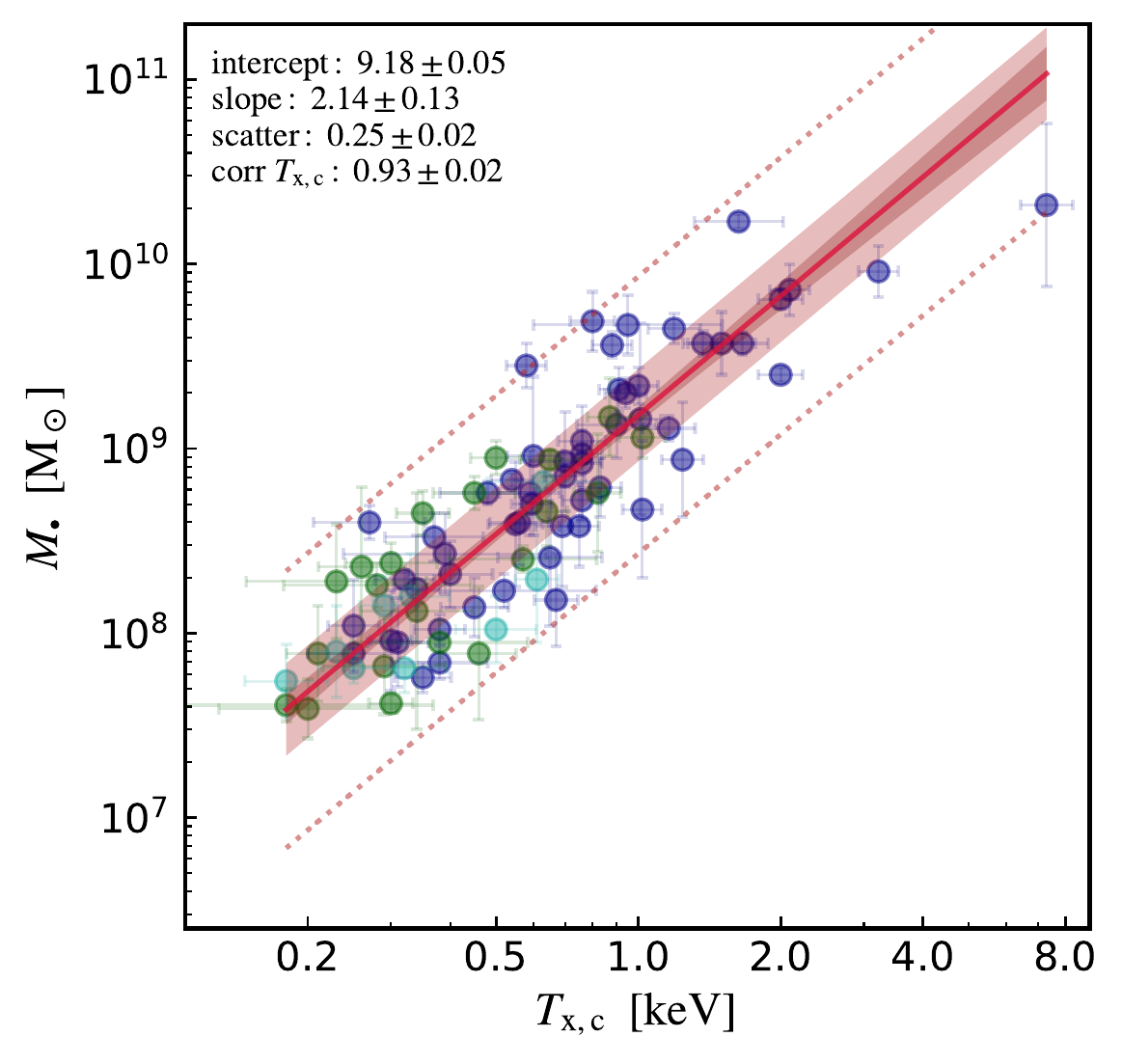}}
 \vskip -0.4cm
\caption{Black hole mass vs.~X-ray temperature (\S\ref{s:xuni}) within the galactic/CGM scale (left; 
$\rxg\approx 2\re\sim 0.03\,\r500$; Eq.~\ref{e:Rxg}) and the group/cluster core (right; $\rxc\approx 0.15\,\r500\sim 10\,\re$; Eq.~\ref{e:Rxc}). We note that $T_{\rm x,c} \approx T_{\rm x, 500}$. The inset shows the mean and errors of all the posteriors from the Bayesian linear fitting (\S\ref{s:corr}), including the intercept, slope, intrinsic scatter (1-$\sigma$ interval plotted as a light red band, 3-$\sigma$ as dotted lines), and linear correlation coefficient. The solid red line and inner dark band are the mean fit and related 15.87\,--\,84.13 percentile interval.
The points are color-coded based on the morphological type (blue: E; green: S0; cyan: S), which is also correlated with $B/T$ and $N_{\rm m}$. 
The key result is the very tight linear correlation retrieved ($\mbh \propto T^{2.5}$), with the lowest scatter found among all analyzed properties, in particular compared with the optical counterpart, the $\mbh -\sige$ relation (\S\ref{s:optuni}).
} 
\vskip +0.3cm 
\label{MbhTx}
\end{figure*}

\subsubsection{X-ray temperature and luminosity} \label{s:xuni}
One of the most fundamental X-ray properties is the plasma temperature (\S\ref{s:xvar}), which is also a measure of the gravitational potential $\phi(r)$. As the gas collapses in the potential wells of the dark matter halos during the formation of the galaxy, group, or cluster, the baryons thermalize (converting kinetic energy into thermal energy mainly via shock heating; \citealt{Kravtsov:2012}) and reach approximate virial equilibrium, $k_{\rm b}T_{\rm x}\propto \phi$ (\S\ref{s:xFP}).\footnote{As is customary, we use interchangeably the temperature in K and keV units (1\,keV\,$ \simeq 1.16\times10^7$\,K), even though the latter technically has the dimensionality of energy.}
Unlike the X-ray luminosity depending on gas mass and thus experiencing evacuation (e.g., via feedback processes), the plasma \Tx\ remains fairly stable in space and time (e.g., \citealt{Gaspari:2014_scalings}). 
Typically, \Tx\ is constant within the galactic region and shows at best 
a factor of 2 variations up to the outer regions (\citealt{Vikhlinin:2006,Diehl:2008b}),
mostly due to radiative cooling.
Given that most of the photons come from the core region, the core temperature is a reasonable proxy for the global/virial temperature ($T_{\rm x,c} \approx T_{\rm x, 500}$), within total uncertainties.

Figure \ref{MbhTx} shows that the (LW) X-ray temperature within the galactic scale $R_{\rm x,g}$
(left; Eq.\,~\ref{e:Rxg}) 
or within the macro-scale group/cluster core $R_{\rm x,c}$ (right; Eq.\,~\ref{e:Rxc}) are very tightly correlated with the BH mass.
The intrinsic scatter $\sim$0.21\,-\,0.25 dex is the smallest found among all the studied correlations (1-$\sigma$ level shown as a filled light red band), in particular compared with all the other optical/stellar properties (\S\ref{s:optuni}) -- including the stellar velocity dispersion which has 0.1\,dex larger scatter.
This is a key result, given the large sample size and diversity of systems, spanning from massive clusters and groups to isolated S0s and spiral galaxies (top-right to bottom-left sectors, or blue to green and cyan color-coding). The best-fit slopes are both consistent with a power-law index of 2.1\,-\,2.7, with correlation coefficients in the very strong regime (0.93\,-\,0.94). 
In terms of normalization, a 0.8\,keV halo corresponds to a $\sim$$10^9\,\msun$ SMBH.

We will dissect in \S\ref{s:disc} what is the role of the potential versus different accretion mechanisms arising from the hot plasma halo. For now, it is worthwhile to understand the notion that a hotter halo leads to substantially more massive black holes ($\mbh\propto T_{\rm x}^{2.5}$), up to even UMBHs in BCGs. This means that the accretion process shall be stimulated by the presence of a larger plasma mass (e.g., in galactic coronae or the more extended IGrM/ICM), rather than hindered by its thermal pressure, which tends to oppose the gas gravitational infall. 
The high end of the $\mbh-T_{\rm x,c}$ hints at a potential saturation, although at present it is unclear whether UMBHs with several tens of billions of solar masses exist in the universe.

\begin{figure}[!ht]
 \vskip -0.05cm
\subfigure{\hspace{+0.06cm}\includegraphics[width=0.941\columnwidth]{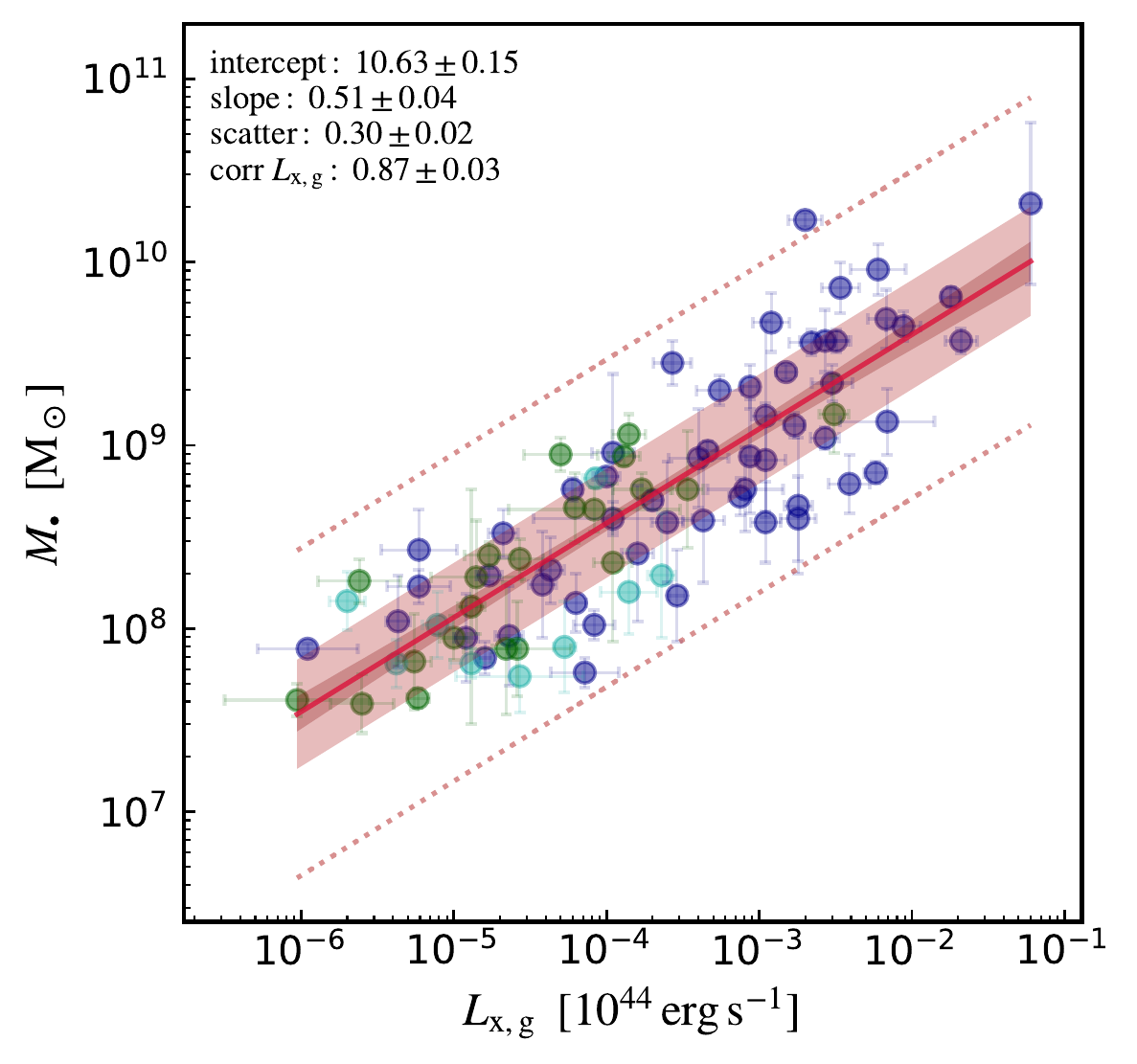}}
 \vskip -0.39cm
\subfigure{\includegraphics[width=0.928\columnwidth]{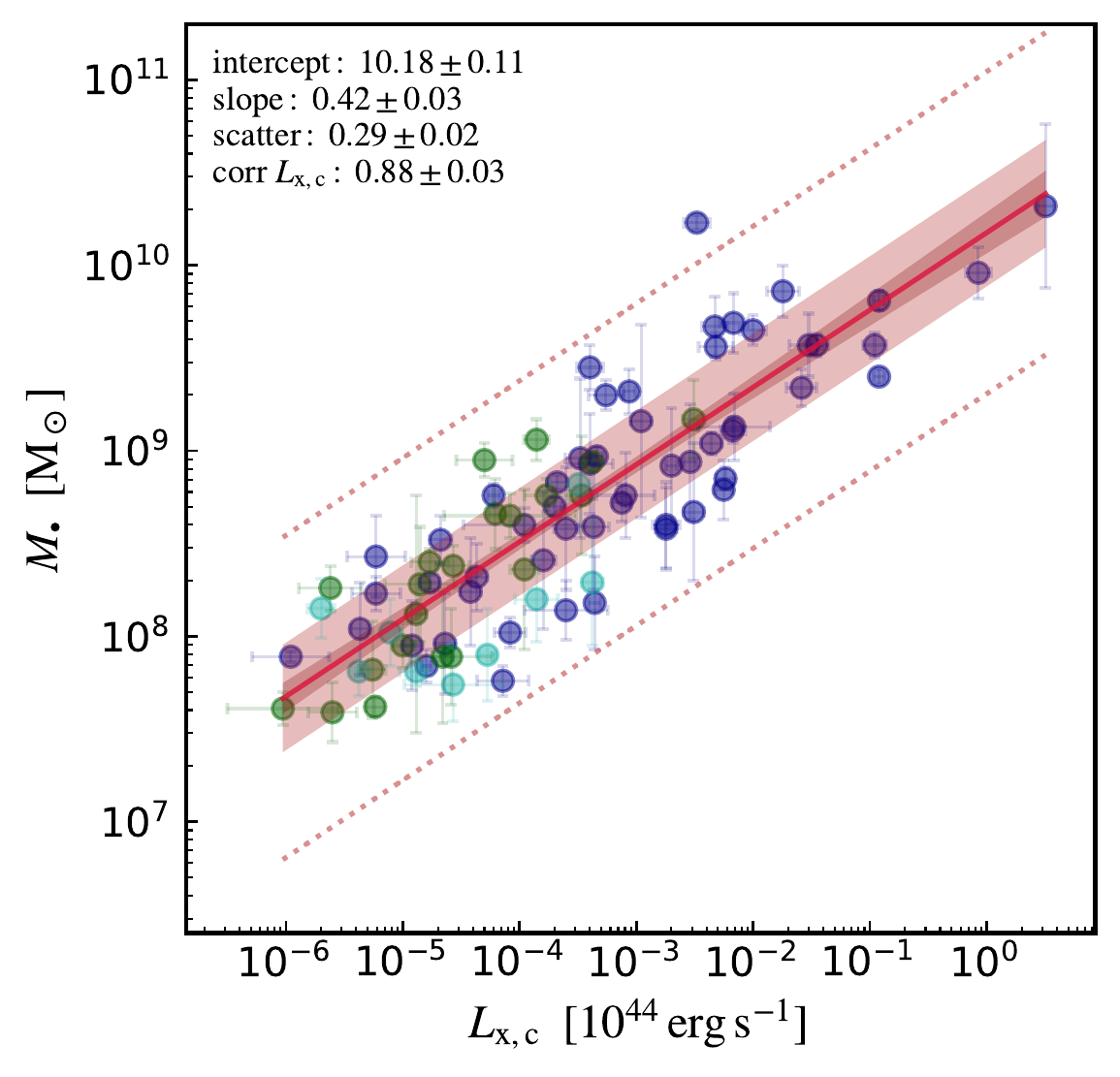}}
 \vskip -0.39cm
\subfigure{\includegraphics[width=0.93\columnwidth]{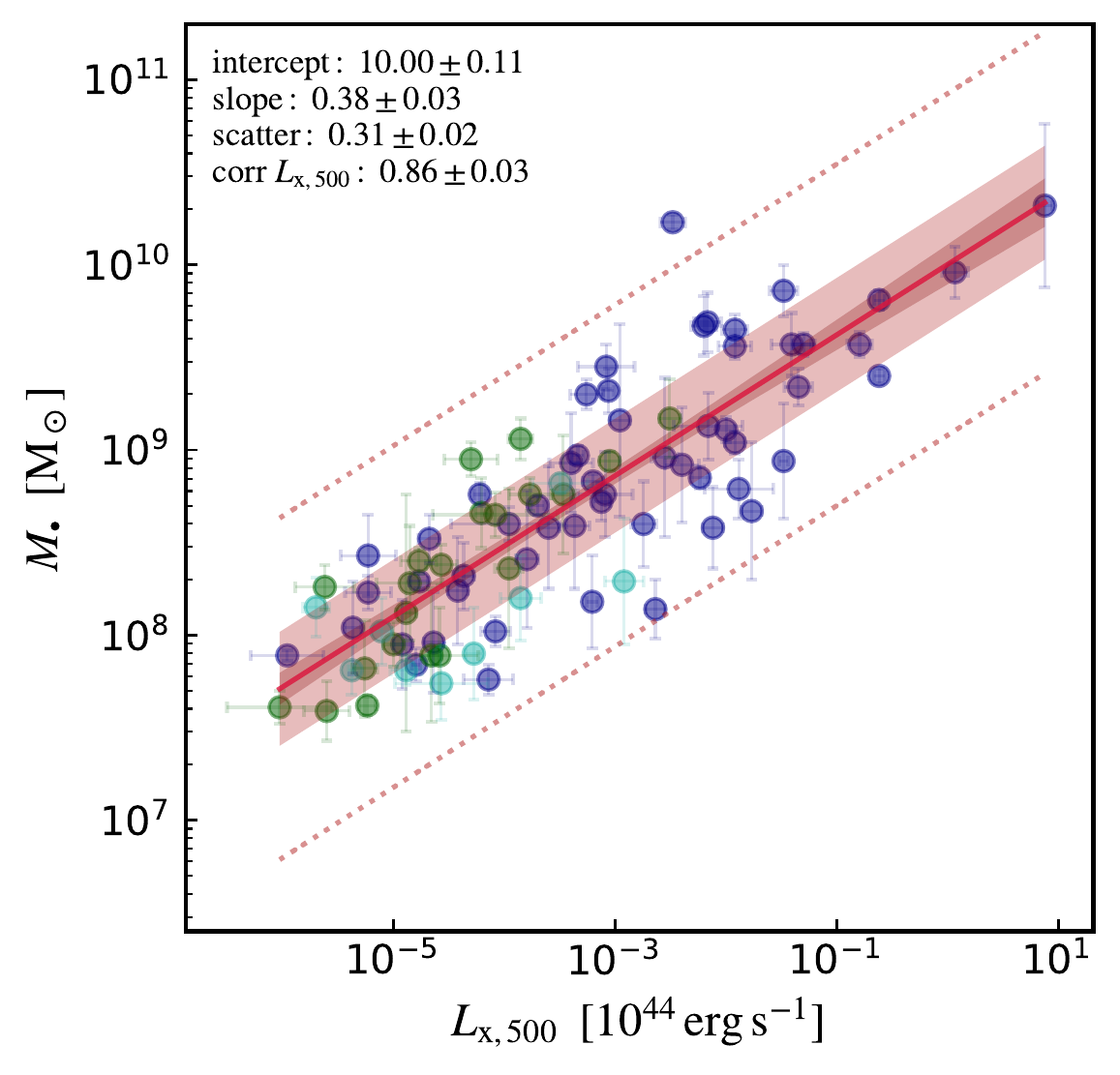}}
 \vskip -0.4cm
\caption{BH mass vs.~X-ray luminosity (\S\ref{s:xuni}) within the galactic/CGM scale ($\rxg \sim 0.03\,\r500$; top), group/cluster core ($\rxc \sim 0.15\,\r500$; middle), and group/cluster $R_{500}$ (bottom). Analog of Fig.~\ref{MbhTx}. 
The X-ray luminosity is tightly correlated with the SMBH mass (better than the optical $L_K$ counterpart), in particular within the core region.
}
 \vskip -0.8cm 
\label{MbhLx}
\end{figure}

\begin{figure}[!ht]
 \vskip -0.05cm
\subfigure{\includegraphics[width=0.92\columnwidth]{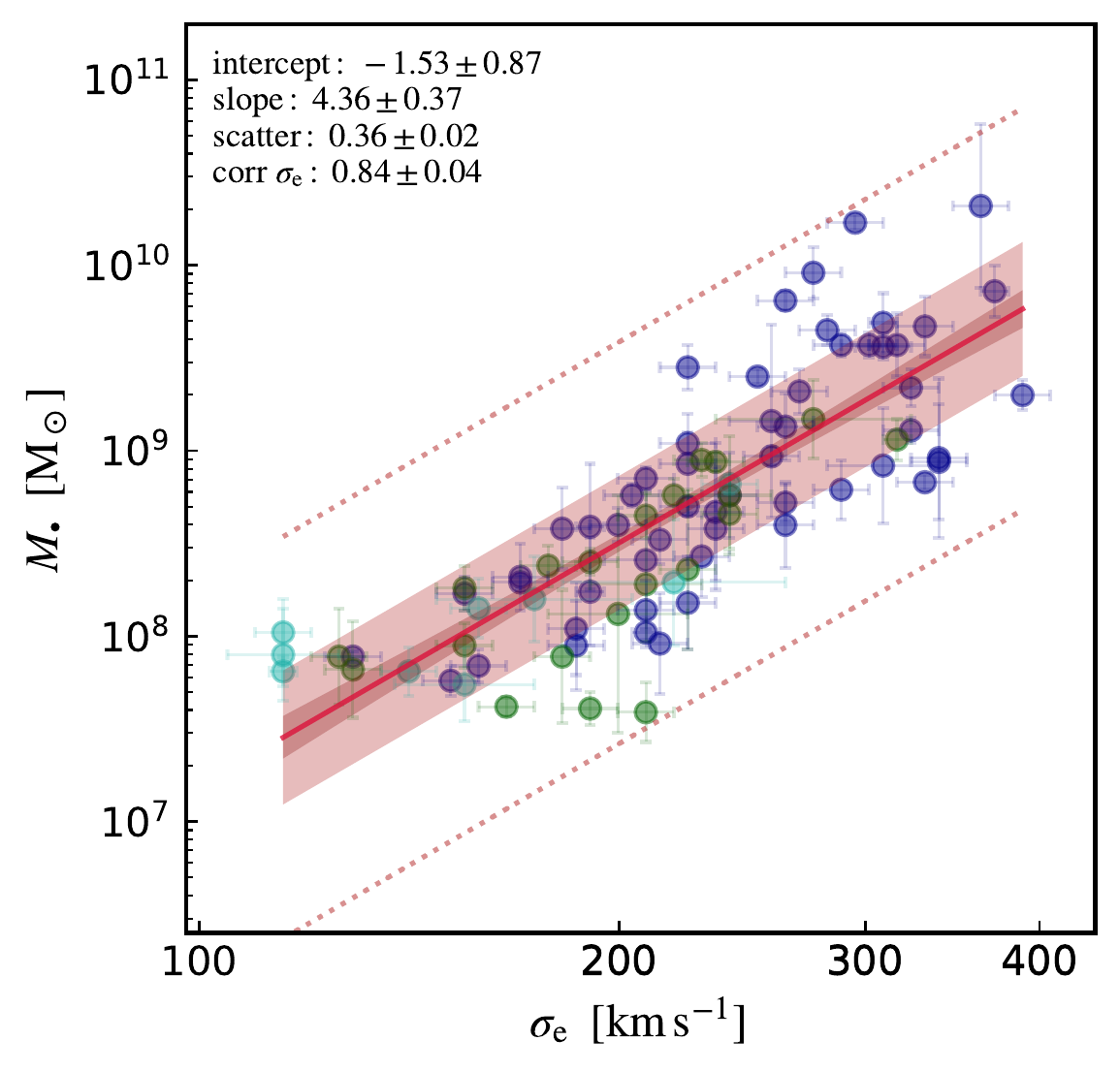}}
 \vskip -0.39cm
\subfigure{\includegraphics[width=0.92\columnwidth]{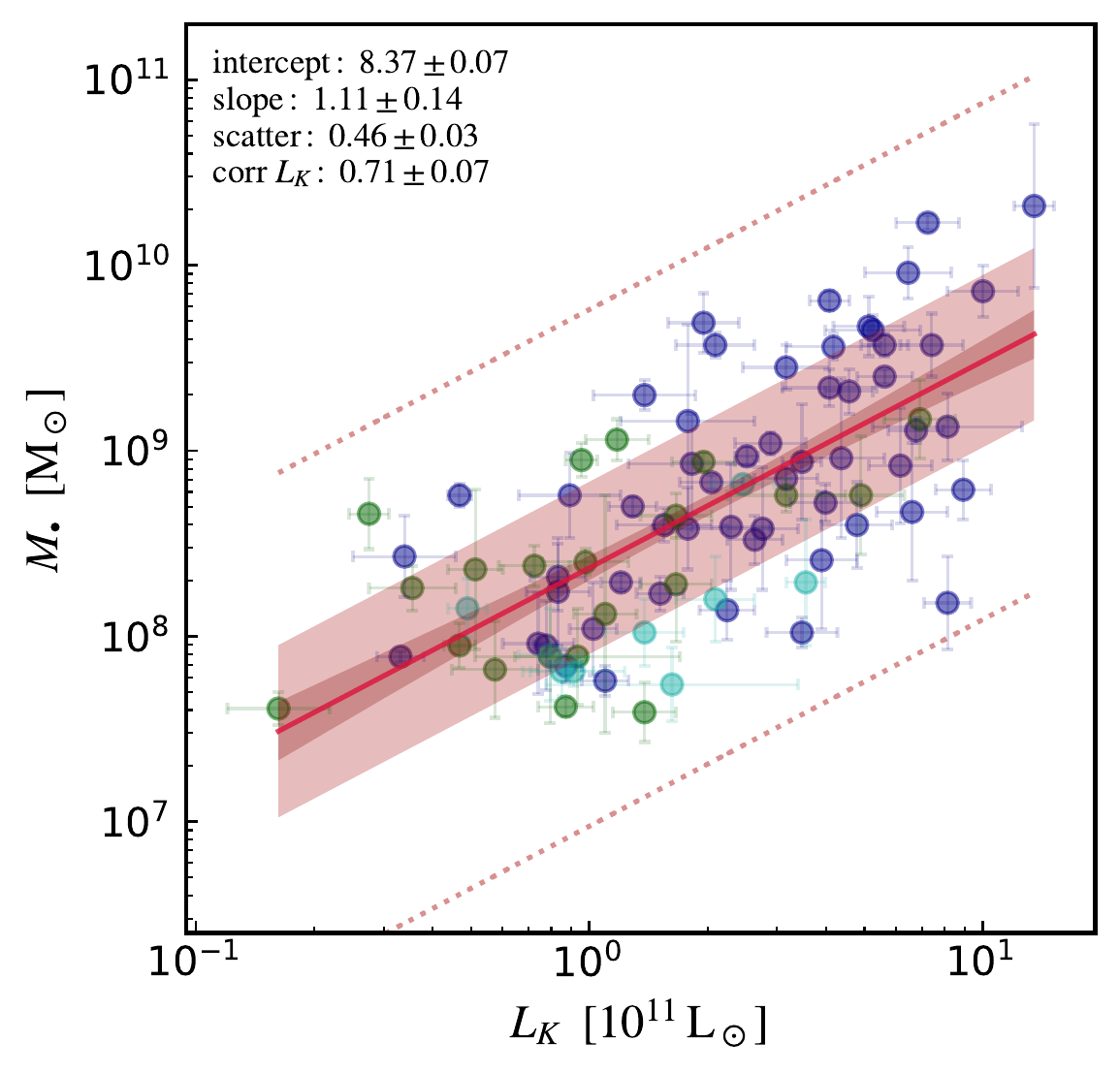}}
 \vskip -0.39cm
\subfigure{\includegraphics[width=0.92\columnwidth]{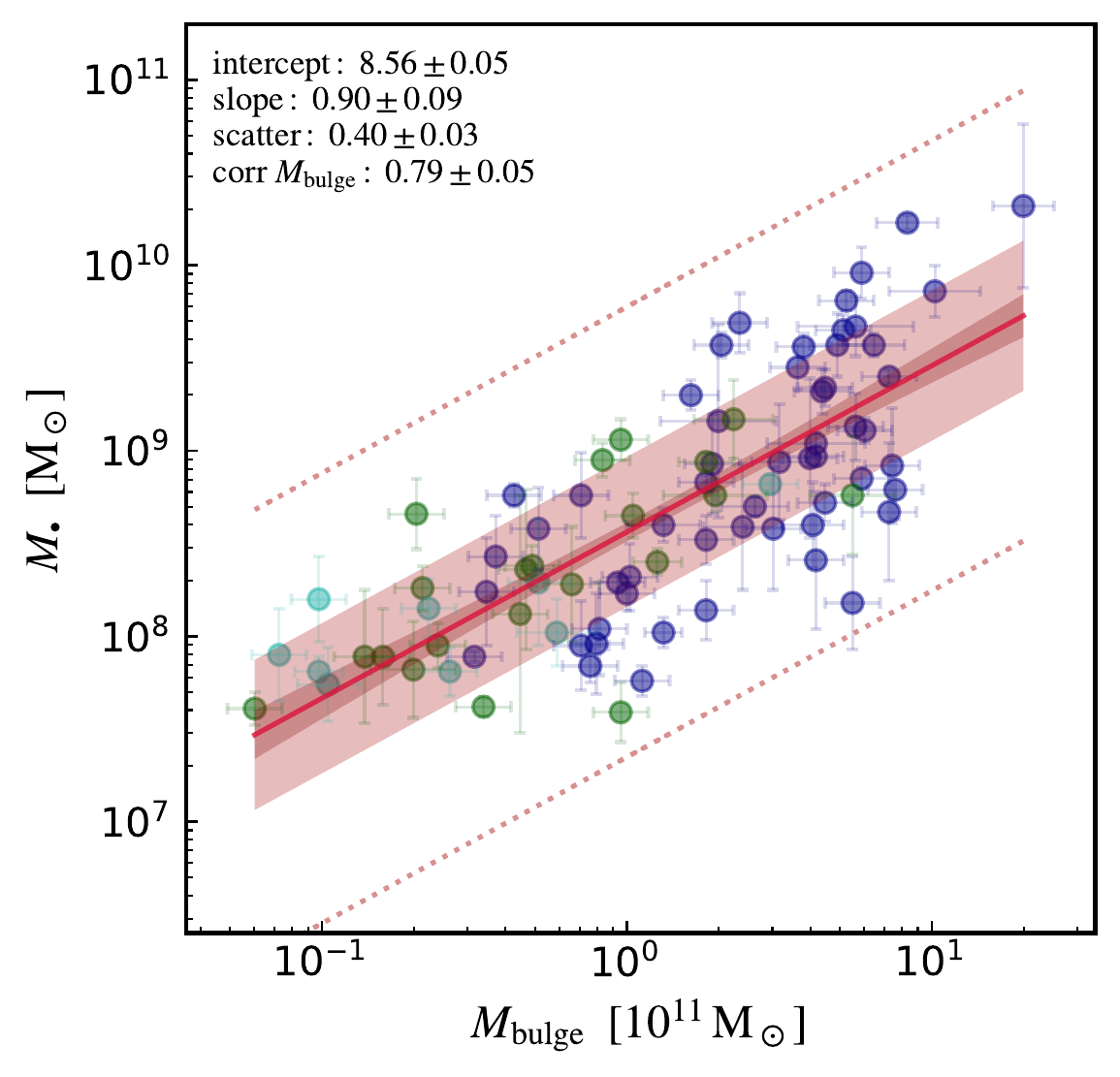}}
 \vskip -0.4cm
\caption{Black hole mass vs.~galactic optical properties (\S\ref{s:optuni}), including the stellar velocity dispersion ($< R_{\rm e}$; top), total (bulge and disk) $K$-band luminosity (middle), and bulge mass (bottom).
Analog of Fig.~\ref{MbhTx}. The optical properties have larger scatter compared with the X-ray ones. The velocity dispersion is the tighter observable, though showing superlinear behavior at the high end. 
}
 \vskip -0.8cm 
\label{Mbhop}
\end{figure}

Figure \ref{MbhLx} shows the second fundamental X-ray variable, i.e., the plasma luminosity, enclosed within our three radial regions (top to bottom panel: \Rxg, \Rxc, and \R500, respectively; Eq.\,\ref{e:Rxg}\,-\,\ref{e:R500}). 
As for the X-ray temperature, \Lx\ shows a very strong positive correlation with the black hole mass (corr $>0.85$). Regarding normalization, a typical SMBH of $10^9\,\msun$ tends to reside in a plasma halo of $\sim$\,$10^{41}\,\es$.
The log slope is now below unity, ranging between 0.5\,-\,0.4, adopting the galactic or cluster/virial scale, respectively (as \Lx\ steadily increases with radius).
However, the range covered by \Lx\ is now over 6 dex, i.e., a $6\times$ range increase compared with the temperature log scaling (we checked that the $\lx-\tx$ relation is consistent with that in other observational works which include low-mass galaxies; see \S\ref{s:xFP} and Fig.~\ref{LxTx}).
It is a major result that the SMBH scaling holds over such a wide range of X-ray luminosities, reflecting very different (poor and rich) environments.
We note the $R_{500}$ scalings are the more uncertain correlations: their contained scatter might be a reflection of the currently low number of available central massive galaxies (i.e., those having macro properties, such as $L_{\rm x,500}$, set by the extended ICM and not the CGM; \S\ref{s:xvar}).

Another important difference with \Tx\ is the relatively larger intrinsic scatter in $\mbh - \lx$, hovering in the range 0.29\,-\,0.31\;dex, though still tighter than any optical scaling relation (Fig.~\ref{Mbhop}). Removing the 3.5-$\sigma$ outlier NGC\,1600 (which could have stochastically suffered a ram-pressure halo stripping or AGN outburst) would reduce the scatter by 0.03\;dex, approaching that of the temperature scalings. Conversely, a case of perfectly matching the mean best fit is provided by M87 ($L_{\rm x,c}\simeq1.2\times10^{43}\;\es$), whose SMBH horizon has been recently imaged via EHT, constraining $\log \mbh/\msun\simeq 9.8$.
Interestingly, the lowest scatter in \Lx\ is found within the core region (though with small significance), where radiative cooling is very effective. 
Since $L_{\rm x}\propto M^2_{\rm gas}(<\,$$r)$, any gas evacuation (or phase transition) is associated with significant luminosity variations; indeed, feedback processes and mergers/cosmic inflows are particularly impactful in the inner and outer regions (e.g., \citealt{Ghirardini:2019}), respectively, while the intermediate region is less affected by them (the $\mbh-f_{\rm gas,c}$ relation will unveil this more clearly; \S\ref{s:fgas}).
In \S\ref{s:CCA}, we will test  the role of gas condensation and CCA-regulated AGN feedback; indeed,
the existence of a tight correlation between BH mass and \Lx/\Tx\ is consistent with first-principle predictions (\citealt{Gaspari:2017_uni}). 

We note that more massive SMBHs are hosted by galaxies with more extended X-ray and effective radii
(e.g., $\mbh-R_{\rm x,g}$ has a mild correlation with slope $1.1\pm0.2$), as more luminous halos have larger atmospheres (Fig.~\ref{sizegal}).
A stronger correlation emerges considering $\mbh-R_{\rm x,c}$ ($\beta=2.4\pm0.3$), although the BH correlations with characteristic radii typically show a significant scatter ($\sim$\,0.4 dex; e.g., Fig.~\ref{Mbhxop}). 
The correlation with $R_{500}$ is tighter and steeper, being a pure reflection of the \Tx\ scaling.
The multivariate, partial correlation analysis in \S\ref{s:multi} combining all the fundamental observables will help to understand any potential X-ray virial relations or deviations from it via non-gravitational processes.

While finishing our five-year project, another short work discussed a correlation between $\mbh$ and cluster halo temperature in 17 BCGs/BGGs (\citealt{Bogdan:2018}; no X-ray value was used here from their paper). They find a best-fit $\log(\mbh/10^9\,\msun)=(0.20\pm0.09)+(1.74\pm0.16)\log(T_{\rm x,c}/{\rm keV})$ (with $T_{\rm x,c}$ as proxy for $T_{\rm x,500}$). The shallower slope is due to a massive-system bias: selecting only central galaxies in our sample leads to $\log(\mbh/10^9\,\msun)=(0.20\pm0.07)+(1.83\pm0.31)\log(T_{\rm x,c}/{\rm keV})$, which is consistent with the above. Given their smaller sample and less robust BCES method (\S\ref{s:corr}), their scatter is $\sim$\,0.1\,dex higher, though still tighter than that of their $\mbh - \sige$.
This marks the importance of collecting a larger and more complete sample covering different morphological, dynamical, and environmental types.

\subsubsection{Optical/stellar velocity dispersion, luminosity, and bulge mass}\label{s:optuni}
We focus now on the counterpart variables in the optical band that are tracing the stellar component, rather than the plasma halo. Given that most stars are confined within a few effective radii, the optical properties can only trace the galactic scale, and not the larger scale core or virial region.

Fig.~\ref{Mbhop} shows the SMBH mass as a function of the three fundamental variables adopted in several previous studies: the 1D velocity dispersion \Sige\ (within an aperture the size of effective radius $R_{\rm e}$), bulge mass, and total (bulge plus disk) galactic luminosity in the NIR ($\sim$\,$2.2\,\mu$m) $K$ band (\S\ref{s:data}).
The first key result is the substantially larger intrinsic scatter of all the optical properties compared with that of the X-ray counterparts (Fig.~\ref{MbhTx}\,-\,\ref{MbhLx}), which can reach values up to 0.5\,dex for $L_K$, with the correlation coefficient dropping to the mild ($\sim$\,0.7) level.

The most reliable optical property is the stellar velocity dispersion \Sige\ (top panel), which represents another tracer of the (inner) galactic potential $\sigma_{\rm e}^2\propto \phi_{\rm g}$ (\S\ref{s:oFP}). 
The retrieved \Sige\ scatter is 0.36\,dex, which is 0.15\,-\,0.11 dex larger than that of the galactic/core X-ray temperature (at over 99.9\% confidence). The $M_\bullet - \sigma_{\rm e}$ log slope is $4.4\pm0.4$, which is consistent with twice that of the $M_\bullet - T_{\rm x}$ scaling, i.e., $\sigma_{\rm e}^2 \propto T_{\rm x}$, as expected in virialized systems\footnote{Specifically, $\sigma_{\rm e}^2 = k_{\rm b} T_{\rm x}/{\mu m_{\rm p}}$ (with $\mu=0.62$ the plasma particle mean weight); while we find a unity slope ($0.93\pm0.08$) for $\sige^2-T_{\rm x,g}$, the normalization is lower than the virial expectation by $\approx$\,40\%, implying extra heating due to feedback processes.}. 
The retrieved $M_\bullet - \sigma_{\rm e}$ slope is similar to that found by previous studies on bulge-dominated galaxies (e.g., \citealt{Kormendy:2013}); it would steepen to a value $\gta 5$, almost doubling the intrinsic scatter, including the more uncertain low-mass BHs and related irregular galaxies (e.g., \citealt{Saglia:2016}). 
The high-mass end of the $M_\bullet - \sigma_{\rm e}$ is a significant source of scatter with increasingly over-massive BHs (five objects are approaching the top 3-$\sigma$ channel), in conjunction with the increased presence of BGGs/BCGs 
(see also Fig.~\ref{BGCCs}; some works interpret this as a nonlinear bend).
The disky (low $B/T$) and spiral galaxies (green/cyan points) start also to show symptoms of a decline below the linear fit (despite measurements of \Sige\ remaining accurate), while $\mbh-\tx$ and $\mbh-\lx$ retain a stable linear behavior regardless of different morphological types and environment (Fig.~\ref{BGCCs}).

The second panel in Fig.~\ref{Mbhop} shows the $M_\bullet - L_K$ scaling. While the total galaxy NIR luminosity is a good proxy for the total stellar mass (since it is not much affected by dust absorption),
the hosted BH mass is only mildly tied to this galactic observable. Unlike the other quantities, the morphological types tend to be significantly mixed from low to large $L_K$ values, corroborating the large $\epsilon$ value.
On the other hand, the slope is consistent with unity, i.e., there is a direct 1:1 conversion in both logarithmic and linear space, with an average $10^{11}\,L_\odot$ galaxy hosting a $\mbh\simeq2\times10^8\,\msun$. Converting $L_K$ to total stellar mass (Eq.~\ref{e:ML})
would show similar correlation slope and scatter, within the 1-$\sigma$ uncertainty (the total stellar mass correlations are thus redundant and not shown).
At variance with temperatures, the X-ray versus optical luminosities (at all radii) show very different scaling with $\mbh$, since $L_{\rm x}$ covers four more orders of magnitude compared with $L_K$. In other words, the X-ray properties allow us to probe more extended regimes and regions than those traced by stars, better separating the loci occupied by LTGs and ETGs. 
The multivariate analysis (\S\ref{s:multi}) will unveil that the X-ray and optical fundamental planes behave differently due to \Lx\ breaking the self-similar gravitational collapse expectation.

A way to reduce the scatter is to consider purely the stellar bulge mass, instead of the total stellar luminosity/mass (which is contaminated by potential disk features). 
The bottom panel in Fig.~\ref{Mbhop} shows the correlation with the stellar bulge mass (known as the `Magorrian relation'; \citealt{Magorrian:1998}).
Translating from a luminosity to a mass is non-trivial since the $M_\ast/L_K$ depends on complex stellar population models (\S\ref{s:sample}). Moreover, the $B/T$ ratio should be taken as approximate, as it can vary significantly between studies. Keeping in mind such hurdles, the $M_\bullet - M_{\rm bulge}$ relation is able to reduce the scatter to 0.40\,dex, albeit not yet reaching the lower level of \Sige. The log slope is slightly shallower than unity ($\beta \approx 0.9$). 
Regarding normalization, $M_{\rm bulge}/\mbh \approx 360$, implying that stars continuously accumulate within the galaxy without substantially feeding the BH during cosmic time, given their collisionless nature. 

The $M_\bullet - \sigma_{\rm e}$ appears to be the most stable optical estimator of BH mass. However, it presents signs of unreliability at the high-mass end, with several galaxies exceeding the 2-$\sigma$ scatter band. 
None of the optical variables shows better correlations than the X-ray counterparts, in terms of intrinsic scatter and correlation coefficient ($>$\,$99$\% confidence level).
Moreover, performing a pairwise correlation analysis on residuals (cf.~\citealt{Shankar:2019}), we find that 
$\log \mbh - \langle \log \mbh | \log \sige \rangle$ versus
$\log \tx - \langle \log \tx | \log \sige \rangle$
has 60\% larger correlation coefficient (0.8) than 
$\log \mbh - \langle \log \mbh | \log \tx \rangle$ versus
$\log \sige - \langle \log \sige | \log \tx \rangle$, suggesting that X-ray properties are more fundamental than optical properties.
There are two reasons that we deem to be important to explain this.
First, the stellar component is tracing purely the inner part of the whole gravitational potential, thus missing the macro group/cluster halo. 
Second, stars are the residual by-product of a more wide-spread top-down multiphase condensation process, which originates in the X-ray plasma atmosphere (particularly in the core, $r \lta 0.1\,R_{500}$; \S\ref{s:disc}).

\subsection{Univariate correlations: composite X-ray variables} \label{s:comp}
We now move on to the univariate correlations of the composite variables, again focusing on their interplay with BH mass.
Indeed, the equations of thermodynamics and hydrodynamics for a diffuse gas/plasma are linked to properties such as pressure and particle number density.
The concept to keep in mind is to derive these properties only from the fundamental observables, i.e., X-ray luminosity and temperature (while propagating the related errors), thus keeping any parameterization and assumption to the minimum.

\subsubsection{Electron number density} \label{s:ne}

\begin{figure}[!ht]
 \vskip -0.05cm
\subfigure{\includegraphics[width=0.908\columnwidth]{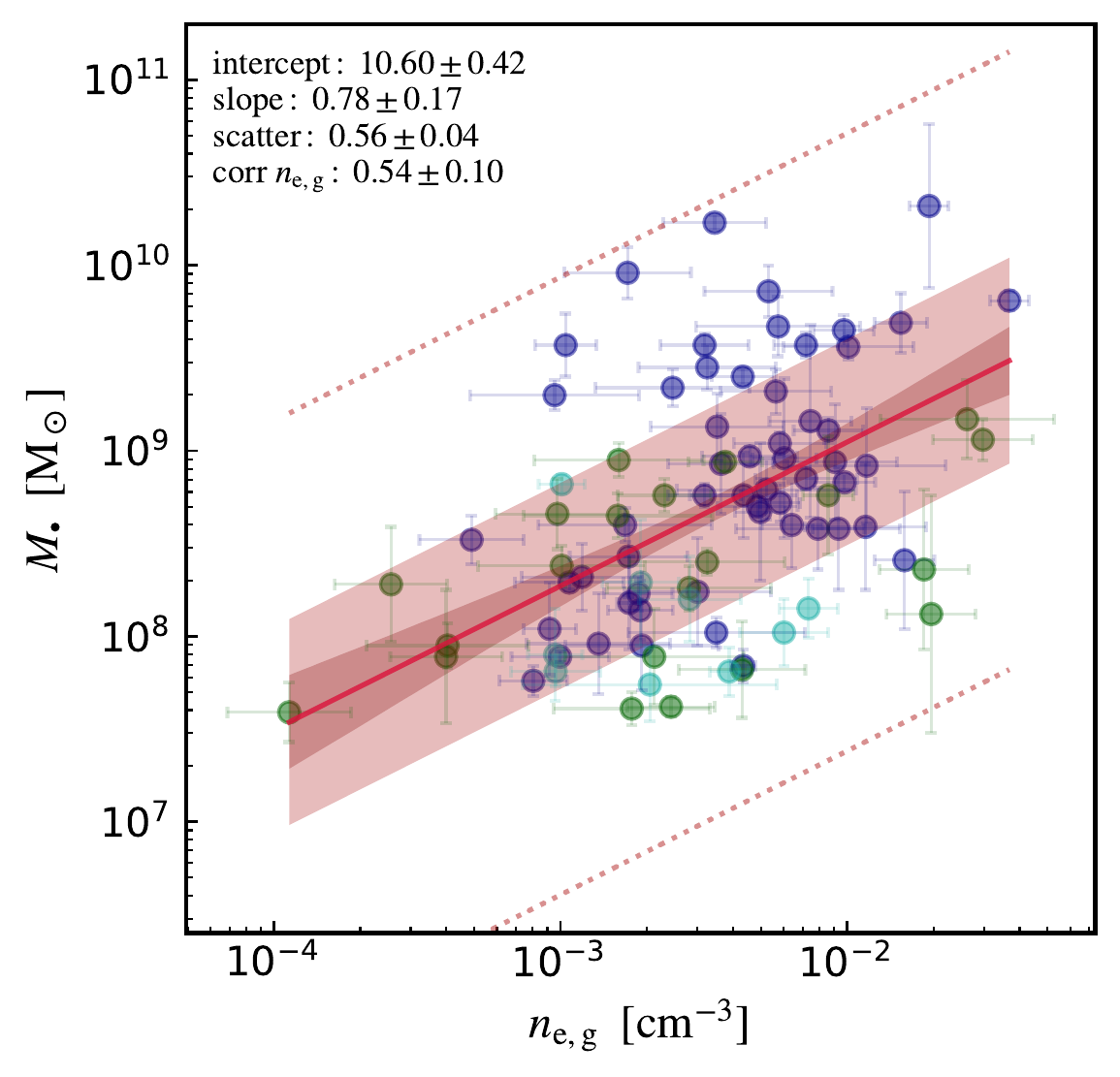}}
 \vskip -0.39cm
\subfigure{\includegraphics[width=0.91\columnwidth]{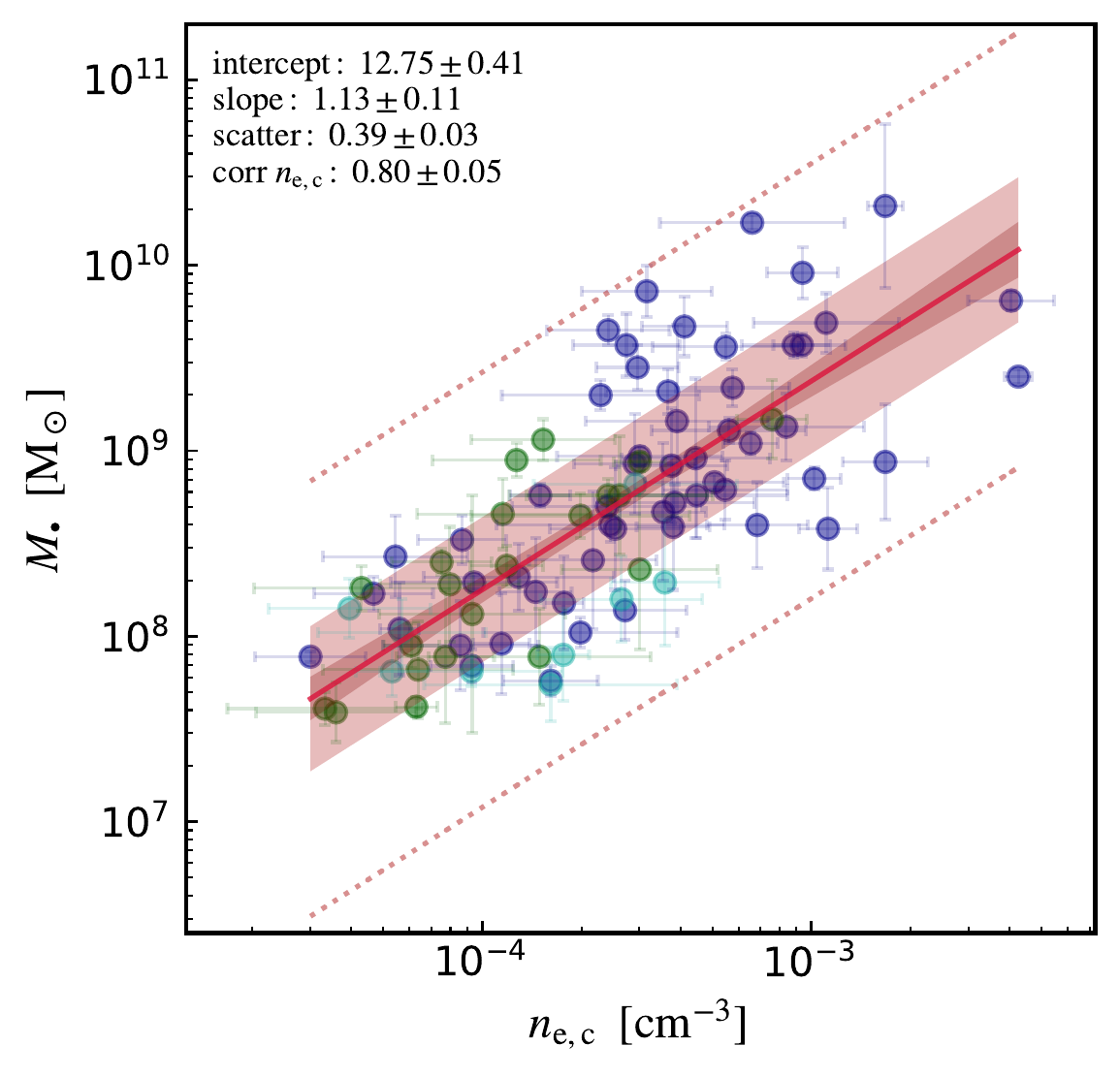}}
 \vskip -0.39cm
\subfigure{\includegraphics[width=0.91\columnwidth]{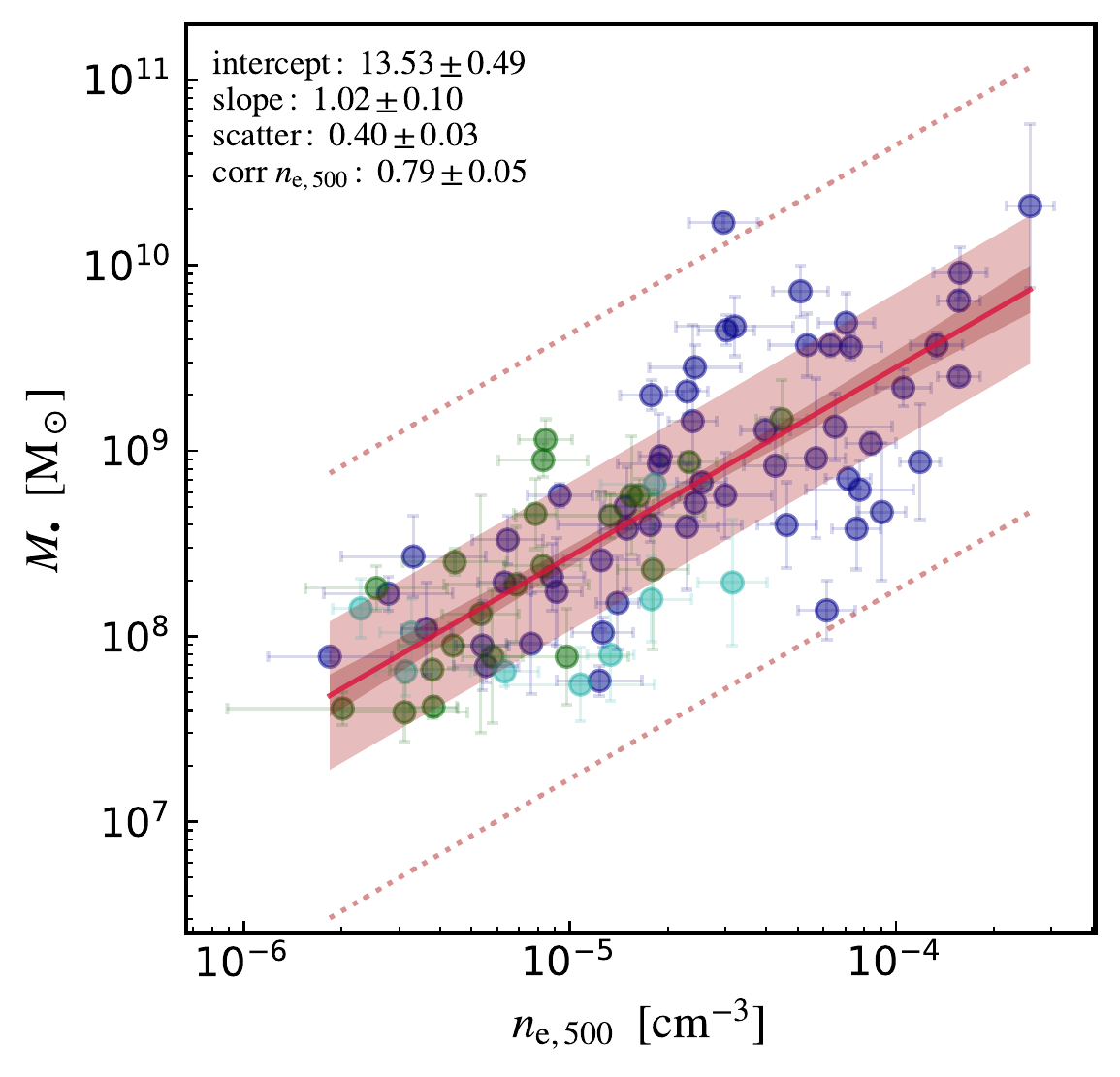}}
 \vskip -0.4cm
\caption{Black hole mass vs.~electron number density (\S\ref{s:ne}) within the galactic/CGM \Rxg\ (top), group/cluster \Rxc\ (middle), and $R_{500}$ (bottom) shells. Analog of Fig.~\ref{MbhTx}. Density is derived from the \Lx\ gradient and $\Lambda(T_{\rm x},Z)$ (propagating the related errors; Eq.~\ref{e:ne}). While the CGM scale displays very weak correlation, the core plasma density is strongly correlated with BH mass.
}
 \vskip -0.5cm 
\label{Mbhne}
\end{figure}

The plasma luminosity is given by 
\begin{equation}\label{e:Lx}
L_{\rm x} = \int{n_{\rm e} n_{\rm i}\, \Lambda(\tx,Z)\, dV}, 
\end{equation}
where $\Lambda(\tx,Z)$ is the radiative plasma cooling function in collisional ionization equilibrium (\citealt{Sutherland:1993})\footnote{The cooling curve includes spectral calculations for H, He, C, N, O, Fe, Ne, Na, Si, Mg, Al, Ar, S, Cl, Ca, Ni, and all related stages of ionization.
We tested different cooling curves (e.g., \citealt{Schure:2009}), finding comparable results.
} adopting metallicity $Z\simeq0.7,\,0.4,\,0.3\;Z_\odot$ for the galactic, core, and $R_{500}$ region, respectively (\citealt{Mernier:2017})\footnote{We tested the full observational scatter of observed abundance values, finding no major change in results. Further, we note that such changes in metallicity alter $\mu_{\rm i}$ by less than 1\%.}. 
By differentiating and discretizing Eq.~\ref{e:Lx} over finite spherical shells, $\Delta V = (4/3)\pi(R_{\rm out}^3 - R_{\rm in}^3$), the plasma electron number density can be retrieved as
\begin{equation}\label{e:ne}
n_{\rm e} \simeq \left(\frac{\Delta\lx}{\Delta V} \frac{{\mu_{\rm i}}/{\mu_{\rm e}}}{\Lambda(\tx,Z)} \right)^{1/2},
\end{equation}
where
$(\mu_{\rm i},\mu_{\rm e}) \simeq (1.30, 1.18)$ are the ion and electron mean weights (for a plasma with solar composition). 
Given finite discretization of Eq.~\ref{e:Lx},
the computed $n_{\rm e}$ should be understood as a mean density inside our three radial shells (0$-$\Rxg, \Rxg$-$\Rxc, \Rxc$-$\R500).\footnote{For non-central galaxies, the core/outskirt gradients are approximated as $dL_{\rm x}/dV\sim L_{\rm x}/V$. }
Over our entire sample and bins, we find a median density gradient $d\log n_{\rm e}/d\log r = -1.6\pm0.3$, which is consistent with that in other works (e.g., \citealt{Babyk:2018} and \citealt{Hogan:2017}).
We remark that gas density is a composite variable given by the combination of X-ray luminosity, temperature, and $R_{\rm x}^3$ (Eq.~\ref{e:ne}).

Figure \ref{Mbhne} shows the correlation between SMBH mass and plasma electron density, inside our three adopted radial shells.
For the inner scale the correlation is weak (absent at the 2-$\sigma$ level), corroborated by the substantial scatter ($\epsilon \simeq 0.6$) and the large errors in the posterior distributions (broad red bands).
Indeed, most of the galaxies have an average galactic/CGM gas number density ranging between $n_{\rm e,g}\sim10^{-3}$\,-\,$10^{-2}$\,cm$^{-3}$ (consistently with other studies, e.g., \citealt{Lakhchaura:2018}).
The correlation enters instead the strong regime (almost halving the scatter) if we consider the core or $R_{500}$ region.
This corroborates the result highlighted by the $\mbh - L_{\rm x,c}$ (\S\ref{s:xuni}), i.e., the halo core region (where the cooling time is typically below the Hubble time) is one of the best predictors for the SMBH growth.
The slope is consistent with unity, with a typical SMBH mass of a few $10^9\msun$ (massive galaxies) linked to $n_{\rm e}\approx 10^{-3}/10^{-4}$\,cm$^{-3}$ in the macro-scale core/outskirt region (in agreement with values retrieved by \citealt{Sun:2012} and \citealt{Hogan:2017} for BGGs and BCGs). Since \Lx\ (via $n_{\rm e}^2\,\Lambda$) is a direct manifestation of the plasma radiative emission, these findings suggest that condensation processes could play a major role in the evolution of SMBHs (\S\ref{s:CCA}).

\subsubsection{Total gas pressure} \label{s:Px}

\begin{figure}[!ht]
 \vskip -0.05cm
\subfigure{\includegraphics[width=0.91\columnwidth]{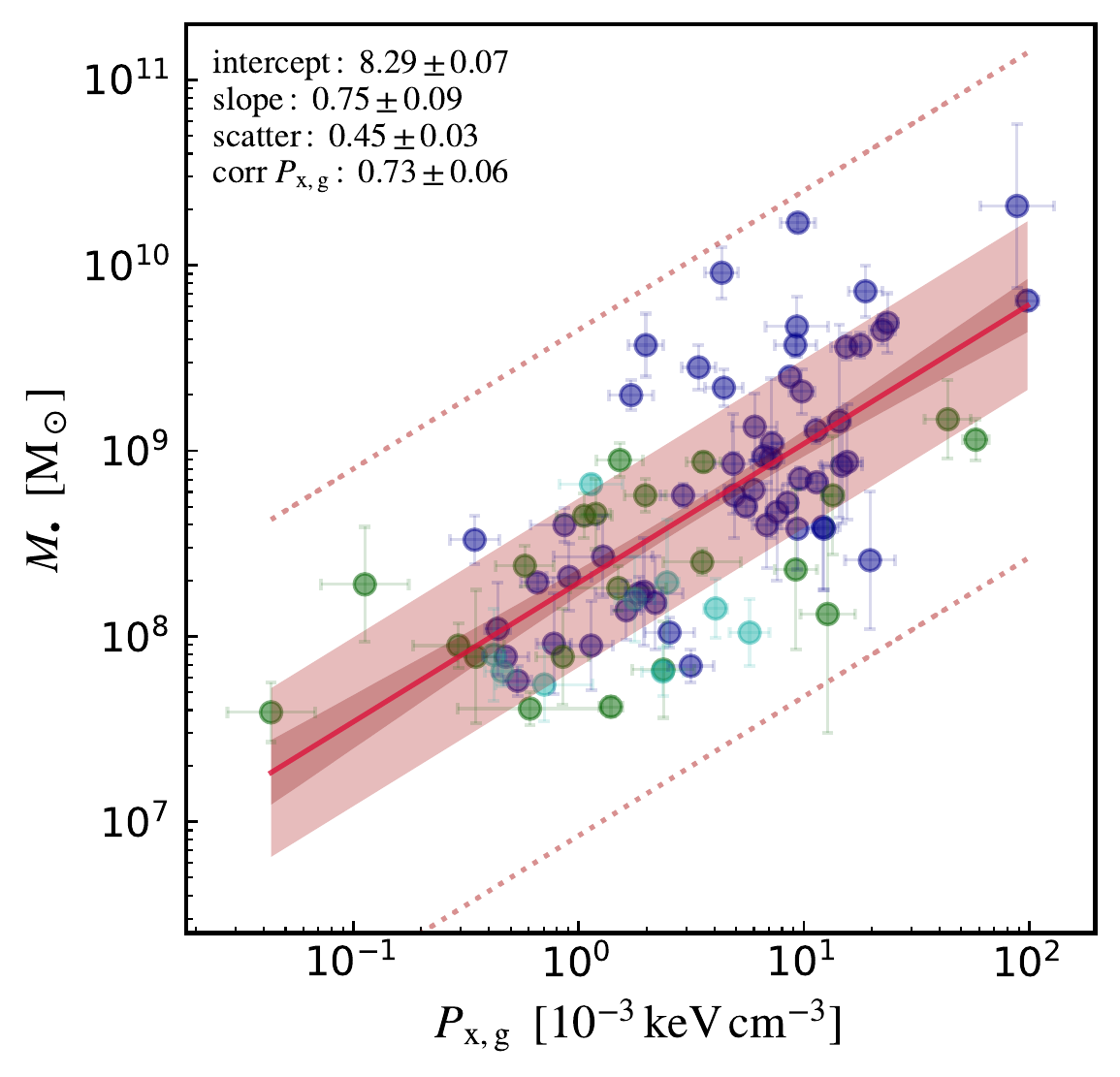}}
 \vskip -0.39cm
\subfigure{\includegraphics[width=0.91\columnwidth]{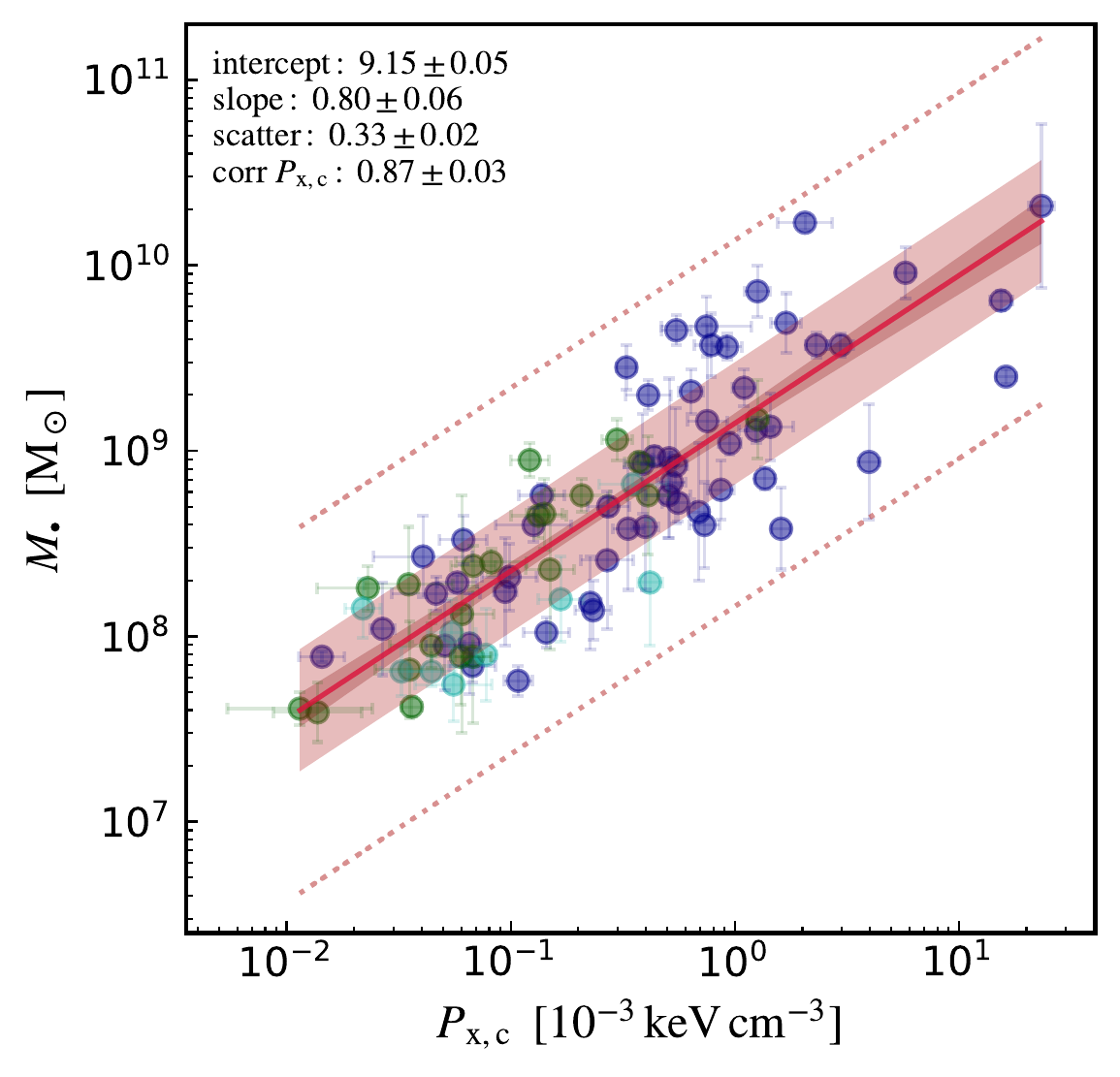}}
 \vskip -0.39cm
\subfigure{\includegraphics[width=0.912\columnwidth]{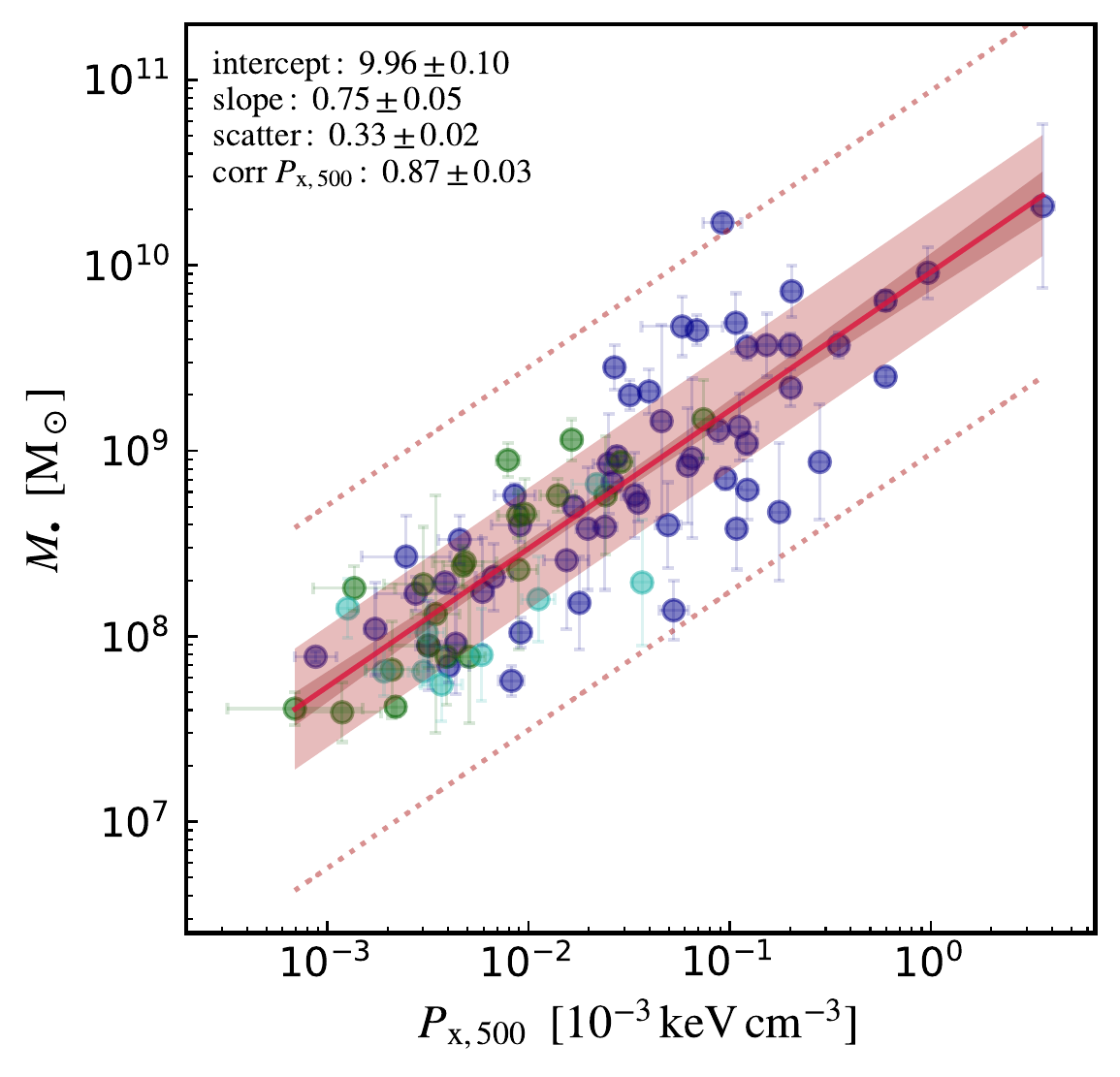}}
 \vskip -0.4cm
\caption{Black hole mass vs.~gas pressure (\S\ref{s:Px}) inside the \Rxg\ (top), \Rxc\ (middle), and $R_{500}$ (bottom) shells. Analog of Fig.~\ref{MbhTx}. 
Similar to the X-ray luminosity, the core plasma pressure is a good and tight indicator of the central BH growth, although it loses efficacy in the galactic region.
}
\vskip -0.4cm 
\label{MbhPx}
\end{figure}

A key thermodynamic variable which determines the hydrostatic balance of a stratified atmosphere is the total gas/plasma pressure ($P=P_{\rm e}+P_{\rm i}$) defined as
\begin{equation}\label{e:pres}
P = n\, k_{\rm b} T_{\rm x}, 
\end{equation}
where $n=n_{\rm e}+n_{\rm i}\simeq 2\,n_{\rm e}$ is the total gas particle number density. 
Figure \ref{MbhPx} shows the $\mbh - P_{\rm x}$ correlation for our three radial shells (\S\ref{s:ne}).
All the gas pressure scalings reside in the regime of strong correlation with BH mass (corr\,$\approx0.7$\,-\,$0.9$). The direct combination of density and temperature seems to ameliorate the galactic scaling, although the scatter remains large at 0.5 dex. 
The best-fit slope is stable at sublinear values, $\beta\approx 0.75$.
As for \Lx, the core region displays the lowest scatter ($\epsilon\simeq0.3$) and highest corr coefficient, 
with a characteristic gas pressure of $\sim$\,$10^{-3}$\,keV\,cm$^{-3}$ for halos hosting a $10^9\,\msun$ BH.

Interestingly, more pressure-supported halos harbor larger SMBHs. If we think in terms of classical hot-mode accretion (Bondi, ADAF, etc.), in which a larger atmospheric pressure suppresses accretion (\S\ref{s:Bondi}), such a trend seems difficult to develop.
However, if accretion proceeds through the cold mode, then a more pressurized gas implies larger available internal energy, $P/(\gamma-1)$,\footnote{The (non-relativistic) plasma adiabatic index is $\gamma=5/3$.} to be radiated away, and thus a larger condensing neutral/molecular gas mass available to rain onto the SMBH (dropping out of the diffuse atmosphere in quasi hydrostatic equilibrium).

\subsubsection{Gas Mass} \label{s:Mgas}

\begin{figure}[!ht]
 \vskip -0.05cm
\subfigure{\includegraphics[width=0.94\columnwidth]{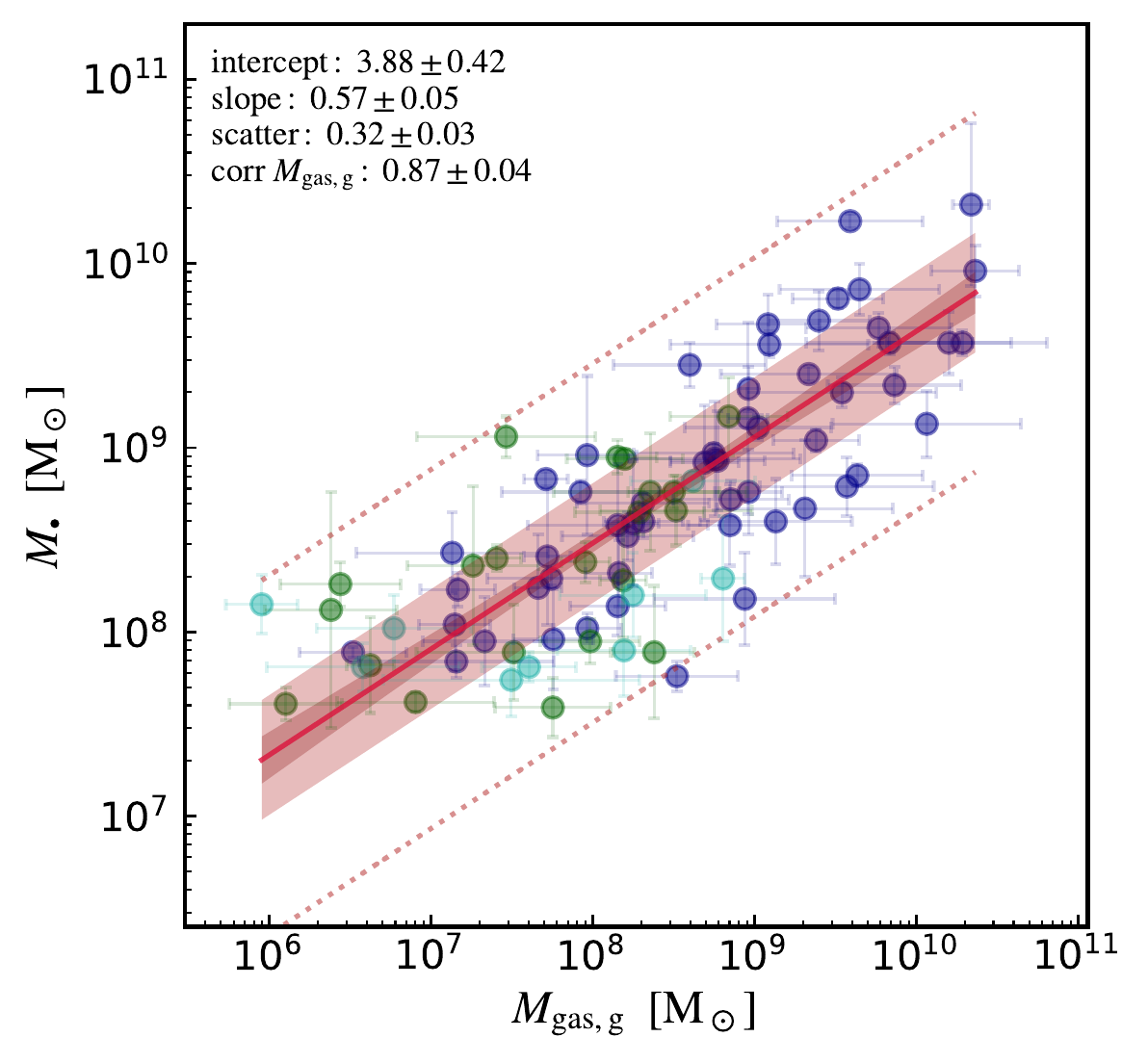}}
 \vskip -0.39cm
\subfigure{\includegraphics[width=0.91\columnwidth]{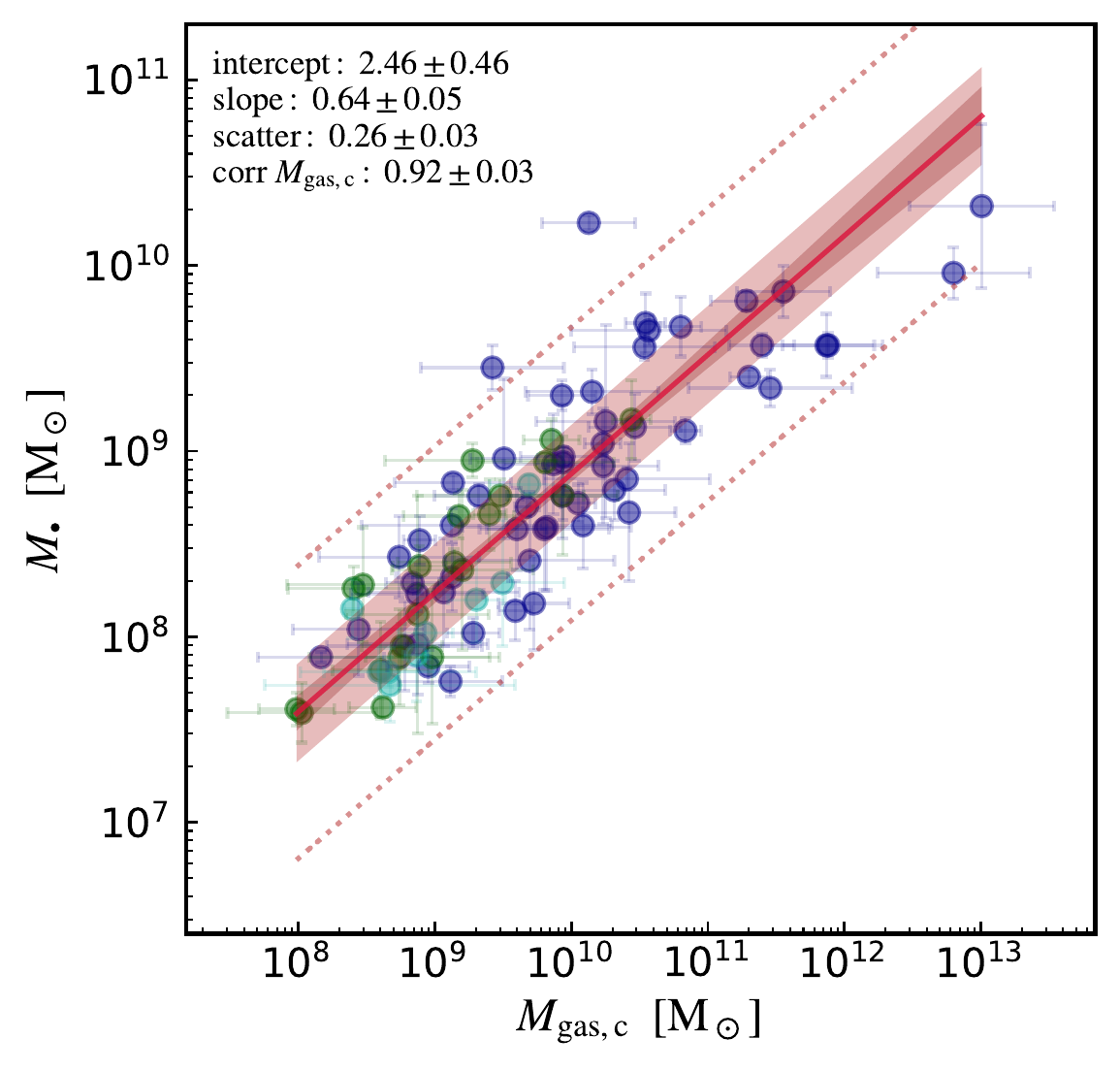}}
 \vskip -0.39cm
\subfigure{\includegraphics[width=0.91\columnwidth]{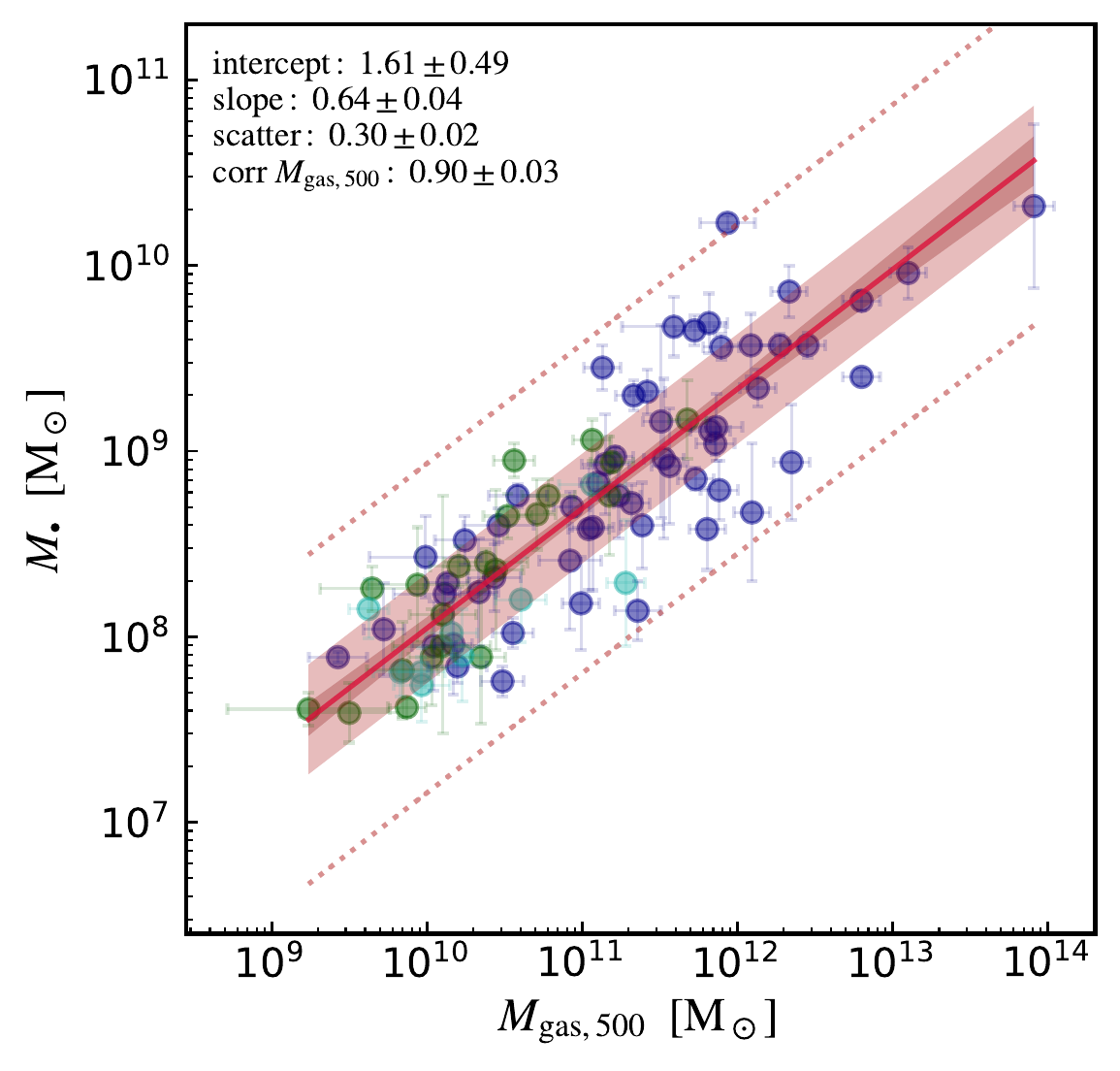}}
 \vskip -0.4cm
\caption{Black hole mass vs.~enclosed gas mass (\S\ref{s:Mgas}) within \Rxg\ (top), \Rxc\ (middle), and $R_{500}$ (bottom).
Although $M_{\rm gas}$ is a derived property, $\mbh - M_{\rm gas}$ is tight across all regions, and shows a nearly maximal positive correlation coefficient in the core region. Compared with the stellar mass (Magorrian) relation, the scatter is halved and is analogous to the fundamental \Tx\ correlations.
}
 \vskip -0.4cm 
\label{MbhMgas}
\end{figure}

The plasma mass within a given enclosed radius can be retrieved (integrating over our bins) via
\begin{equation}\label{e:mgas}
M_{\rm gas} (< r') = \int^{r'}_0 \rho \;4\pi r^2 dr,
\end{equation}
where the total gas density is $\rho = n_{\rm e}\mu_{\rm e} m_{\rm p} = n \mu m_{\rm p}$.
Figure \ref{MbhMgas} shows the $\mbh - M_{\rm gas}$ correlations. In spite of being (here) a derived variable, $M_{\rm gas}$ is very well correlated with the BH mass, lowering the intrinsic scatter to $\epsilon\simeq0.26\pm0.03$, comparable to that of the fundamental X-ray temperature scalings. Being an integrated quantity, gas mass has a smoothening advantage compared with local properties. 
In the core region (middle panel), the Bayesian posterior of the corr coefficient shows a value $0.92\pm0.03$, which is consistent with a maximal positive correlation at the 3-$\sigma$ level. 
Evidently, the gas mass plays a key role in the evolution of SMBHs.
The slope is similar across all radial bins, with values $\beta\approx 0.6$. The galactic (top panel) and virial (bottom panel) relations bound the locus of optimal correlation.

Regarding normalization, the median $10^9\,\msun$ BHs occupy core halos that have $\sim$10$\times$ more gas mass. 
As a ratio, this is over an order of magnitude lower compared with that involving the stellar bulge mass (the Magorrian relation; \S\ref{s:optuni}). We note that the retrieved range of core gas masses $M_{\rm gas,c}\sim10^9-10^{11}\,\msun$ is consistent with that of similar samples (e.g., \citealt{Babyk:2018}), corroborating our derivation method.
Considering the galactic scale (top panel), the median SMBH has $M_{\rm gas}$ roughly equivalent to $\mbh$ (at least at $z\sim0$).
This suggests that, while a major fraction of the BH mass can be built up in time via gas accretion due to collisional processes (e.g., CCA inelastic collisions, hydrodynamical instabilities, viscosity, shocks, turbulent mixing), stars remain largely unaffected by the accretion process being collisionless systems. This may also explain why most stellar properties present substantial scatter as estimators of $\mbh$, being linked to the BH growth via secondary/indirect effects.

\subsubsection{$Y_{\rm x}$ Compton parameter -- Thermal energy} \label{s:Yx}

\begin{figure}[!ht]
 \vskip -0.05cm
\subfigure{\includegraphics[width=0.91\columnwidth]{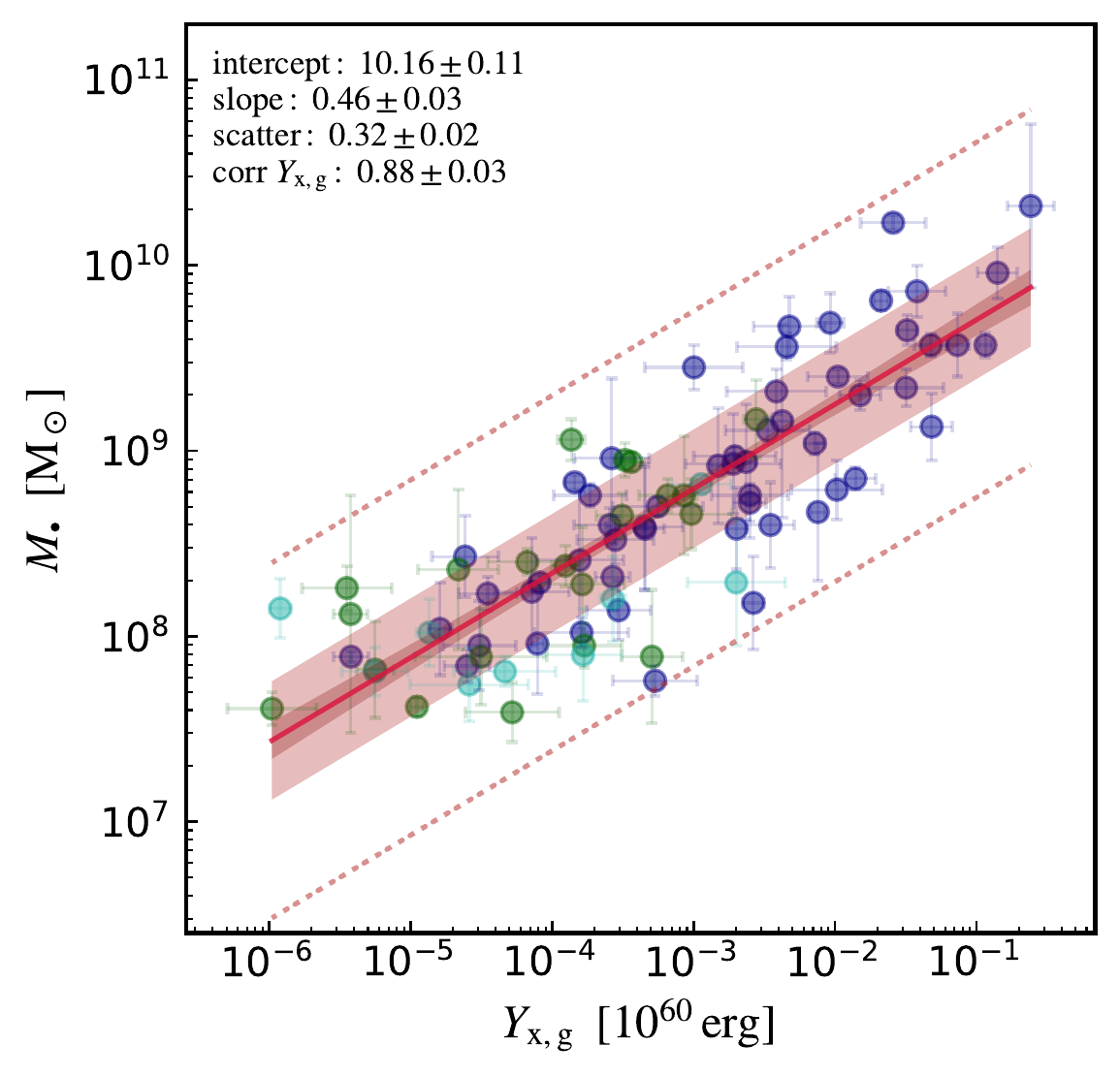}}
 \vskip -0.39cm
\subfigure{\includegraphics[width=0.91\columnwidth]{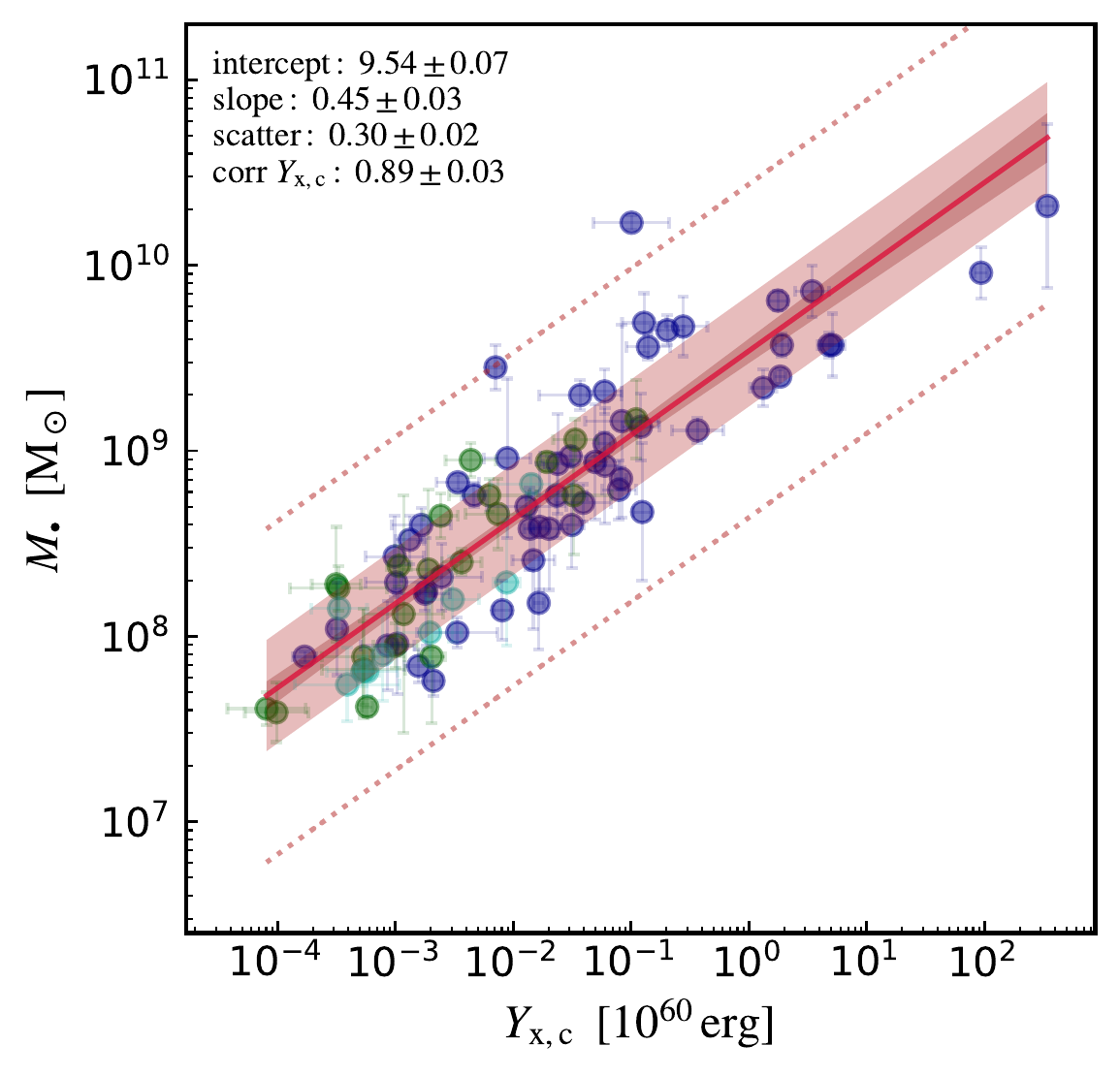}}
 \vskip -0.39cm
\subfigure{\includegraphics[width=0.91\columnwidth]{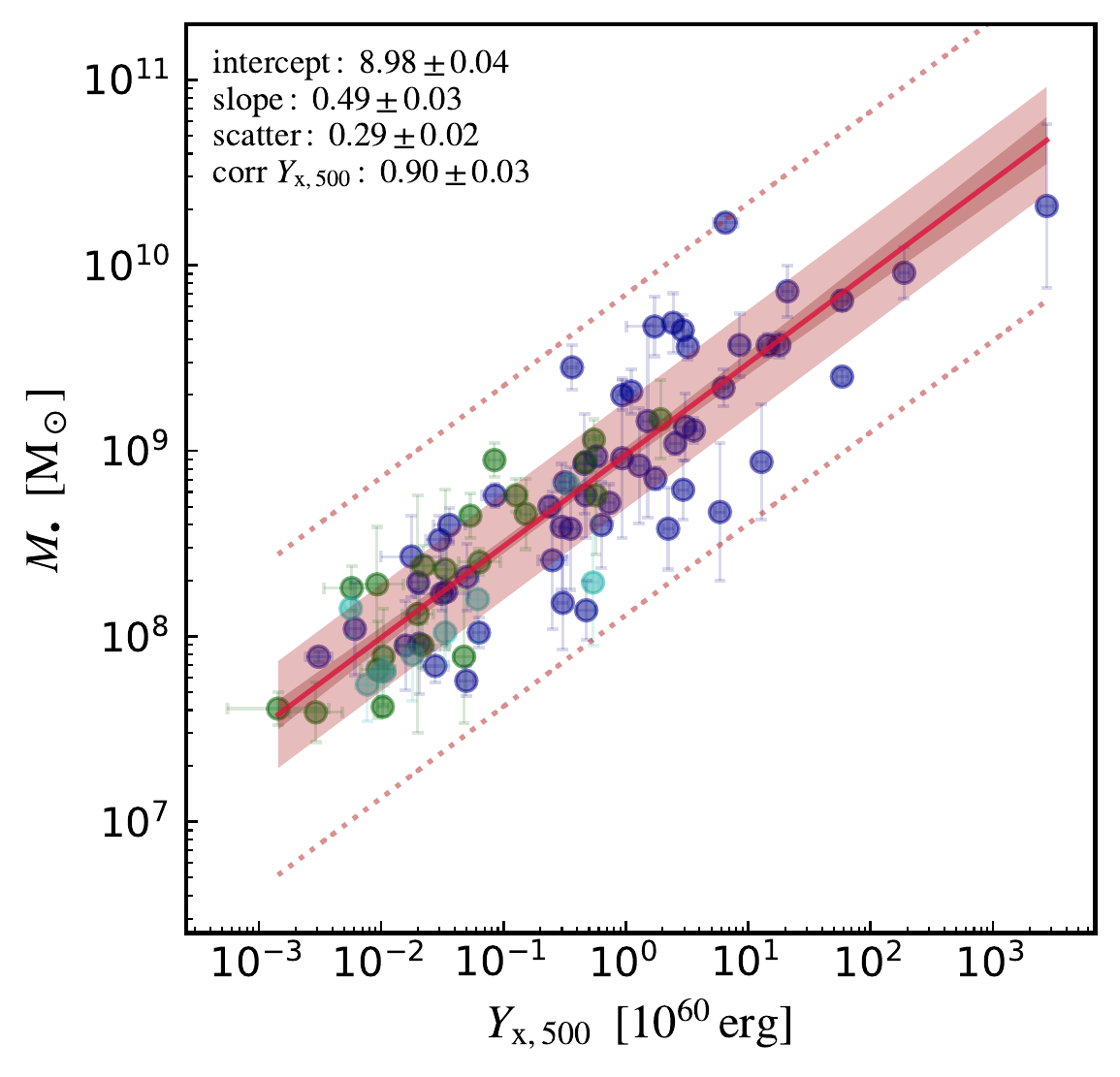}}
 \vskip -0.4cm
\caption{BH mass vs.~$Y_{\rm x}$ parameter (gas thermal energy; \S\ref{s:Yx}) within \Rxg\ (top), \Rxc\ (middle), and $R_{500}$ (bottom). Unlike $P_{\rm x}$ and $n_{\rm e}$,
the Compton parameter shows very stable behavior among all the inner and outer regions. The strong and tight correlation with $\mbh$, particularly over the whole cluster/group, can be leveraged by the next-generation radio/SZ telescopes.
}
 \vskip -0.7cm 
\label{MbhYx}
\end{figure}

Another key quantity that has become central to cluster and cosmological studies is the X-ray analog of the Compton $Y_{\rm sz}$ parameter (\citealt{Kravtsov:2006}), which describes the strength of the thermal Sunyaev-Zel'dovich (SZ) effect\footnote{The distortion of the cosmic microwave background spectrum via inverse Compton scattering by the hot plasma electrons (e.g., \citealt{Khatri:2016}).}. 
We define this X-ray analog as follows:
\begin{equation}\label{e:yx}
Y_{\rm x} \equiv E_{\rm th} = \frac{3}{2}\frac{k_{\rm b}T_{\rm x}}{\mu m_{\rm p}} M_{\rm gas}.
\end{equation}
Multiplying by the relevant thermodynamic constants, $Y_{\rm x}$ represents another form of the plasma thermal energy content (or integrated pressure). Recent studies (e.g., \citealt{Planelles:2017} and refs.~within) agree that $Y_{\rm x,500}$ is a good proxy for the total mass, being relatively insensitive to feedback processes (e.g., the diffuse atmosphere is heated while being evacuated at the same time).

Figure \ref{MbhYx} shows the $\mbh - Y_{\rm x}$ relation within our three X-ray radii. 
The Compton parameter correlation is able to reduce the scatter compared with the linked (punctual) $\mbh - P_{\rm x}$: the galactic scaling indeed reduces the scatter down by 30\%, which is near $\epsilon\simeq0.30$ across all regions.
This is analogous to that of the gas mass scalings, except for the core region.
The correlation coefficient remains in the very strong regime (${\rm corr }\approx 0.9$). The slope is shallow, down to a value of $\sim$1/2; indeed, thermal energy covers a wide range of values from small spirals to massive BCGs, $E_{\rm th,c}\approx 10^{56}$\,-\,$10^{62}$\,erg. 

Interestingly, an SMBH of $10^9\,\msun$ has an available rest-mass feedback energy of $E_\bullet = \eta\,M_\bullet c^2 \sim 2\times10^{60}$\,erg (using a median mechanical efficiency $\eta\sim10^{-3}$; \citealt{Gaspari:2017_uni}) 
which can potentially unbind the galactic/core region if released in short time. 
However, such ejective (quasar-like) feedback matching the gas gravitational binding ($\sim$\,$M_{\rm gas}\phi \simeq 2 E_{\rm th}$) energy would induce
$\mbh \sim {2\,Y_{\rm x}}/(\eta c^2) \propto M_{\rm gas} \sige^2$ 
(the latter via $\sigma_{\rm e}^2 \propto \tx$; \S\ref{s:optuni}), 
leading to much steeper scaling than that found in Fig.~\ref{MbhYx} or \ref{Mbhop}.
Thereby a gentler AGN feeding/feedback self-regulation and gradual deposition is required (\S\ref{s:disc}; see also \citealt{Gaspari:2014_scalings}).

Overall, the stability and tightness across largely different radial regions (varying each by one order of magnitude; \S\ref{s:xvar}) corroborates the importance of using $Y_{\rm x}$ over other thermodynamic observables (such as pressure or entropy).
Such strong and tight correlations with the Compton parameter, particularly for the large-scale \R500\ region, imply that we can use the thermal SZ signal from hot halos to probe or trace SMBHs. 
This novel approach presents several advantages over the X-ray counterpart, since we can fully leverage the new ground-based radio facilities (instead of the more expensive X-ray space telescopes), which have recently entered a golden age (e.g., ALMA, MUSTANG-2, NIKA-2, SPT).

\subsubsection{Total mass (dark matter, gas, stars)} \label{s:mtot}

\begin{figure}[!ht]
 \vskip -0.05cm
\subfigure{\includegraphics[width=0.945\columnwidth]{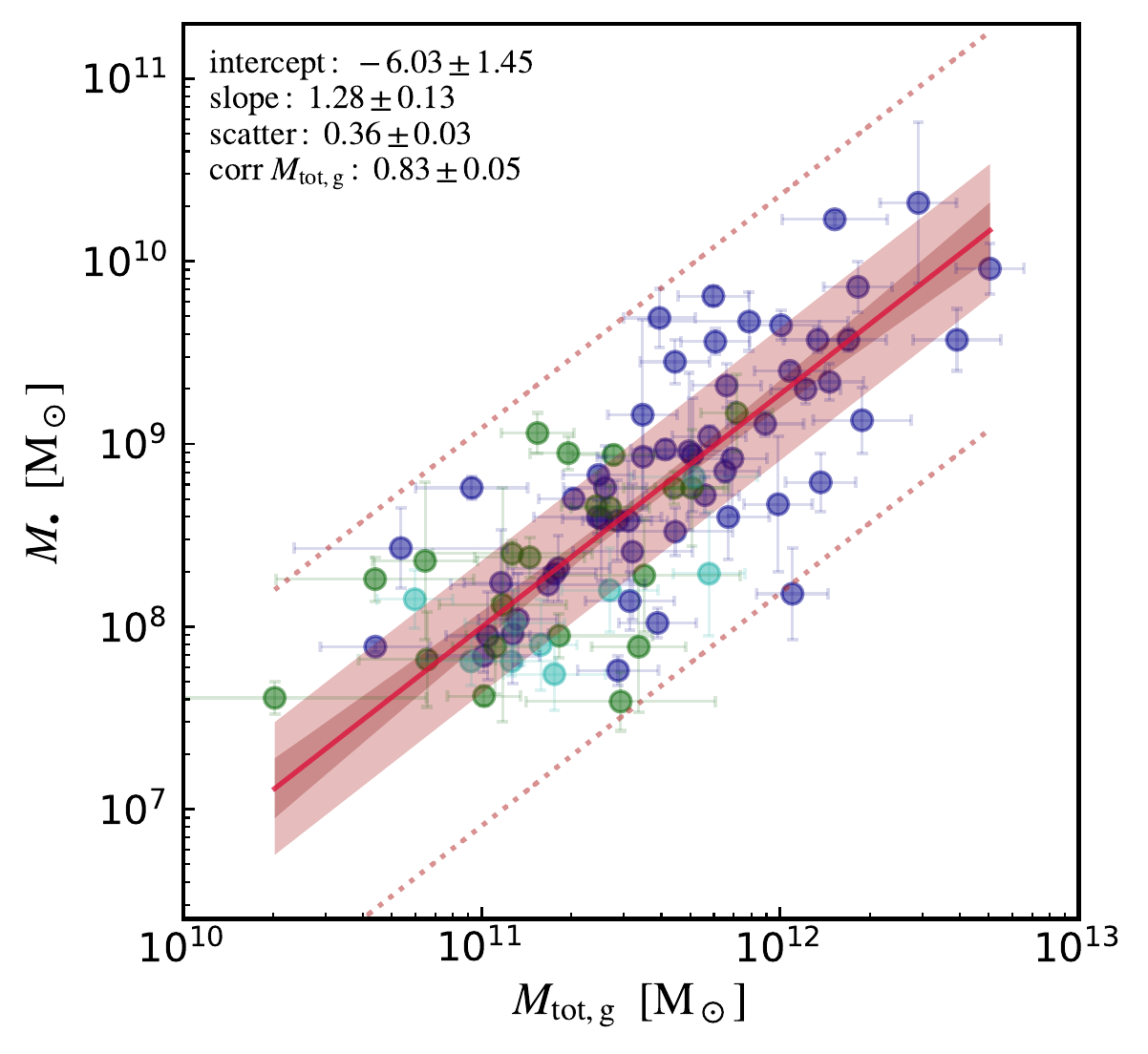}}
 \vskip -0.39cm
\subfigure{\includegraphics[width=0.91\columnwidth]{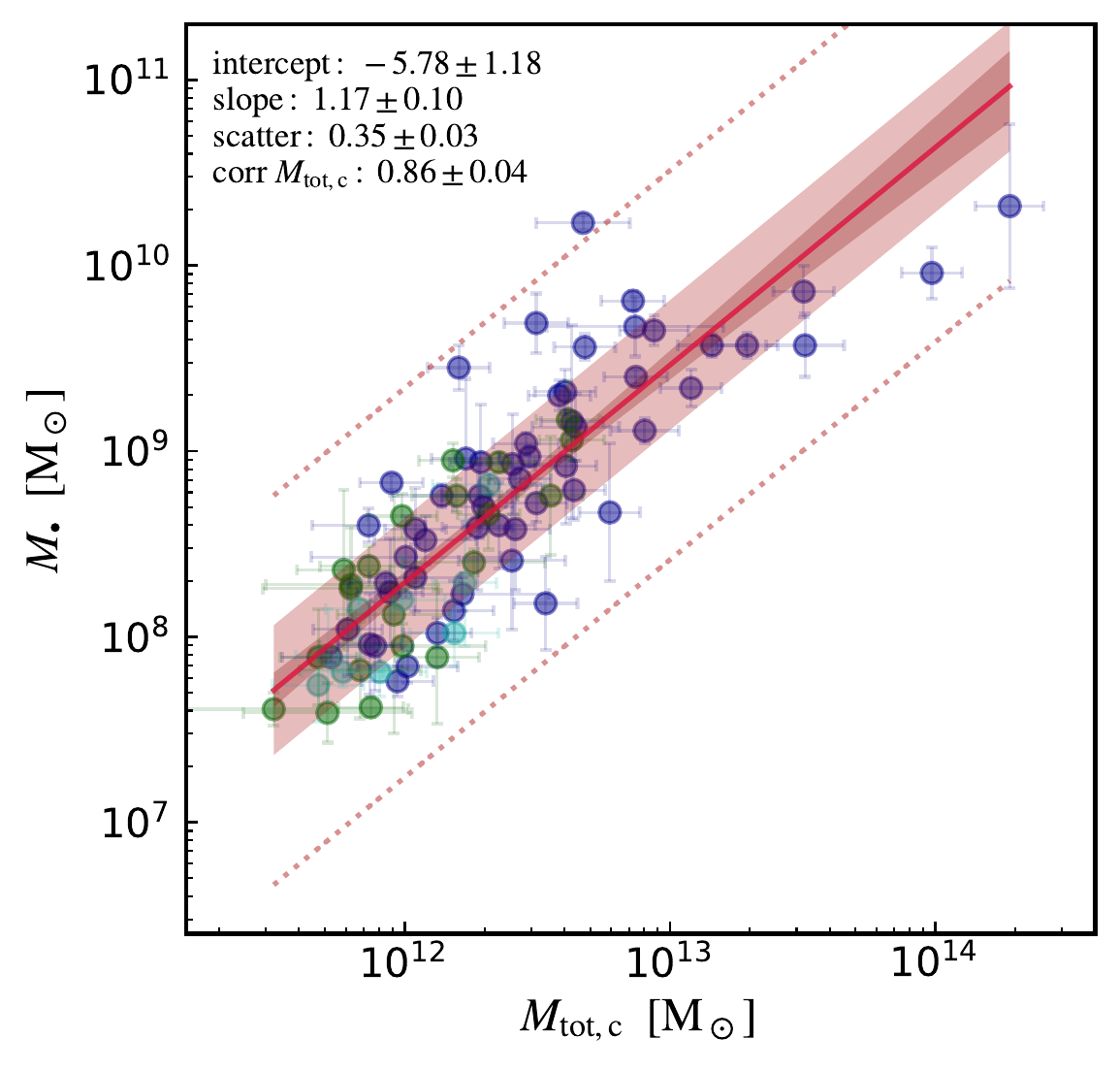}}
 \vskip -0.39cm
\subfigure{\includegraphics[width=0.91\columnwidth]{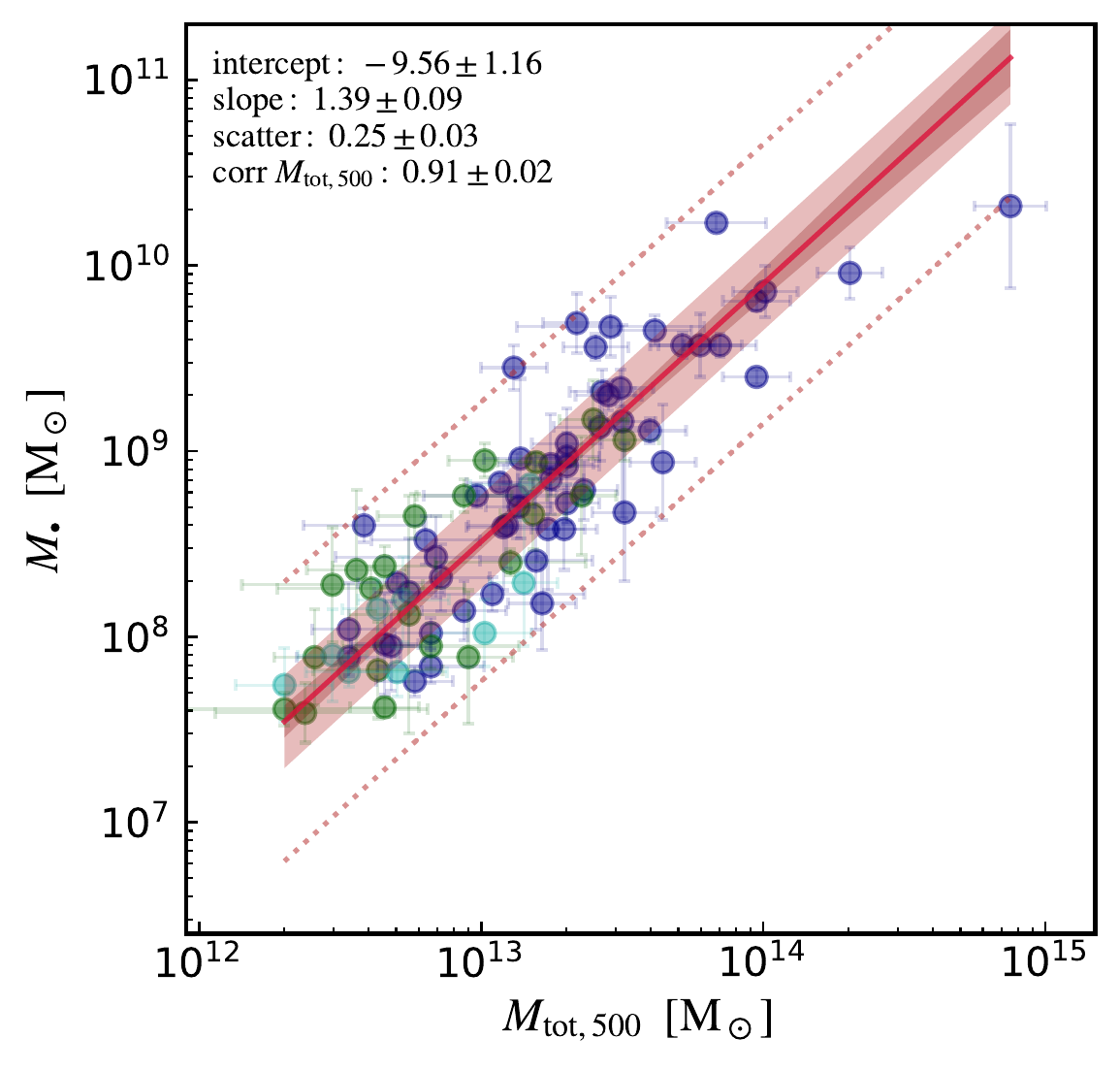}}
 \vskip -0.4cm
\caption{Black hole mass vs.~total mass (dark matter, gas, stars; \S\ref{s:mtot}) within \Rxg\ (top), \Rxc\ (middle), and $R_{500}$ (bottom). 
Besides $M_{\rm tot, 500}$ (that is a simple reflection of $T_{\rm x, 500}$), the total mass scalings still show a significant correlation with $\mbh$, albeit less tight than most gas scalings, in particular $M_{\rm gas}$. This suggests that gaseous halos may play a more central role than DM halos in the growth of SMBHs.
}
\vskip -0.5cm 
\label{MbhMtot}
\end{figure}

We analyze now the total (gravitational) mass, which is the sum of the baryonic (gas and stars) and dark matter (DM) component. The latter dominates ($>$\,90\%) the total matter content, particularly in the core and outskirt regions of the group/cluster. Studies show that the dark matter distribution can be well described by a NFW profile (\citealt{Navarro:1996}) in most galaxies and clusters (e.g., \citealt{Humphrey:2006b,Ettori:2019}), which is shaped by one minimal parameter. 
While the global mass is given via \R500\ (\S\ref{s:xvar}), $M_{\rm tot, 500}=(4/3)\pi R_{500}^3\,500 \rho_{\rm c}$, 
the dark matter mass enclosed within smaller radii can be retrieved via the NFW profile
\begin{equation}\label{e:Mnfw}
M_{\rm DM}(< r) = 4\pi\rho_{\rm s}\,R^3_{\rm s}\left[\ln{\left(\frac{R_{\rm s}+r}{R_{\rm s}}\right)} - \frac{r}{R_{\rm s}+r} \right], 
\end{equation}
where the scale radius is $R_{\rm s}\equiv R_{500}/c_{500}$ and the concentration parameter is given by observations of galaxies and groups (\citealt{Sun:2009a}),
$c_{500} \simeq 5.0 \left({M_{\rm tot, 500}}/{10^{13}\,\msun}\right)^{-0.09}$ (with 0.1\;dex scatter).
The characteristic DM density is defined as 
\begin{equation}\label{e:rhos}
\rho_{\rm s} = \frac{500\rho_{\rm c}}{3} \frac{c_{500}^3}{\ln(1+c_{500})-c_{500}/(1+c_{500})}.
\end{equation}
As secondary components of the total matter content, we add the enclosed gas mass (\S\ref{s:Mgas}) and galaxy stellar mass (\S\ref{s:optuni})\footnote{Albeit having minor role, we adopt a stellar (\citealt{Hernquist:1990}) profile $M_\ast(<r)=M_{\ast,\rm tot}\,[r^2/(r+a)^2]$, where $a=R_{\rm e}/(1+\sqrt{2})$.} to the above DM mass. 
The retrieved total masses within \R500\ span a range of $M_{\rm tot,500}\sim10^{12}$\,-\,$10^{15}\,\msun$ from isolated galaxies to massive clusters (consistently with \citealt{Forbes:2016} and \citealt{Lovisari:2015}, respectively), and decrease by one/two orders of magnitude in the core/galactic regions. 
In agreement with analogous samples (\citealt{Babyk:2018}), most of the objects have $M_{\rm tot,c}\sim2\times10^{11}$\,-\,$2\times10^{13}\,\msun$ within $r \lta $\,0.15\,\R500 (see \citealt{Humphrey:2009} for comparable $M_{\rm tot,g}$).
Moreover, we retrieve a $M_{\rm tot,c} \propto Y_{\rm x,c}^{0.40\pm 0.03}$ scaling which is consistent with that found by \citet{Babyk:2018} with slope ${0.38\pm 0.05}$.\footnote{We note that as our sample extends down to isolated/low-mass galaxies, our scaling relations show a larger departure from self-similarity than those solely including massive ETGs and clusters (e.g., \citealt{Vikhlinin:2009}; see also Appendix~\ref{a:uextra}).}  
Additional permutations of the halo properties can be retrieved via the total mass-to-light ratios shown in Fig.~\ref{MtotdLx}.

\begin{figure}[!ht]
 \vskip -0.05cm
\subfigure{\includegraphics[width=0.91\columnwidth]{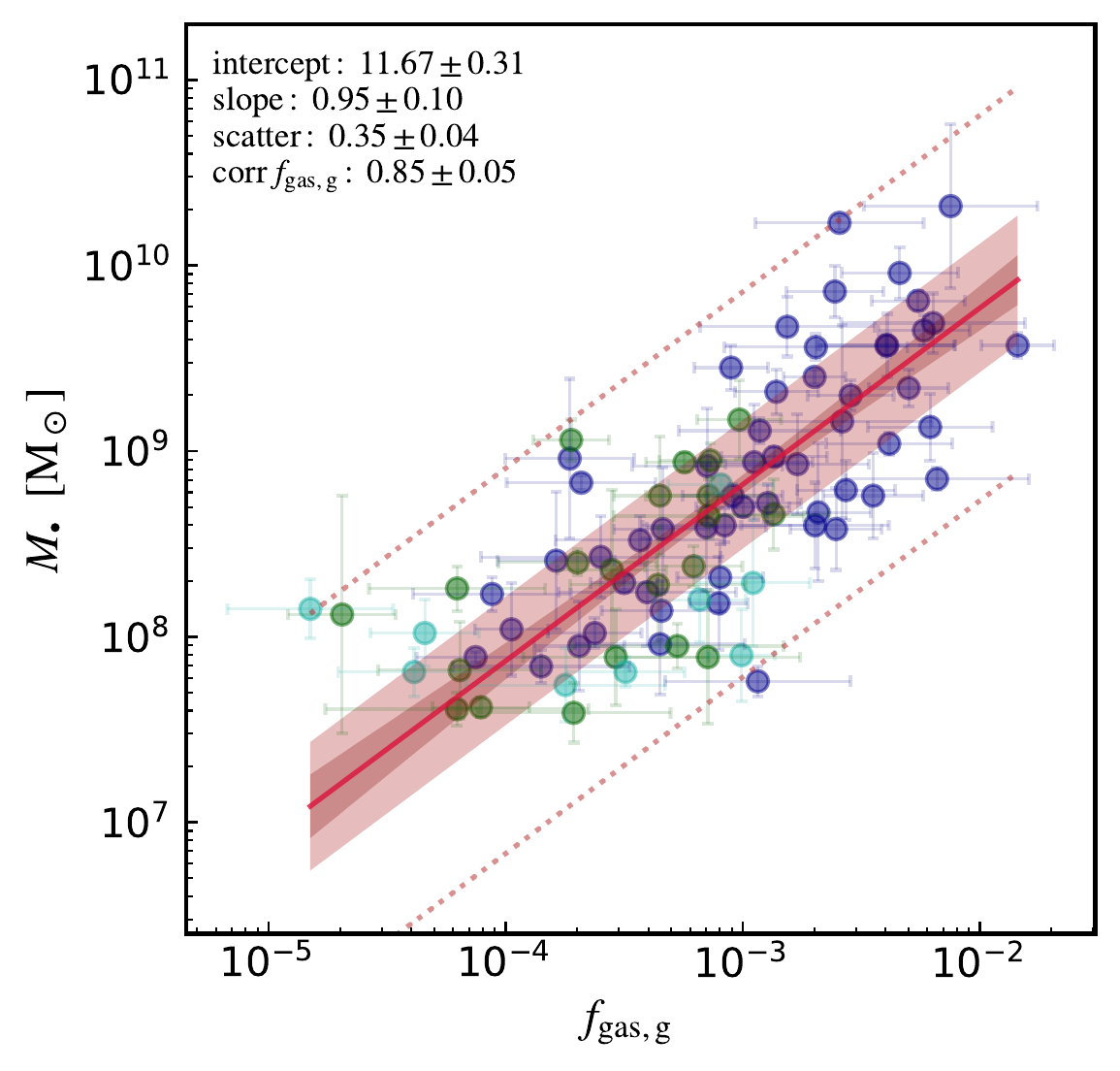}}
 \vskip -0.39cm
\subfigure{\includegraphics[width=0.91\columnwidth]{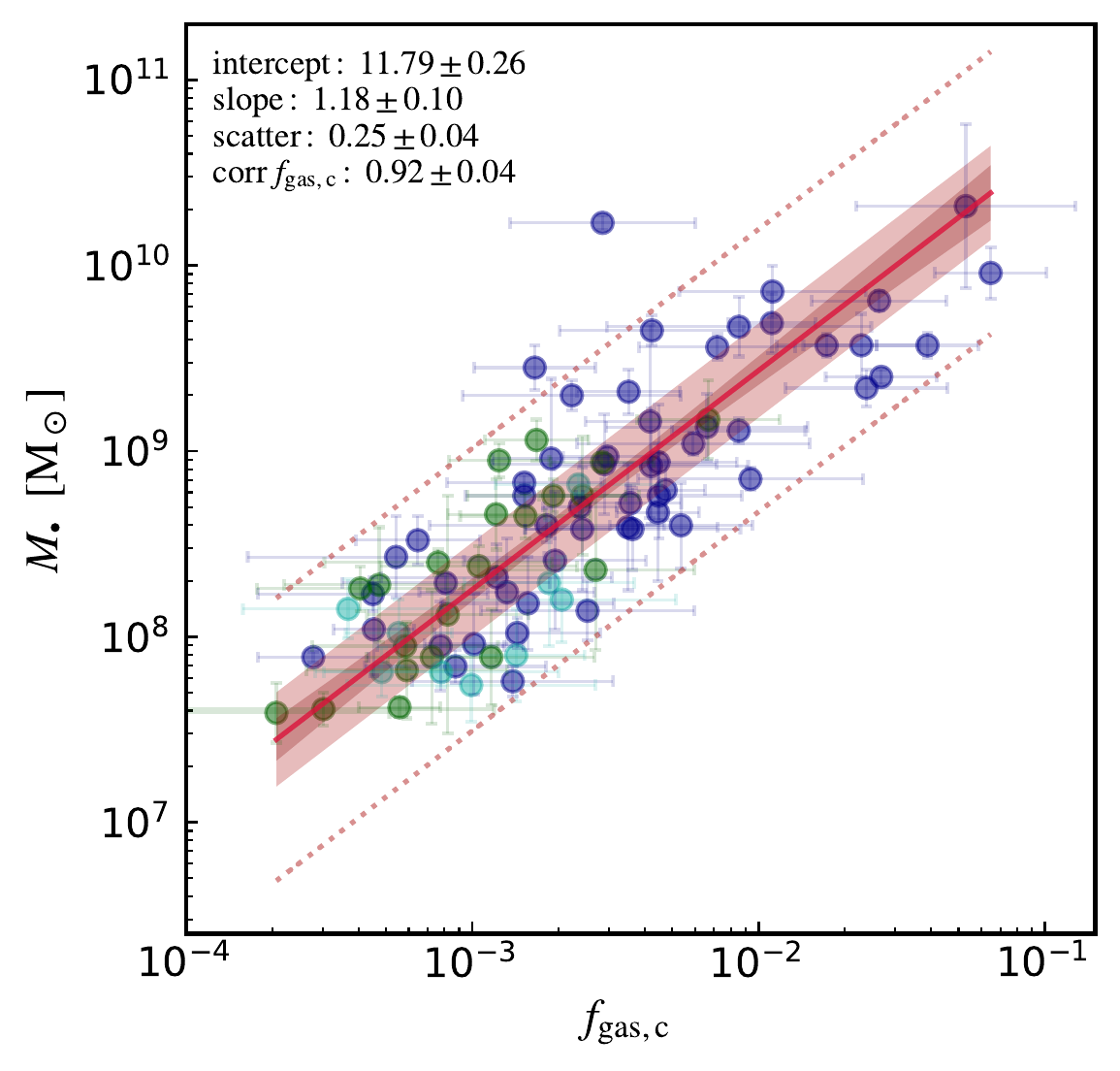}}
 \vskip -0.39cm
\subfigure{\includegraphics[width=0.91\columnwidth]{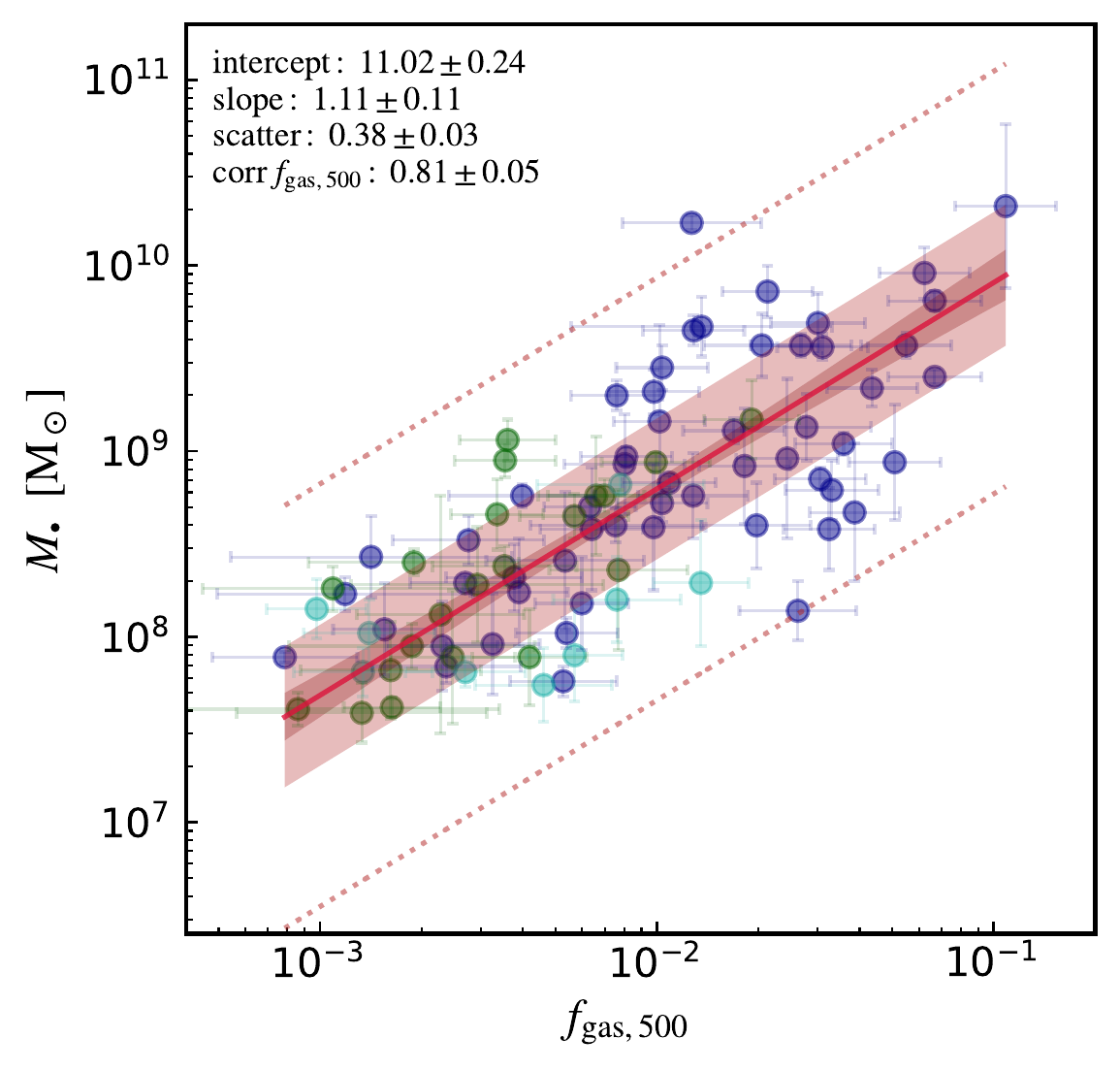}}
 \vskip -0.4cm
\caption{Black hole mass vs.~gas fraction within \Rxg\ (top), \Rxc\ (middle), and $R_{500}$ (bottom). 
A strong correlation is again found in the core, while the extremal regions show substantial intrinsic scatter. The inclusion of total mass seems to weaken the stronger relation with gas mass. Overall, SMBHs seem to grow faster in halos which host relatively larger amount of diffuse plasma.
}
 \vskip -0.5cm 
\label{Mbhfgas}
\end{figure}

Figure \ref{MbhMtot} shows the $\mbh - M_{\rm tot}$ correlations retrieved via our customary Bayesian analysis (\S\ref{s:corr}), together with the 1- to 3-$\sigma$ scatter bands.
As expected, $M_{\rm tot, 500}$ is purely a reflection of the X-ray temperature via $R_{500}\propto T_{\rm x,500}^{1/2}$ (\S\ref{s:xvar}), thus preserving the small scatter of $\mbh - \tx$.
On the other hand, the correlation between $\mbh$ versus total mass within the core and galactic scale shows a significant intrinsic scatter, which is comparable to that of the $\mbh - \sigma_{\rm e}$ and larger than that of most gas scalings. 
In particular, compared with the gas mass relation (\S\ref{s:Mgas}), adding the DM component does not improve the mean corr coefficient and induces it to drop to a lower level. 
Under our assumptions, these results suggest that the plasma halos, and related baryonic properties, may play a more central role than the sole gravitational/DM potential in growing SMBHs.
A positive correlation with $M_{\rm tot}$ is nevertheless established because hotter plasma halos are created in larger potential wells, as they get shock heated during the primordial halo formation. While a correlation cannot probe causation, we devote \S\ref{s:disc} to testing the BH mass growth via either gas accretion or mergers (which purely increase the gravitational potential), finding that the latter channel is sub-dominant over most of the cosmic time (\S\ref{s:mergers}). 

DM halos still represent a reasonable, useful proxy to predict the central SMBH mass. The correlation slopes are mildly superlinear, $\beta\simeq1.2$\,-\,1.4 (with outliers becoming more frequent at the high-mass end). Such simple total mass scalings can be used by large-scale cosmological simulations (including both LTGs and ETGs) and semi-analytic models (SAMs) to either test their results or calibrate the subgrid parameters on the $\mbh - M_{\rm tot}$ (instead of the more complex stellar scalings; \S\ref{s:subgrid}).
Another potential application is the inclusion of the AGN feedback power modeled directly from the DM mass; the latter is one of the best resolved and convergent properties in cosmological simulations (\citealt{Sembolini:2016}).

Interestingly, equating the total BH feedback energy $E_\bullet=\eta M_\bullet c^2$ (\S\ref{s:Yx}) to the DM gravitational binding energy $\sim$\,$M_{\rm tot}\phi \simeq M_{\rm tot} [3\,k_{\rm b}T_{\rm x}/(\mu m_{\rm p})]$ ameliorates the above superlinear scaling into a quasi-linear mass scaling with $\beta\sim0.8$\,-\,0.9 (outskirt to galactic scale; not shown). The retrieved BH mass (normalization) is similar to the observed one at the galactic scale, but is overestimated for the outer regions. 
In other words, $\mbh \sim M_{\rm tot,g} \phi /(\eta c^2)$ is a better proxy for BH mass than purely $M_{\rm tot,g}$.

\subsubsection{Gas fraction}  \label{s:fgas}
Now that we have $M_{\rm tot}$, it is possible to analyze the gas fraction, which is the ratio between the gas mass and total mass, within the three enclosed radii, $f_{\rm gas}=M_{\rm gas}/M_{\rm tot}$.
Figure \ref{Mbhfgas} shows that the inclusion of total mass weakens the likely more fundamental correlation with $M_{\rm gas}$.
The core region (where radiative cooling plays a key role) still preserves a low intrinsic scatter ($\approx$\,$0.25$), with the two enclosing regions showing 35\% larger $\epsilon$. We note that the propagated errors have now become substantial, given that $f_{\rm gas}$ is a highly composite variable. 
Remarkably, the slope is consistent with unity (considering 1-$\sigma$ uncertainty), which allows for a straightforward linear-space conversion between gas fractions and BH masses.

Overall, it is evident that larger BHs prefer to grow in halos which host larger amount -- both in absolute and relative sense -- of diffuse gas, despite this being only a small fraction of the whole matter budget. Indeed, the retrieved core gas fraction for galaxies/groups hosting a few $10^9\,\msun$ BHs is a few percent (consistent with \citealt{Sun:2009a,Babyk:2018}). Such fractions tend to approach the cosmic baryon fraction ($\sim$\,0.1) when we consider the $R_{\rm 500}$ of the cluster halo and the most massive BCGs (e.g., NGC\,4889, NGC\,3842; see also \citealt{Eckert:2019}). In this regime, feedback processes cannot easily evacuate the gas mass due to the large binding energy (\citealt{Gaspari:2014_scalings}). Conversely, the inner galactic/CGM region (strongly affected by AGN and stellar feedback) show values below the percent level (e.g., as found by \citealt{Humphrey:2008,Humphrey:2009}), in particular for isolated galaxies, thus requiring deeper and more challenging observations.

Moving forward, it will be crucial to obtain observations via X-ray telescopes with significantly improved sensitivity and resolution, for both imaging and spectroscopy, to test the faintest hot halos at the low-mass end and thus extending the sample to more late-type objects.
A series of dedicated X-ray missions with such characteristics will operate in the upcoming decade, such as 
eROSITA, XRISM, {\it Athena}, and possibly AXIS and {\it Lynx} (see \S\ref{s:cav} for more details on the future developments).
\\

\subsection{Multivariate correlations} \label{s:multi}

A further key investigation angle that is worth dissecting is the Bayesian multivariate correlation analysis (\S\ref{s:corr}) of the fundamental X-ray/optical variables. We limit the analysis here to a three-dimensional (3D) space, i.e., a correlation plane (with thickness given by $\epsilon$) with some inclination and position angle. In principle, higher-dimensional hyperplanes can be explored; however, the free parameters also increase substantially, thus diminishing the physical and predicting value of the fitting. This is also why the univariate correlations on the composite variables are in general preferred, given that the composite variables are set by physical intuition, rather than by a statistical random parameter search. 

While the Bayesian prior/posterior procedure and MCMC analysis is essentially identical, an important difference with the univariate fitting is that, with {\tt mlinmix} (Eq.~\ref{e:mlinmix}), we are carrying out a dual {\it partial} (conditional) correlation analysis, implying that we will retrieve two partial correlation coefficients (related to $X_1$ via the control variable $X_2$, and vice versa).
Before dissecting the multivariate relations with BH mass, it is essential to first analyze the `fundamental' planes in the optical and X-ray bands, in order to understand the major differences between the stellar and hot halos, and in which kind of environment the SMBHs reside and grow. Such planes are also crucial probes for competing evolution models of galaxies, groups, and clusters of galaxies.\\

\subsubsection{Optical/stellar fundamental plane (oFP)} \label{s:oFP}

\begin{figure}[!ht]
\subfigure{\includegraphics[width=0.95\columnwidth]{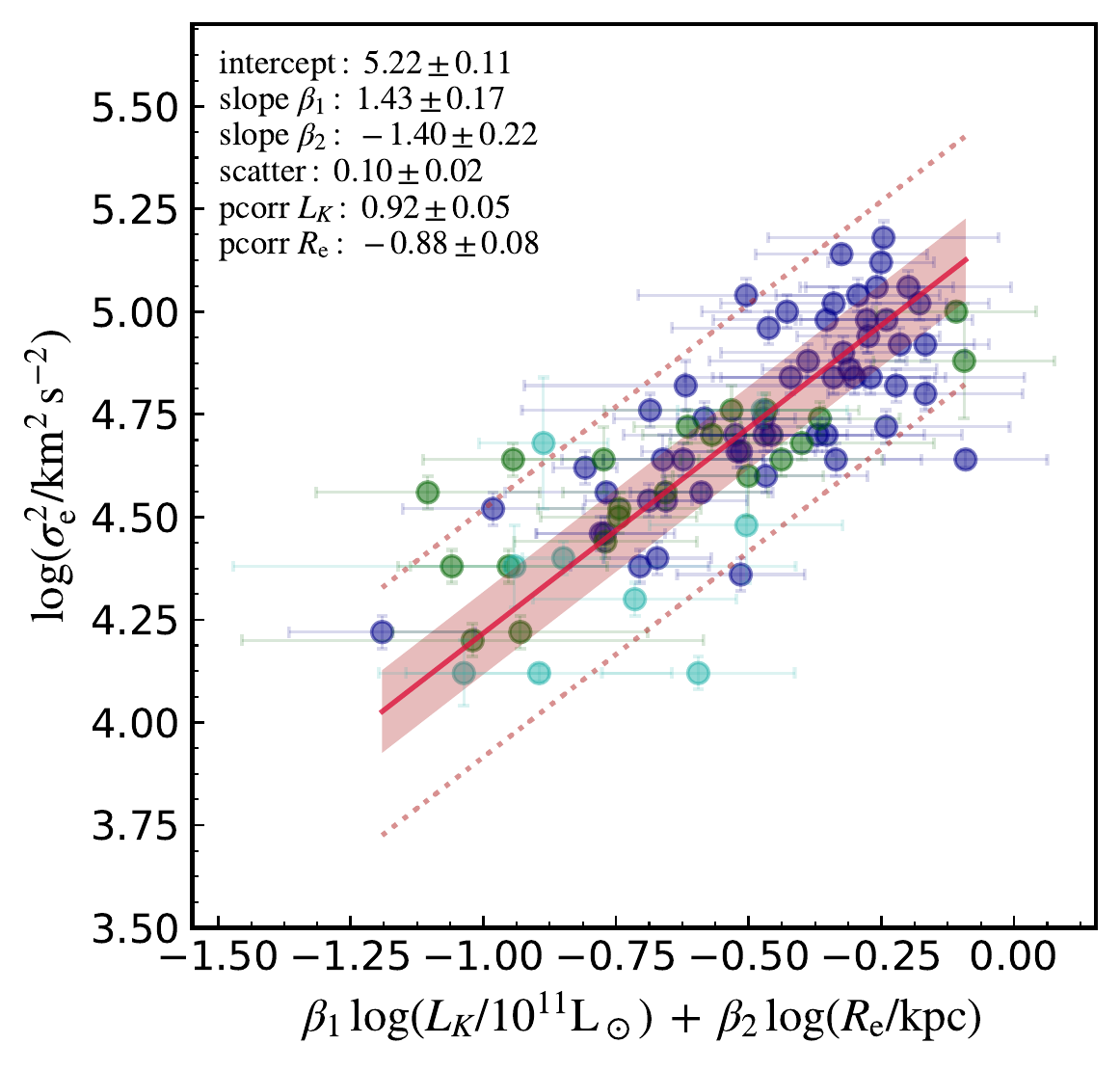}}
\caption{Optical/stellar fundamental plane (edge-on view): 
multivariate correlation between the stellar velocity variance, total $K$-band luminosity, and effective radius. 
The inset lists the mean and errors of all the posteriors from the {\tt mlinmix} analysis, including the intercept, slope, intrinsic scatter (1-$\sigma$ interval plotted as a filled red band, 3-$\sigma$ as dotted lines), and the two partial correlation coefficients. 
The points are color-coded as per morphological type (blue: E; green: S0; cyan: S).
The retrieved galactic oFP can be understood via the VT (Eq.~\ref{e:virstar}) plus quasi homology and a mild tilt due to $M_\ast/L_K$, as supported by the strong correlation between $\sige^2$ and $L_K/\re$ and the very tight scatter.
} 
\label{oFP}
\end{figure}

A key motivation to analyze multivariate correlations resides in the {\it virial theorem} (VT): a stationary system of particles bound by gravity is expected to have an average kinetic energy $\mathcal{T}$ directly related to the average gravitational potential energy $U$ such that
\begin{equation}\label{e:virial}
2\langle \mathcal{T} \rangle = -\langle U \rangle.
\end{equation}
For a virialized stellar system with $\sigma_{\rm e,tot}^2=3\sige^2$, we can write
\begin{equation}\label{e:virstar}
\sige^2 = \frac{\kappa}{3}\,\frac{GM_\ast(< \re)}{\re} \propto \frac{L_K}{\re},
\end{equation}
where $\kappa/3\approx0.1$\,-\,0.3\footnote{The normalization factor of the potential energy is tied to the detailed geometry of the particle spatial distribution; e.g., for a sphere with uniform density, $\kappa = 3/5$.} is a structural parameter and the second conversion mainly depends on the stellar mass-to-light ratio $M_\ast/L_K$ 
(note that within \Re\ the stellar mass dominates over the other mass components).
It is important to note that the optical observables are only proxies for the intrinsic VT properties; the existence of an optical fundamental plane (oFP) among galaxies requires also significant (structural and dynamical) homology and tight $M_\ast/L_K$ (e.g., \citealt{Ciotti:1997_VT} and refs.~within).

Figure \ref{oFP} shows the edge-on view of the best-fit plane correlating (in logarithmic space) the three key stellar observables $\sige^2 - L_K - \re$. Since the virial theorem is centrally important, in this section we will adopt the velocity variance instead of the velocity dispersion.
Notice that the plot abscissa implies a rotation about the $Y$ axis, given by the two non-zero slopes. As is customary, the error bars are obtained by propagating the single errors weighted by $\beta_1$ and $\beta_2$.
The top-left inset lists the mean and standard deviation of all the posterior distributions of the Bayesian {\tt mlinmix} analysis
(Eq.~\ref{e:mlinmix}).

Two are the key results. First, the measured multivariate optical properties are consistent with the VT prediction of a plane, as both $\beta_1$ and $\beta_2$ slopes have identical value but opposite signs. Second, the intrinsic scatter is very small ($\epsilon\simeq0.1$ dex). If we consider only the univariate correlations,
the scatter increases up to $3\times$ and the corr coefficient decreases to the weak regime (as for the size versus velocity variance; Fig.~\ref{sizegal}); some of these univariate scalings indeed represent highly inclined projections of the best-fit oFP.
In more detail, the multivariate result indicates that, over our whole sample, the virial relation $\sigma_{\rm e}^2 \propto L_K/\re$ 
holds tightly but with a mild tilt (1.4) in the slope, which departs from unity (below 2 standard deviations). The pcorr coefficients (0.9) further reflect the very strong positive/negative partial correlation related to $L_K$ and $\re$, respectively. On the other hand, propagated error bars can reach relatively large uncertainty for lower mass galaxies.

The thin oFP is a well-known property, in particular for ETGs (e.g., \citealt{Djorgovski:1987,Dressler:1987}). The tilt in the observed plane can mainly be attributed to the dependence of the stellar mass-to-light ratio on the velocity dispersion, $M_\ast/L_K \propto \sige^{0.3-0.4}$ (\citealt{Kormendy:2013}). 
Further minor variations are due to DM, non-homology, and projection effects (cf.~\citealt{vBosch:2016}). 
Applying our stellar mass-to-light conversion (Eq.~\ref{e:ML}), we find a univariate correlation $\sige^2 = (0.20\pm0.02)\,(GM_\ast/\re)^{1.16\pm0.11}$, which is close to Eq.~\ref{e:virstar} with $\kappa\sim3/5$.

Overall, the optical scalings can be filtered down to a single key variable, $\sige^2$, or its virial analog $L_K/\re$.
Using $M_{\rm bulge}$ instead of $L_K$ would show similar results (Fig.~\ref{MbuoFP}), although with significantly larger scatter due to the larger presence of disk-dominated galaxies at the low-mass end. 
In Appendix~\ref{a:uextra}, we include additional variants of the oFP, which may be of interest for other observational and theoretical studies.

\subsubsection{X-ray/plasma fundamental plane (xFP)} \label{s:xFP}

\begin{figure}[!ht]
 \vskip -0.1cm
\subfigure{\includegraphics[width=0.9\columnwidth]{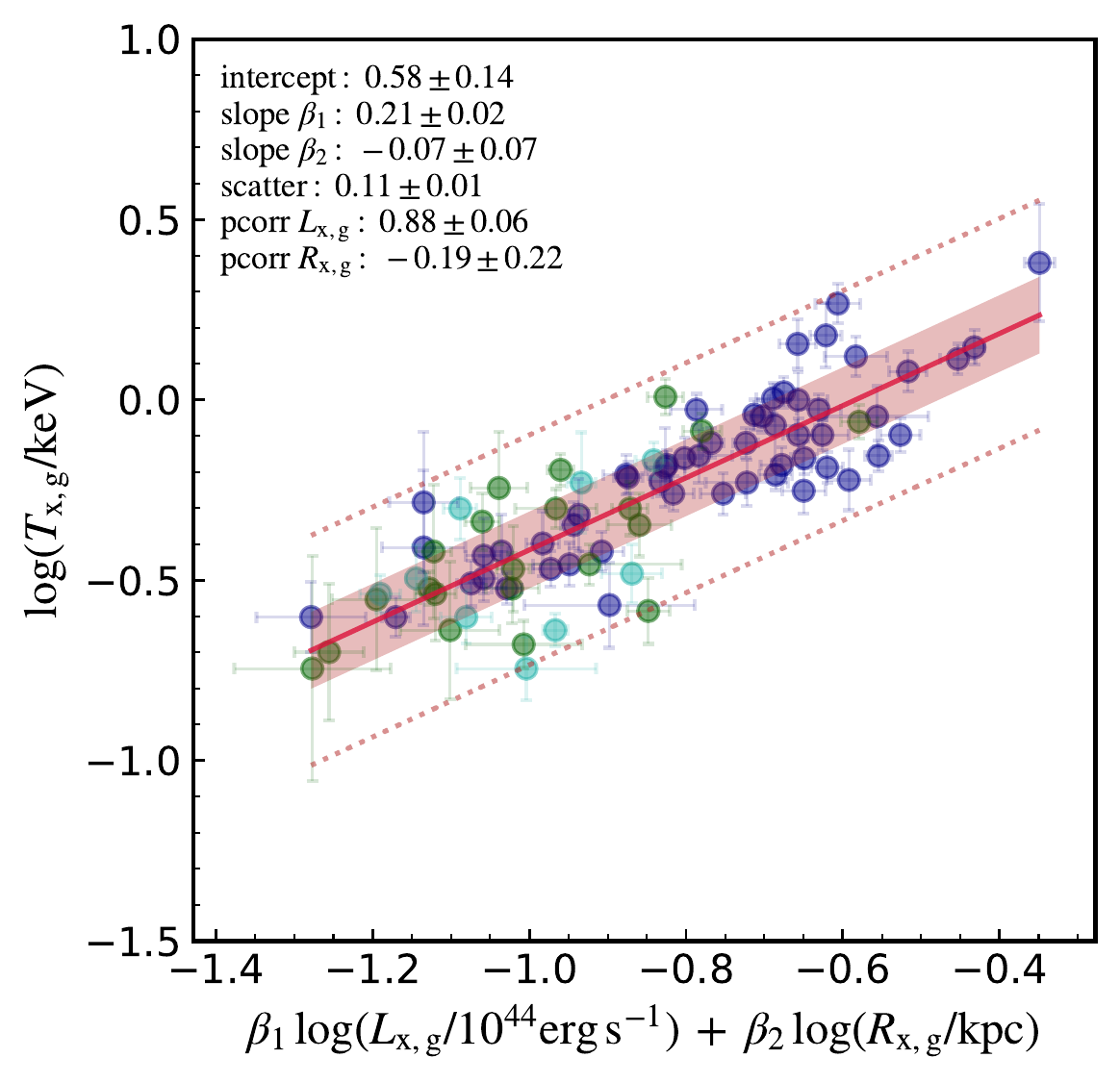}}
 \vskip -0.38cm
\subfigure{\includegraphics[width=0.9\columnwidth]{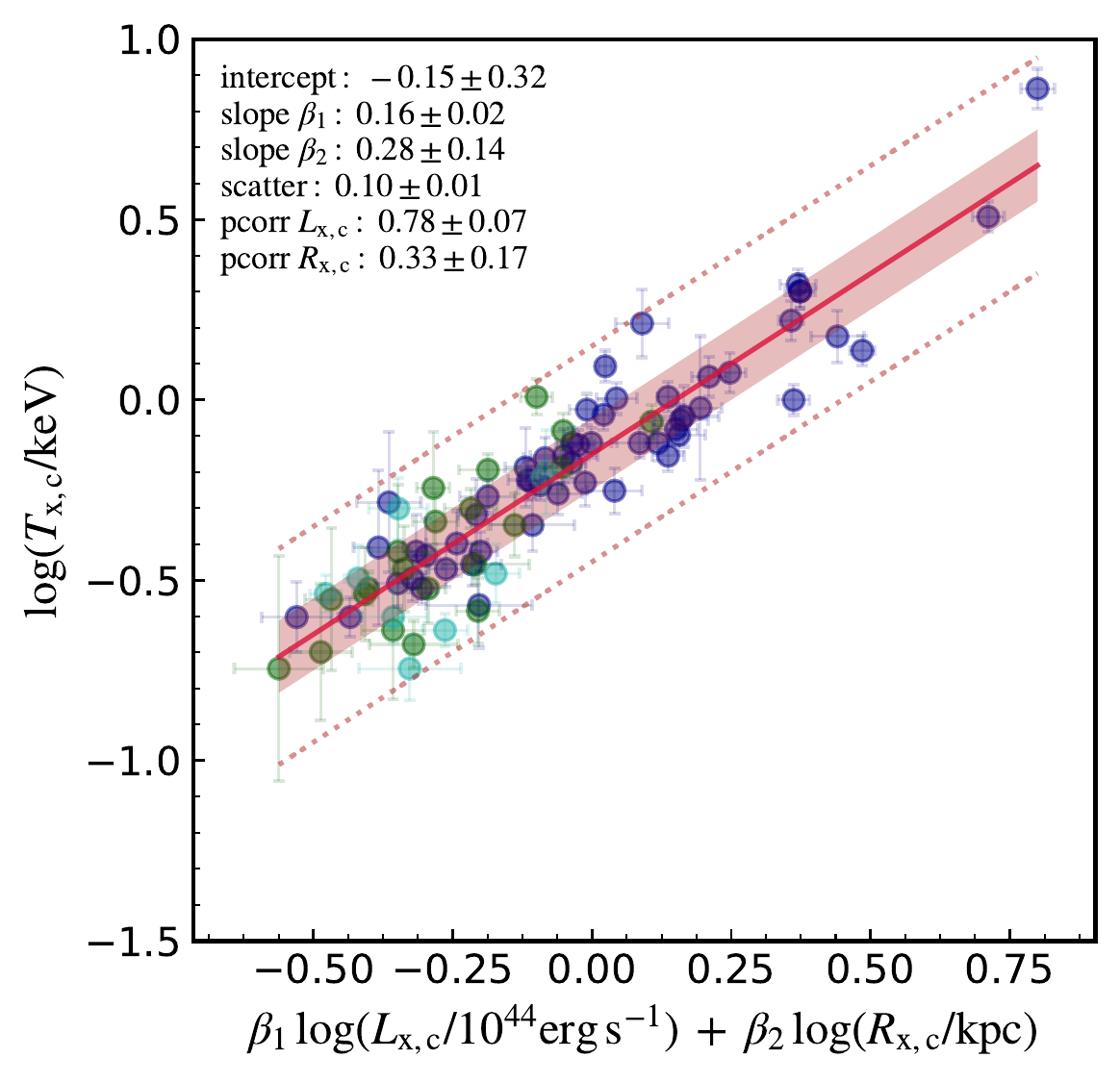}}
 \vskip -0.4cm
\subfigure{\includegraphics[width=0.90\columnwidth]{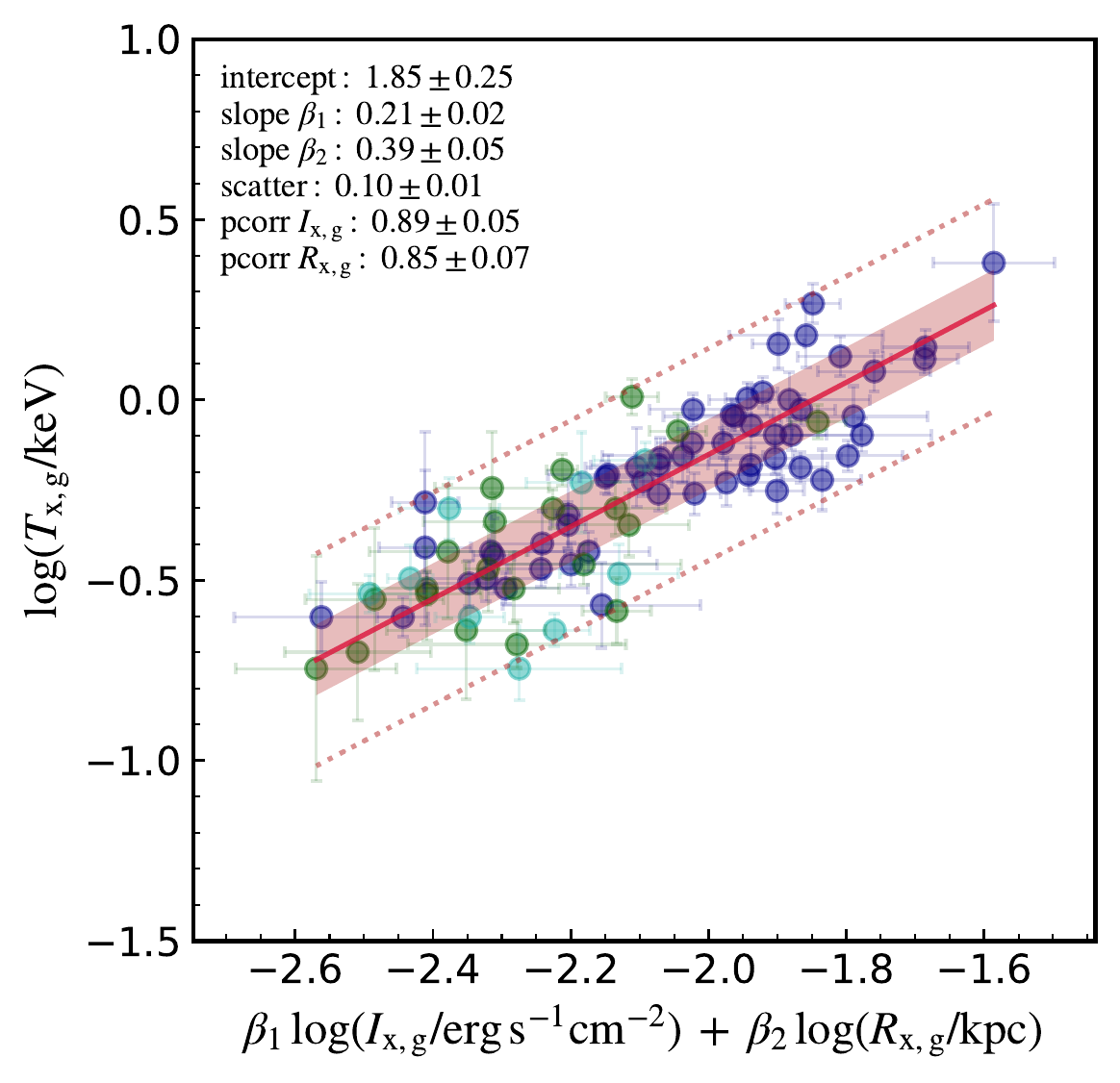}}
 \vskip -0.4cm
\caption{X-ray/plasma fundamental plane (edge-on view): 
multivariate correlation between the X-ray temperature, luminosity, and radius, for the galactic (top) and core region (middle). The bottom panel shows the galactic X-ray intensity scaling.
The $R_{500}$ scaling is not shown, since $R_{500}$ is redundant with $T_{\rm x,c}$.
The xFP substantially reduces the intrinsic scatter, but
shows a dominant correlation with $L_{\rm x}$, departing from the VT expectation. Unlike the oFP, the xFP is shaped by thermo- and hydrodynamical collisional processes 
rather than experiencing a pure virialization. 
} 
 \vskip -1.4cm
\label{xFP}
\end{figure}

We now dissect the X-ray fundamental plane (xFP) of hot halos, testing if the virial relation manifests so evidently in the plasma atmosphere too. This plane should not be confused with the AGN X-ray fundamental plane 
(\citealt{Merloni:2003_FP}), which focuses on the {\it nuclear} X-ray luminosity of the central point source, instead of the ISM/IGrM/ICM $\lx$ and \Tx. 
The analog of the virial relation (Eq.~\ref{e:virstar}) for a thermal plasma can be retrieved by using the specific thermal energy for the average kinetic energy such that
\begin{equation}\label{e:virgas}
c^2_{\rm s,i} \equiv \frac{k_{\rm b}\tx}{\mu m_{\rm p}} = \frac{\kappa}{3}\,\frac{GM_{\rm tot}(< \rx)}{\rx}\propto \frac{\lx}{\rx},
\end{equation}
where the first term is the gas isothermal sound speed. Since X-ray halos cover more extended regions than the stellar $\re$ (Fig.~\ref{sizegal}), the probed total mass is no longer dominated by the stellar component, but by the DM. Interestingly, the hydrostatic equilibrium equation (neglecting the non-thermal pressure support term) is
\begin{equation}\label{e:HSE}
\frac{GM_{\rm tot}(<\rx)}{\rx} = -c^2_{\rm s,i}\left(\frac{d\ln n_{\rm e}}{d\ln r} + \frac{d\ln\tx}{d\ln r}\right),
\end{equation}
which is akin to a virial relation with the normalization $\kappa$ given by the gas density and temperature log slopes\footnote{Typical large-scale gradients observed in groups/clusters give reasonable consistency between Eq.~\ref{e:virgas} and \ref{e:HSE} normalizations.}.

Figure \ref{xFP} shows the edge-on view of the $\tx - \lx - \rx$ best-fit plane. 
As for the optical properties, introducing a multivariate fitting substantially reduces the intrinsic scatter, showing a notable $\epsilon\approx0.1$ dex for both the galactic (top) and macro-scale core region (middle panel). 
At variance with the oFP,
the $\beta_{1,2}$ slopes are significantly different and 
much shallower than for a VT relation. 
The $\beta_2$ slope is also consistent with null (within 1-2 standard deviations), meaning that the multivariate correlation reduces to the simple $L_{\rm x}-T_{\rm x}$ relation. 
If we inspect pcorr, the conditional correlation is strongest for \Lx, for both the galactic and macro scale.
We note that this is different from the univariate, non-conditional analysis, which indicates that the characteristic radii positively correlate with \Lx\ or \Tx\ (corr $\approx$\;0.7\,-\,0.8, $\epsilon\approx0.2$ dex; e.g., Fig.~\ref{sizegal}).
Adopting the intensity $I_{\rm x} = L_{\rm x}/(4\pi \rx^2)$ better equilibrates the pcorr coefficients and lowers the scatter (bottom panel). 

The xFP deviates significantly from the simple virial expectation, $\lx \propto \tx\,\rx$ (Eq.~\ref{e:virgas}; also called `self-similarity' in cluster studies, modulo the cooling function $\lx \propto \tx\,\rx\,\Lambda$). While stars are strongly collisionless systems solely driven by gravitational effects, the plasma halos are complex systems shaped also by thermodynamical processes (e.g., radiative cooling and feedback heating) as well as hydrodynamical/collisional features (e.g., turbulence, shocks, Kelvin-Helmholtz and Rayleigh-Taylor instabilities). 
Indeed, on top of the virialization process within the DM halo, the hot halos continuously experience multiphase condensation and feedback heating (from both stars and AGN; e.g., \citealt{Gaspari:2014_scalings,Gaspari:2017_cca}), which evacuate and induce circulation throughout the macro atmosphere. The evacuation process is particularly important to reduce the density and thus the X-ray emission ($\propto n^2$) in less bound objects, ultimately leading to $\lx \propto \tx^{4.5}$ (Fig.~\ref{LxTx}). This observed steep scaling is consistent with other studies extending the luminosity\,--\,temperature relation down to low-mass and satellite galaxies (\citealt{Diehl:2005,Kim:2013_obs,Kim:2015,Goulding:2016,Babyk:2018}).

We can investigate in more detail the main reason for the difference between the oFP and xFP. Is the characteristic radius scaling the main culprit? If we analyze the univariate $\re - \sige^2$ (Fig.~\ref{sizegal}), the optical log slope is $0.6\pm0.1$, which is consistent with the X-ray slope of $\rxc - T_{\rm x,c}$ ($0.7\pm0.1$). The culprit is mainly the major difference between the optical and X-ray mass-to-light ratios.
The optical $M_\ast/L_K$ shows only very minor variations as a function of optical `temperature' ($T_\ast \propto \sige^2$), with a log slope $\approx$\,0.2 (Eq.~\ref{e:ML}). The observed\footnote{Assuming simple cluster self-similarity, the predicted X-ray mass-to-light ratio would be $M_{\rm tot}/L_{\rm x}\propto \tx^{-1/2}$.} X-ray counterpart instead shows a steep anticorrelation with X-ray temperature, $M_{\rm tot}/L_{\rm x} \propto T_{\rm x}^{-3}$
(Fig.~\ref{MtotdLx}). 
There are crucial differences between the X-ray and stellar emission. First, while $L_K$ is essentially the sum of many black-body spectra with a given stellar age and metallicity, \Lx\ is instead given by plasma collisional ionization processes ($\propto n_{\rm e}\,n_{\rm i}\,\Lambda(\tx, Z)$; \S\ref{s:ne}). In addition, the above-mentioned heating processes break self-similarity, introducing a steep dependency between $f_{\rm gas}$ and halo mass (which otherwise would remain constant; Fig.~\ref{Mbhfgas} and \citealt{Sun:2012}). 
In sum, the observed fundamental planes are a composition of more than three VT variables, including a mass-to-light ratio, such as
\begin{equation}\label{e:MLvirial}
\frac{L}{R}\propto \frac{T}{(M/L)}\propto T_\ast^{0.8}\ {\rm (stellar)}\ |\ T_{\rm x}^{4}\ {\rm (gas)},
\end{equation}
i.e., while galaxies with larger stellar velocity dispersion emit less optical light relative to mass, hotter plasma halos emit increasingly more X-ray photons.
Since the shallow X-ray radius scaling is swamped by the stronger $(M_{\rm x}/\lx)-\tx$ correlation, the observed xFP tends to closely approach the $\lx - \tx$ projection (Fig.~\ref{xFP}). 
Aggravating the difference is the several times larger intrinsic scatter of the X-ray (Fig.~\ref{MtotdLx}) versus optical ($\sim$\,0.1\,dex) mass-to-light ratio, which can be interpreted as a form of non-homology.

Recently, \citet{Fujita:2018} showed a variant of the xFP for 20 massive clusters ($M_{\rm tot,vir}\gta10^{15}\,\msun$; $T_{\rm x,vir} \gta 8$\,keV) by analyzing lensing masses. While their mass/temperature range is far beyond that of our sample, it is interesting to note that they also find a tight (0.05\,dex) xFP, involving $\tx - R_{\rm s} - M_{\rm s}$ (where $M_{\rm s}$ is the mass within the NFW scale radius $R_{\rm s}$; Eq.~\ref{e:Mnfw}). The significant thinness of the xFP is analogous to that in our Fig.~\ref{xFP}, albeit $2\times$ larger likely due to the inclusion of galactic X-ray halos. 
Their plane substantially deviates from the simple virial expectation too, although it is unfeasible to compare absolute values, given the different observables and more pronounced self-similarity break of low-mass systems via non-gravitational processes.

\subsubsection{Black hole mass versus oFP} \label{s:MbhoFP}

\begin{figure}[!ht]
 \vskip -0.05cm
\subfigure{\includegraphics[width=0.89\columnwidth]{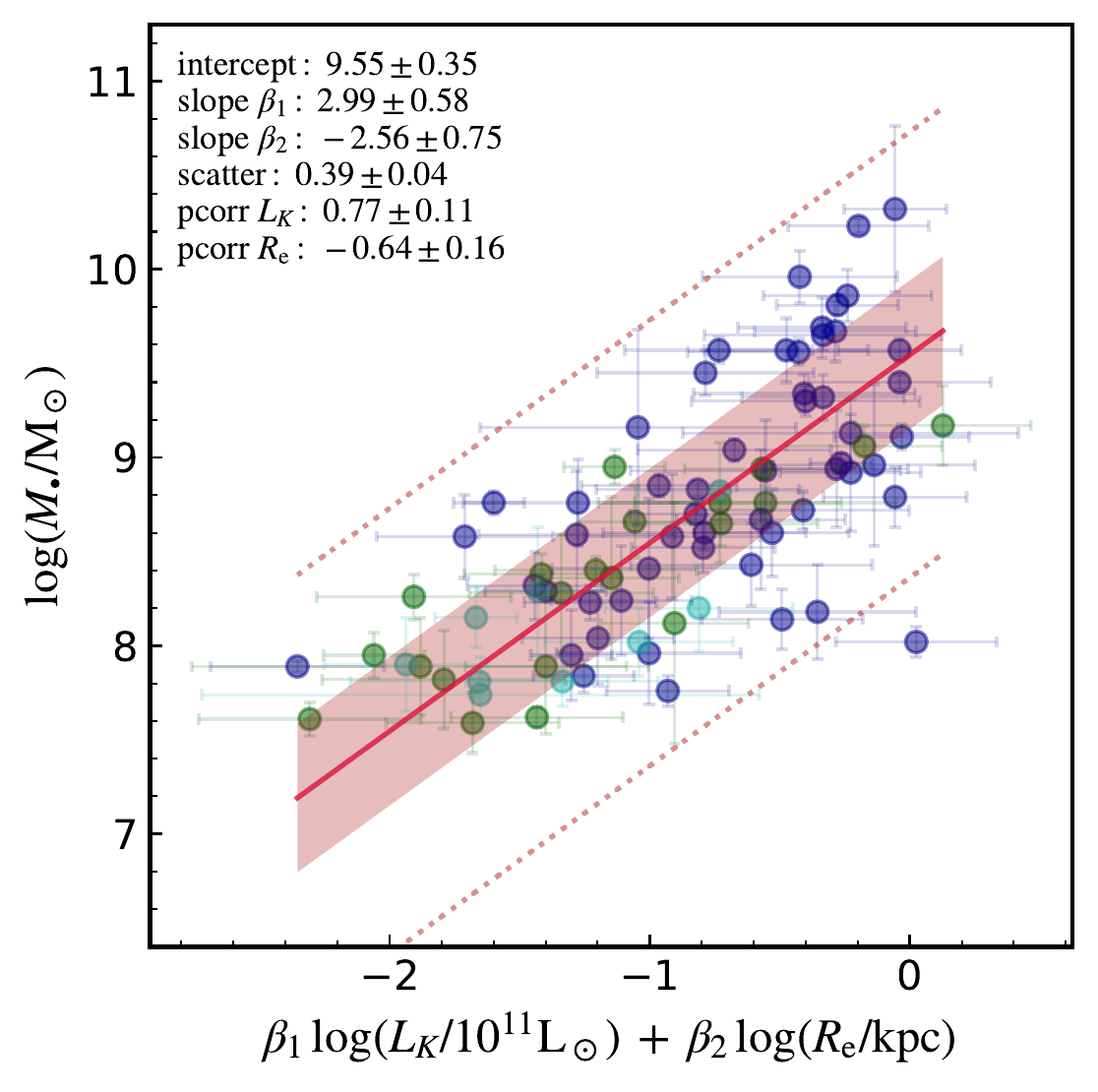}}
 \vskip -0.39cm
\subfigure{\includegraphics[width=0.89\columnwidth]{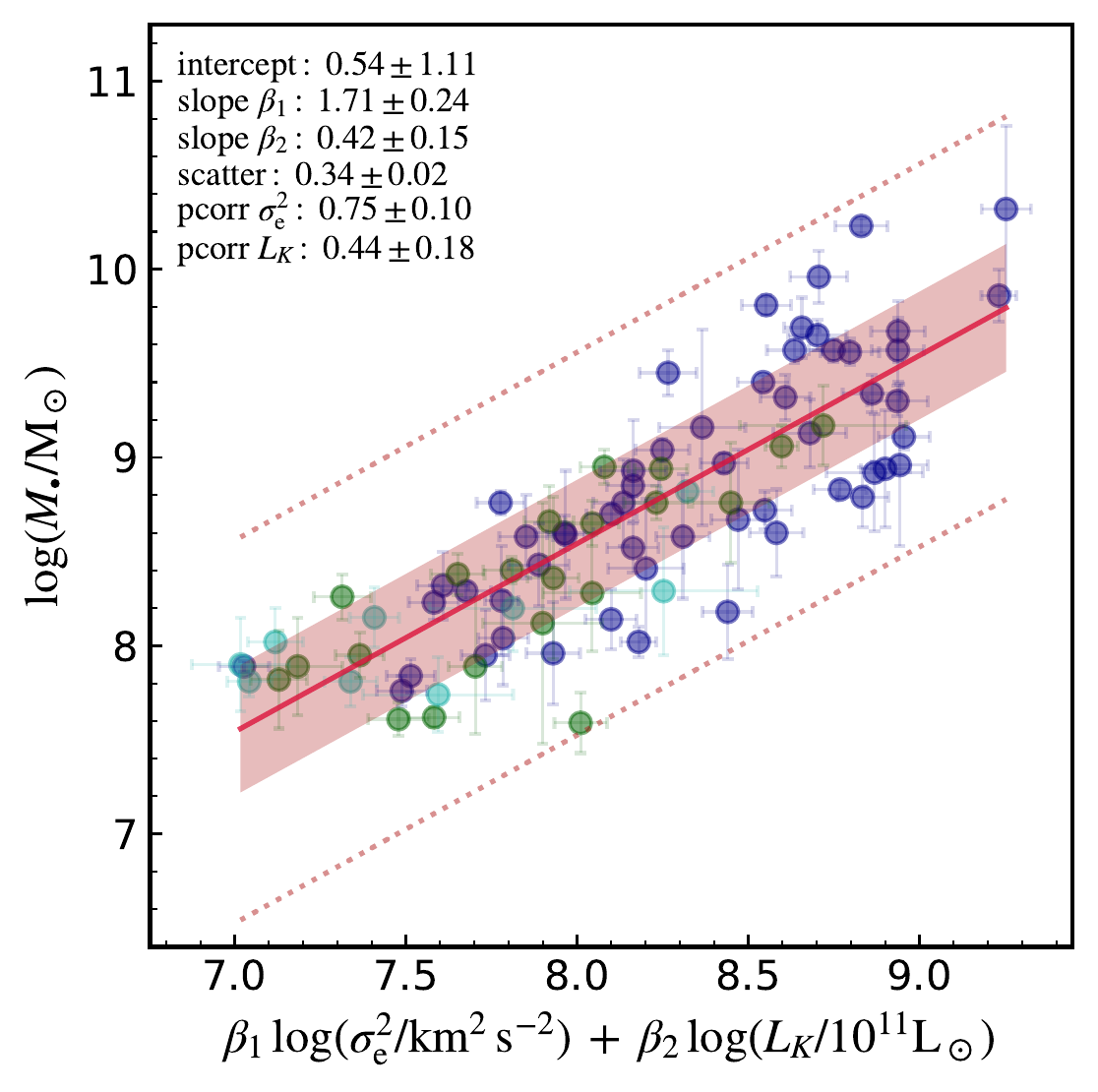}}
 \vskip -0.39cm
\subfigure{\includegraphics[width=0.9\columnwidth]{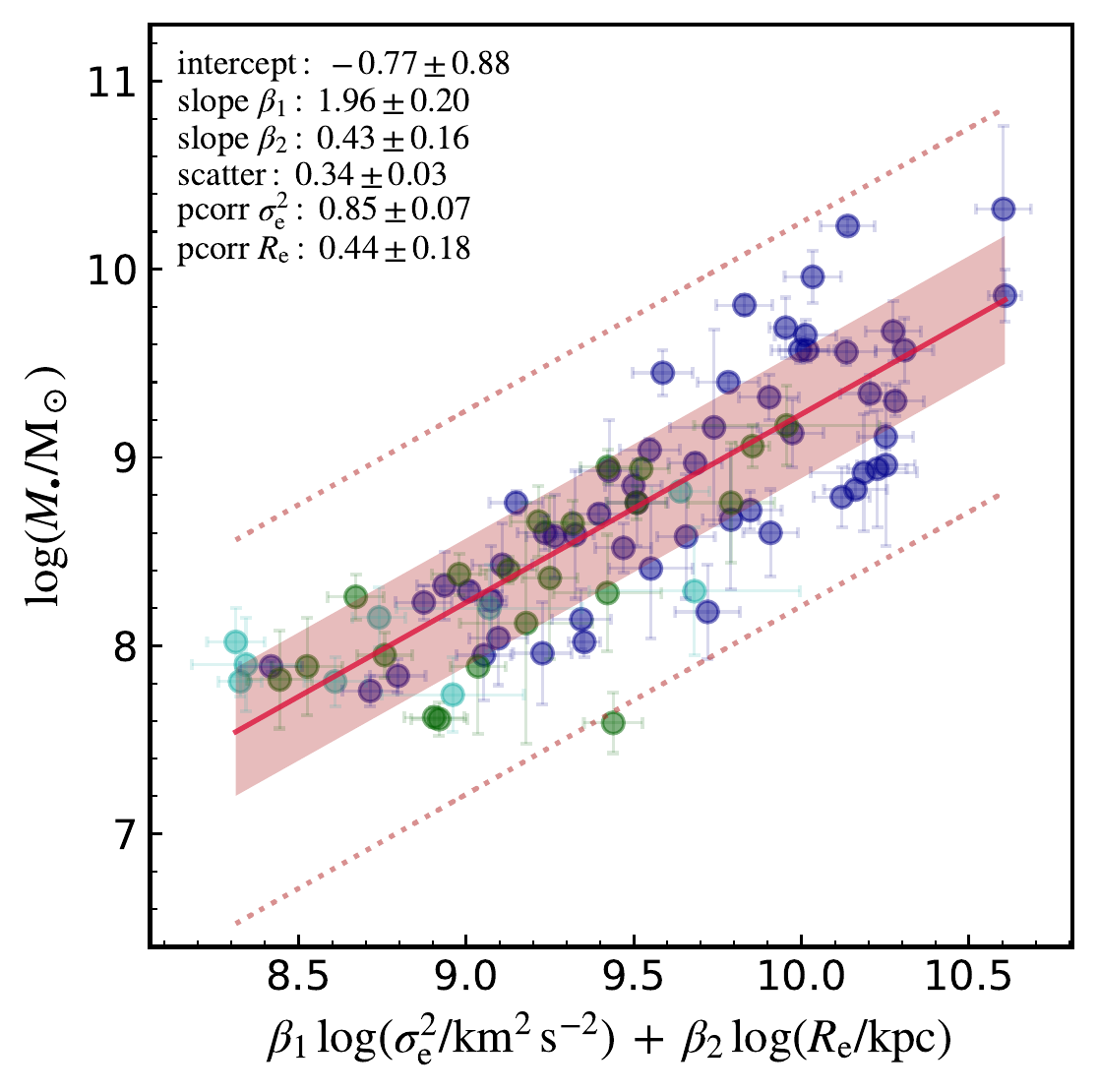}}
 \vskip -0.4cm
\caption{BH mass vs.~dual properties of the oFP (edge-on view), permutating the stellar luminosity, size, and velocity variance. Analog of Fig.~\ref{oFP}. 
The $\mbh - \sige$ can be directly converted into the $\mbh - L_K/\re$, given the tightness of the oFP, which follows very well the virial prediction. 
} 
 \vskip -0.4cm
\label{MbhoFP}
\end{figure}

We are now able to test the role of the SMBH mass in relation to the oFP and xFP.
Fig.~\ref{MbhoFP} shows $\mbh$ versus at least two of the oFP variables, with the usual posterior results of the Bayesian {\tt mlinmix} analysis. The top panel shows that the BH mass\,--\,luminosity\,--\,size in the optical band is essentially equivalent to the $\mbh - \sige$ relation (Fig.~\ref{Mbhop}; see also \citealt{Beifiori:2012,vBosch:2016}). Both have scatter consistent within 1-$\sigma$. As for the oFP (\S\ref{s:oFP}), the stellar luminosity/size shows pcorr strongly correlated/anti-correlated, with $\beta_{1,2}$ slopes being specular at a value of approximately $\pm 3$. 
Indeed using the oFP, $\mbh \propto (\sige^2)^{2.2} \propto [(L_K/\re)^{1.4}]^{2.2}\propto(L_K/\re)^3$, as retrieved here.
Overall, given the tight correlation between the stellar velocity variance and $L_K/\re$, we can on average convert from one to the other, making stellar `temperature' the unique fundamental variable for the optical component.

The middle and bottom panels show instead the multivariate correlations between $\mbh$ and the other two combinations of optical variables. Given the always higher pcorr (0.8 vs.~0.4) and steeper $\beta_1>\beta_2$, it is clear that the dominant variable is $\sige^2$. However, compared with the $L_K/\re$ correlation, the scatter is reduced slightly, even below that of the $\mbh - \sige$. The major improvement is in comparison with the univariate $\mbh - L_K$ (Fig.~\ref{Mbhop}), reducing its scatter by 30\%.
By using the bulge mass instead of $L_K$, similar results would apply (Fig.~\ref{MbhMbuop}).
Overall, this shows that the multivariate optical correlations can improve the scatter, although only by a mild amount, and yet not below the level of most X-ray correlations.

\subsubsection{Black hole mass versus xFP} \label{s:MbhxFP}

\begin{figure}[!ht]
 \vskip -0.05cm
\subfigure{\includegraphics[width=0.89\columnwidth]{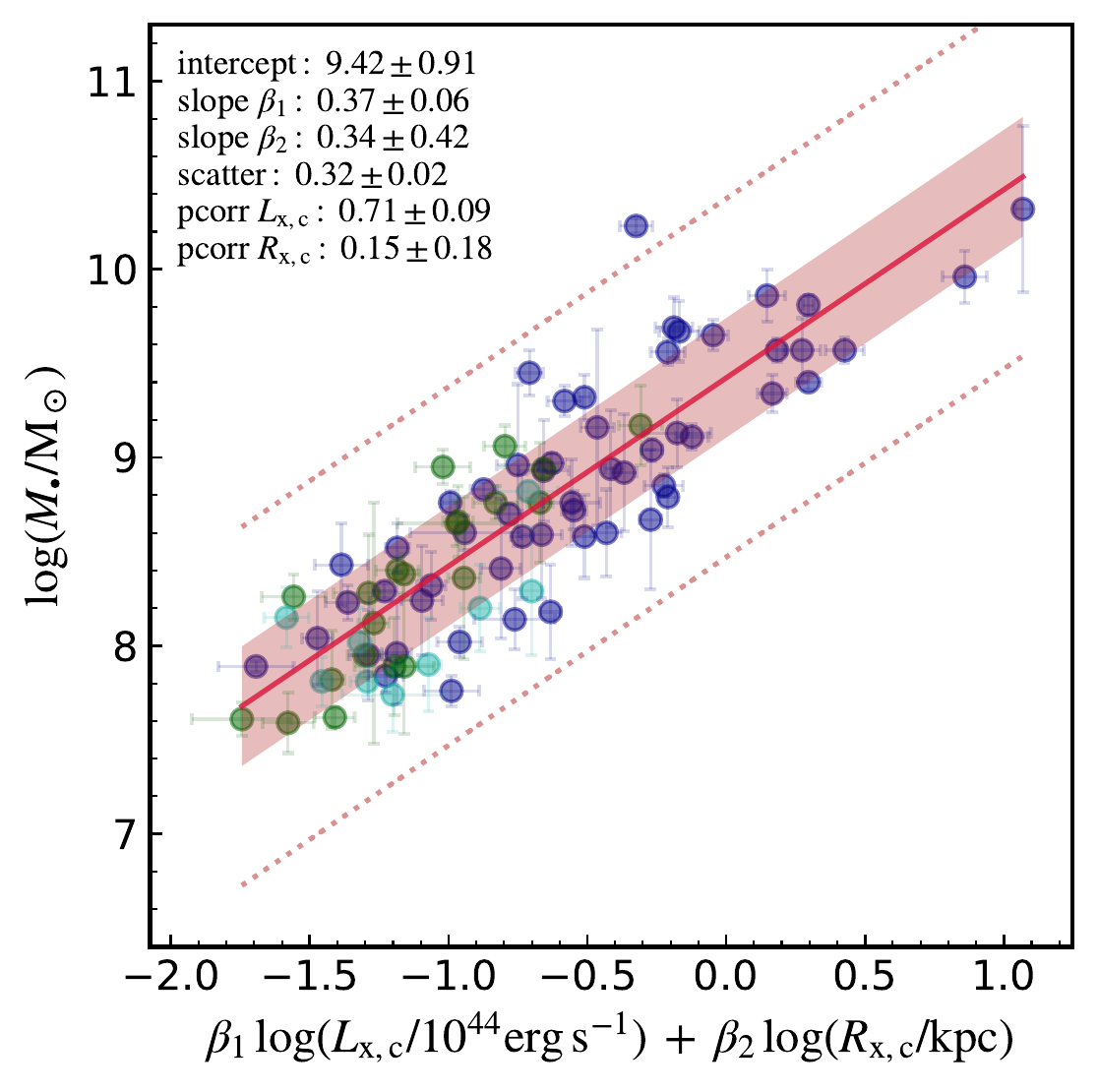}}
 \vskip -0.39cm
\subfigure{\includegraphics[width=0.89\columnwidth]{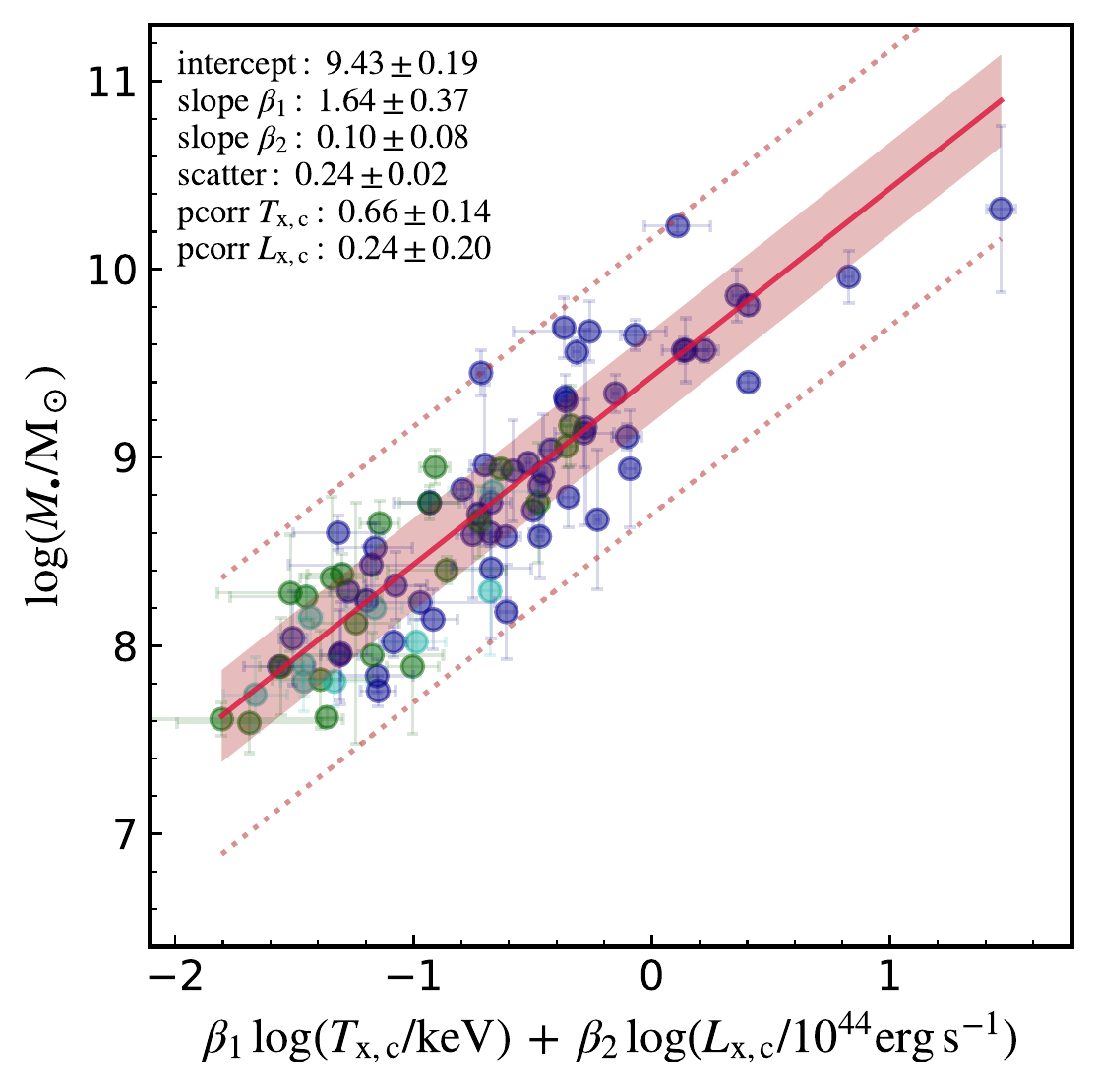}}
 \vskip -0.39cm
\subfigure{\includegraphics[width=0.89\columnwidth]{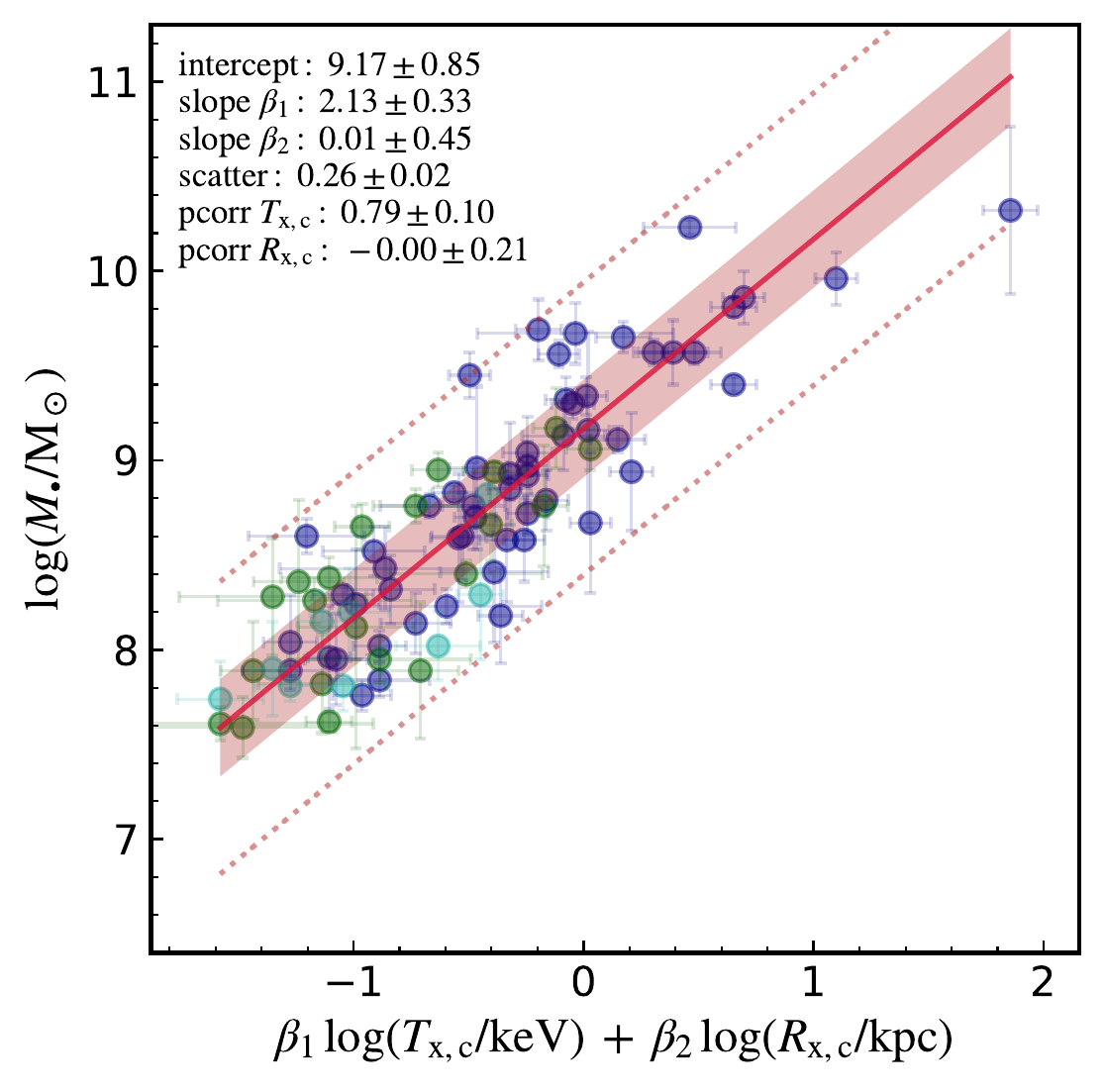}}
 \vskip -0.3cm
\caption{BH mass vs.~dual properties of the xFP (edge-on view), permutating the X-ray luminosity, size, and temperature for the core region. Analog of Fig.~\ref{oFP}. The multivariate fitting on the X-ray properties does not significantly reduce the intrinsic $\epsilon$, although it remains lower than for the optical counterparts.
} 
\vskip -0.6cm
\label{MbhxFPc}
\end{figure}

\begin{figure}[!ht]
 \vskip -0.05cm
\subfigure{\includegraphics[width=0.89\columnwidth]{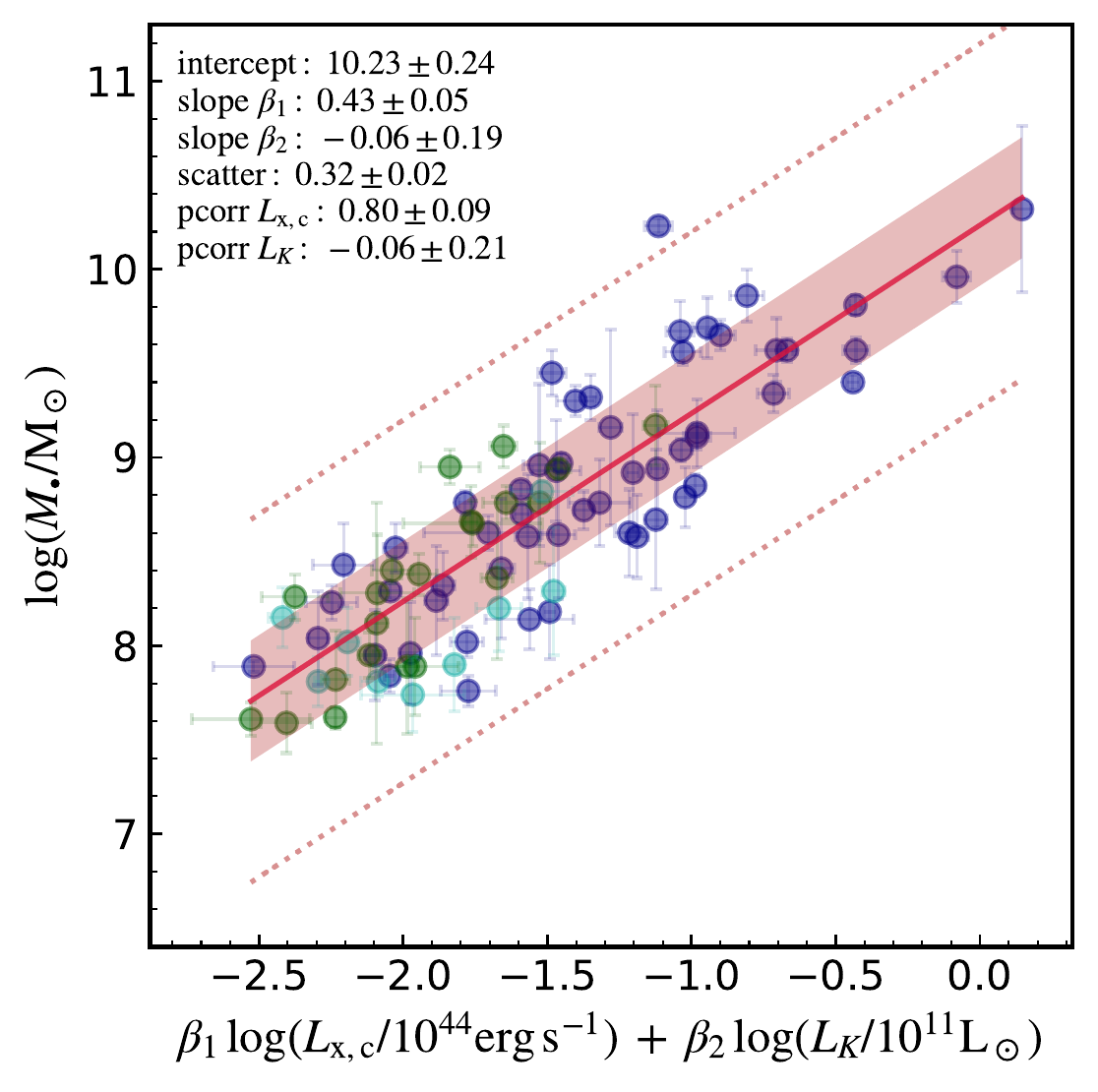}}
 \vskip -0.39cm
\subfigure{\includegraphics[width=0.89\columnwidth]{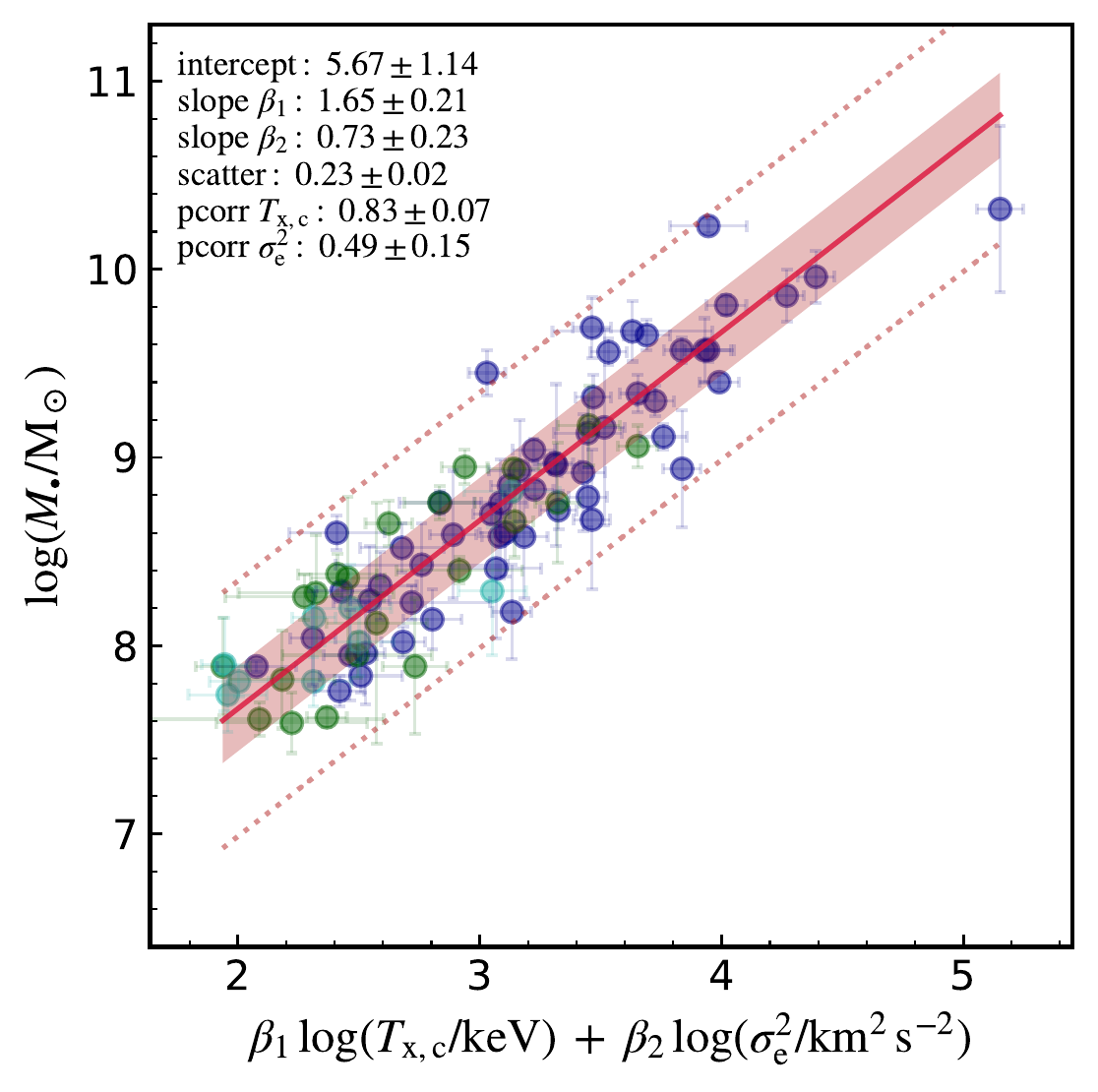}}
 \vskip -0.39cm
\subfigure{\includegraphics[width=0.89\columnwidth]{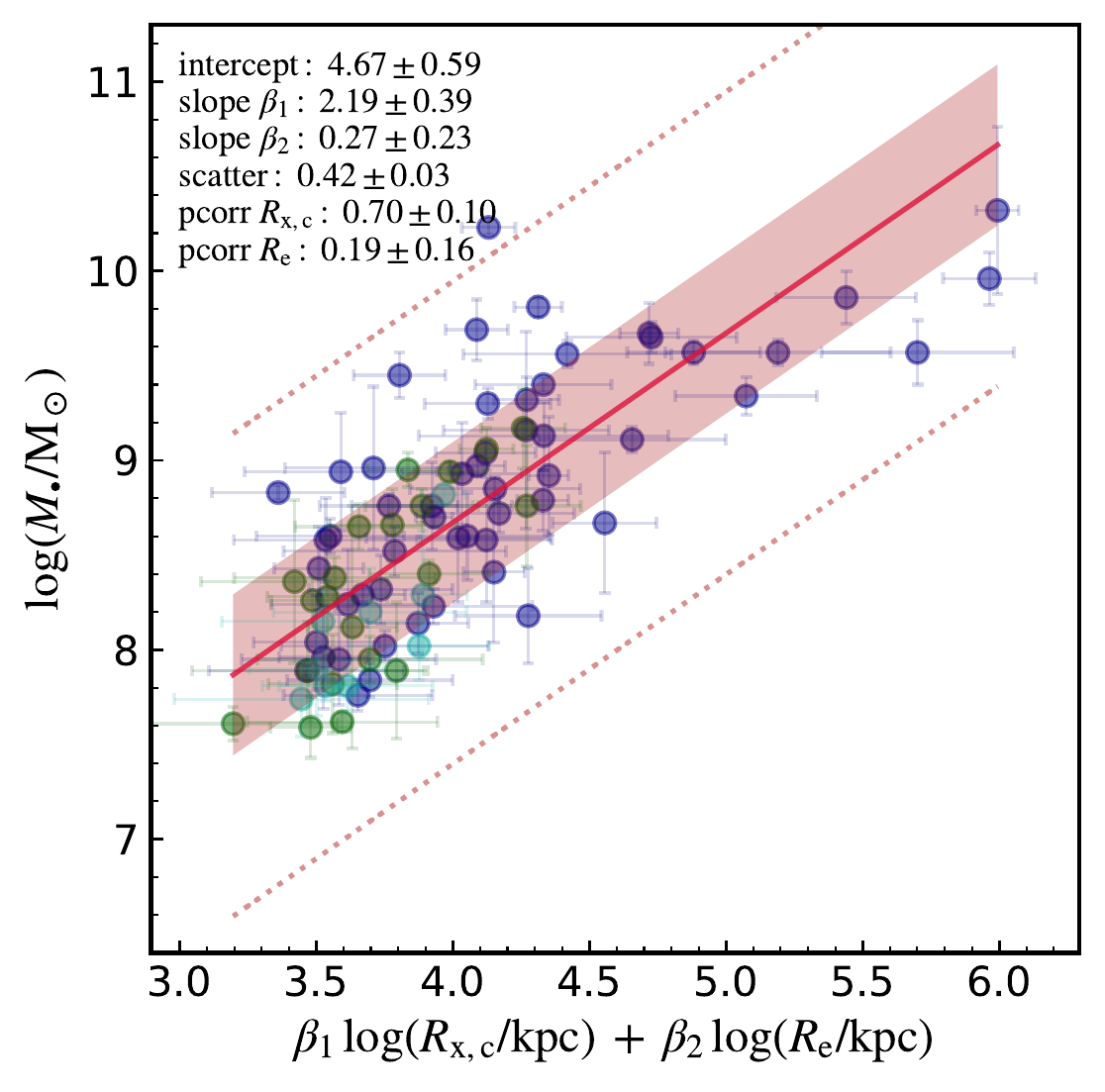}}
 \vskip -0.4cm
\caption{BH mass vs.~counterpart properties of the xFP and oFP, including luminosities, temperatures, and sizes. Analog of Fig.~\ref{oFP}. In all cases, the X-ray properties are more deeply linked to the BH mass, with dominant partial correlation coefficients. Similar results apply considering the X-ray galactic region (not shown).
} 
\vskip -0.3cm
\label{Mbhxop}
\end{figure}

Figure \ref{MbhxFPc} shows the BH mass as a function of two other fundamental X-ray properties, varying between luminosity, temperature, and size, for the core region (the galactic region leads to similar results; Fig.~\ref{MbhxFPg}).
 In all cases, the scatter remains on an almost identical level compared with the X-ray univariate correlations. When the size is involved (top and bottom panels), the \Lx\ or \Tx\ always dominates the pcorr coefficient. Indeed, as discussed in \S\ref{s:xFP}, the xFP is mostly driven by $\lx - \tx$ (and related plasma processes).
The thinnest plane involving the SMBH mass is achieved with $\mbh - \tx - \lx$ (middle panel).
The intrinsic scatter is significantly below any other optical multivariate scaling (\S\ref{s:MbhoFP}).

Looking at the slopes, temperature displays the largest values.
A drawback of the multivariate fitting is that it tries to statistically optimize the parameters, regardless of a potential physical meaning (this becomes progressively more severe with increasing number of dimensions). For instance, the linear slopes retrieved for $\mbh - T_{\rm x, c} - L_{\rm x, c}$ are polar opposite, $\beta_1\simeq1.6$ and $\beta_2\simeq0.1$; 
combining $\tx^{3/2}\,\lx^{1/10}$ does not lead to any evident thermodynamic property. 
Inspecting all the univariate scalings as a function of temperature, the closer composite variables are $f_{\rm gas,c}$ and
$R_{\rm cond}$ (\S\ref{s:CCA}), 
which uncoincidentally are among the properties with the lowest scatter. On a similar note, the normalization values do not evidently relate to physical constants. 
By investigating the alternative $\mbh - I_{\rm x} - T_{\rm x}$ scaling (not shown), we find again that temperature dominates the partial correlations, for all the considered regions. 

Overall, the fact that $\beta_1 \gg \beta_2$ and ${\rm pcorr_1 \gg pcorr_2}$ implies that the X-ray univariate fitting is minimally sufficient and better physically motivated. The tightness and simplicity of the xFP suggest that we can adopt either \Tx\ or \Lx\ as the key driver for the black hole mass growth, in a more confident way than the optical counterparts.
To better quantify the last statement, we show in Fig.~\ref{Mbhxop} the multivariate scalings (for the core region) between the xFP and specular oFP variable.
While these planes are not significantly tighter than the pure $\mbh -$\,xFP scalings, they are instructive in showing how, in all cases, the X-ray property has a much deeper link to the SMBH mass (${\rm pcorr_1 \gg pcorr_2}$), even when $\sige$ is involved, which is the key driver of the oFP (middle panel).
Nevertheless, while the univariate scalings (\S\ref{s:uni}) lead to a minimal and tighter interpretation (from the statistical and physical point of view), the presented multivariate (pure or mixed) scalings 
are additional stringent tests for theoretical/numerical models, which need to be passed to achieve a full theory of co-evolving stars, diffuse gas, and SMBHs in galaxies and groups of galaxies.

\vspace{+0.3cm}
\section{Discussion -- Physical Interpretation} \label{s:disc}
In \S\ref{s:res}, we focused on the observed statistical correlations and comparison between X-ray and optical properties, at face value. Here, we discuss potential physical interpretations, caveats, and future developments.

\subsection{Testing SMBH growth mechanisms}
By now, it is clear that the X-ray gaseous atmospheres play some relevant role in the growth of SMBHs. 
The models concerning the feeding of SMBHs arising from macro-scale properties\footnote{
Since we are not probing relations with the AGN luminosity and given that we are concerned with the integrated BH mass over long timescales, accretion models dealing with the micro-scale of a few 10 $R_{\rm S}$ are beyond the scope of this work (see also \S\ref{s:CCA}).}
can be grouped into two major categories, hot/smooth plasma accretion (\S\ref{s:Bondi}) versus cold/chaotic gas accretion (\S\ref{s:CCA}), which show different correlations between $\mbh$ and X-ray properties. 
Binary SMBH mergers are a third viable growth channel, which we probe in \S\ref{s:mergers}.

\subsubsection{Hot gas accretion} \label{s:Bondi}

\begin{figure}[!t]
 \vskip -0.05cm
 \subfigure{\includegraphics[width=0.89\columnwidth]{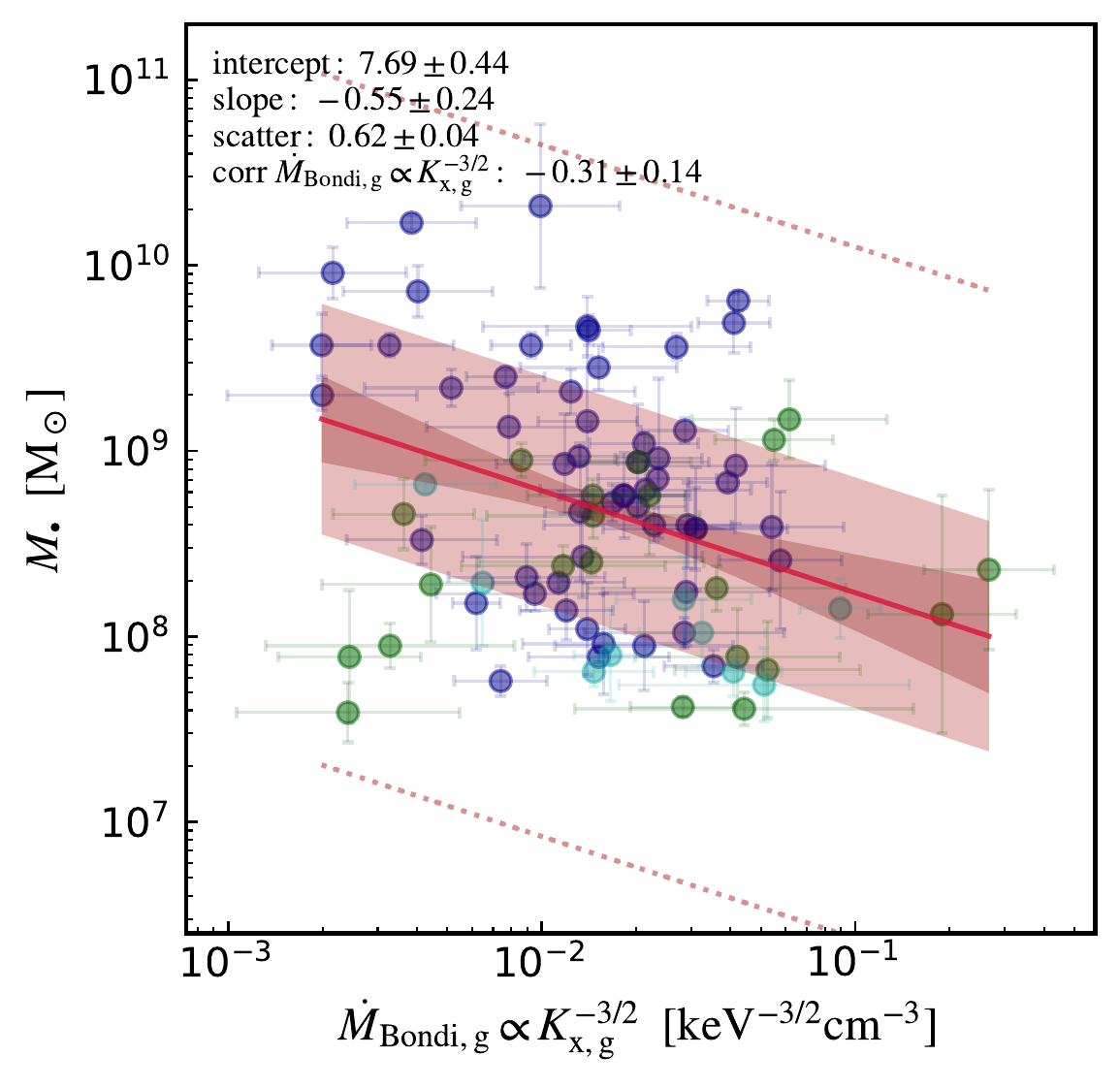}}
 \vskip -0.39cm
\subfigure{\includegraphics[width=0.892\columnwidth]{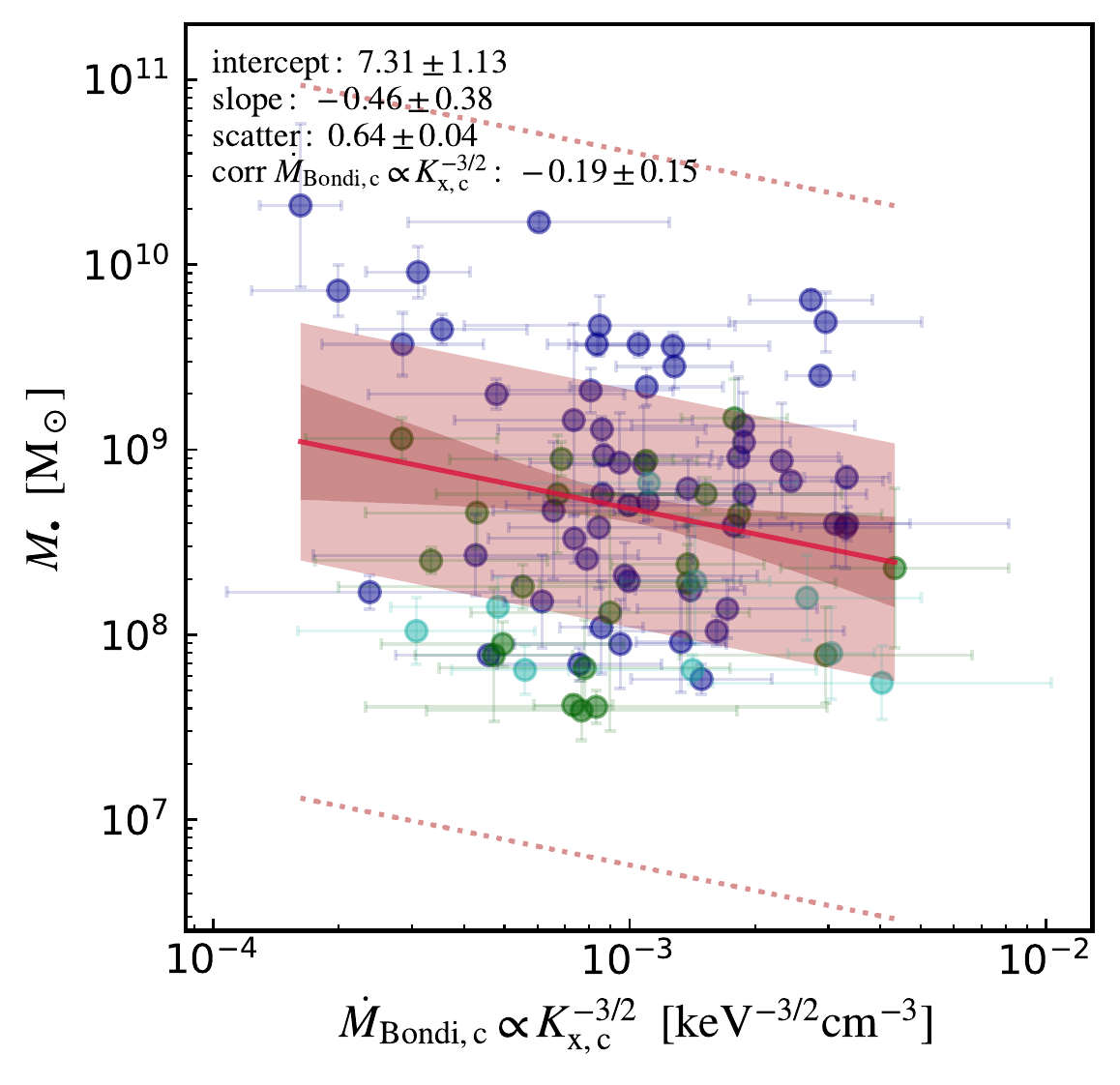}}
 \vskip -0.39cm
\subfigure{\includegraphics[width=0.89\columnwidth]{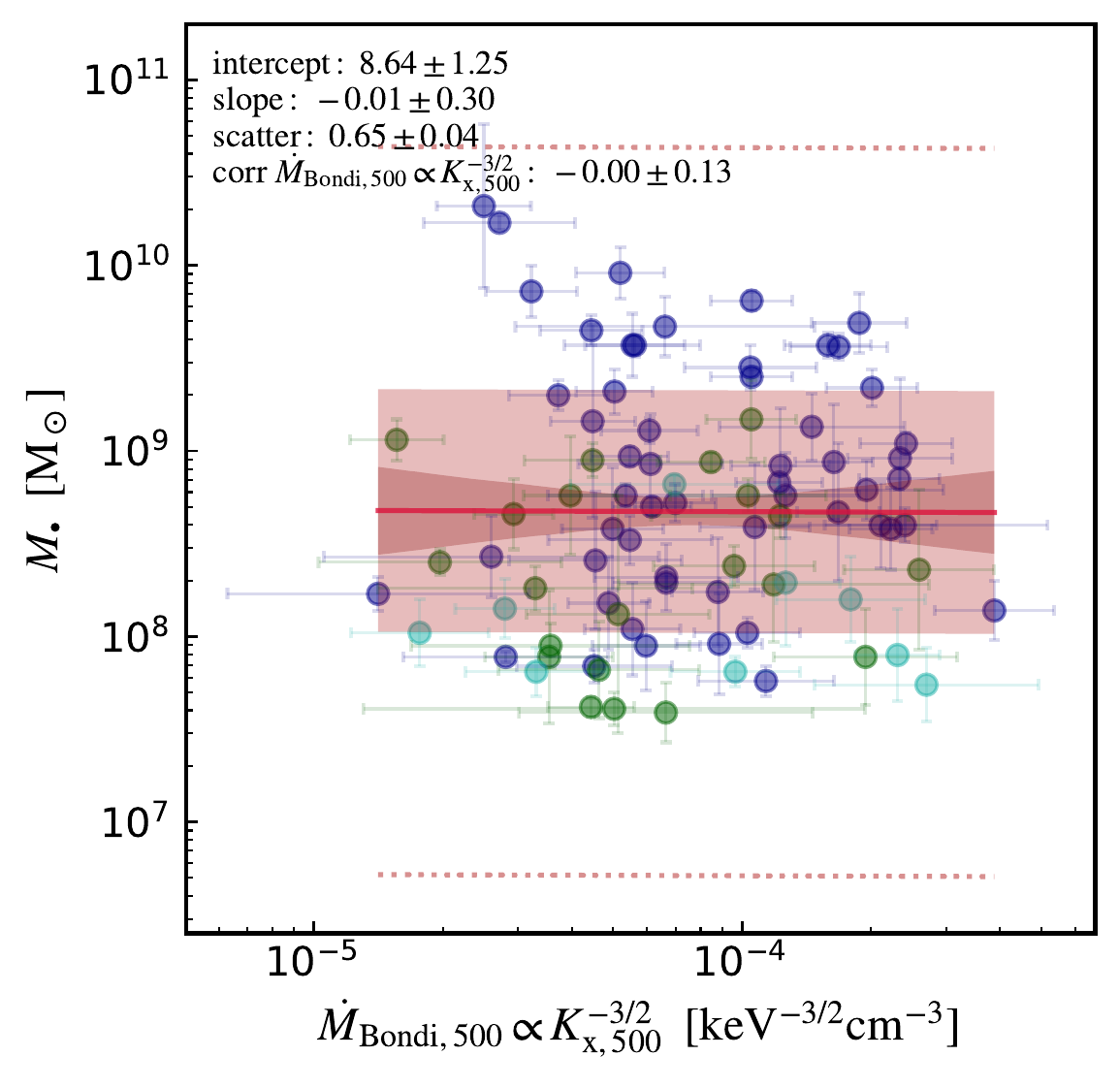}} 
 \vskip -0.4cm
\caption{BH mass vs.~hot halo scaling (X-ray plasma entropy) of the Bondi rate for the CGM scale (top), group/cluster core (middle), and $R_{500}$ (bottom). 
The black hole mass poorly correlates with the Bondi rate ($\propto K_{\rm x}^{-3/2}$; \S\ref{s:Bondi}), even showing a weak negative trend. Both the slope and nearly null corr coefficient rule out hot accretion models as major drivers of BH growth. 
} 
\vskip -0.4cm
\label{MbhMdbondi}
\end{figure}

The majority of hot accretion models are directly (or indirectly) based on the seminal work led by \citet{Bondi:1952}. In a spherically symmetric, steady, and adiabatic gaseous atmosphere, the equations of hydrodynamics reduce to a simple formula for the accretion rate onto the central compact object (as per classical \citealt{Bondi:1952}):
\begin{equation}\label{e:MdBondi}
\dot M_{\rm B} = \lambda\,4\pi(G\mbh)^2\frac{\rho_{\infty}}{c^3_{s,\infty}}, 
\end{equation}
where $\lambda(\gamma)$ is a normalization factor of order unity
($\lambda(5/3)=1/4$), and
with the gas density and adiabatic sound speed, $c^2_{\rm s}=\gamma\,c^2_{\rm s,i}=\gamma k_{\rm b}T_{\rm x}/(\mu m_{\rm p})$, taken at large radii from the accretor, $r_\infty \gg r_{\rm B}\equiv G\mbh/c^2_{\rm s}$. The first drawback of the Bondi rate is that, as absolute value, it produces a very low accretion rate. Even assuming a fully formed SMBH for a median 1\,keV galaxy, then $\dot M_{\rm B} \sim 10^{-4}\,\msun\,{\rm yr^{-1}}$, i.e., a maximal accretion for 10 Gyr would grow the BH only by $10^6\,\msun$.
The inclusion of additional physics breaking the steady-state assumption (e.g., turbulence), spherical geometry (e.g., rotation), or adiabaticity (e.g., non-thermal support via radiation or magnetic fields), each leads to a further suppressed Bondi rate by over 1 order of magnitude (\citealt{Proga:2003, Park:2012, Gaspari:2013_cca,Gaspari:2015_cca,Ciotti:2017}). 
Similar low/suppressed values apply to analogous hot accretion models, such as ADAF (advection dominated accretion flow) and related variants (e.g., \citealt{Narayan:2011}). A key property characterizes all hot, single-phase models: in order to accrete, the flow has to overcome the large thermal pressure support of the hot halo that strongly counterbalances (with negative radial gradient) the gravitational pull of the SMBH, galaxy, and cluster core.
Eq.~\ref{e:MdBondi} is thus a firm upper limit for hot gas accretion models.

The last term in Eq.~\ref{e:MdBondi} is the key dependency tied to the hot X-ray halo. 
Combining the plasma density and sound speed, it can be rewritten as a steep inverse function of plasma entropy:
\begin{equation}\label{e:MdBondiK}
\dot M_{\rm B} = \lambda\,4\pi(G\mbh)^2\,\frac{(\mu m_{\rm p})^{5/2}}{\gamma^{3/2}}\,K_{\rm x,\infty}^{-3/2},
\end{equation}
where the X-ray plasma entropy (related to the thermodynamic entropy as $S\propto \ln K$) is defined as
\begin{equation}\label{e:Kx}
K_{\rm x} \equiv \frac{k_{\rm b}T_{\rm x}}{n^{\gamma-1}} = \frac{k_{\rm b}T_{\rm x}}{n^{2/3}}.
\end{equation}

\begin{figure}[!t]
 \vskip -0.05cm
 \subfigure{\includegraphics[width=0.91\columnwidth]{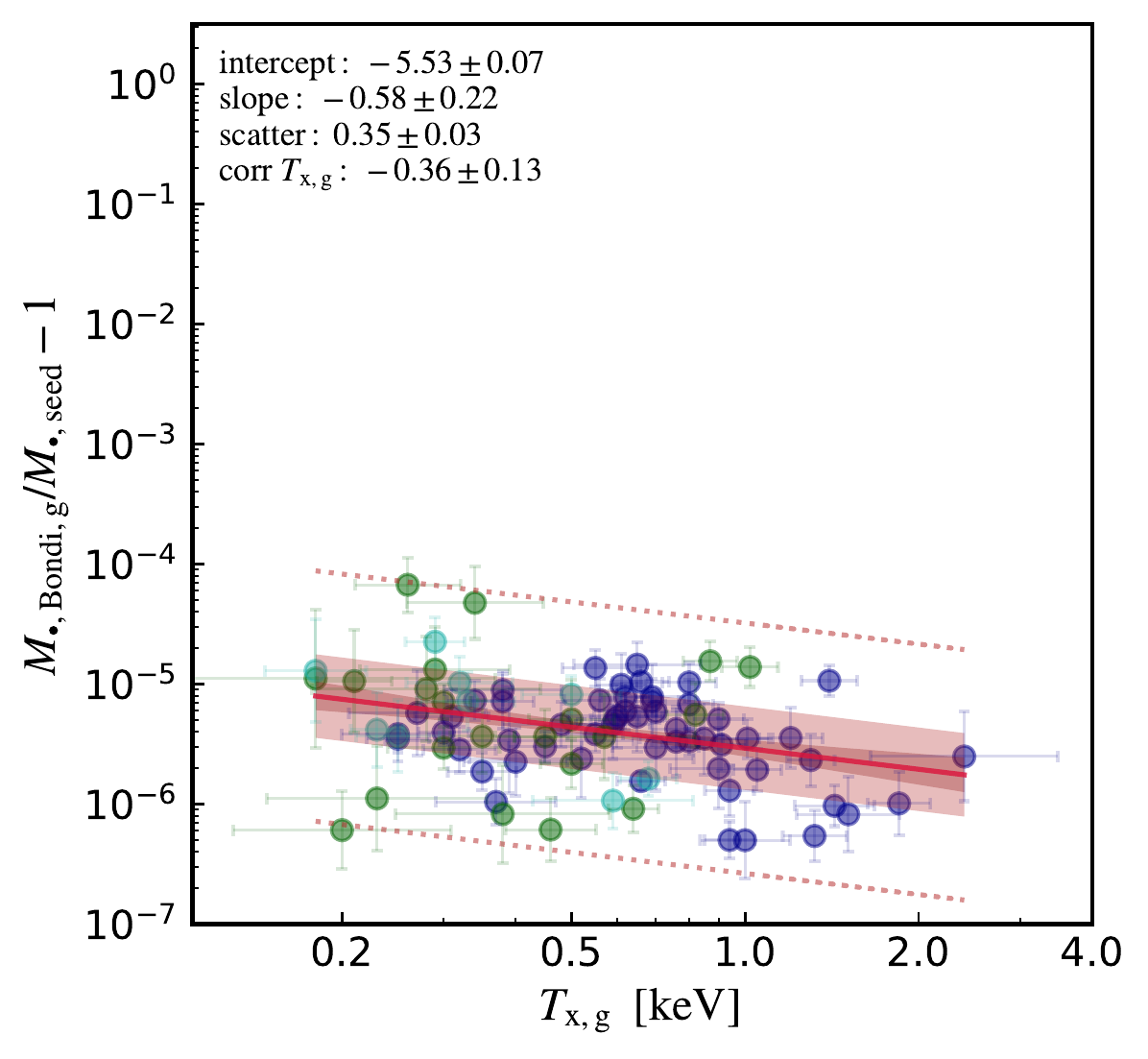}}
 \vskip -0.35cm
 \subfigure{\includegraphics[width=0.91\columnwidth]{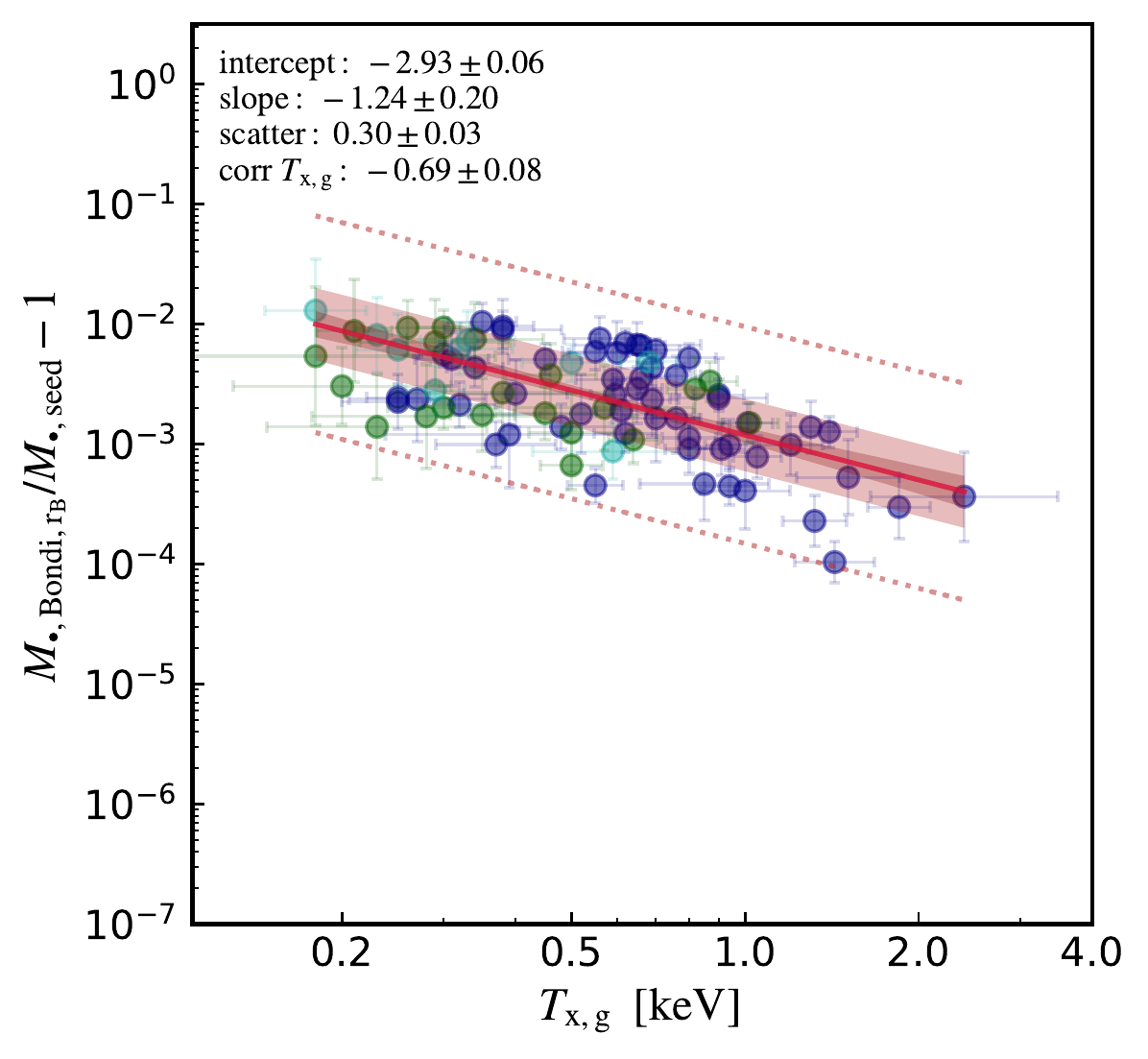}}
 \vskip -0.35cm
\caption{Bondi-driven BH mass growth as a fraction of the high-$z$ seed (Eq.~\ref{e:MBondi}; $M_{\rm \bullet, 0}=10^5\,\msun$), adopting as boundary condition the galactic scale (top) and the extrapolation down to the Bondi radius $r_{\rm B}$ (bottom). 
During the entire cosmic time, pure Bondi accretion can only grow a tiny percentage of the initial seed ($< 1$\%), proving a minor impact of hot-mode accretion in the evolution of SMBHs.
}
\vspace{+0.2cm} 
\label{MbhMbondi}
\end{figure}

Figure \ref{MbhMdbondi} shows the BH mass versus the gas scaling of the Bondi rate (i.e., the plasma entropy\footnote{The $\mbh^2$ dependence is a trivial self-correlation, which is unrelated to the X-ray halo. Eq.~\ref{e:MBondi} also shows the key role of entropy.} for a non-relativistic gas with $\gamma=5/3$). It is evident that the SMBH mass does not correlate well with the plasma entropy (thus hot-mode accretion), adopting any radial bin.
All the corr coefficients reside in the absent or weak regime, even within the 1-$\sigma$ level. 
The scatter is one of the largest reported in this study, $\epsilon\sim0.6$ dex, which is 3$\times$ that of the tightest X-ray relations, essentially spanning the entire observed BH mass range within the 3-$\sigma$ channel.
Even for the galactic case with mild slope (top), the weak correlation is negative, meaning hotter halos accrete relatively less gas mass, as stifled by the plasma pressure support (conversely to cold models which condense more heavily in more massive halos; \S\ref{s:CCA}).

We can further test whether the integrated Bondi rate is capable of growing an SMBH to the current level throughout the cosmic history of the galaxy, group, or cluster.
Integrating Eq.~\ref{e:MdBondiK} (nonlinear ODE), $\int^{t_{\rm H}}_{t_0} \dot M_{\rm B}\, dt$, and adopting a decreasing entropy with redshift
$K_{\rm x}(t) = K_{\rm x, now}(t/t_{\rm H})^\theta$, yields
\begin{equation}\label{e:MBondi}
M_{\bullet, \rm{B}}(t_{\rm H}) = \frac{M_{\bullet, 0}}{1-M_{\bullet, 0}\,K_{\rm x, now}^{-3/2}A\left(t_{\rm H}^{1-3\theta/2}-t_0^{1-3\theta/2}\right)},
\end{equation}
with integration constant given by the following: $A\equiv\lambda 4\pi G^2 (\mu m_{\rm p})^{5/2}\,t_{\rm H}^{3\theta/2}/[\gamma^{3/2}(1-3\theta/2)]$.
We note there is no duty cycle for hot-mode accretion, since the diffuse atmosphere can not be removed and accretion remains continuous.
Fig.~\ref{MbhMbondi} shows the fractional increase of the BH mass given a typical high-$z$ seed of $M_{\bullet, 0}=10^5\,\msun$ and $\theta=4/3$ entropy evolution (\citealt{Voit:2005a}). 
Assuming as boundary condition the galactic scale (top panel), 
the BH grows only a tiny mass fraction during the cosmic evolution ($t_{\rm H}-t_0\simeq13$\;Gyr), and it would worsen by adopting the large-scale $R_{\rm x,c}$ and $R_{500}$. 
The bottom panel shows the extrapolation of the Bondi rate down to the inner $r_{\rm B}$ of each galaxy ($<100$ pc) by using the average observed CC entropy profile ($K\propto r^{2/3}$; \citealt{Panagoulia:2014})\footnote{By using the non-CC profile, the entropy would remain flat in the core, with no variation in $\dot M_{\rm B}$.}. Even under such a best-case scenario the fractional increase reaches 1\% and decreases toward the (hotter) BCGs due to the entropy dependence.

Overall, all the above tests rule out (even adopting favorable parameter values) hot-mode accretion as the primary mechanism of SMBH growth.
This is consistent with high-resolution hydrodynamical simulations finding hot accretion sub-dominant compared with cold accretion (e.g., \citealt{Gaspari:2012a,Gaspari:2013_cca}).
Moreover, applying the macro AGN mechanical efficiency ($\eta=10^{-3}$; \S\ref{s:Yx}), the driven kpc-scale jet/outflow power for our median system is $P_{\rm mech} = \eta \dot M_{\rm B} c^2 \sim 10^{39}$ erg\,s$^{-1}$, several orders of magnitude below the core X-ray cooling luminosity. This implies an inefficient AGN feedback mechanism (see also \citealt{McNamara:2011,Russell:2013}), together with a low-variability behavior inconsistent with most of the observed (X-ray) AGN light curves (e.g., \citealt{Peterson:2001,LaMassa:2015}).\\

\subsubsection{Chaotic cold accretion (CCA)} \label{s:CCA}
We test now the other main BH feeding theory, chaotic cold accretion (CCA), i.e., the raining of warm optical filaments and cold molecular clouds condensing out of the X-ray atmosphere via nonlinear thermal instability (\S\ref{s:intro}).
CCA behaves in a different manner from the above hot accretion models, displaying two distinct properties.
First, the accretion rates can be boosted intermittently up to several orders of magnitude compared with the Bondi rate, given the recurrent chaotic inelastic collisions between the cold and warm clouds or filaments.
Second, CCA displays large variability, with power spectral density described by a flicker noise, 
as shown by other natural chaotic and fractal processes, including AGN, quasars, meteorological data, and semiconductors (cf.~\citealt{Gaspari:2017_cca}). Moreover, this duty cycle is more frequent toward low-mass systems.
In CCA-driven BH growth, the feeding and feedback processes are closely intertwined.
While the rain recurrently triggers the AGN down at the horizon scale\footnote{Once CCA funnels the clouds within $\sim$\,20 Schwarzschild radii ($r_{\rm S}\equiv2\,G\mbh/c^2$; \citealt{Gaspari:2013_cca}), the Maxwell and Reynolds stresses generated via MRI turbulence in the inner torus induce the gas to radially accrete within $\sim$10 orbital periods (\citealt{Sadowski:2017,Jiang:2019}), i.e., $10\times2\pi[(20\,r_{\rm S})^3/G\mbh]^{1/2}\sim$\,2\;yr (for $\mbh=10^9\,\msun$), which is negligible compared with the macro-scale plasma halo timescales.}
(\citealt{Sadowski:2017}), the AGN feedback quickly responds by injecting back a substantial amount of energy in the form of massive outflows and jets, which deposit their energy at the macro scale, in a gentle self-regulated feedback loop (e.g., \citealt{Gaspari:2012b,Gaspari:2012a,Li:2014,Barai:2016,Yang:2016a,Yang:2016b,Yang:2019}). Such mechanical AGN feedback, on the one hand quenches cooling flows, on the other hand induces over the long term an irreducible level of subsonic turbulence that shapes the halo weather (\citealt{Lau:2017,Hitomi:2018,Simionescu:2019}).

\begin{figure}[!ht]
 \vskip -0.05cm
 \subfigure{\includegraphics[width=0.91\columnwidth]{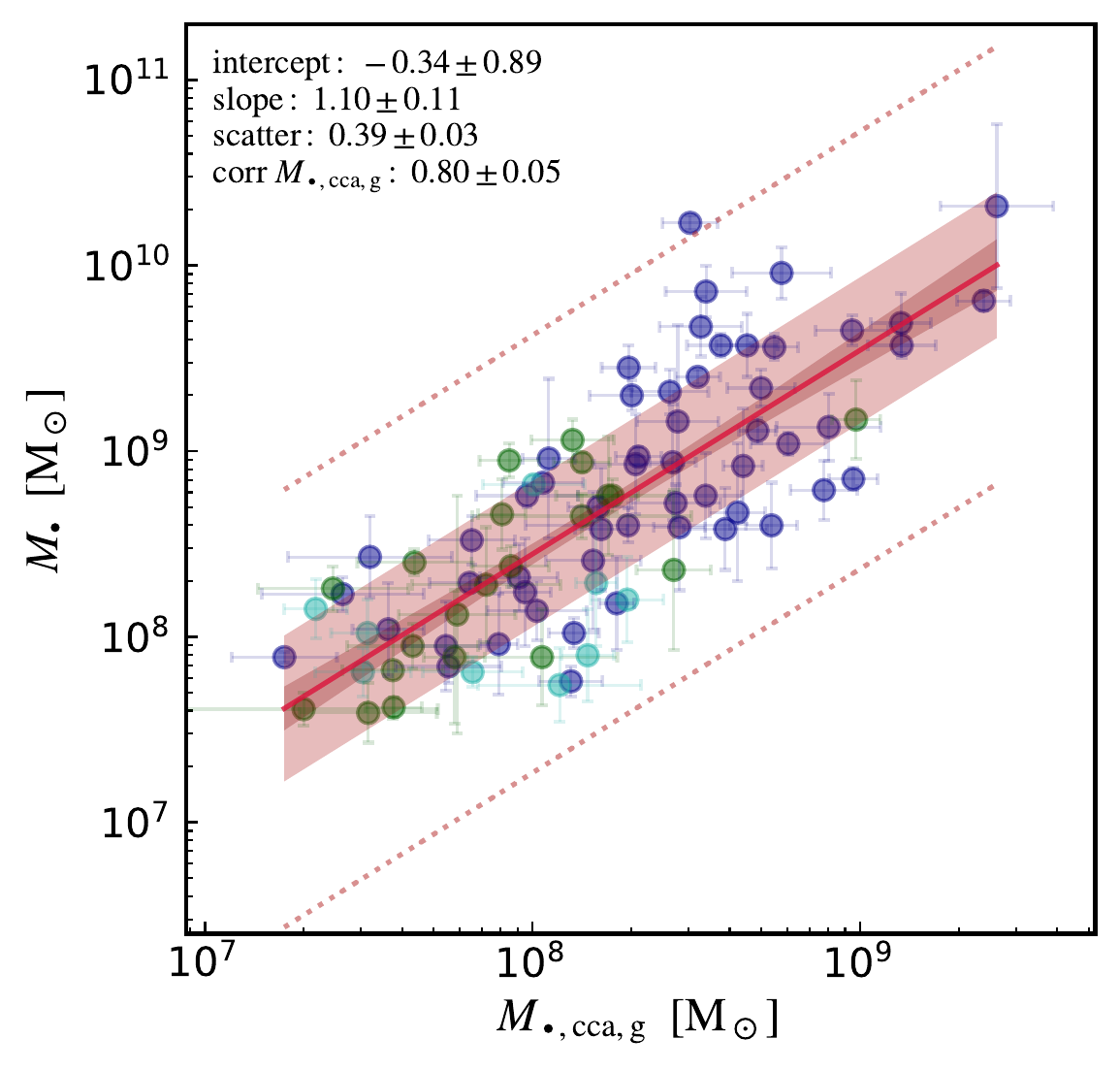}}
 \vskip -0.35cm
\subfigure{\includegraphics[width=0.91\columnwidth]{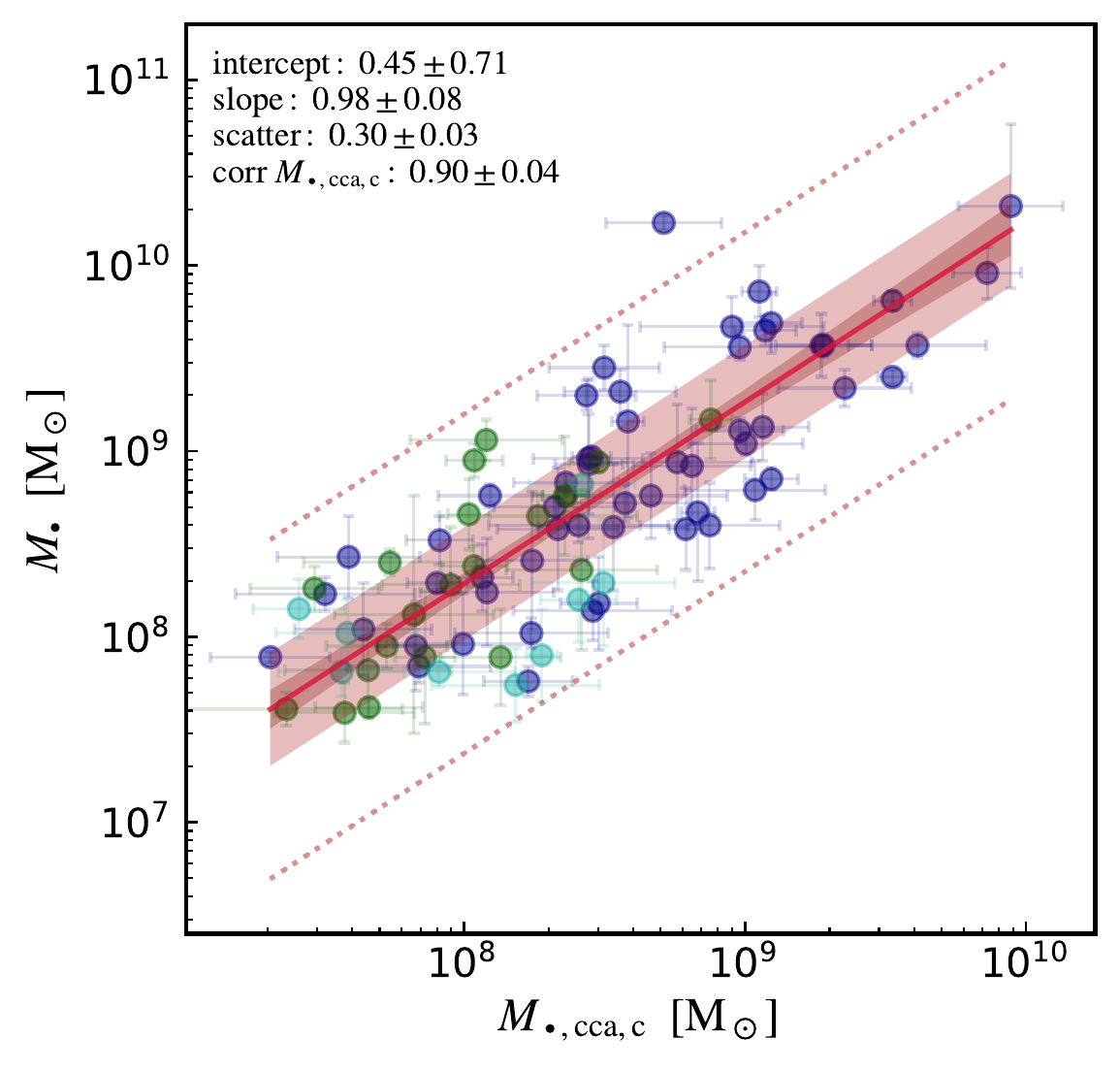}}
 \vskip -0.35cm
\subfigure{\includegraphics[width=0.91\columnwidth]{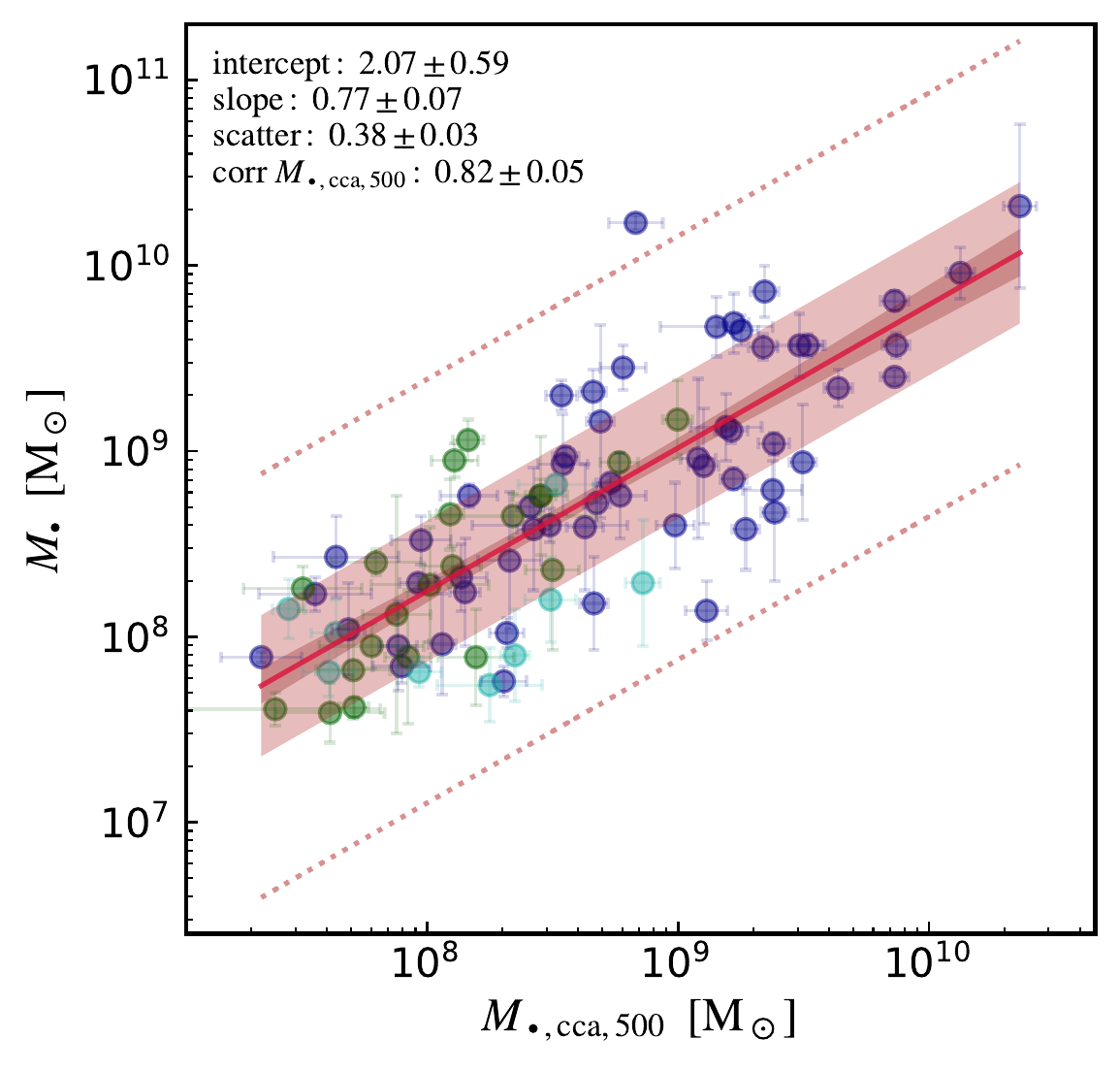}} 
 \vskip -0.4cm
\caption{CCA-driven BH mass growth (Eq.~\ref{cca}). 
The substantial boosting of the accretion rate via CCA (due to the direct link to the halo \Lx\ and related cooling rate, in particular in the core region), coupled with the lower CCA frequency toward more massive halos, leads to a consistent build-up of SMBHs throughout the Hubble time, generating also $10^{10}\,\msun$ UMBHs in BCGs.
} 
\vskip -0.3cm
\label{MbhCCA}
\end{figure}

While expensive hydrodynamical simulations are required to track the detailed chaotic process (e.g., \citealt{Gaspari:2013_cca,Prasad:2017}), analytic predictions can be retrieved from the macro-scale properties (\citealt{Gaspari:2017_uni}). 
CCA is tightly linked to the cooling rate of the X-ray plasma:
\begin{equation}\label{Mdcool}
\dot M_{\rm cool} = \frac{q\,\mu m_{\rm p}}{(3/2)\,k_{\rm b}}\frac{L_{\rm x}}{T_{\rm x}},
\end{equation} 
where $q\approx10\%$ is the evidence that most pure cooling flows are significantly quenched via AGN feedback (\citealt{Peterson:2006,McNamara:2012,Gaspari:2012a}). Since $L_{\rm x}/T_{\rm x}\propto t^{-1}$ (\citealt{Maughan:2012}) and accounting for the CCA variability, the integral of Eq.~\ref{Mdcool}
over cosmic time 
($t_{\rm H}-t_0\simeq13$\;Gyr)
leads to 
\begin{equation}\label{cca}
M_{\rm \bullet,cca}(t_{\rm H}) \simeq  \left[\dot M_{\rm cool,now} \, t_{\rm H} \ln({t_{\rm H}}/{t_0})\right]\frac{\nu_{\rm cca}}{\tilde{\nu}_{\rm cca}},
\end{equation} 
where the last term is related to the CCA rain active cycle, with a characteristic frequency given by the turbulent eddy turnover frequency (\citealt{Gaspari:2018}), $\nu_{\rm cca} = (2\pi r^{2/3} L^{\rm 1/3}/\sigma_{v, L})^{-1}$.\footnote{We note that for $r > L$, the eddy time reduces to $\propto L/\sigma_{v,L}$, since the injection scale is the maximum driving scale.} 
Indeed, turbulence is the key physics enabling the generation of significant overdensities in the stratified hot halo, which then nonlinearly condense in multiphase filaments and clouds. By using the ensemble warm-gas single spectra for 72 galaxies in groups and clusters, the above authors show that the (3D) turbulent velocity dispersion is contained in a fairly narrow range, $\sigma_{v,L}\approx 240\pm30$\;km\,s$^{-1}$, with a minor halo scaling $\propto (T_{\rm x}/\,{\rm 2\,keV})^{1/3}$. The injection scale $L$ is related to the AGN feedback influence region and can be retrieved from the observed size and distance covered by the pair of inflated AGN bubbles, scaling as $L\approx{\rm 10\,kpc}\,(T_{\rm x}/{\rm 1\,keV})^2$ 
(\citealt{Shin:2016}) and reaching up to 200 kpc for massive clusters (e.g., MS\,0735.6+7421).
Further, the frequency of the CCA rain increases from massive clusters to low-mass galaxies due to the relatively stronger radiative cooling (\citealt{Gaspari:2011a,Gaspari:2012b,Sharma:2012,Prasad:2015}).
While for low-mass systems the AGN feeding/feedback events are so frequent to be nearly continuous, massive clusters experience longer duty cycles, with powerful AGN outbursts (up to $10^{45}$\,erg\,s$^{-1}$) followed by quiescent periods of a few 100 Myr. The last term in Eq.~\ref{cca} models such CCA duty cycle, with normalization $\tilde{\nu}_{\rm cca}$ taken as the eddy frequency at the low-mass end of the halo distribution ($\tilde{\nu}_{\rm cca}\approx0.05$\;Myr$^{-1}$). 
The retrieved median CCA frequency (with 1 RMS) over the whole sample is $\log \nu_{\rm cca}/{\rm Myr^{-1}}\simeq-2.1\pm0.5$.

The tight correlation between BH mass and X-ray luminosity (Fig.~\ref{MbhLx}) already implies an important connection with the plasma cooling (and thus CCA), since its rate $\dot M_{\rm cool}$ is a strong function of $L_{\rm x}$. At variance with hot-mode accretion models (\S\ref{s:Bondi}), the CCA driven accretion rate is capable of generating SMBHs, and even UMBHs, over the whole Hubble time.
Indeed, for a typical ETG cooling rate of 0.1\,-\,1\,$\msun$\,yr$^{-1}$, steady
feeding would imply BHs with $\mbh \sim$\,$10^9$\,-\,$10^{10}\,\msun$.
While in hot accretion models the hotter the halo the lower the accretion rate, in CCA hotter -- thus more luminous -- halos drive stronger radiative emissivity and condensation. 
Similarly, the tight correlations found for $\mbh - M_{\rm gas}$ (Fig.~\ref{MbhMgas}) and $\mbh - f_{\rm gas}$ (Fig.~\ref{Mbhfgas}), particularly in the core region ($\epsilon\simeq0.25$), support the key role of the gas mass
in feeding the BH.
On the other hand, the fact that $\mbh - \lx$ deviates from a pure linear scaling ($\beta\sim1/2$) rules out cold accretion models that are perennially turned on over the entire mass range. Indeed, for massive clusters, such a model would generate UMBHs with masses in excess of $10^{11}\,\msun$ during the Hubble time. The intrinsic \textit{chaotic} variability of CCA, with retrieved lower frequency toward more luminous/extended halos, makes it a compelling model to solve such a hurdle.

Fig.~\ref{MbhCCA} shows the quantitative CCA-driven BH mass growth computed via Eq.~\ref{cca} and the X-ray properties of our sample. The best-consistent region to extract the CCA variables (as the cooling rate) is the core region ($\sim$0.1\,$R_{\rm 500}$; middle panel). For the core region, the observed and predicted BH masses match in a linear way within 1-$\sigma$ ($\beta\simeq0.98$), covering the full range of detected masses from $3\times10^7\,\msun$ to $10^{10}\,\msun$. The intrinsic scatter is the lowest among the three regions and tight ($\epsilon\simeq0.30$), with the correlation coefficient in the very strong regime (corr\,$\simeq$\,0.9).
The galactic region instead shows a superlinear slope (top), since the $\mbh$ at the high-mass end are underestimated; the correlation is also weaker, displaying 30\% larger scatter.
Conversely, using the outskirt properties (bottom) leads to an overestimate of the BH masses in BCGs, inducing a too shallow slope; however, the correlation is still in the strong regime.
Overall, the X-ray cooling rate initiating from the halo core region and then feeding the central BH appears to be the optimal predictor for the long-term\footnote{The instantaneous accretion rates in one cycle can be quickly estimated via $L_{\rm x}$; the total BH mass requires the integrated cooling rate, which is essentially the raining mass through several cycles.} BH growth. This is further consistent with the core region correlating with the drop in temperature that differentiates CC versus non-CC systems (\citealt{Vikhlinin:2006,Ghirardini:2019}). 
It is also worth noting that the CCA average Eddington ratio is $\log \dot M_{\rm \bullet, cca,c}/\dot M_{\rm Edd} = -2.7\pm0.4$, i.e., significantly sub-Eddington at the present time (but expected to slowly increase at higher $z$; e.g., to $-2$ at $z\sim2$), as found by X-ray AGN surveys (e.g., \citealt{Aird:2018}).

\begin{figure}[!t]
 \vskip -0.05cm
 \subfigure{\includegraphics[width=0.91\columnwidth]{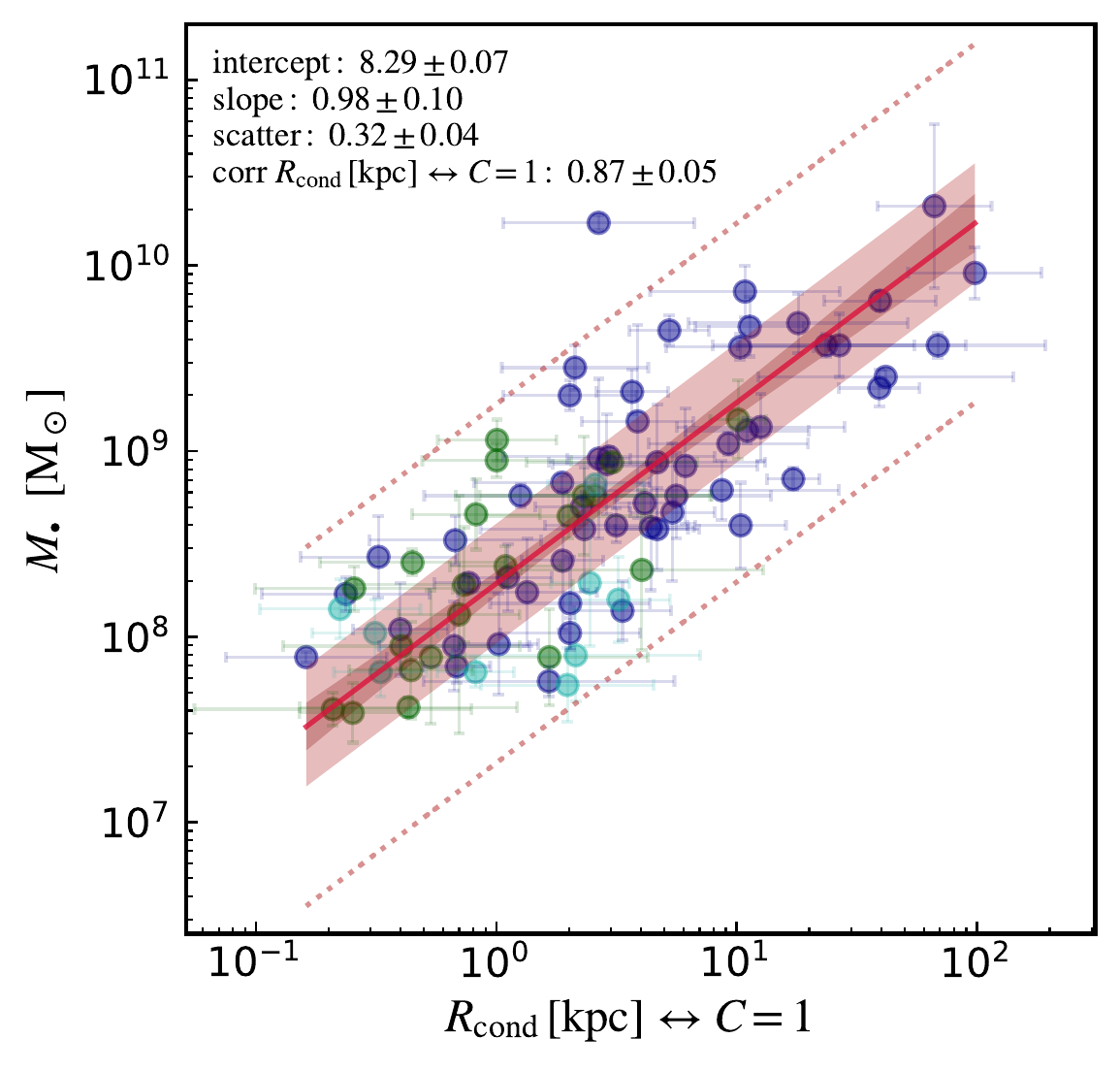}}
 \vskip -0.35cm
 \hskip -0.1cm
 \subfigure{\includegraphics[width=0.95\columnwidth]{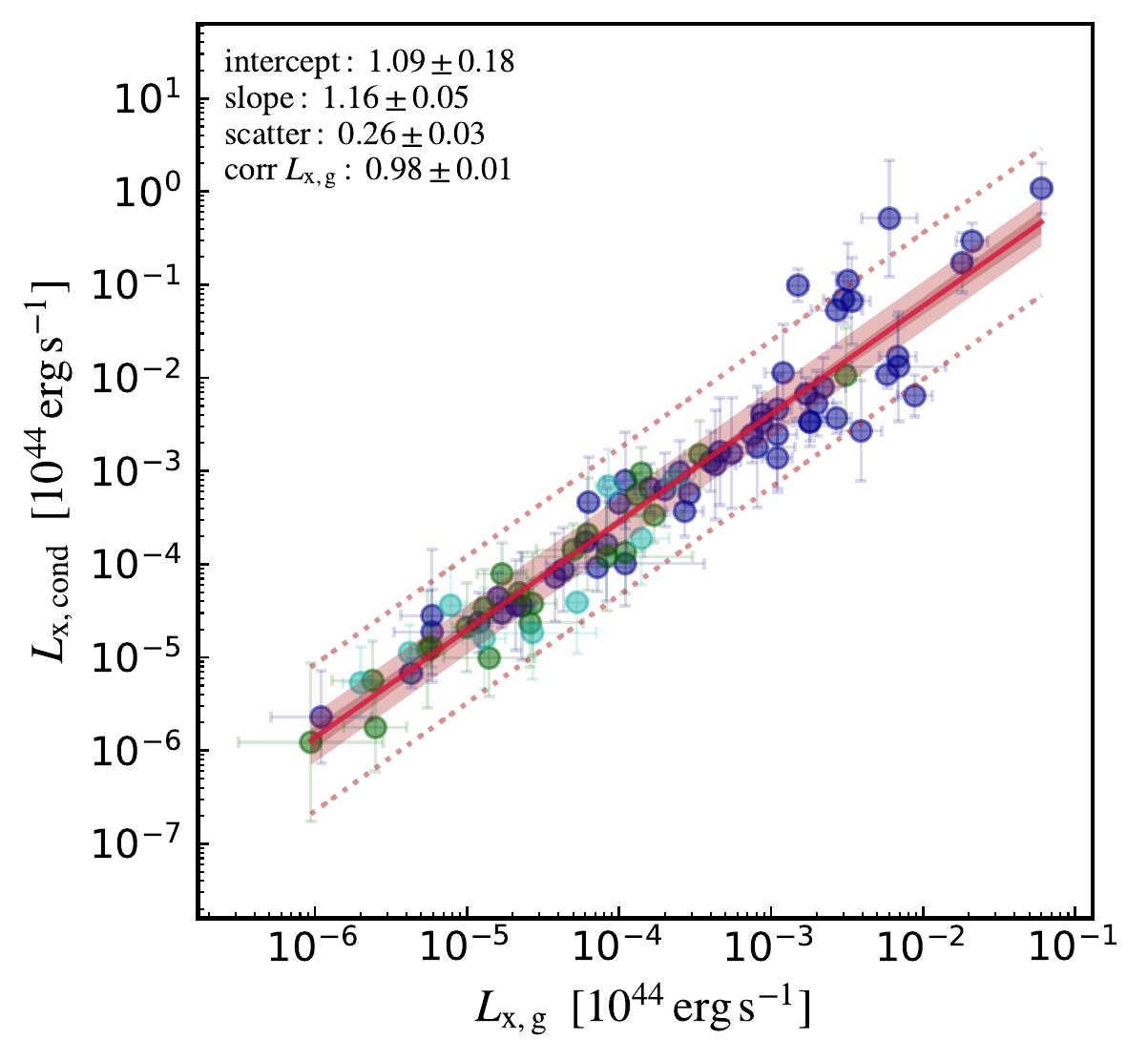}}
\caption{\textit{Top}: Condensation radius for the precipitating warm optical filaments and cold molecular clouds, provided by the $C$-ratio reaching unity inside the core region. 
The BH masses are linearly and strongly linked to the condensation region, suggesting a key interplay between CCA and the secular growth of SMBHs.
\textit{Bottom}: X-ray luminosity predicted by the CCA condensation criterion $C\equiv t_{\rm cool}/t_{\rm eddy}=1$ (Eq.~\ref{Lx_cond}), compared with the observed galactic X-ray luminosity of the sample. The CCA prediction is consistent with the data over a similar extent radius, supporting the key role of the $C$-ratio in describing the diffuse gas properties.
} 
\label{MbhRcond}
\end{figure}

Another key property that is worth testing is the CCA condensation radius, $R_{\rm cond}$.
While significant X-ray cooling initiates in the core, the inner region where the condensed ionized (H$\alpha$+[NII]) filaments and molecular (CO) clouds end up precipitating is given by $R_{\rm cond}$, the radius at which the eddy time ($t_{\rm eddy}=\nu_{\rm cca}^{-1}$) and cooling time $t_{\rm cool} \simeq 3\,k_{\rm b}T_{\rm x}/(n_{\rm e}\Lambda)$ become comparable, $C\equiv t_{\rm cool}/t_{\rm eddy}=1$ (\citealt{Gaspari:2018}). 
Given the electron density profile, $n_{\rm e}(r)=n_{\rm e,c}\,(r/\rxc)^{-a}$ (with median $a\simeq1.6$ found in \S\ref{s:ne}; see also \citealt{Hogan:2017,Babyk:2018}), a $C$-ratio of unity translates into 
\begin{equation}\label{Rcond}
R_{\rm cond} = \left(\frac{3}{2\pi}\frac{k_{\rm b}T_{\rm x}}{\Lambda\,n_{\rm e,c}\,\rxc^a}\frac{\sigma_v}{L^{1/3}}\right)^{\frac{3}{2-3a}}.
\end{equation}
Adopting the core properties, the $\mbh - R_{\rm cond}$ scaling shown in Fig.~\ref{MbhRcond} (top) displays a tight correlation and linear slope ($\epsilon\simeq0.32$ and $\beta\simeq0.98$) with corr coefficient in the very strong regime. This corroborates the above result (Fig.~\ref{MbhCCA}) that the growth of SMBHs is tightly linked to the CCA physics and related multiphase rain occurring in galaxies, groups, and clusters (see also \citealt{Voit:2015_nat,Voit:2015_gE,Soker:2016,Voit:2018}). 
In terms of normalization, ETGs in groups have $R_{\rm cond}\sim0.6$\,-\,6\;kpc, while typical BCGs have $R_{\rm cond}\sim7$\,-\,40\;kpc, which is in good agreement with the extent of H$\alpha$ nebulae observed in low- to high-mass halos  (\citealt{McDonald:2010,McDonald:2011a,Werner:2014}; NGC\,5044/A1795 a notable group/cluster with nebular warm gas). 
Using instead the galactic/$R_{500}$ variables (not shown) leads to an underestimate/overestimate of the nebular radius by $\sim$1\,dex; the correlation with \Mbh\ is also weaker, with 40\% larger scatter, signaling that we are moving away from the main thermally unstable plasma source (the core region). 

It is important to note that $R_{\rm cond}$ is a function of predominantly \Tx\ (log slope $\sim$2), 
thus, the tight $\mbh - \tx$ correlation found in Fig.~\ref{MbhTx} may be thought of as a reflection of the residual condensed 
phase recurrently feeding the central BH. This correlation is also linked to the gravitational potential (or $R_{\rm 500}$/total mass): hotter halos have larger gas mass, and thus more vigorous precipitation; in other words, larger, hotter halos have a more extended raining region (the role of mergers, also affecting the potential, is tested in \S\ref{s:mergers}).
Conversely, the characteristic radius of Bondi/hot models (\S\ref{s:Bondi}) is $r_{\rm B}\propto T_{\rm x}^{-1}$, i.e., anti-correlated with hotter halos, hence going against the observed positive trend with BH masses, as well as being disconnected from the group/cluster properties (given its size of a couple orders of magnitude smaller than $R_{\rm cond}$). 

We can further test whether the CCA theory can self-consistently predict the $\lx - \tx$ scaling relation based on the $C$-ratio criterion. As the $C$-ratio becomes unity, this yields a characteristic electron number density
\begin{equation}\label{ne_cond}
n_{\rm e,cond} = \frac{3}{2\pi}\frac{k_{\rm b}\tx}{\Lambda}\frac{\sigma_{v,L}}{r^{2/3}L^{1/3}}.
\end{equation}
The X-ray luminosity can then be retrieved via the integral shown in Eq.~\ref{e:Lx}, adopting $n^2_{\rm e,cond}$ (Eq.~\ref{ne_cond}) in the emission measure,
thus yielding
\begin{equation}\label{Lx_cond}
L_{\rm x,cond}  = \frac{27}{5\pi}\,\frac{(k_{\rm b}\tx)^2}{\Lambda}\frac{\sigma_{v,L}^2}{L^{2/3}}\,R_{\rm cond}^{5/3},
\end{equation}
with $\sigma_{v,L}$ and $L$ given by the observational scalings (as a function of \Tx) discussed after Eq.~\ref{Mdcool}.
Figure \ref{MbhRcond} (bottom) shows the CCA prediction, by using the above $R_{\rm cond}$
in Eq.~\ref{Lx_cond}, 
compared with the independently observed $L_{\rm x,g}$, with the condensation radius typically being $\pm0.3$\,dex from the galactic/CGM scale.
The match with the observed X-ray luminosity is good, with both normalization 
and scaling
reproduced up to $\sim$\,$10^{42}\,\es$, above which the scatter increases due to $R_{\rm cond}$ exceeding the CGM region. By extrapolating the density profile, Eq.~\ref{ne_cond} and \ref{Lx_cond} may be further generalized and applied in observational or theoretical studies to interpret or model the thermodynamic properties of diffuse media at different radii in a (semi)analytic framework.
Summarizing, the predictions of CCA in terms of both the hot halo properties and SMBH masses are well in agreement with the analyzed observational data.

\subsubsection{Stars}

As introduced in \S\ref{s:res}, the tightness of the X-ray temperature/luminosity/gas mass scalings ($\epsilon\sim0.2$\,-\,0.3), compared with the optical correlations ($\epsilon\sim0.4$\,-\,0.5) is an indicator that the gaseous atmospheres play a key role in the co-evolution of SMBHs. Besides the above quantitative tests, first principles suggest the stellar component is not the main source of fueling. Once rapidly collapsed from the progenitor molecular cloud, stars become collisionless, with negligible dynamical friction to feed the BH (this drag force is proportional to the square of the moving object mass). While  SMBHs can efficiently accrete the collisional gas (in particular via chaotic collisions), the stellar component represents the residual, unaccreted mass that is progressively stored in the galactic potential. 

The stellar mass is still linked to the gas condensation (hence the $\mbh - M_{\rm bulge}$), however, the link with the BH growth is progressively washed out during the recurrent CCA cycles.
Taking the potential ($\propto \sigma_{\rm e}^2$ or $L_K/\re$) as reference improves the BH mass optical correlations, since the potential is less affected by the baryonic physics. 
While the stellar component can be well described via the VT and homology (as proven by the oFP; \S\ref{s:oFP}), X-ray halos are primarily affected by thermo-hydrodynamical processes (as found via the xFP; \S\ref{s:xFP}), rather than solely experiencing a virialization in a gravitational potential.
The above results (and cosmological simulations in \S\ref{s:mergers}) point toward gas accretion and condensation as the dominant mechanism establishing the relations with $T_{\rm x}$ and $L_{\rm x}$, instead of such scalings being a pure passive tracer of the total mass/DM (\citealt{Bogdan:2018}).
Needless to say, the secondary connection with the potential will be always present to some degree, as it sets an upper dynamical limit to any gas accretion%
\footnote{For an isothermal sphere $\rho_{\rm gas}=f_{\rm gas}\,\sige^2/(2\pi G r^2)$,
the dynamical limit is given by
$\dot M_{\rm dyn} = 4\pi\rho_{\rm gas}\,r^2\sige \approx 10^3 \sigma_{\rm e,300}^3\,\msun\,{\rm yr^{-1}}$ 
for $f_{\rm gas}=0.1$. The cooling rate is much lower than this, as the X-ray halo is not collapsing in freefall.
Integration by a timescale $\propto \sige^{1.5}$ would lead to the observed $\mbh - \sige$ relation (e.g., \citealt{King:2003}).},
as well as being linked to the hierarchical merger growth -- the focus of the next section.

\subsubsection{SMBH mergers} \label{s:mergers}
Another potentially important channel for BH growth is the hierarchical merger build-up throughout the cosmological evolution of galaxies and clusters. Let us discuss first a simple statistical approach, and then the more detailed realistic cosmological simulation.
In the concordance $\Lambda$CDM universe, the first galaxies form after $\sim$1 Gyr after the Big Bang and start to stochastically merge into large-scale structures, such as proto-groups and then proto-clusters. 
Beyond a scale of tens kpc, the hosted SMBHs slowly sink (in a few Gyr) 
toward the new, merged potential center via dynamical friction onto stars, until the SMBH pair forms a hard binary. 
At the parsec scale, 3-body scattering with nuclear stars leads to the ejection of the stellar bodies and shrinking of the binary orbit to the milliparsec scale, where gravitational-wave radiation drives the final coalescence of the two SMBHs (e.g., \citealt{Begelman:1980}). 

It is thus expected that, in a hierarchical universe, SMBH masses and any halo property positively correlate, including those of the X-ray plasma (which is heated up via the virialization shock during the infall in the DM potential). However, the key differentiating point is how well they correlate, in terms of slope and scatter.
Idealized SAMs (\citealt{Peng:2007,Jahnke:2011}) show that, starting from an uncorrelated random distribution of seeded intermediate BHs,
a sequence of mergers will naturally lead to a linear correlation ($\beta = 1$) between SMBH mass and bulge/total mass, being both the results of a large number of summation events, which are independent from any gas feeding/feedback physics.
In this `central-limit-theorem' (CLT)\footnote{We preserve the CLT nomenclature adopted in the astrophysics literature; however, a better term would be to refer to this effect as an \textit{attractor} in the evolution of a stochastic process.} view, 
the ensemble averaging ensures that the fractional dispersion in both coordinates should decrease with increasing halo mass  as $\propto N^{-1/2}$, assuming a very large number of experienced mergers $N$.
Rule of thumb for `large' statistical samples is at least $N=30$, hence a relative drop of the scatter down to $1/\sqrt{30} \simeq 0.18$.

Fig.~\ref{Mbhop} shows that the $\mbh - M_{\rm bulge}$ has a slope below unity ($\beta=0.90$); the dispersion (from the best fit) for the bottom/top half of the bulge masses is 0.39/0.46\,dex, which is inconsistent with the CLT prediction of a decreasing scatter. 
Fig.~\ref{MbhMtot} shows that the $\mbh - M_{\rm tot}$ has a slope above unity for all the three main regions ($\beta=1.2$\,-\,1.4); 
moreover, the dispersion in the top half of the total masses within $R_{\rm x,g}$, $R_{\rm x,c}$, or \R500 is in all cases 1.1$\times$ that in the bottom half, i.e., opposite to the CLT prediction. 
This is supported by the below-described cosmological simulations (see also Fig.~4 in \citealt{Bassini:2019}), which further show that starting from a purely random uncorrelated distribution at large redshift is unrealistic, invalidating the CLT assumption.
Analogous non-decreasing dispersion applies to the $\mbh - M_{\rm gas}$ correlations, which also have slopes deviating below unity ($\beta \approx 0.6$).
All these results suggest that, while mergers are part of the evolutionary process, they are likely not the overwhelming force in the build-up of the SMBH -- X-ray halo scaling relations. 

\begin{figure}[!t]
 \vskip -0.4cm
 \hskip -0.2cm
 \subfigure{\includegraphics[width=1.04\columnwidth]{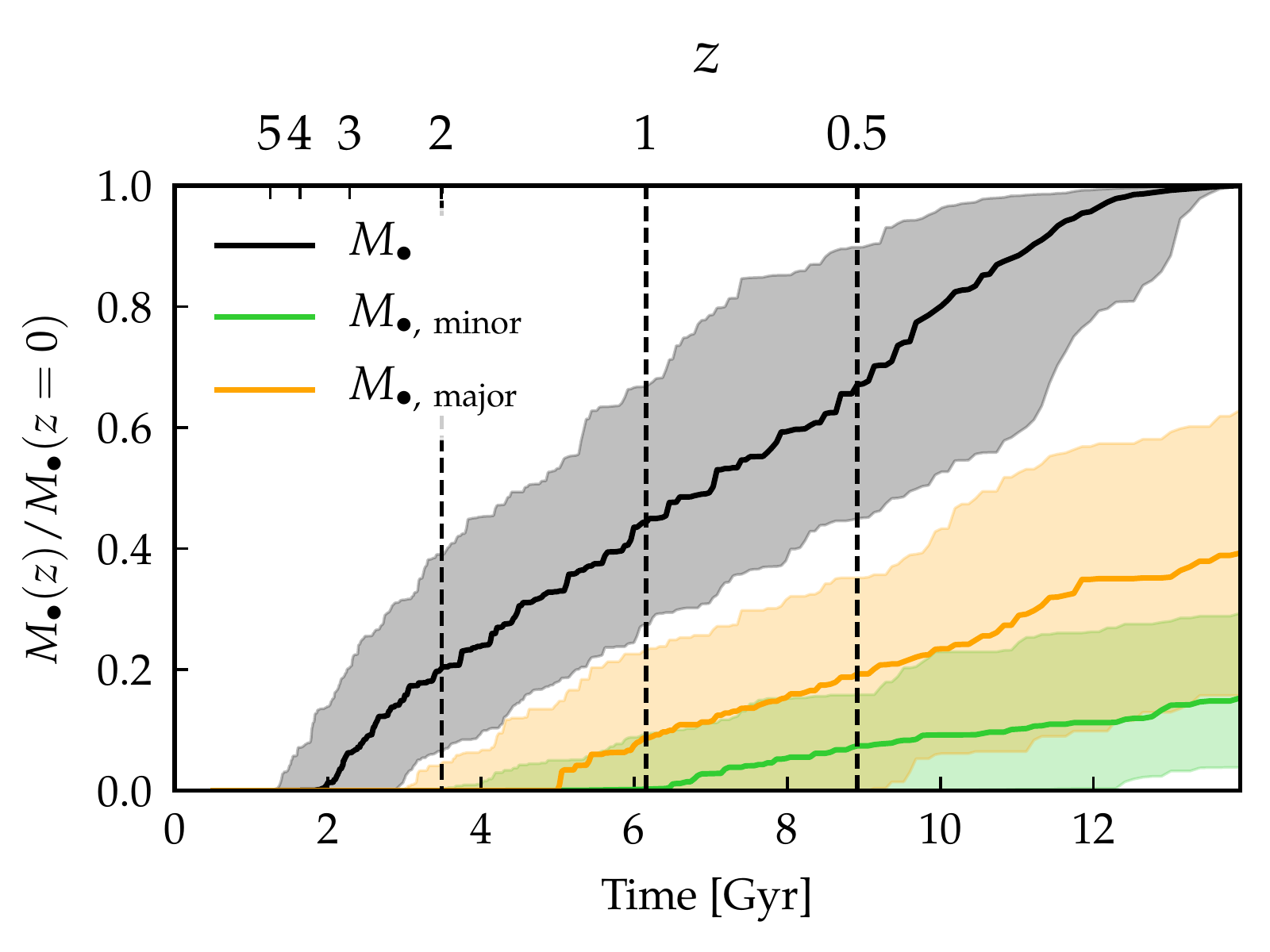}}
 \vskip -0.4cm
\caption{Cosmological hydrodynamical zoom-in simulation of 135 groups/clusters, showing the median and 1-$\sigma$ of the fractional SMBH mass increase during the Hubble time, for central galaxies only. 
The redshift/linear time is depicted on the top/bottom $x$-axis.
The black line indicates the total SMBH mass (due to gas accretion and mergers), while the orange and green lines show the relative contribution of major and minor mergers (above/below a mass ratio of 1:4), respectively.  
Even for central galaxies (which experience the largest amount of major mergers; Fig.~\ref{mergers_mag}),
merger-driven BH growth is sub-dominant over most of the cosmic time and driven by infrequent major mergers. } 
\vskip +0.05cm
\label{mergers_frac}
\end{figure}

\begin{figure}[!t]
 \vskip -0.3cm
 \hskip -0.32cm
 \subfigure{\includegraphics[width=1.06\columnwidth]{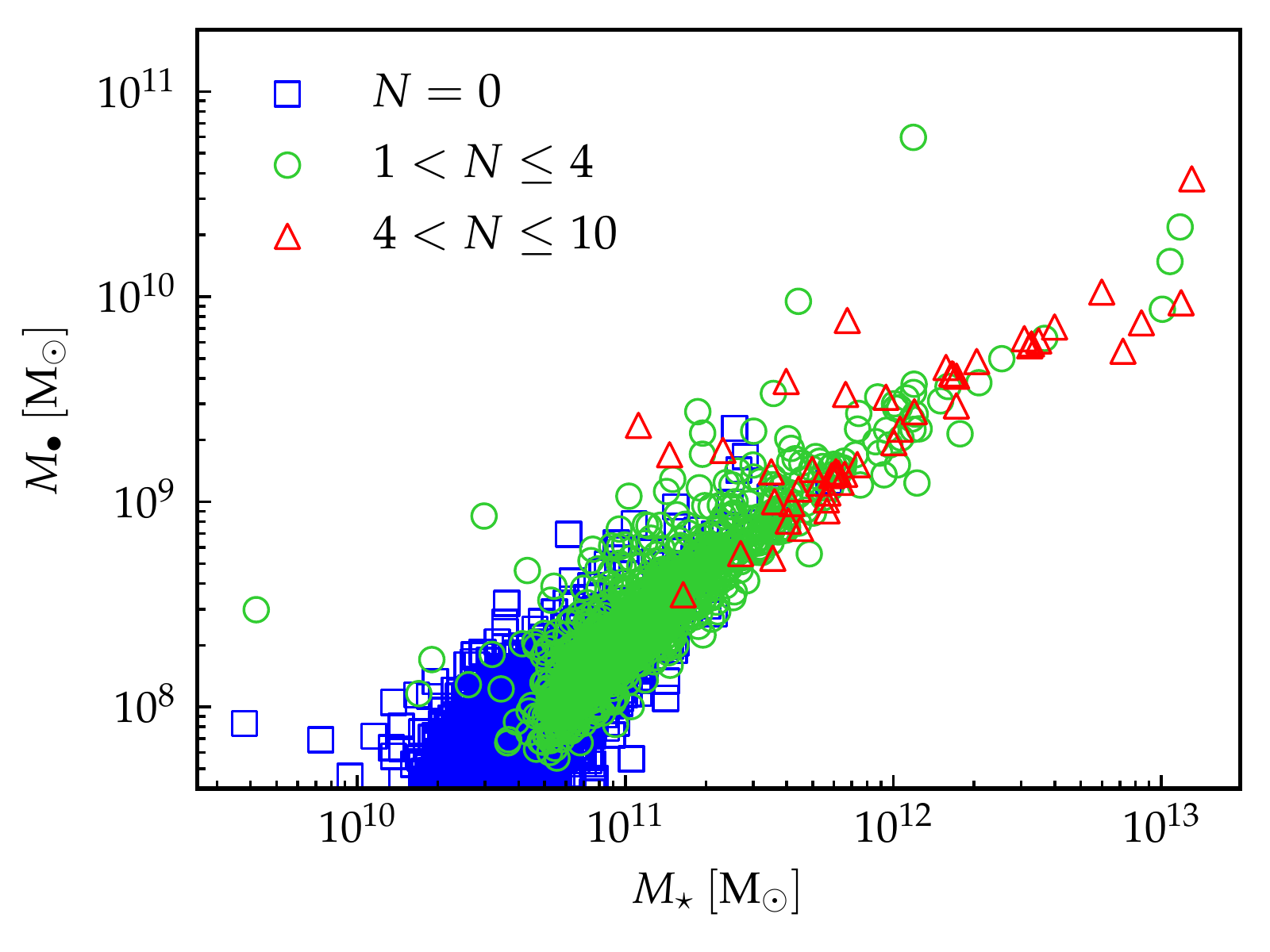}}
 \vskip -0.4cm
\caption{Magorrian relation developed in our cosmological simulation (\S\ref{s:mergers}) at $z=0$, color-coding the number of major ($>$\,1:4 mass ratio) BH mergers experienced by each final SMBH. It is clear that major mergers are rare events: the majority of objects experience only a few major mergers, while most of the mass is built via gas accretion. Moreover, the dispersion for SMBHs above $10^8\msun$ with $N=0$, $N<2$, or $N>2$ remains similar at 0.2\;dex.
} 
\label{mergers_mag}
\end{figure}

To better quantify the above finding in a more realistic setup, we use state-of-the-art cosmological hydrodynamical simulations with subgrid models for baryon astrophysics, including radiative cooling, star formation, metal enrichment, and stellar/AGN feedback, which well reproduce the observed properties of X-ray hot halos (\citealt{Rasia:2015, Planelles:2017, Biffi:2018,Bassini:2019} for the simulation setup).
Such Lagrangian (SPH) zoom-in simulations of 135 groups/clusters of galaxies ($M_{\rm tot,500}\sim10^{13}$\,-\,$10^{15}\,\msun$) are particularly useful to track the evolution and merging of the SMBHs ($\sim$\,6000)%
\footnote{Softening/merging length of the sink particles is $\sim3\,h^{-1}$\,kpc.}
hosted by each galaxy,
in a large and robust statistical sample. 
Moreover, the simulations include subgrid modeling for both hot and cold gas accretion, as well as ejective (quasar-mode) and maintenance (radio-mode) AGN feedback (\citealt{Ragone-Figueroa:2018} for the numerical details).
Fig.~\ref{mergers_frac} shows the median growth of the SMBH masses (normalized to the final mass) for the central galaxies over the $\sim$\,14 Gyr cosmological evolution (black), involving a combination of (predominantly cold) gas accretion and BH mergers. The orange/green line indicates the contribution of mergers, separating them into major/minor events, defined as having a mass ratio above/below 1:4, respectively. The first key result is that mergers contribute a sub-dominant fraction of the growing $\mbh$ over most of the cosmic evolution; 
only in the last Gyr they start to catch up with the level of gas accretion. 
It is important to note that Fig.~\ref{mergers_frac} shows the SMBHs solely grown in central galaxies, thus experiencing the largest amount of mergers. As shown below (Fig.~\ref{mergers_mag}), over 70\% of the total BHs do not experience a major merger and are entirely driven by (cold) gas accretion.
Indeed, even with no mergers, the seeded BHs (starting at $\sim$\,$10^5\,\msun$) evolve toward the Magorrian relation.
Overall, gas accretion is the main channel of SMBH growth for most of the population and over the long-term evolution,
even in the best-case merger scenario of massive BCGs.

The second key result from Fig.~\ref{mergers_frac} is that infrequent major mergers dominate the BH growth, rather than frequent small sub-structure accretion.
The low rate of major galaxy mergers ($\sim$0.03\,-\,$0.5\ {\rm Gyr^{-1}}$ for $z\sim0$\,-\,3; see also \citealt{Rodriguez-Gomez:2015}) means that for each final massive galaxy we expect $\sim$2 major events during its evolution. 
In a similar vein, Fig.~\ref{mergers_mag} shows our simulated Magorrian relation at $z=0$ (for the entire galaxies), with the number of major BH-BH mergers highlighted with different colors for each final object. The SMBHs with a final mass of $\sim$\,$5\times10^7, 5\times10^8, 5\times10^9\,\msun$ have on average experienced 0, 2, 4 total major mergers (along the full tree), respectively.  
If we weight each major merger by the fraction of mass contributed to the final SMBH mass, such $N$ values are roughly halved.
These results shows that binary BH major mergers are rare events, hence breaking down the CLT requirement of a significantly large $N>30$.
Moreover, computing the scatter in the simulated Magorrian relation for the sub-sample with $N=0$, $N<2$, or $N>2$ leads to  very similar dispersions of 0.2\;dex,
in contrast with the CLT expectation. 
The above combined results rule out hierarchical BH mergers and related CLT averaging as dominant drivers of the $\mbh$ versus (X-ray) halo scaling relations, making gas accretion the preferred -- although not unique -- channel.
It is also important to point out that 75\% of the simulated major mergers contain or bring a significant amount of gas very near the BH (`wet' mergers), meaning that they further enhance the gas accretion channel rather than the pure SMBH binary merging.

We note that another evolutionary formation scenario could be the full growth of SMBHs at very high redshift ($z \gg 3$). However, as shown by our cosmological hydrodynamical simulations (Fig.~\ref{mergers_frac}), over 99\% of the BH mass is built at $z<3$, since at very early cosmic times mergers are extremely rare and gas accretion is inefficient.
Moreover, high-resolution simulations have still major hurdles in achieving the build-up of high-$z$ intermediate BHs with $\mbh<10^5\,\msun$, e.g., via direct gas collapse or supermassive/Pop III stars (e.g., \citealt{Latif:2013,Luo:2018,Wise:2019}), 
well before the SMBH regime. The handful of SMBHs observed at $z\gta7$ (e.g., \citealt{Mortlock:2011}) might thus represent the far tail of the random BH population; in addition, such measurements have highly uncertain masses due to the lack of direct dynamical measurements and large systematic errors in the modeling of the broad-line region.
Instead of requiring all BHs to accrete at high-$z$ via unrealistically high gas accretion and/or mergers, 
the models and data suggest a gentler, long-term co-evolution of SMBHs (\S\ref{s:CCA}) and host galaxies during the entire Hubble time.

\vspace{+0.15cm}
\subsection{Relic galaxies, galactic coronae, environment, and morphology} \label{s:env}
Interesting astrophysical laboratories for testing the growth of SMBHs are compact relic galaxies,
which appear to be the local analogs of high-redshift ($z\sim2$) `red nuggets',
the progenitors of massive ETGs
(e.g., \citealt{Ferre:2017,Werner:2018,Buote:2019}).
These compact elliptical galaxies are thought to have formed 13 Gyr ago through early dissipative processes; however, at variance with other ETGs, they managed to randomly avoid the subsequent series of merging events, remaining fairly isolated until $z=0$ with a passively evolving stellar population. They thus represent a direct probe for pure gas accretion models. 
Mrk\,1216 and NGC\,1277 are exemplary cases discovered in the local universe ($D\sim70$\,-\,90\;Mpc) with dynamically detected BH masses (\citealt{Graham:2016,Walsh:2017}).
Both compact relic galaxies are among the major outliers in $\mbh-\re$ and $\mbh-M_{\rm bulge}$ (up to 1\,dex from the mean relations). Since they have not experienced the slow $z<3$ non-dissipative phase that enlarges the stellar envelopes, their effective radii and bulge masses have remained relatively low, $\re \sim1$\,-\,$3$\;kpc and $M_{\rm bulge}\sim10^{11}\;\msun$, respectively (see Tab.~\ref{tabop}). On the other hand, such compact galaxies host SMBHs with significant masses ($\mbh\sim1$\,-\,$5\times10^9\,\msun$); indeed, they possess galactic X-ray emitting atmospheres with $\tx\sim1$\;keV. 
As they have been untouched by mergers for 13 Gyr, the only available source for accretion is the plasma atmosphere. 
Bondi hot-mode gas accretion drives too low accretion rates (\S\ref{s:Bondi}). Conversely, CCA mode is directly linked to the X-ray luminosity and can drive substantial accretion rates via the multiphase condensation of cold/warm clouds out of the persistent hot halo (\S\ref{s:CCA}) from high to low $z$, hence leading to the steady growth of BHs up to several $10^9\,\msun$ at the present time (regardless of the stellar component).
Both relic galaxies consistently fit within $\sim$1-$\sigma$ of the BH mass versus $L_{\rm x,g}$ relation, indistinguishable from other normal ETGs in our sample.

\begin{figure}[!t]
 \hskip -0.1cm
 \subfigure{\includegraphics[width=0.944\columnwidth]{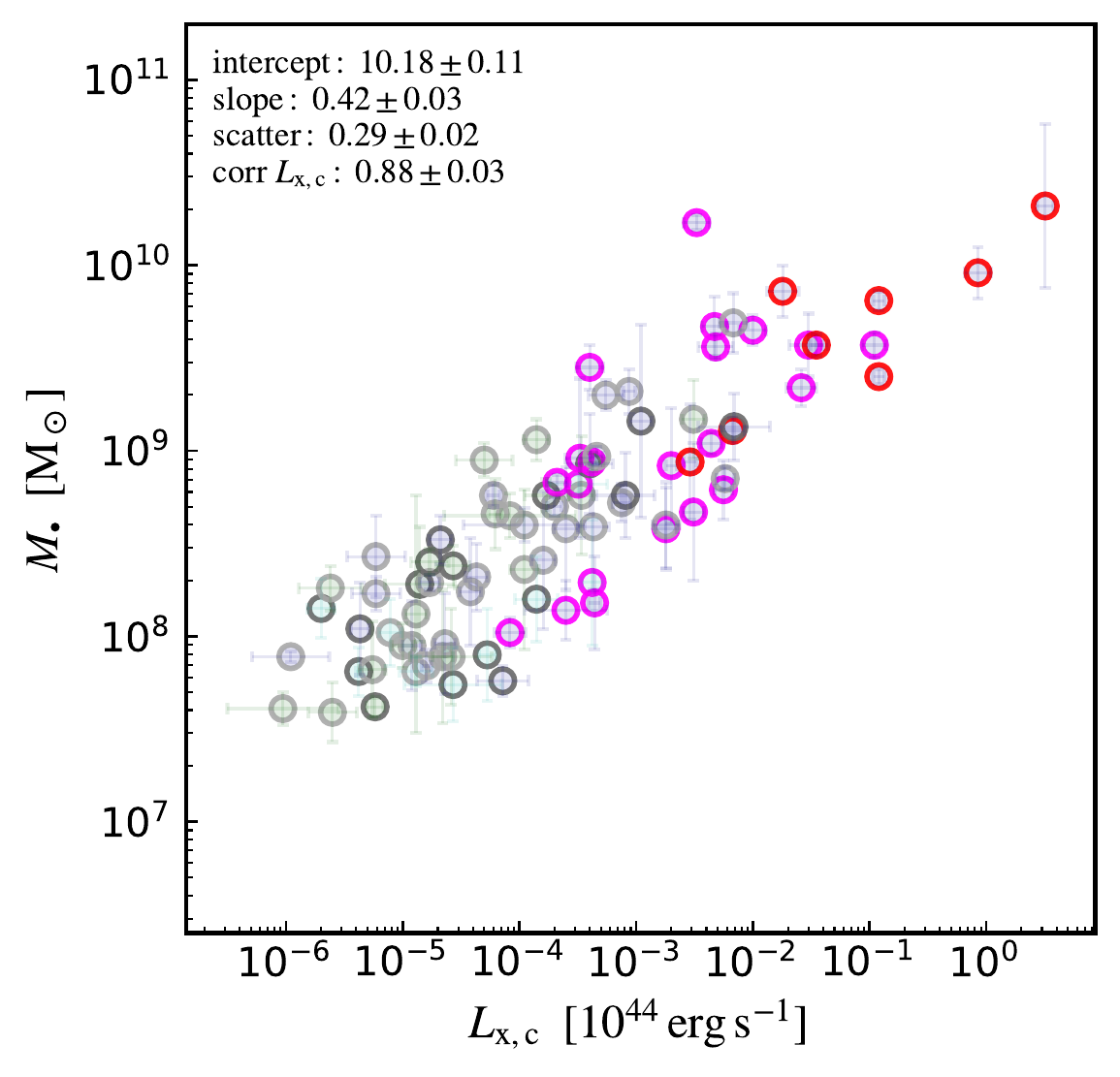}}
 \vskip -0.5cm 
 \hskip -0.11cm  
 \subfigure{\includegraphics[width=0.948\columnwidth]{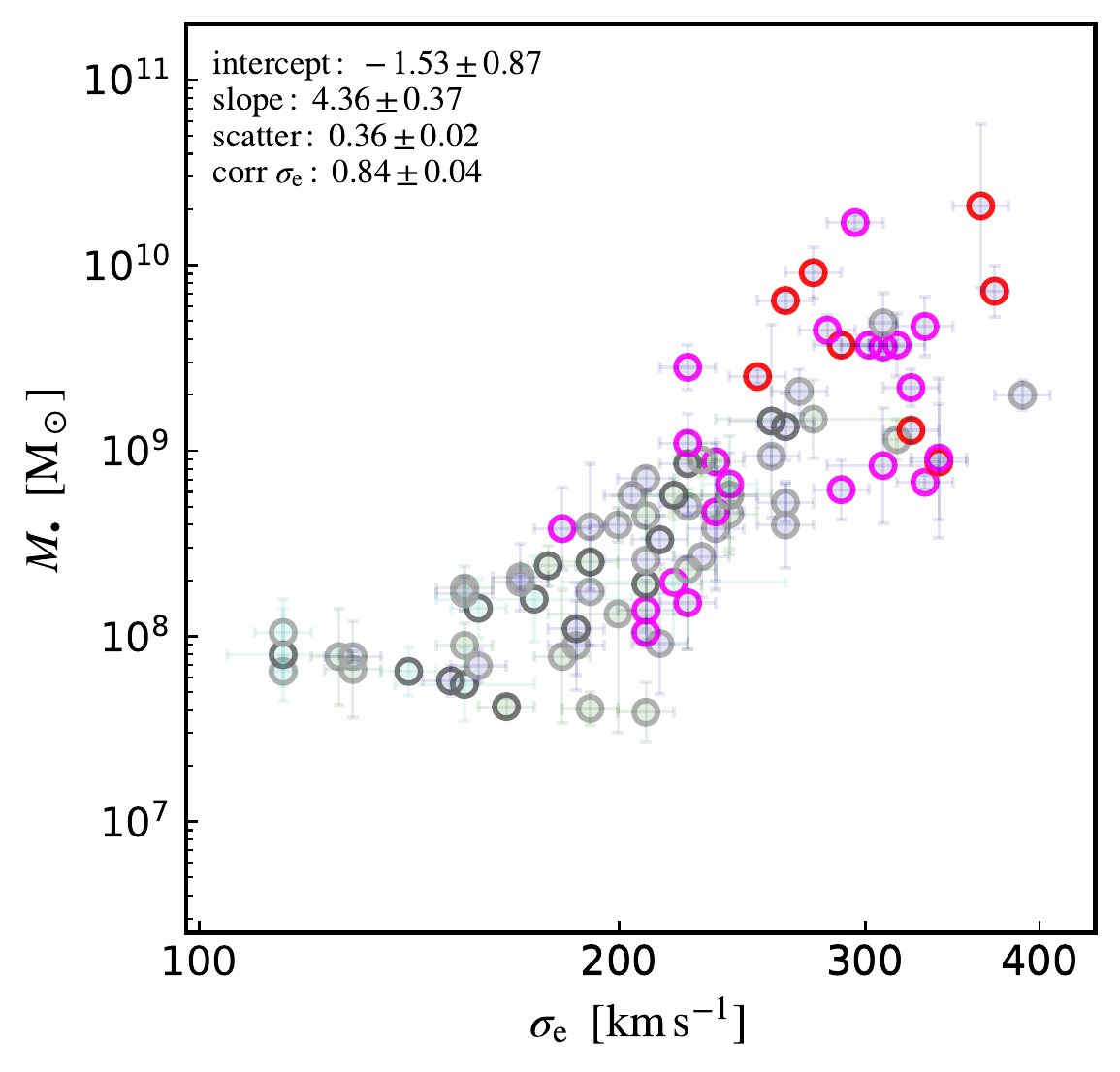}} 
 \vskip -0.4cm
\caption{BH mass vs.~core X-ray luminosity (top) and stellar velocity dispersion (bottom), differentiating brightest central group/cluster galaxies (BGGs with magenta circles; BCGs in red) and other field or isolated/satellite galaxies (gray; BFGs have bold gray circles).
Analog of Fig.~\ref{MbhLx} and \ref{Mbhop}. 
More massive SMBHs (e.g., UMBHs) reside at the center of hotter and more luminous clusters/groups (and in E types -- pale blue). Conversely low-mass BHs tend to be hosted by galaxies in poor environments (correlated with late morphological types, such as S0s/spirals -- pale green/cyan).}
\label{BGCCs}
\end{figure}

Coronae are another class of dense X-ray atmospheres shrinked to $\sim$1\,-\,5 kpc scale. They appear to be ubiquitous in ETGs (\citealt{Sun:2007}). Analogs of the above compact systems, they represent the irreducible hot X-ray atmosphere that co-evolves with the central SMBH; indeed, the relatively high gas density makes them survive stripping, evaporation, and AGN outflows.
Thereby, even central ETGs in non-CC systems or satellites in poor environments
can feed the SMBH via the condensation rain stimulated in the corona (for a very long time).
Two examples of residual coronae are NGC\,4889 and 3842 in the Coma and A1367 non-CC clusters, in which the galactic (corona) temperature seems to provide a better fit than $T_{\rm x,c}$. We note that non-CC systems might have been CCs at higher $z$, thus experiencing long periods of past macro rain.

Focusing on the environment, Fig.~\ref{BGCCs} (top panel) shows that satellite/isolated
galaxies (gray) tend to have both low BH masses and low X-ray luminosity/temperature. Such galaxies are also correlated with late-type galaxies, such as S0s and spirals (with $B/T < 0.5$), mostly being in poor environments ($N_{\rm m} < 10$).
While central galaxies sitting at the bottom of the group/cluster potential can feed from the macro plasma halo, satellite (and merging) galaxies are left to feed from the inner weather/corona, as they move at several 100\;$\kms$ relative to the external weather. If the stripping is, however, substantial, even the corona could be lost leading to undermassive BHs (perhaps the fate of dwarf galaxies as NGC\,4486A with $\mbh \lta 10^7\,\msun$).
Similarly, isolated/field\footnote{Brightest field galaxies (BFGs; bold gray circles) can be considered analog of isolated galaxies, since the other field galaxies typically have a negligible contribution to the hot halo.} galaxies are naturally starved of the large-scale gas reservoir.
BCGs (red circles) host instead the largest BH masses up to $10^{10}\,\msun$ of UMBHs, 
while BGGs (magenta) occupy the intermediate $10^9\,\msun$ SMBH locus (with NGC\,1600 as an outlier, likely due to pure stochasticity). 
About 95\% of such BCGs/BGGs are E-type galaxies ($B/T\approx1$) and essentially all in rich environments ($N_{\rm m} \sim 10$\,-\,500).
The more central the galaxy in a group/cluster potential, the more massive the X-ray halo and the larger the condensation radius (\S\ref{s:CCA}), hence, with raining clouds and filaments that can be quickly ($\lta 100$\,Myr) funneled toward the center of the group/cluster from larger BH distances. 
Conversely, as shown in \S\ref{s:mergers}, group/cluster mergers are rare, with the formation of a binary SMBH via dynamical friction onto stars requiring several Gyr%
\footnote{The timescale of BH dynamical friction onto stars is (\citealt{Combes:2002})
$t_{\rm df} \sim 2.5\,{\rm Gyr}\,(v_{\rm circ}/300\,\kms)(\re/{\rm 5\,kpc})^2(\mbh/10^8\,\msun)$.
}.
Adopting instead the optical properties as \Sige\ (bottom panel) shows that BCGs drive a major scatter and rough upward trend at the high end of the distribution (see also \S\ref{s:optuni}); thereby, the large-scale environment plays an important role in the evolution of SMBHs.
In other words, the BH physics at the micro scale is substantially linked to the macro gaseous atmosphere thermodynamics, thus creating a unified symbiotic system (which may be referred to as `BH weather') covering over $\sim$10 orders of magnitude in spatial (and temporal) scale.\\

\subsection{Scaling relations as test/calibration for cosmological simulations} \label{s:subgrid}
In \S\ref{s:res}, we presented a large set of scalings related to each of the thermodynamic quantities of hot halos, including the X-ray luminosity, temperature, electron density, pressure, thermal energy, and gas/total mass, both as univariate and multivariate correlations.
Such a study is vital (i) to advance our observational constraints on the observable universe; (ii) to test/falsify models of BH accretion/growth; (iii) to carefully calibrate and check the predictive power of cosmological simulations.
While the first two points have been amply dissected in the previous sections, let us focus on the last one.

Granted the relevant progresses in the past decade, state-of-the-art cosmological simulations still lack substantial resolution to implement all the baryonic physics in a self-consistent manner, thus having to rely on a simplified analytic approach known as `subgrid' modeling (e.g., \citealt{Sijacki:2007,Rasia:2015,Schaye:2015,Tremmel:2017,Weinberger:2018}). This is particularly challenging for AGN feeding/feedback physics as it has to operate on scales spanning 10\,dex. Most -- if not all -- subgrid models tend to calibrate the impact of AGN feedback and related SMBH growth on the optical scalings, in particular the $\mbh-\sige$ or Magorrian relations. However, \S\ref{s:optuni}\,-\,\ref{s:xuni} show that the optical scalings have significantly larger scatter than the X-ray halo properties and tend to steepen at the high-mass end due to UMBHs. 
As discussed in \S\ref{s:CCA}, stars are indeed the final by-product of the top-down multiphase gas condensation that occurs throughout the cosmic time. Instead, the plasma halos are the primordial reservoir out of which the raining matter will feed recurrently the central SMBHs. 

It is crucial to calibrate first the AGN feedback/feeding parameters of cosmological simulations on a few selected X-ray halo relations (e.g., $\mbh-\tx$). 
The next step would be to check the predictions for all other thermodynamic properties (e.g., $\mbh-M_{\rm gas}$) and/or stellar counterparts. 
The calibration should be as minimal as possible, in order to avoid overfitting. 
We note that calibrating first on $\mbh - M_{\rm tot}$ might lead to a large parameter-space uncertainty; as shown in \S\ref{s:mergers}, the total (DM dominated) potential is likely a secondary element of the BH growth, while it is primary to model the hot gas physics first (e.g., galaxies without hot halos appear to show significant decorrelation from BH properties, regardless of DM mass).
The dozen scaling relations presented in this paper should allow for multiple independent calibration tests to check the robustness of the subgrid model (see the recent work by \citealt{Bassini:2019} for such an example). In parallel, subgrid prescriptions should gradually move toward physically-motivated -- rather than fitting-oriented -- models. E.g., a minimal subgrid approach based on a realistic AGN raining/outflow self-regulation is provided in \citet{Gaspari:2017_uni}. Needless to say, the same scaling relations can be readily applied in purely SAMs of galaxy (e.g., \citealt{Hirschmann:2012}) and cluster (e.g., \citealt{Flender:2017}) formation.\\

\subsection{Caveats, selection effects, and future prospects} \label{s:cav}
Before concluding, it is important to remark the limitations of the current work, selection effects, and expansions worth pursuing in future investigations.

The selection is first done in optical, as it is based on the available direct dynamical BH masses (\S\ref{s:data}). 
Since optical telescopes have to resolve the BH influence region, this limits our analysis to systems in the local universe ($z<0.04$). Given the tight co-evolution of SMBHs and hot halos retrieved in simulations (e.g., \citealt{Bassini:2019}), we expect the retrieved $\mbh$ -- X-ray halo correlations to hold at least up to $z\sim2$ with similar scatter, albeit with steeper slope given the relatively slower growth of macro-scale cluster halos (\citealt{Bassini:2019} find a $\mbh - \tx$ slope increase of $2\times$ from $z=0$ to 2).

Regarding the X-ray selection, we only included direct BH masses with X-ray halo detections, except for a handful of contaminated systems (\S\ref{s:xvar}). 
The final sample of 85 systems includes a very diverse mix of morphological, dynamical, and environmental types (Tab.~\ref{tabop}-\ref{tabx}), hence we do not expect major biases in these directions. 
The vast majority of the galaxies with direct BH masses without X-ray detections are low-mass galaxies, which are expected to fall below a halo temperature of 0.2 keV and $\mbh < 3\times10^7\,\msun$, i.e., the IMBH regime. 
This is also related to $\lx <10^{38}$\;erg\,s$^{-1}$, which current X-ray instruments have severe difficulties in constraining.
Extrapolating the $\mbh - \tx$, these IMBHs should reside in the UV regime (e.g., NGC\,2787 and NGC\,7582 at 0.18\,keV show already signs of being unconstrained in X-rays), which is not sufficiently covered by any current telescope.
This IMBH regime corresponds to low-mass galaxies with $M_\ast < 5\times10^9\,\msun$. 
Bulgeless disk galaxies have been shown to start to deviate from the mean optical scalings (\citealt{Kormendy:2013}), and might present a challenge for those involving hot halos too. 
Such departures can be due to the low binding energy of small halos and thus higher susceptibility to AGN feedback evacuation.
Moreover, low-mass galaxies experience rotation as dominant physics, which changes the formation and frequency of the condensation rain (\citealt{Gaspari:2015_cca}), reducing the accreted gas mass via the stronger centrifugal barrier (as a function of the turbulent Taylor number ${\rm Ta_t}\equiv v_{\rm rot}/\sigma_v$, linked to $B/T$; e.g., \citealt{Juranova:2019}).
Overall, given the above effects, we envision an increased scatter and steeper/lower slope/normalization in the correlations including such unprobed low-mass regime.
Further, rapid rotation triggers different accretion mechanisms such as disk shearing and bar/ring instabilities (\citealt{Hopkins:2010}); it may be thus worthwhile to explore other correlations, as that between $\mbh$ and the spiral-arm pitch angle (\citealt{Davis:2017}).

Another relevant problem moving toward IMBHs and low-mass galaxies/halos is the fact that X-ray binaries (XBs) start to swamp the diffuse hot gas emission. As shown by the stacking of the ROSAT X-ray emission of over 250\,000 galaxies (\citealt{Anderson:2015}; their Fig.~5), at $\log M_\ast/\msun < 10.7$, XBs dominate the X-ray luminosity. 
Similarly, X-ray AGN/quasars with high Eddington ratio ($\gta 0.05$) can introduce significant uncertainty in the X-ray halo detection, especially in Seyfert galaxies (conspicuously contaminating AGN can even be hosted in BCG, as in NGC\,1275).
Active starbursts drive plasma energized via stellar winds and supernovae, thus creating a marked soft X-ray excess (\citealt{LaMassa:2012}; e.g., NGC\,7582) and often require uncertain 2-$T$ models (the hotter component tied to the starburst/spiral arms, the cooler component to the diffuse gas; \citealt{Li:2013_halo}).
Interestingly, for one of the closest dwarf galaxies M32 (NGC\,0221; $D\simeq0.8$\,Mpc), it is possible to put a crude constraint on $\log \lx/{\rm 10^{44}\,erg\,s^{-1}} \approx -7.9 \pm 0.2$ (\citealt{Boroson:2011}), which is consistent within 1-$\sigma$ from the extrapolated mean $\mbh - L_{\rm x,g}$ (Fig.~\ref{MbhLx}); however, such measurement remains tentative considering the 2\,dex larger $\lx$ by XBs (plus ABs and CVs).

While the X-ray properties are in general a more robust indicator of the BH mass growth compared with the stellar properties, extreme events driving shocks can cause the hot halo to become overheated up to $R_{500}$. 
In particular, ongoing wet major mergers can drive strong shocks (${\rm Mach > 10}$) over large regions of the group/cluster (e.g., \citealt{Vazza:2009_shock}), thus temporarily biasing the halo $\tx$ toward larger values. 
Violent AGN jet cocoon shocks and hotspots lead to analogous bias (enhancing $\lx$ too via major density compressions), albeit being contained mostly in the CGM/core region. 
We have retained in our sample the large majority of such systems (e.g., NGC\,5128, NGC\,1316, NGC\,7626, UGC\,12064, NGC\,1399), 
except for the dramatic case of 3C405 (Cygnus\,A), whose X-ray emission is fully dominated by the bipolar lobes driven via the 100\,kpc FRII jets.
All the above objects increase the scatter. To achieve tighter $\mbh$ -- X-ray halo correlations, future studies may thus aim to mask the strong anisotropic merger/jet features (e.g., as done for NGC\,3801) or to separate the shocked medium from the diffuse halo via a 2-$T$ fit analysis (e.g., as carried out for NGC\,2110).

Given the above limitations, future investigations should be aimed to better probe the low-mass, disk ($B/T \ll 0.3$) regime, in terms of direct dynamical BH mass measurements (e.g., by using ALMA sub-arcsec resolution to detect circumnuclear molecular disks; \citealt{Barth:2016}), while moving toward the very soft X-ray (e.g., \Athena) or UV emission regime (e.g., HST/COS). 
At the same time, the community should aim to expand the \textit{direct} BH mass measurements in the BCGs of massive clusters with $T_{\rm x,c}> 2$\;keV, of which there are only a handful available at the present. 
The very-high-end regime is important to fill: can the universe develop UMBHs with masses in excess of 20 billion $\msun$ or will a saturation develop? 
Enlarging the cluster sample is also crucial to test whether the scatter of the (more uncertain) $R_{500}$ scalings remains smaller than that of the optical scalings.
In terms of evolution, we should aim to push BH mass detections at higher $z$, understanding whether the presented $\mbh$ -- X-ray scaling relations develop already with the formation of the first CC clusters (e.g., \citealt{McDonald:2017}).

Future X-ray observations -- leveraging next-generation telescopes like 
\Athena\footnote{\url{https://www.the-athena-x-ray-observatory.eu}} (\citealt{Nandra:2013}, \citealt{Ettori:2013}),
XRISM\footnote{\url{https://heasarc.gsfc.nasa.gov/docs/xrism}} (\citealt{Kitayama:2014}),  
eROSITA\footnote{\url{https://www.mpe.mpg.de/eROSITA}} (\citealt{Merloni:2012}), 
and possibly AXIS\footnote{\url{http://axis.astro.umd.edu}} (\citealt{Mushotzky:2019})
and {\it Lynx}\footnote{\url{https://www.lynxobservatory.com}} (\citealt{Gaskin:2018}) --
should aim to improve the exposure, not only of hot halos in LTGs, but also in the outer regions of the host group and cluster (e.g., as carried out with the X-COP program; \citealt{Eckert:2019}). Expanding the sample toward isolated galaxies ($N_{\rm m} =1$; e.g., NGC\,7457) and compact relic galaxies is also crucial, albeit very challenging, requiring very deep X-ray observations (with exposure $\gg 300$\,ks); however, such systems are excellent laboratories to test the differences with BCGs/BGGs and disentangle the gas accretion versus merger channel.
Further, achieving a finer radial profiling of the hot halo properties for the whole sample would help to better elucidate the variation of the BH scaling relations over more homogeneous extraction radii. 
A more accurate determination of the X-ray half-light radius would also facilitate the comparison with $\re$, and better determine the virialization of hot halos in the multivariate xFP.
Finally, direct detections of total masses (e.g., via lensing) and other gas mass properties as $Y_{\rm sz}$ (e.g., via CMB) would help to relax our derived properties assumptions and better probe the large scales, in particular at and beyond $R_{500}$.

\vspace{+0.3cm}
\section{Conclusions} \label{s:conc}
We thoroughly probed the thermodynamical correlations between SMBHs and X-ray emitting plasma halos for 85 systems (Tab.\,\ref{tabsum}), through a large literature search and Bayesian analysis approach (\S\ref{s:res}), as well as from a theoretical perspective testing different theories and mechanisms (\S\ref{s:disc}). We probed univariate and multivariate (FPs) correlations over three major radial extraction regions, galactic, group/cluster core, and outskirt regions ($r\sim$\,0.03, 0.15, 1.0\,\R500), including a wide range of systems, spanning from massive galaxies to isolated S0s and spiral galaxies.
The main results are as follows.

\begin{itemize}
\item
We found key novel BH mass versus X-ray halo (univariate) correlations, 
first in terms of fundamental variables,
with the tightest relation being the $\mbh - \tx$ (slope 2.1\,-\,2.7), followed by $\mbh - \lx$ (slope 0.4\,-\,0.5) 
The intrinsic scatter is significantly low, $\epsilon = 0.2$\,-\,0.3 dex, in particular adopting the 
galactic/core region for the temperature/luminosity.
The correlation coefficients are in the very strong regime (corr\,$\gta0.9$). 
X-ray halos are thus excellent indicators for SMBH masses hosted by diverse types of galaxies (BCGs/satellites or ETGs/LTGs) with luminosities spanning over 6 dex.
On average, a 0.8\;keV or $10^{41}\;\es$ hot halo is expected to host an SMBH with $10^9\,\msun$.

\item 
We compared the X-ray/plasma scalings with the optical/stellar counterparts ($\mbh - \sige$, $\mbh - L_K$, Magorrian), finding that the stellar scalings have significantly larger scatter, $\epsilon \approx 0.4$\,-\,0.5.
This and the pairwise residual correlation analysis
suggest a more fundamental role of gaseous halos in growing and tracing SMBHs. 
The UMBH regime is well fitted via the X-ray scalings (including the cluster/group halo), while it increases the scatter in the stellar scalings; indeed, unlike satellites and field/isolated galaxies, central BCGs/BGGs can efficiently feed from the macro plasma halo.
A median $10^9\,\msun$ SMBH is hosted by galaxies with $\sige\approx260\,\kms$, $L_K\approx4\times10^{11}\,L_\odot$, and $M_{\rm bulge}\approx3\times10^{11}\,\msun$.

\item 
We presented new BH mass (univariate) correlations in terms of the composite/derived X-ray variables: 
$\mbh - M_{\rm gas}$ (slope 0.5\,-\,0.6) is among the tightest and strongest correlations ({corr\,$\sim$\,0.9}), in particular within the group/cluster core region, with half the scatter of the Magorrian; remarkably, a $10^9\,\msun$ SMBH is hosted by a hot halo with same $M_{\rm gas,g}\approx10^9\,\msun$.
$\mbh-P_{\rm x}$ and $\mbh-n_{\rm e}$ have significantly larger scatter (especially in the galactic region), the latter approaching a quasi linear scaling.
$\mbh - Y_{\rm x}$ is instead stable (slope $\simeq 1/2$) and tight ($\epsilon \simeq 0.3$) over all regions, and can be leveraged by SZ observations up to \R500 (hot halos with $Y_{\rm x,c}\approx6\times10^{58}\,{\rm erg}$ typically host $10^9\,\msun$ SMBHs).

\item
Besides the trivial $\mbh-M_{\rm tot, 500}$ inherited from \Tx,
the total mass $M_{\rm tot}$ is less correlated with BH mass than most gas relations, showing superlinear slopes ($\beta\simeq1.2$\,-\,1.4) and significant scatter comparable to that of the $\mbh - \sige$ (a better alternative is to use the binding energy $\mbh \propto M_{\rm tot}\phi$). 
The $\mbh-f_{\rm gas}$ unveils a tighter connection with the BH mass (as for $M_{\rm gas}$),
establishing a linear relation ($\beta \simeq 1$) over all regions.
Under our assumptions, this and above gas scalings
suggest that $\mbh$ correlates better with X-ray halos rather than DM halos.

\item
The multivariate correlation analysis of the optical (oFP) and X-ray fundamental plane (xFP)
shows that, while the stellar observables can be well described via the virial theorem (plus homology) in a multivariate fundamental plane ($\sige^2\propto L_K/\re$), X-ray halos are better described by the univariate scalings (e.g., $L_{\rm x}\propto T_{\rm x}^4$), with major deviations from self-similarity and primarily driven by plasma physics, rather than a pure virialization. 
The main difference between the optical and X-ray component resides in the mass-to-light ratios ($M_\ast/L_K \propto T_\ast^{0.2}$ vs.~$M_{\rm tot}/L_{\rm x} \propto T_{\rm x}^{-3}$).  

\item
Given the existence of the oFP/xFP, multivariate correlations between $\mbh$ and (statistical) combinations of X-ray/optical properties leads to a minor scatter improvement; the above univariate correlations (especially involving \Tx\ and $\sige^2$) are a better minimal interpretation of BH growth. 
The partial correlation analysis of the mixed X-ray and optical properties (e.g., $\mbh-\tx-\sige^2$ and $\mbh-\lx-L_K$) corroborates that hot halos are more tightly linked to $\mbh$ rather than the stellar component.

\item
We tested the three major channels for $\mbh$ growth: hot gas (Bondi-like) accretion, chaotic cold accretion (CCA), and mergers. Hot/smooth accretion models are rejected by the data being anti-correlated with plasma-halo properties and inducing too low feeding levels. Hierarchical binary BH mergers are sub-dominant during most of the cosmic time, with CLT predictions and assumptions inconsistent with our data and cosmological simulation; the latter shows that major mergers are rare and do not substantially decrease the scatter at the high-mass end.
The X-ray scaling relations and simulations indicate a key role of CCA (with predictions consistent with the observed BH mass growth, halo dependence, and condensation radius), whose rain is induced by the cooling of the turbulent plasma halo (tied to \Lx/\Tx\ and $M_{\rm gas}$) and balanced by the AGN feedback cycling through the Hubble time. 

\item Relic galaxies (descendant of high-$z$ red nuggets) and galactic coronae, both having sizes of a few kpc, are vital astrophysical laboratories to test the SMBH growth via purely gas accretion, since the former have been untouched by mergers for 13 Gyr and the latter can survive extreme or poor environments, thus enabling the recurrent inner feeding via gas condensation. On the other hand, galaxies at the center of rich environments (large $N_{\rm m}$ or $B/T$, and early types) can feed from the macro-scale rain forming UMBHs. 

\item
The new X-ray halo correlations can be leveraged to calibrate and test large-scale cosmological simulations with AGN feeding/feedback subgrid schemes, as well as SAMs of galaxy and cluster formation. These can be now carried out from multiple angles not only in terms of $\lx$ and $\tx$ (plasma studies), but also via gas pressure/thermal energy (SZ studies), gas fractions (cosmological studies), and total/DM masses (lensing studies).

\end{itemize}

\vspace{-0.07cm}
In sum, the BH physics at the micro scale is tightly linked to the macro gaseous atmosphere thermodynamics, thus creating a unique symbiotic system (which we refer to as `black hole weather') spanning over $\sim$10 orders of magnitude in space and time (as predicted by first-principle arguments; \citealt{Gaspari:2017_uni}).
This study highlights the importance of  combining both observational analysis (and exquisite single studies produced in the literature) with a deep theoretical/numerical interpretation.
We live exciting times for multiphase SMBH feeding/feedback observations and modeling,
with new missions directly probing and discovering the multiphase halo rain and AGN outflows ({\Chandra}, XMM, ALMA, MUSE, MUSTANG-2, LOFAR, SOFIA)
and with next-generation telescopes that might be able to open new windows on multi-messenger processes related to SMBHs and gaseous halos (\Athena, eROSITA, XRISM, EHT, JWST, SPT-3G, LSST, SKA, LISA). These programs will lead to order-of-magnitude leaps in sample size, sensitivity, and accuracy, and thus, in our understanding of BHs and their co-evolution with the host galaxies, groups and clusters of galaxies.

\section*{\bf \scriptsize Acknowledgements}
\noindent
M.G.\ is supported by the \textit{Lyman Spitzer Jr.}~Fellowship (Princeton University) and by NASA \textit{Chandra} grants GO7-18121X, GO8-19104X, and GO9-20114X. 
S.E.\ acknowledges contribution from ASI 2015-046-R.0 and ASI-INAF 2017-14-H.0.
L.B., S.B., and E.R acknowledge the ExaNeSt and EuroExa projects funded by the EU Horizon 2020 program (grant 671553 and 54337) and financial contribution from ASI-INAF 2017-14-H.0.
S.B. acknowledges financial support from PRIN-MIUR 2015W7KAWC and the INFN INDARK grant.
S.D.J.\ is supported by NASA \textit{Hubble} Fellowship (HST-HF2-51375.001-A).
H.-Y.K.Y.\ acknowledges support from NASA ATP NNX17AK70G and NSF AST\,1713722.
F.T.\ is supported by the `Programma per Giovani Ricercatori \textit{Rita Levi Montalcini}' (2014).
HPC resources were in part provided by the NASA/Ames HEC Program (SMD-18-7305/7320/7321/7251). 
We are thankful for the `Multiphase AGN Feeding \& Feedback' conference 
({\small \url{http://www.sexten-cfa.eu/event/multiphase-agn-feeding-feedback}})
held by the Sexten Center for Astrophysics in Sesto (BZ, Italy),
which has stimulated insightful interactions. 
We thank G.\,L.~Granato, F.~Gastaldello, M.~Rossetti, N.~Caplar for constructive feedback.
This work made use of public databases: HyperLEDA (\citealt{Paturel:2003})\footnote{\url{http://leda.univ-lyon1.fr}}, SAO/NASA Astrophysics Data System (ADS)\footnote{\url{http://www.adsabs.harvard.edu}}, and NASA/IPAC Extragalactic Database (NED)\footnote{\url{https://ned.ipac.caltech.edu}}. 

\bibliographystyle{biblio}
\bibliography{biblio}

\begin{appendix}
\renewcommand{\theHfigure}{A\arabic{figure}} 
\renewcommand{\theHtable}{A\arabic{table}}
\section{A. Complementary scalings} \label{a:uextra}
We include here additional univariate and multivariate scalings complementary to the discussion in \S\ref{s:res}\,-\,\ref{s:disc}.
Such scalings can be useful for other studies that intend to probe and/or calibrate their parameters on a wider range of observational constraints. Being secondary and mentioned in the above sections, we do not dissect each scaling, although we provide insights in each caption. 
The interested reader can assess the quantitative properties of the correlations from the inset, listing the mean and standard deviations for all the posterior distributions (\S\ref{s:corr}). 
Additional permutations of the scalings can be easily computed by combining the below correlations and/or those in Tab.~\ref{tabsum}.

\vspace{-0.42cm}
\begin{figure*}[!h]
\hskip -0.375cm
\subfigure{\includegraphics[width=0.358\columnwidth]{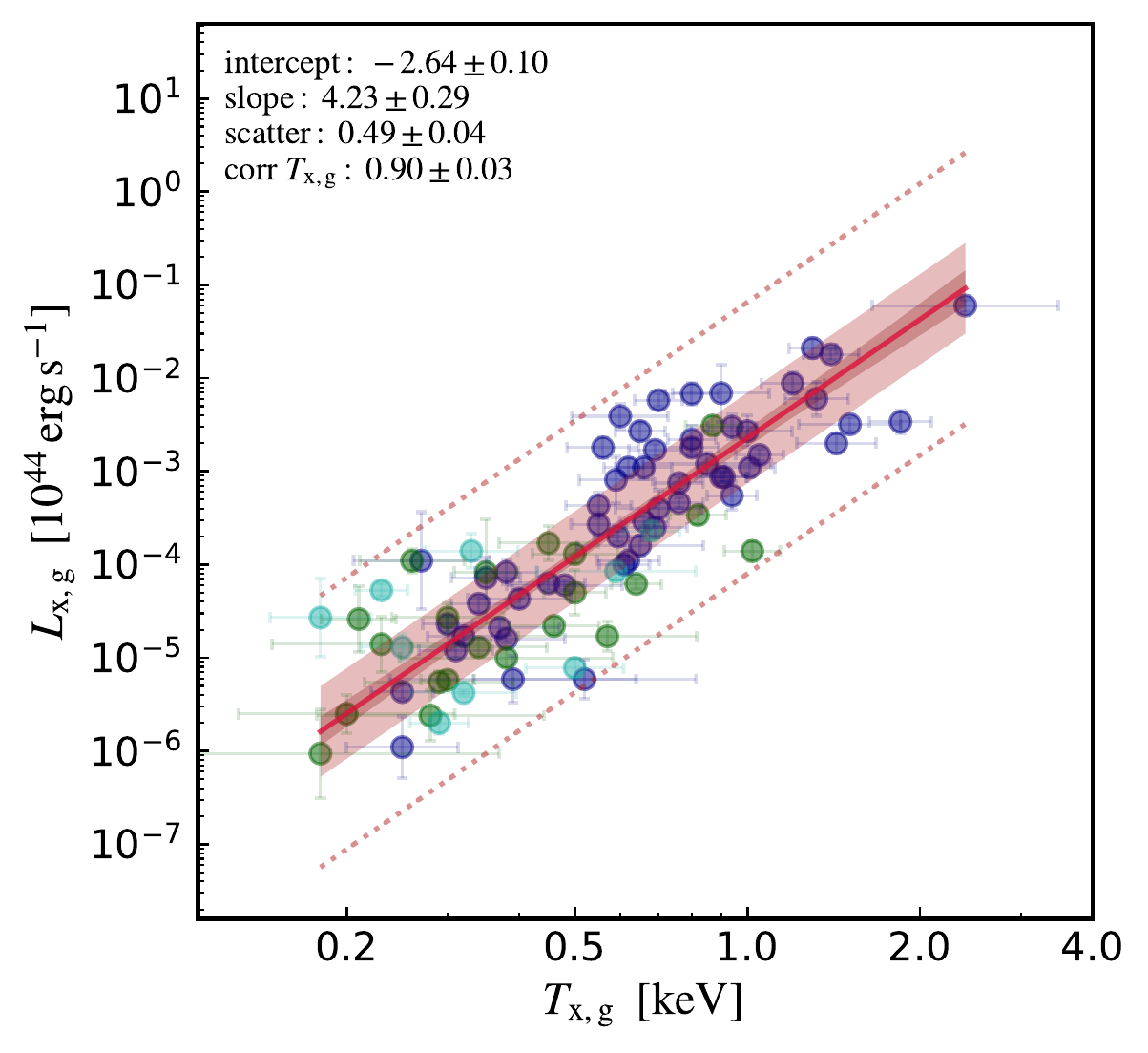}}
\hskip -0.22cm
\subfigure{\includegraphics[width=0.35\columnwidth]{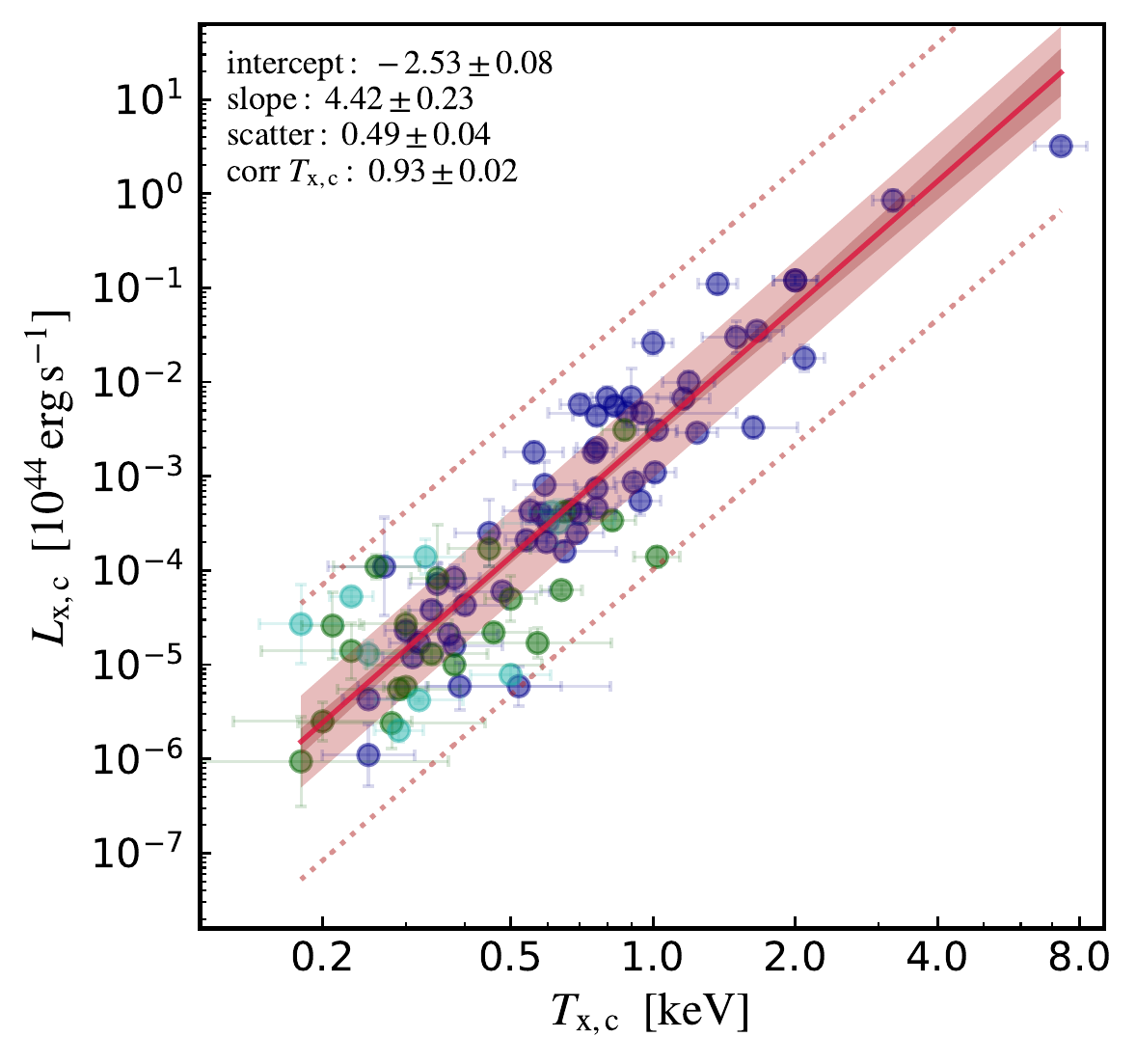}}
\hskip -0.22cm
\subfigure{\includegraphics[width=0.35\columnwidth]{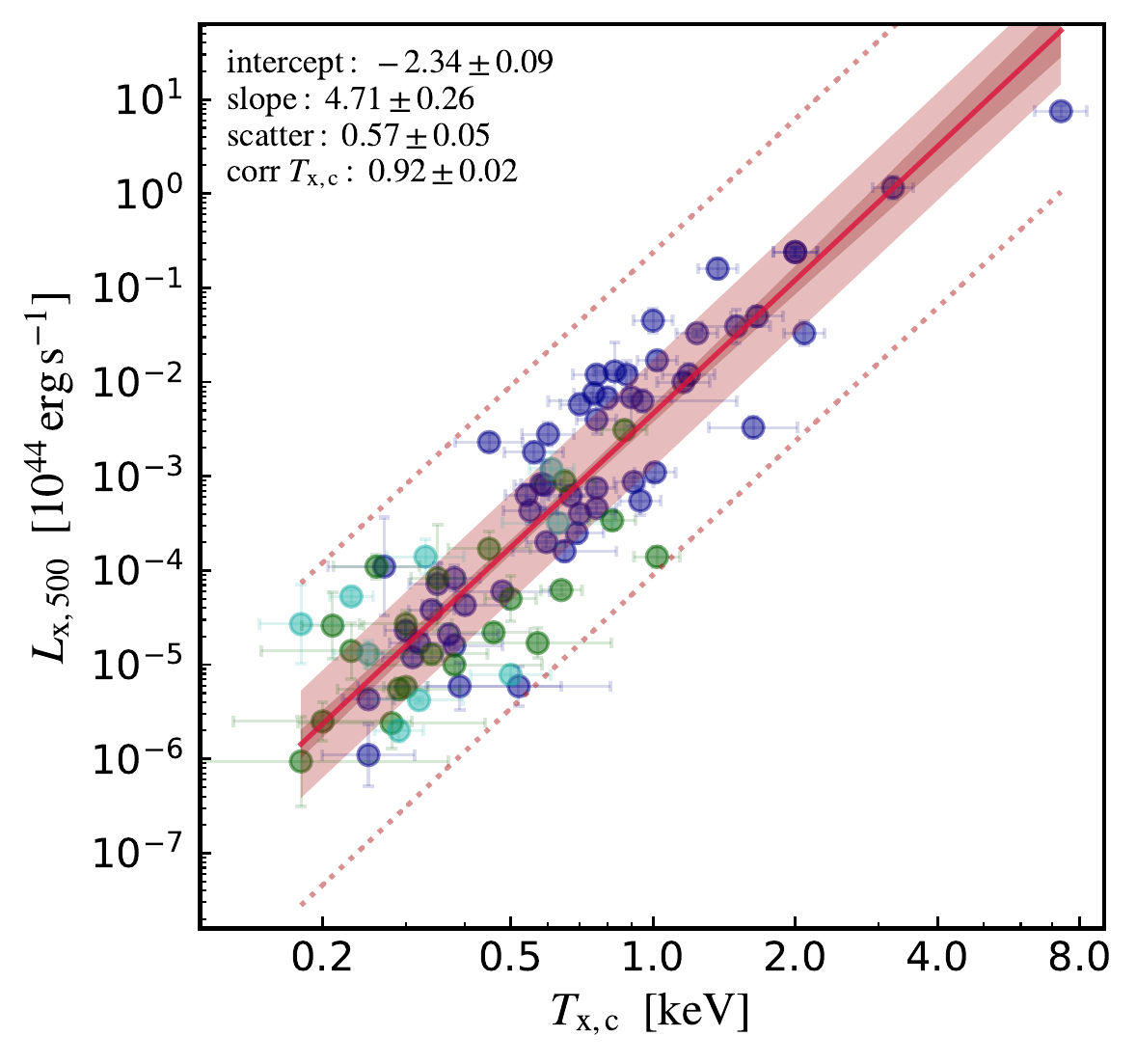}}
\vskip -0.4cm
\caption{X-ray luminosity vs.~temperature within the galactic/CGM \Rxg\ (left), group/cluster \Rxc\ (middle), and $R_{500}$ scale (right; note that $T_{\rm x, 500}\approx T_{\rm x,c}$). The inset lists the posterior mean and errors for the intercept, slope, intrinsic scatter (1-$\sigma$ interval plotted as a filled light red band, 3-$\sigma$ as dotted lines), and correlation coefficient. The solid red line and inner dark band show the mean fit and related 15.87\,--\,84.13 percentile interval.
The blue, green, and cyan points correspond to E, S0, and S morphological types, respectively.
The steep, non-self-similar scalings are consistent with that found by other works including a significant number of isolated galaxies down to the 0.3\;keV regime (e.g., \citealt{Kim:2015,Goulding:2016,Babyk:2018}). 
}
\vskip -0.1cm 
\label{LxTx}

\hskip -0.375cm
\subfigure{\includegraphics[width=0.358\columnwidth]{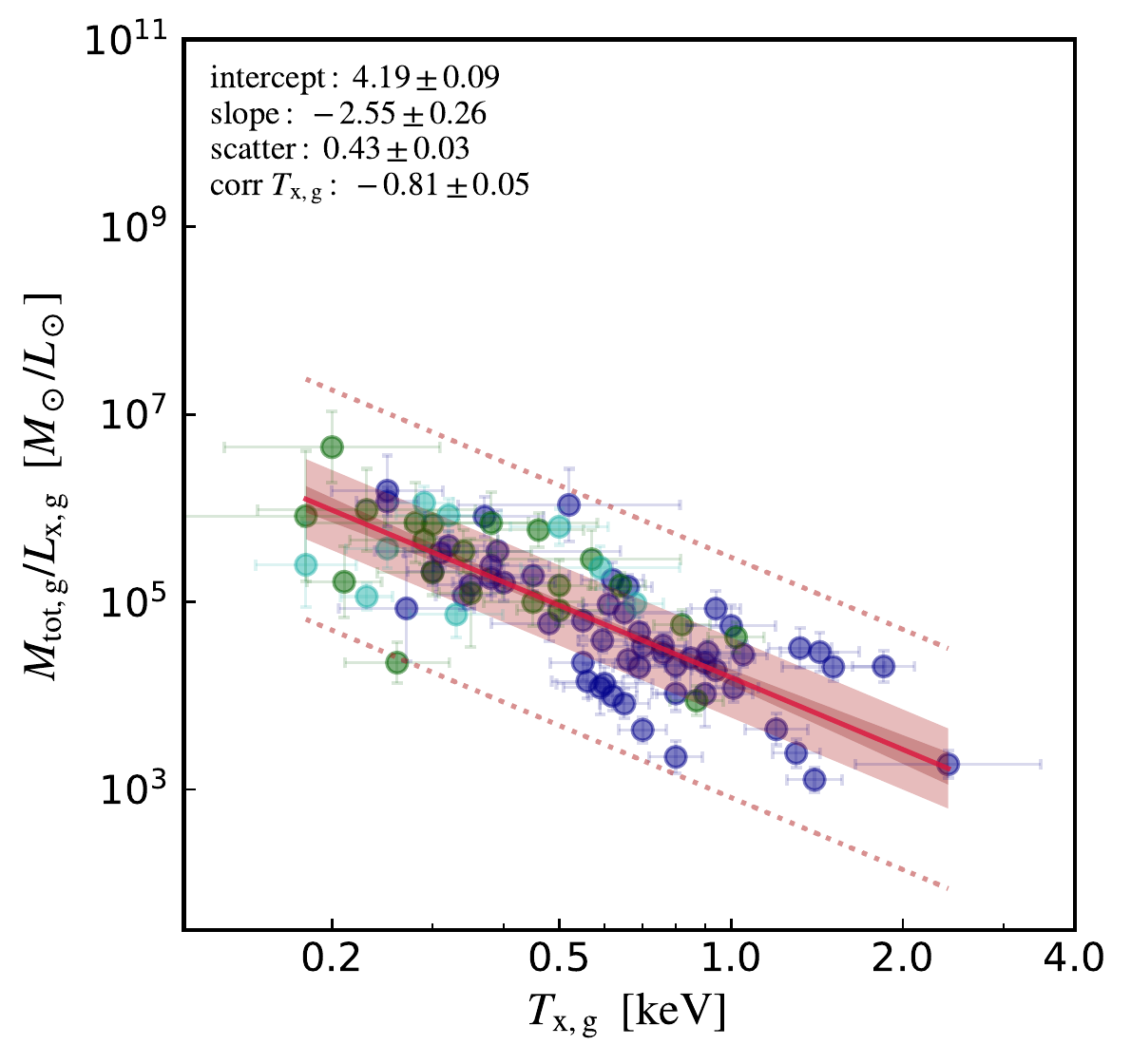}}
\hskip -0.22cm
\subfigure{\includegraphics[width=0.35\columnwidth]{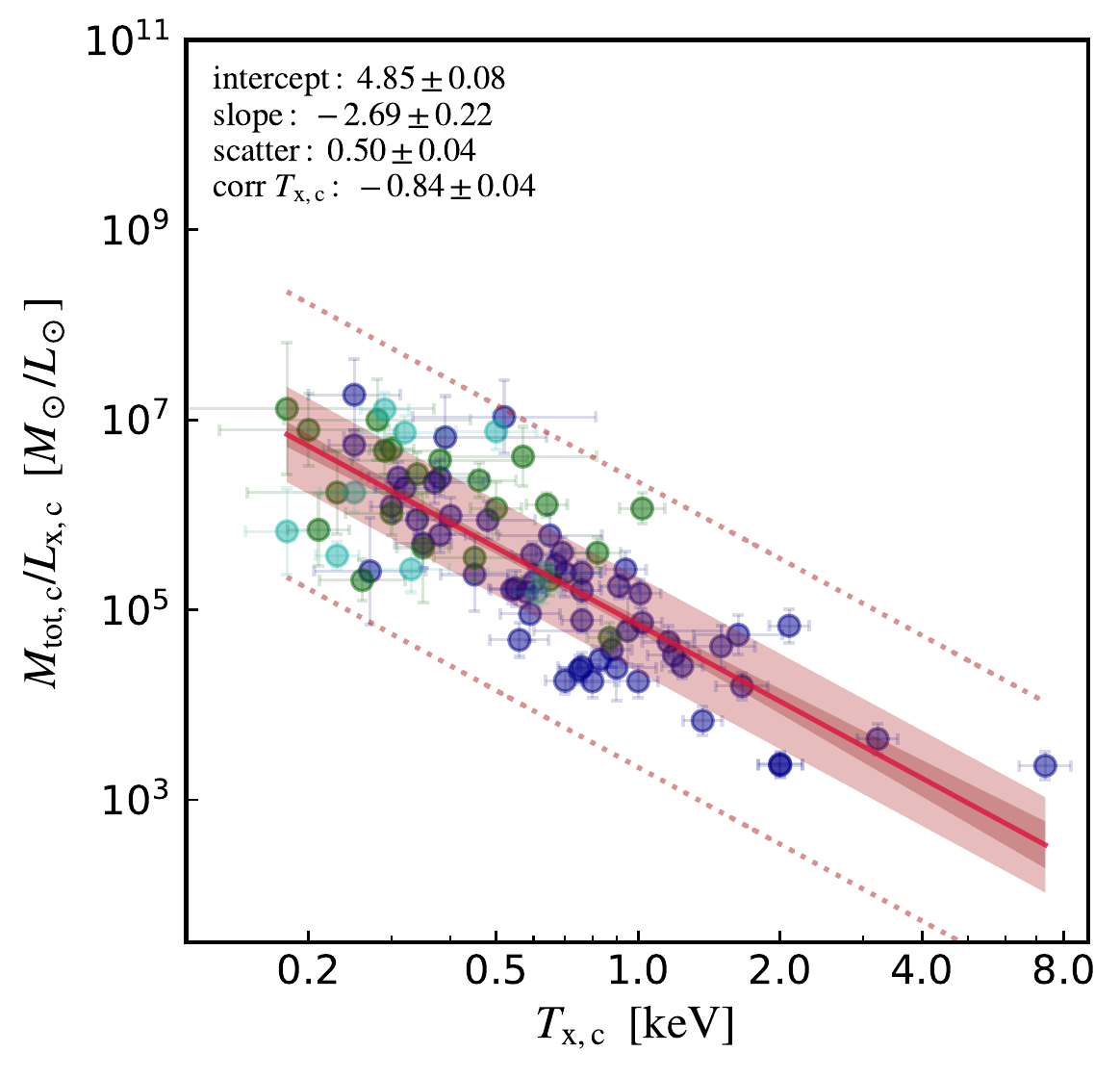}}
\hskip -0.22cm
\subfigure{\includegraphics[width=0.35\columnwidth]{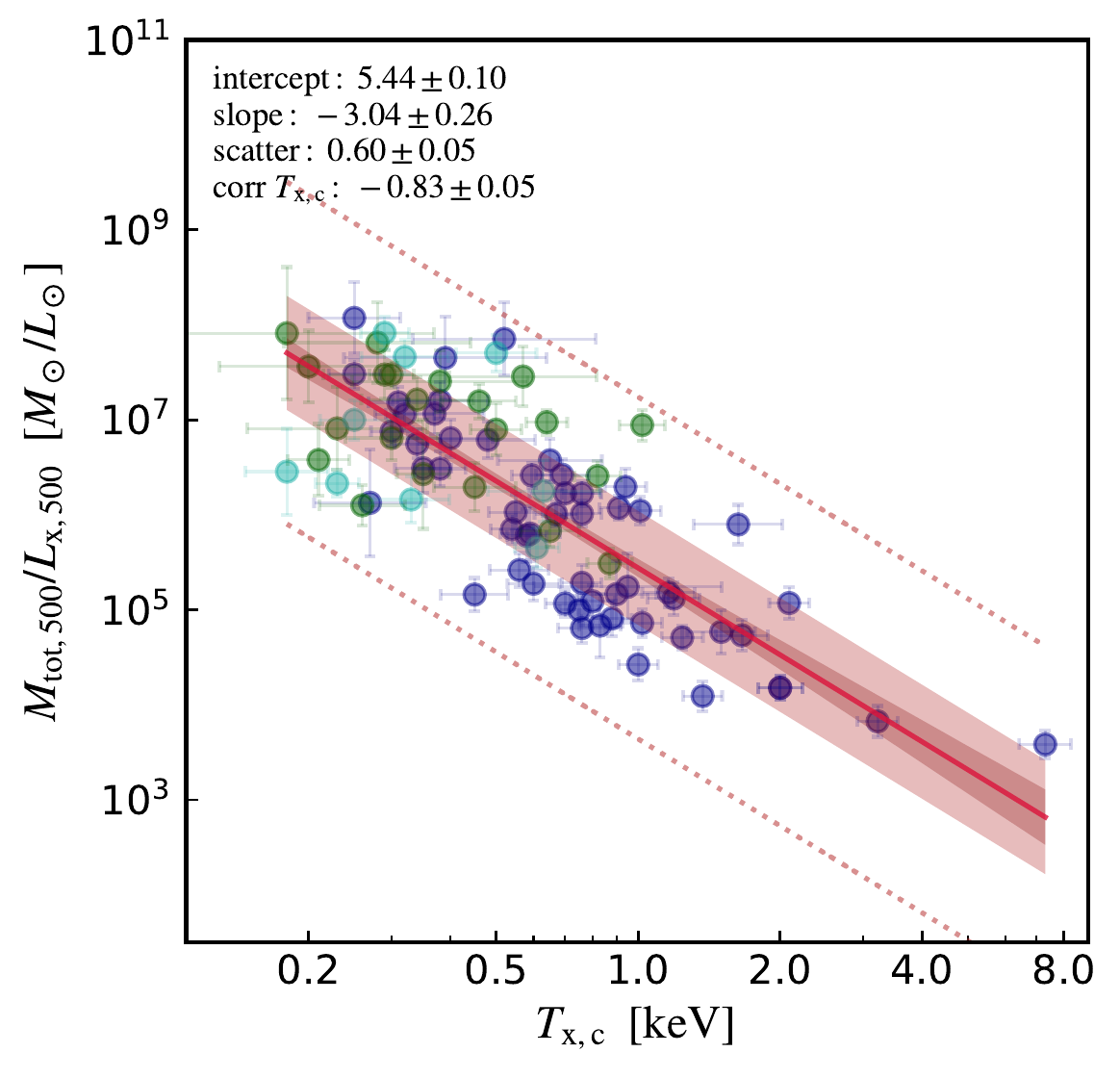}}
\vskip -0.4cm
\caption{Total mass (dominated by DM) to X-ray light ratio vs.~X-ray temperature contained within \Rxg\ (left), \Rxc\ (middle), and $R_{500}$ (right). Analog of Fig.~\ref{LxTx}, which we can also use to convert to an average $M_{\rm tot} \propto  \tx^{1.7}$ and $\lx \propto M_{\rm tot}^{2.5} $ scalings; such relations are steeper than the self-similar expectations (3/2 and 4/3, respectively), particularly in the latter case (as found in other studies, e.g., \citealt{Anderson:2015}). We note that, due to the larger photon counts, the majority of previous investigations on X-ray scalings focused on massive/central BCGs (thus leading to shallower slopes for relations involving \Lx\ and $M_{\rm gas}$), while here we include a large variety of lower mass objects, such as gas-poor galaxies, isolated Es, S0s, and disky LTGs. See \S\ref{s:xFP} for further discussion on the mass-to-light ratios.
}
\label{MtotdLx}

\hskip -0.375cm
\subfigure{\includegraphics[width=0.35\columnwidth]{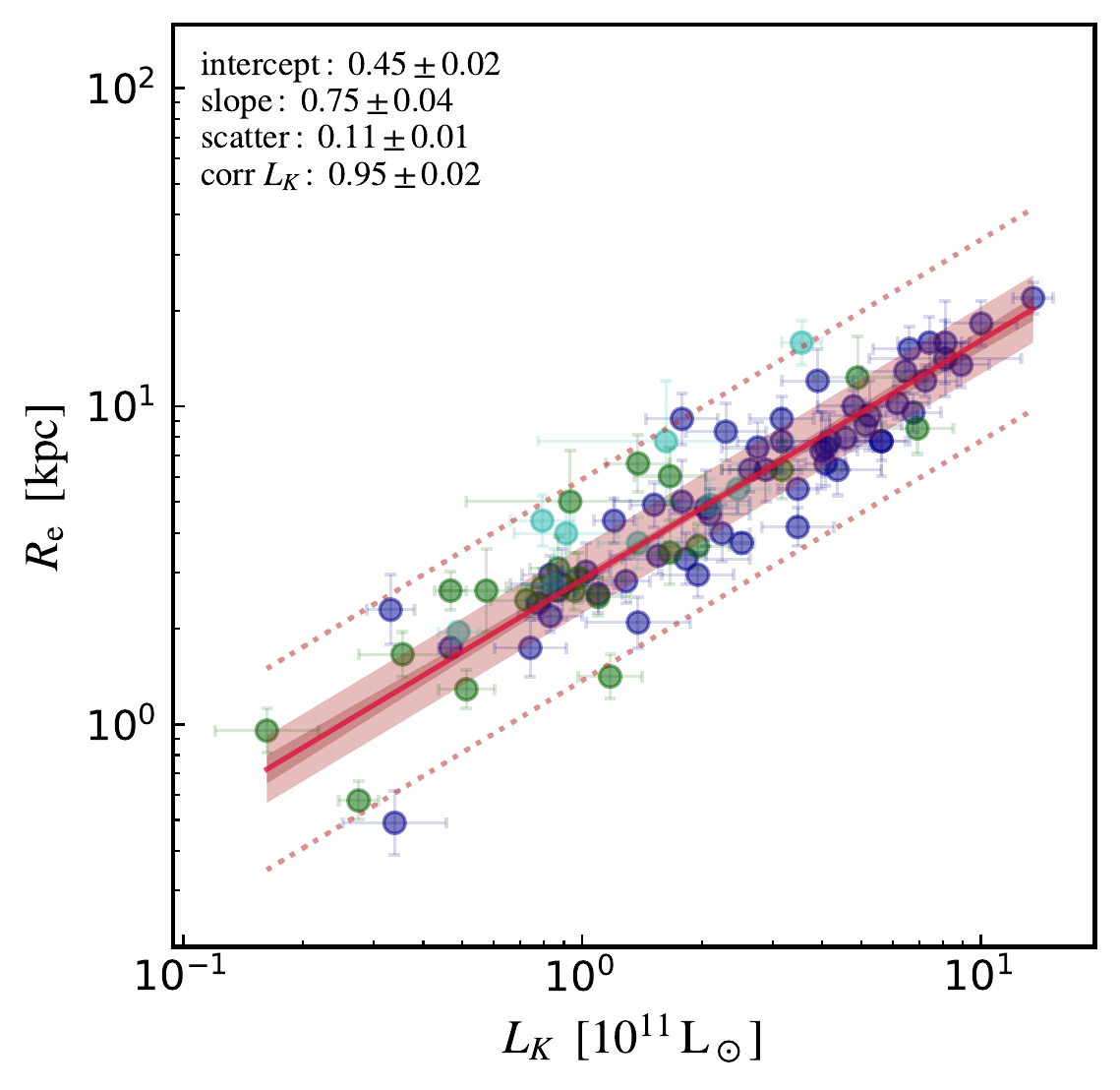}}
\hskip -0.21cm
\subfigure{\includegraphics[width=0.35\columnwidth]{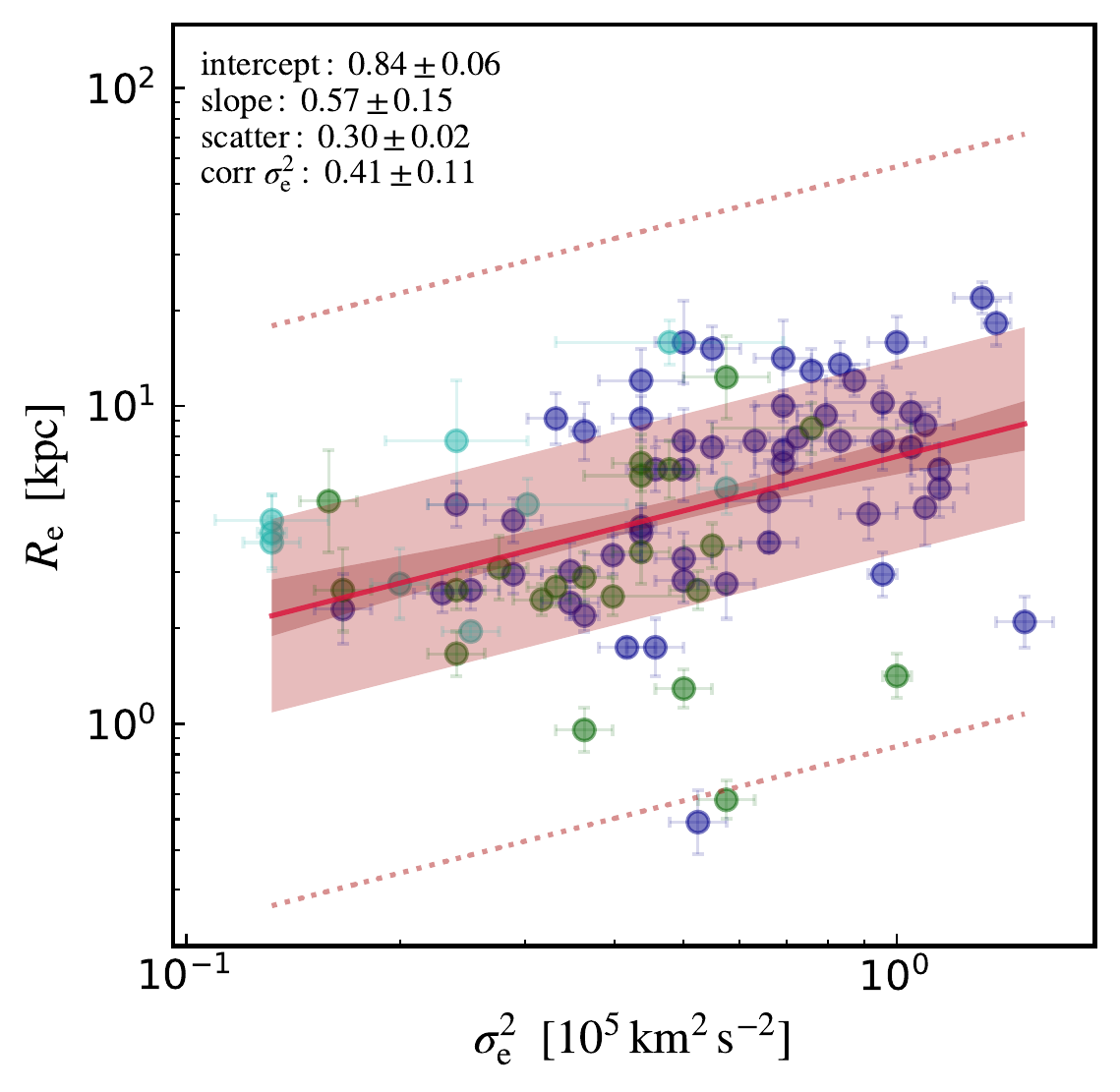}}
\hskip -0.21cm
\subfigure{\includegraphics[width=0.359\columnwidth]{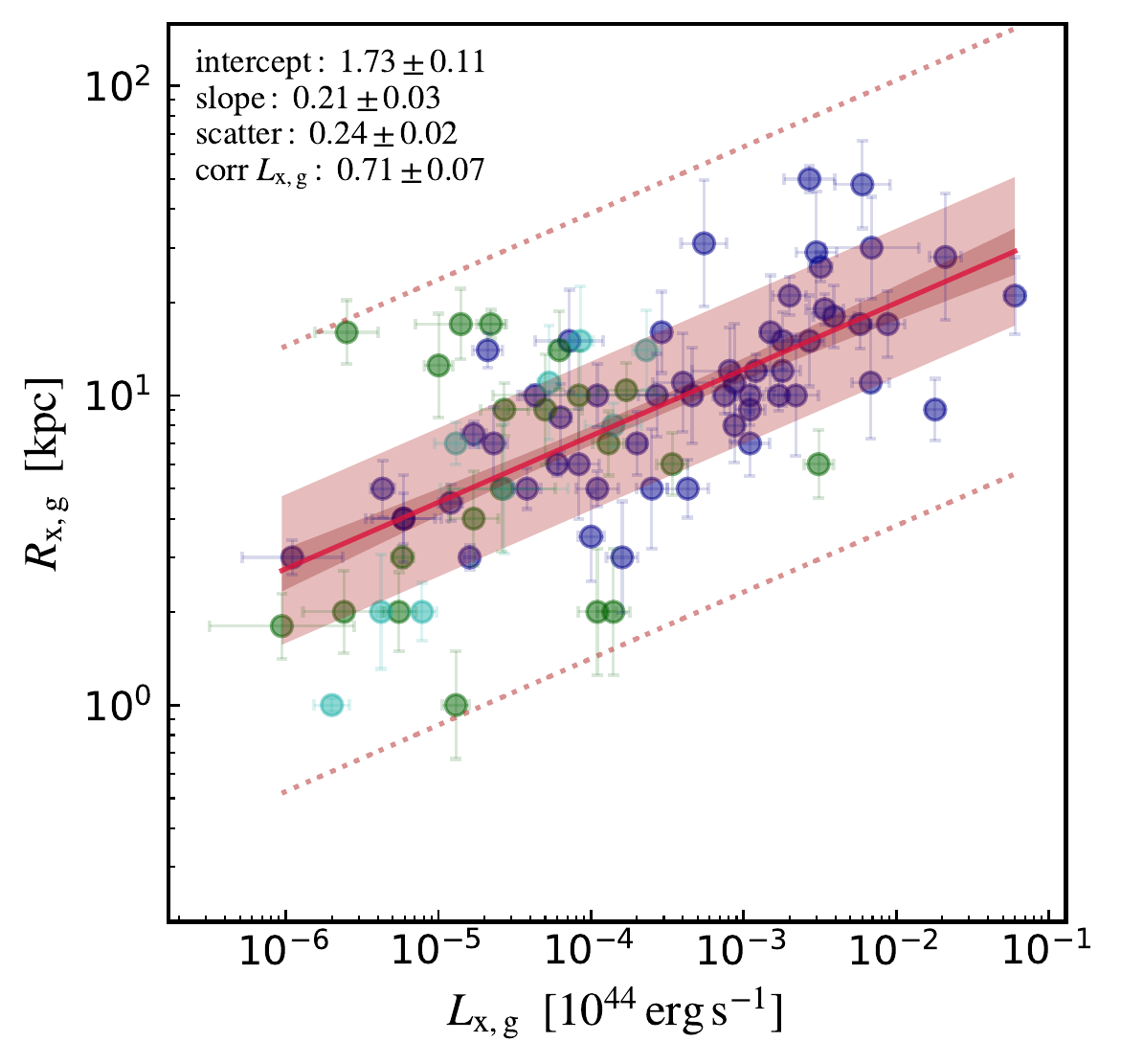}}
\vskip -0.4cm
\caption{Size vs.~luminosity or temperature scaling relations for the considered galaxies in optical (left and middle) and X-ray band (right). Analog of Fig.~\ref{LxTx}. 
Note the strong and tight stellar size-luminosity scaling, which is consistent with the recent version of the Kormendy relation for bulge-dominated galaxies (e.g, \citealt{Lasker:2014}). LTGs have instead a major impact on the correlation between stellar size and $\sige^2$, making it fairly shallow and weaker, with $3\times$ larger scatter. 
The X-ray size-luminosity scaling shows scatter in between, but substantially lower slope than the optical counterpart, although now probing three more orders of magnitude in luminosity. It is clear from the normalization that hot halos are generally more extended even at the galactic scale.
}
\vskip -5.5cm 
\label{sizegal}
\end{figure*}

\begin{figure*}[!h]
\hskip -0.375cm
\subfigure{\includegraphics[width=0.35\columnwidth]{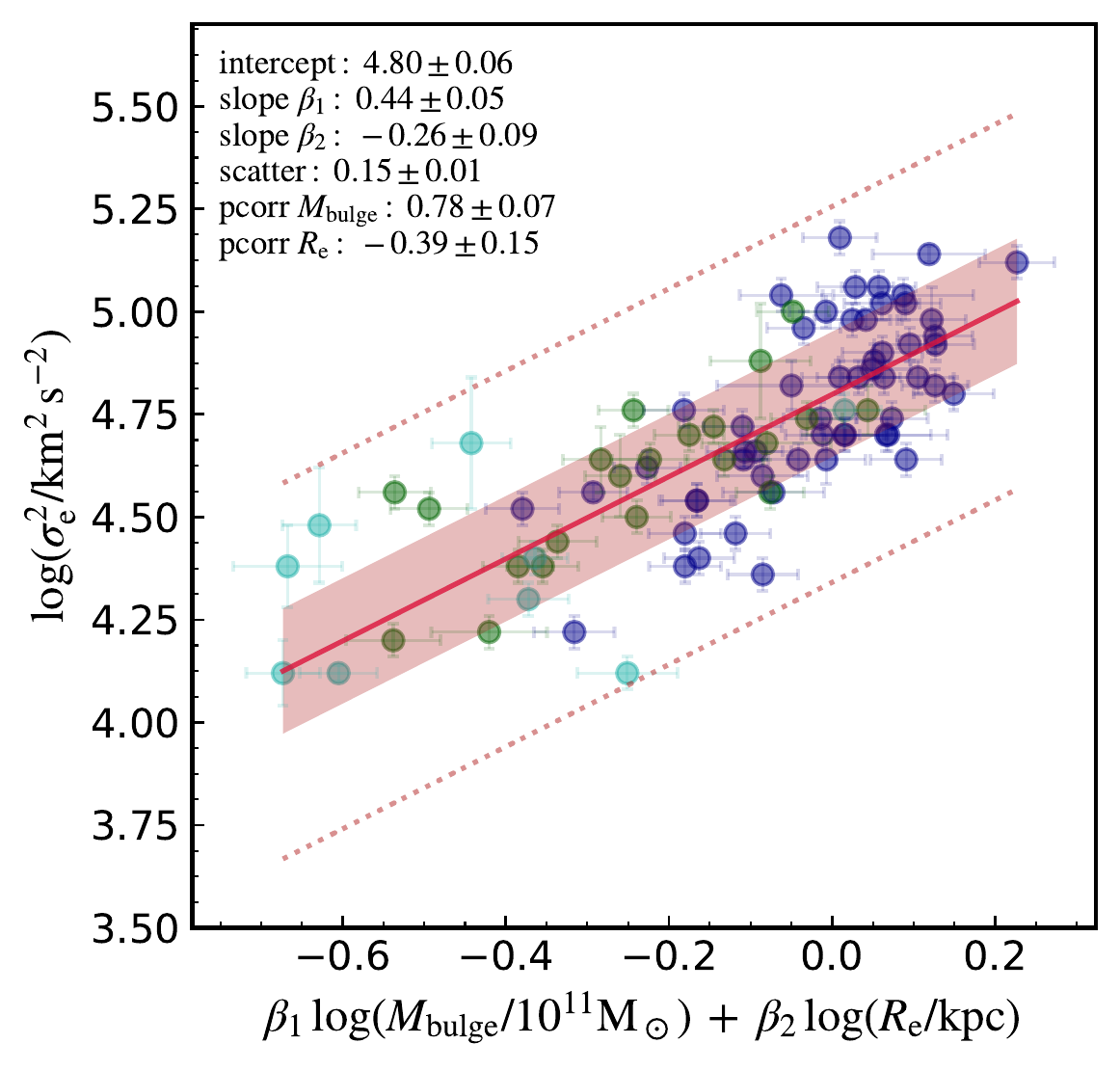}}
\hskip -0.21cm
\subfigure{\includegraphics[width=0.35\columnwidth]{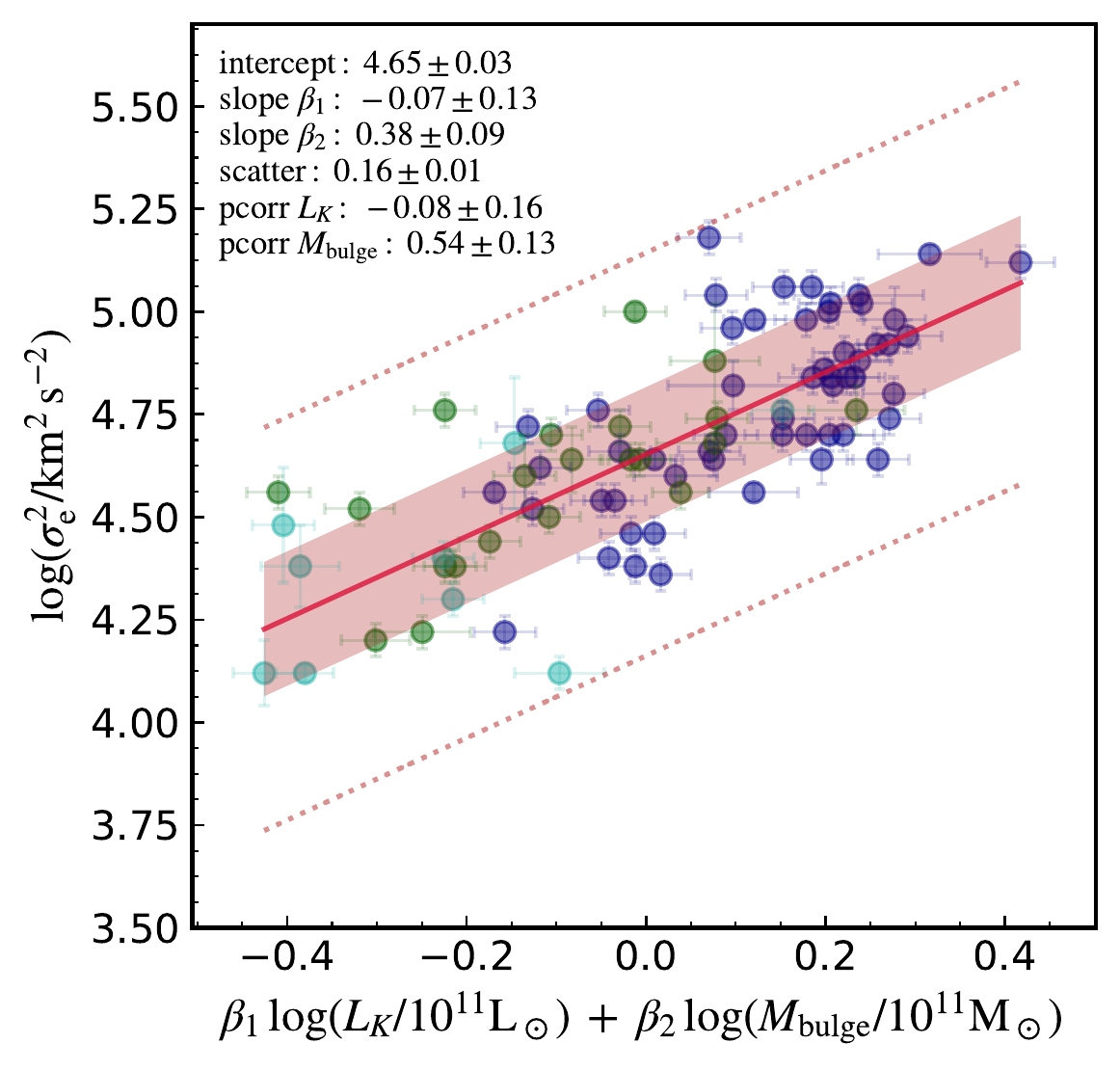}}
\hskip -0.21cm
\subfigure{\includegraphics[width=0.352\columnwidth]{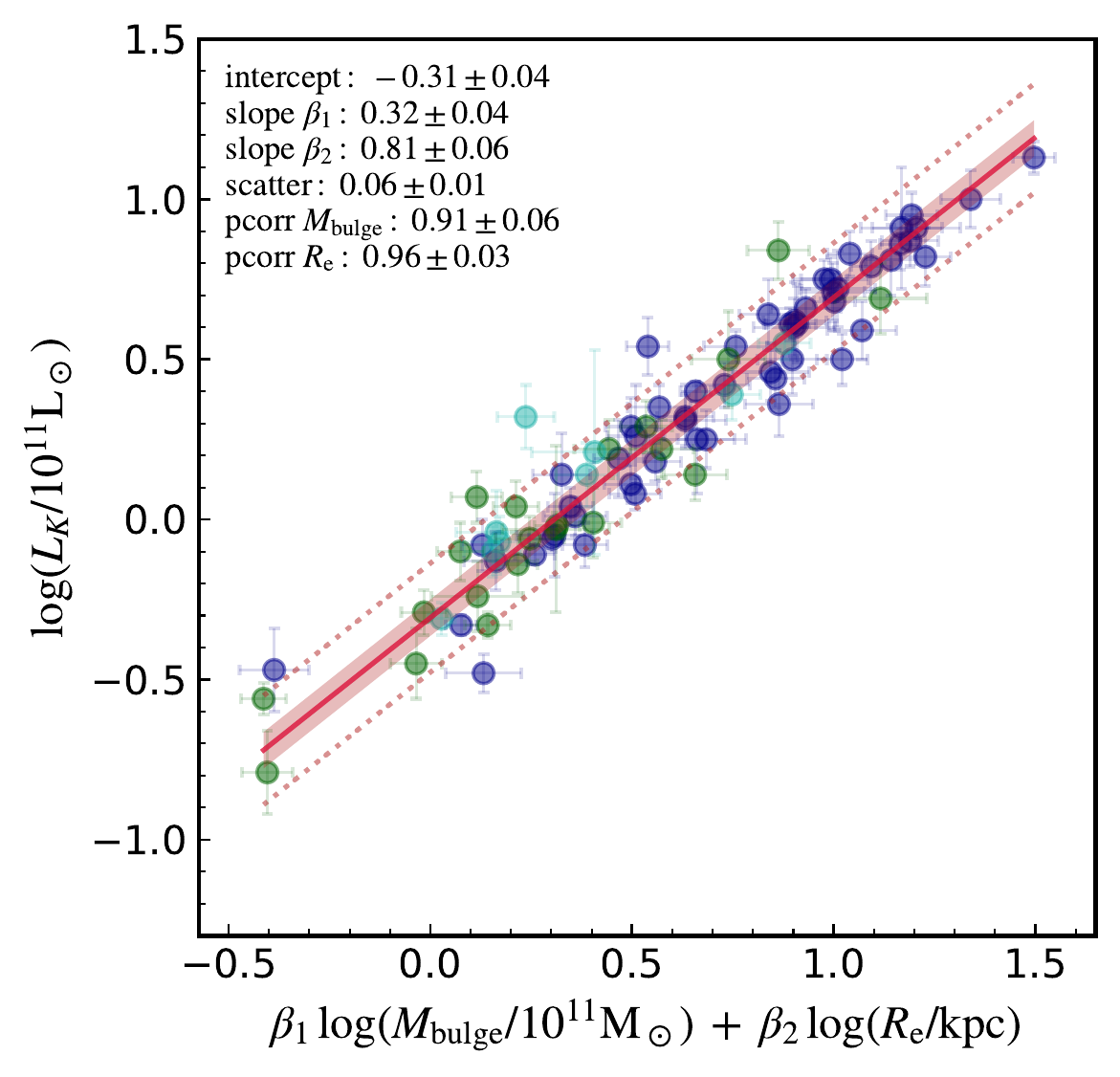}}
\vskip -0.4cm
\caption{Variants of the optical fundamental plane including the bulge mass. Edge-on view of the multivariate correlation plane (red line is the best-fit with the 1-$\sigma$ and 3-$\sigma$ intrinsic scatter bands overplotted). The inset shows all the retrieved {\tt mlinmix} Bayesian posterior parameters (mean and standard deviation; \S\ref{s:corr}).
Although not necessarily linked to the VT, all the above multivariate correlations show a small scatter.
The first two panels show 50\% larger scatter compared with the standard oFP (Fig.~\ref{oFP}), while the scatter of the luminosity\,--\,$M_{\rm bulge}$\,--\,size relation is 40\% smaller  (with high significance). This is partly due to the fairly stable optical mass-to-light ratios, although both pcorr$_{1,2}$ are in the very strong regime, signaling that \Re\ also plays a role.
}
\vskip +0.4cm 
\label{MbuoFP}

\hskip -0.375cm
\subfigure{\includegraphics[width=0.35\columnwidth]{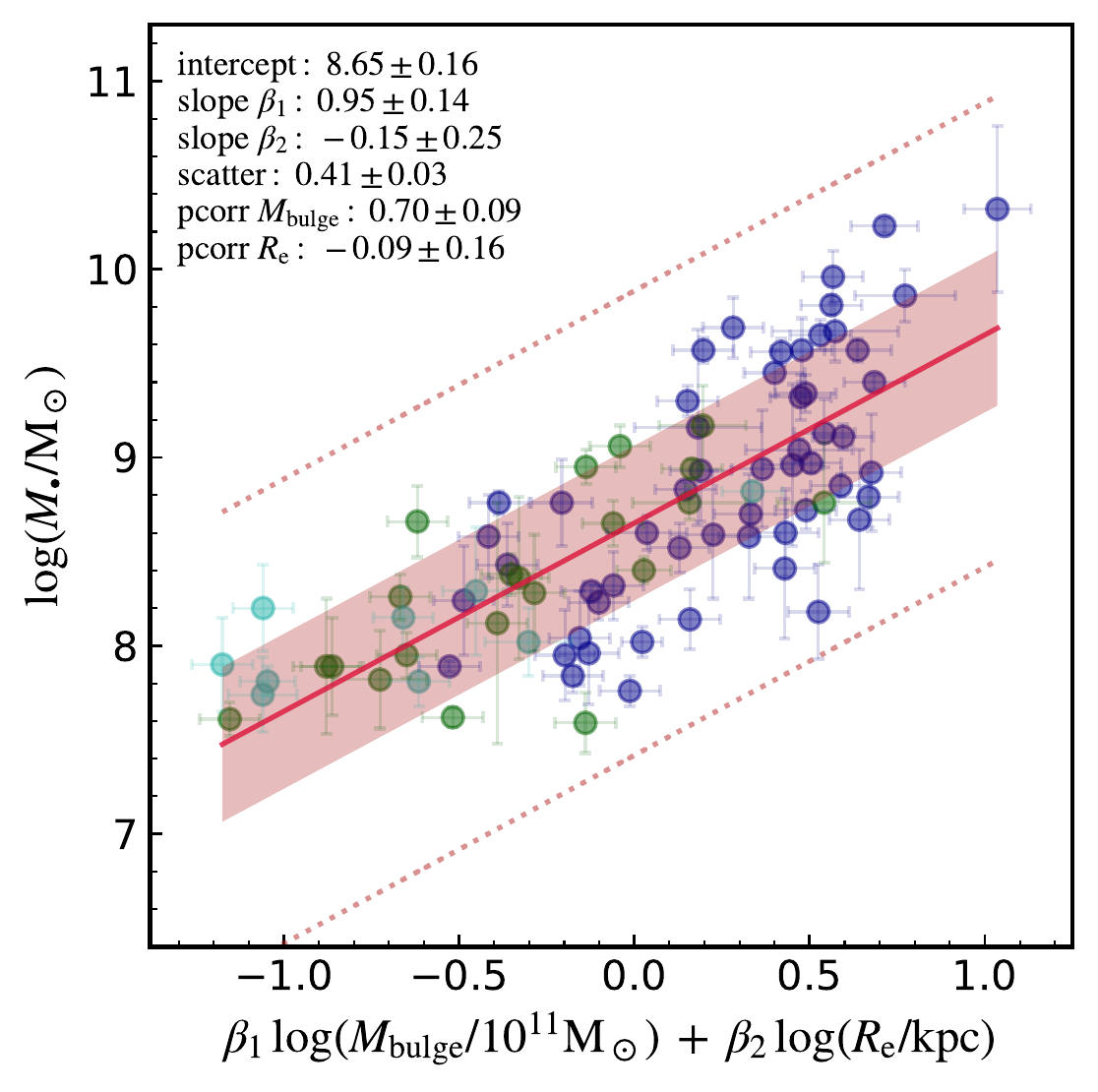}}
\hskip -0.21cm
\subfigure{\includegraphics[width=0.35\columnwidth]{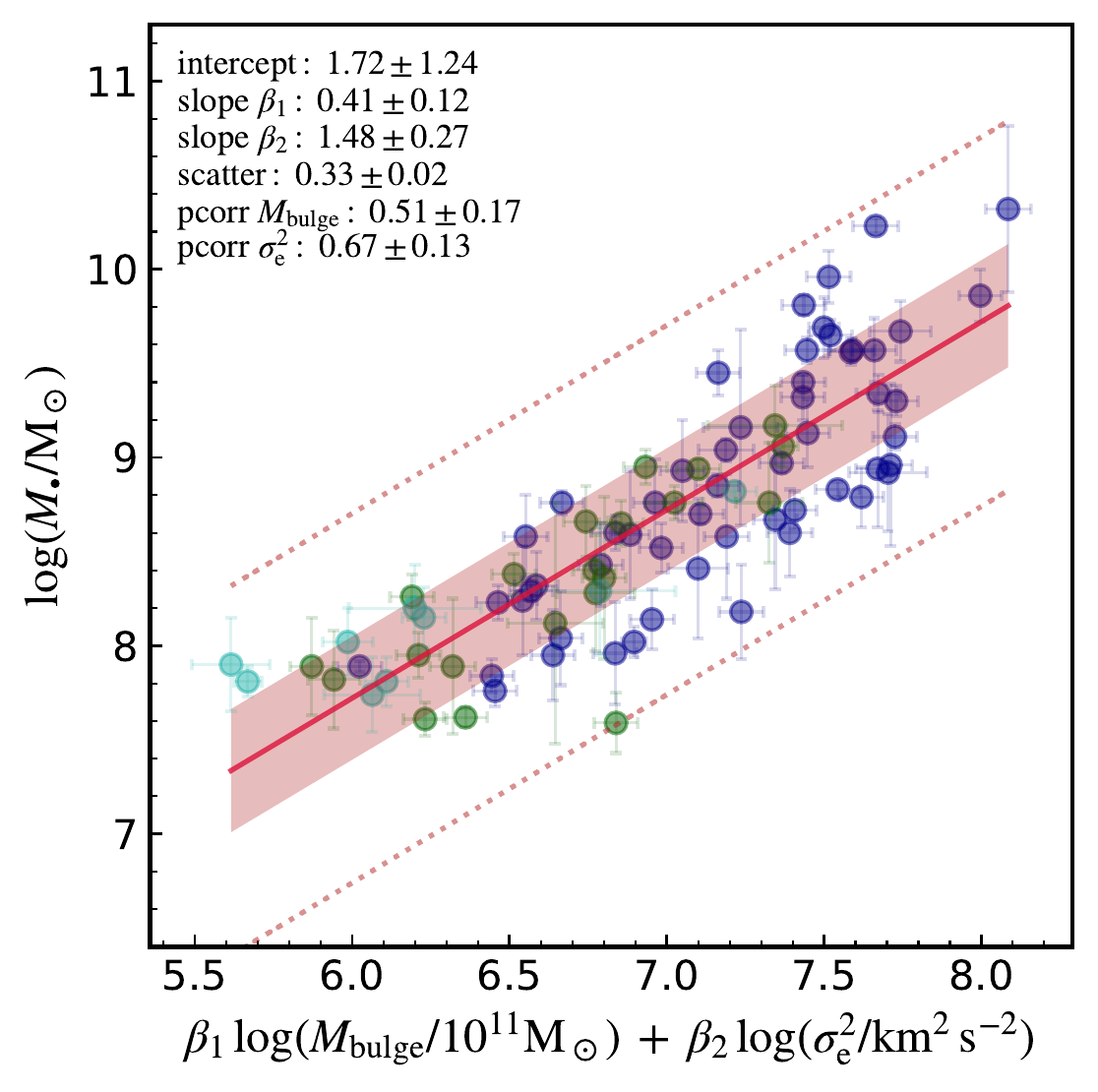}}
\hskip -0.21cm
\subfigure{\includegraphics[width=0.35\columnwidth]{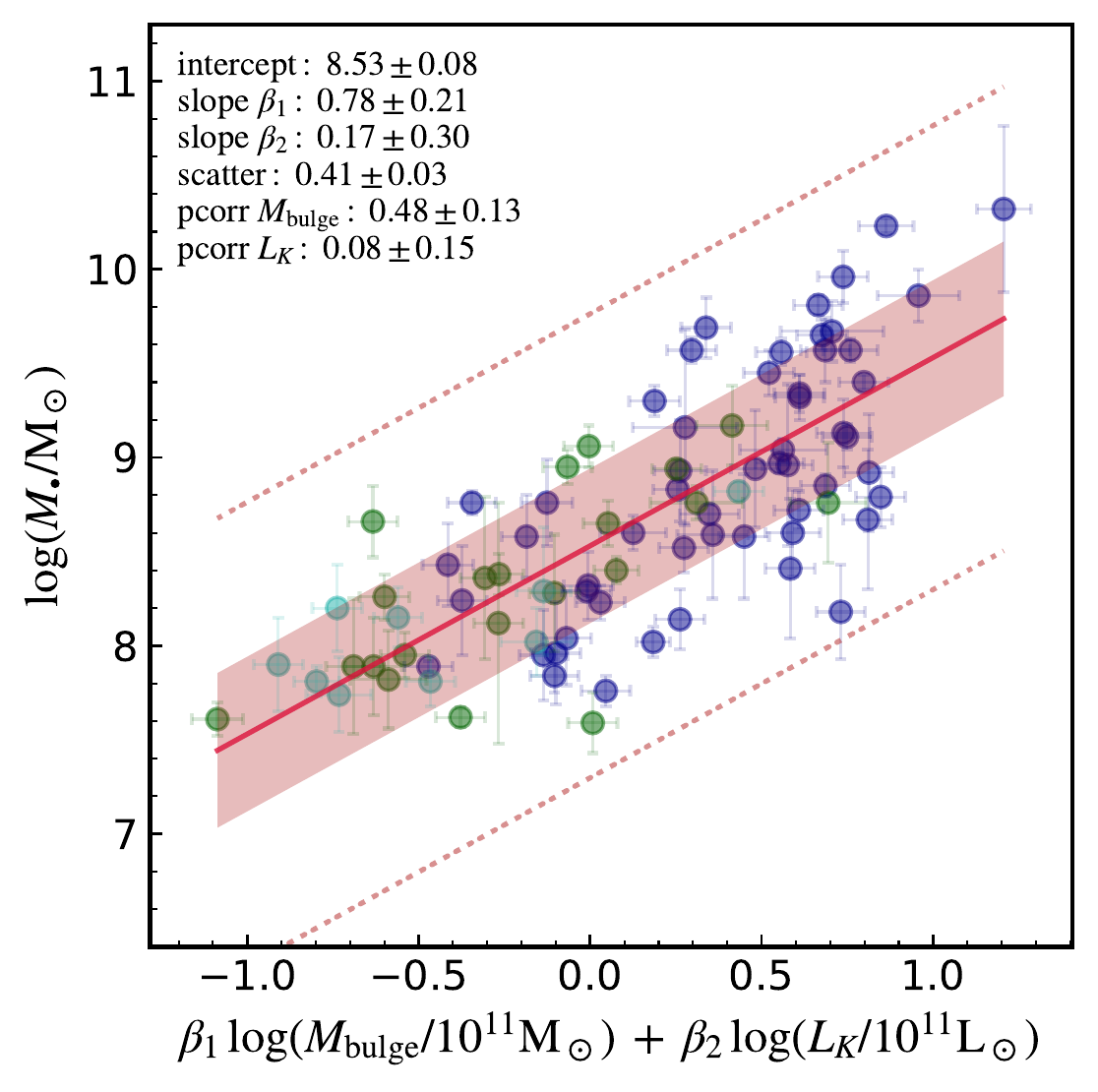}}
\vskip -0.4cm
\caption{BH mass vs.~bulge mass and another fundamental optical variable, including stellar size (left), velocity variance (middle), or luminosity (right). Analog of Fig.~\ref{MbuoFP}. Using the bulge mass instead of $L_K$ shows similar results for the multivariate correlations (\S\ref{s:MbhoFP}), albeit $M_{\rm bulge}$ tends to emerge even more over the second independent variable (${\rm pcorr_1 \gg pcorr_2}$; except for the $\sige^2$ case).
}
\vskip -0.8cm 
\label{MbhMbuop}
\end{figure*}

\begin{figure*}[!h]
\hskip -0.375cm
\subfigure{\includegraphics[width=0.361\columnwidth]{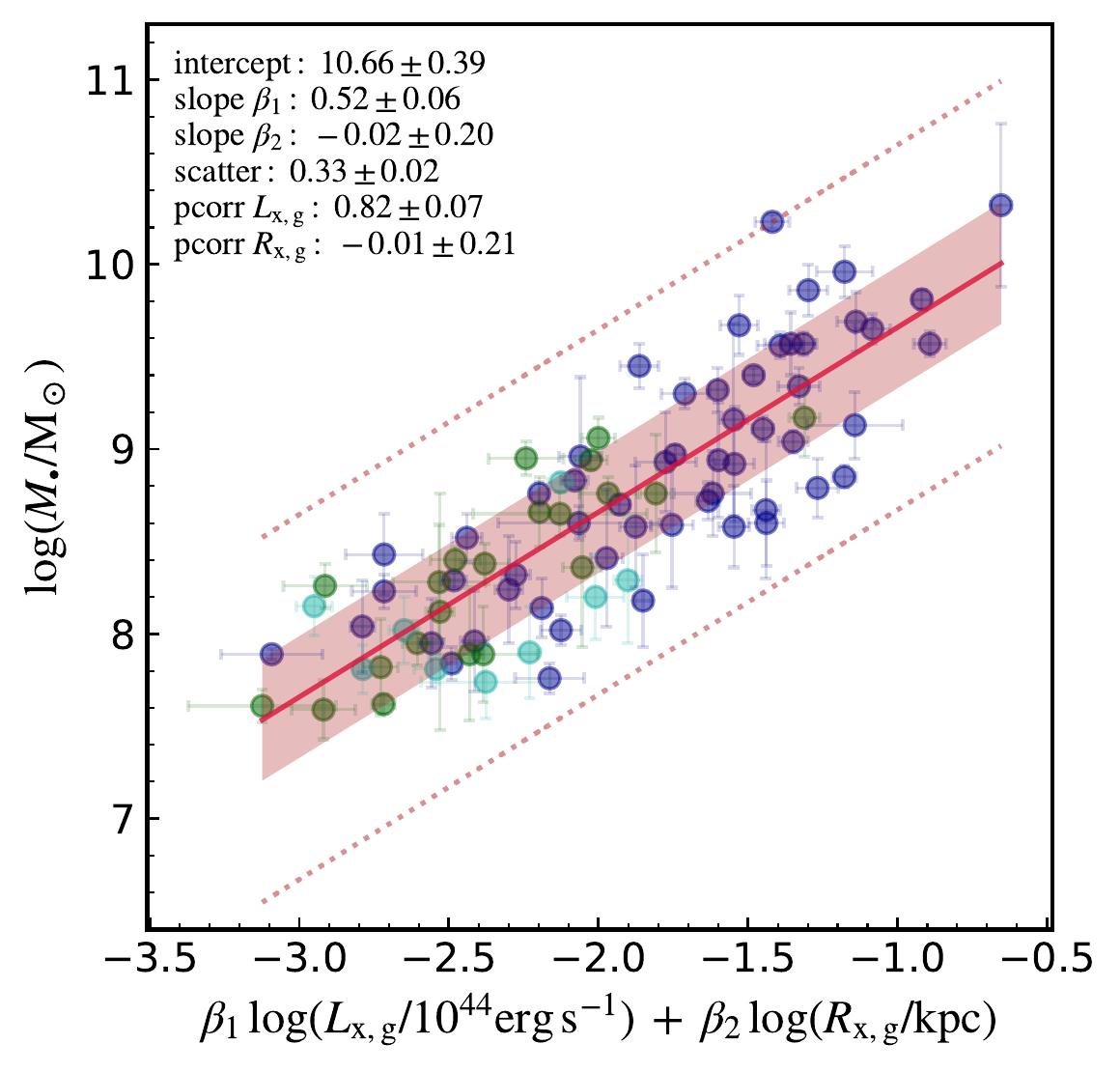}}
\hskip -0.37cm
\subfigure{\includegraphics[width=0.347\columnwidth]{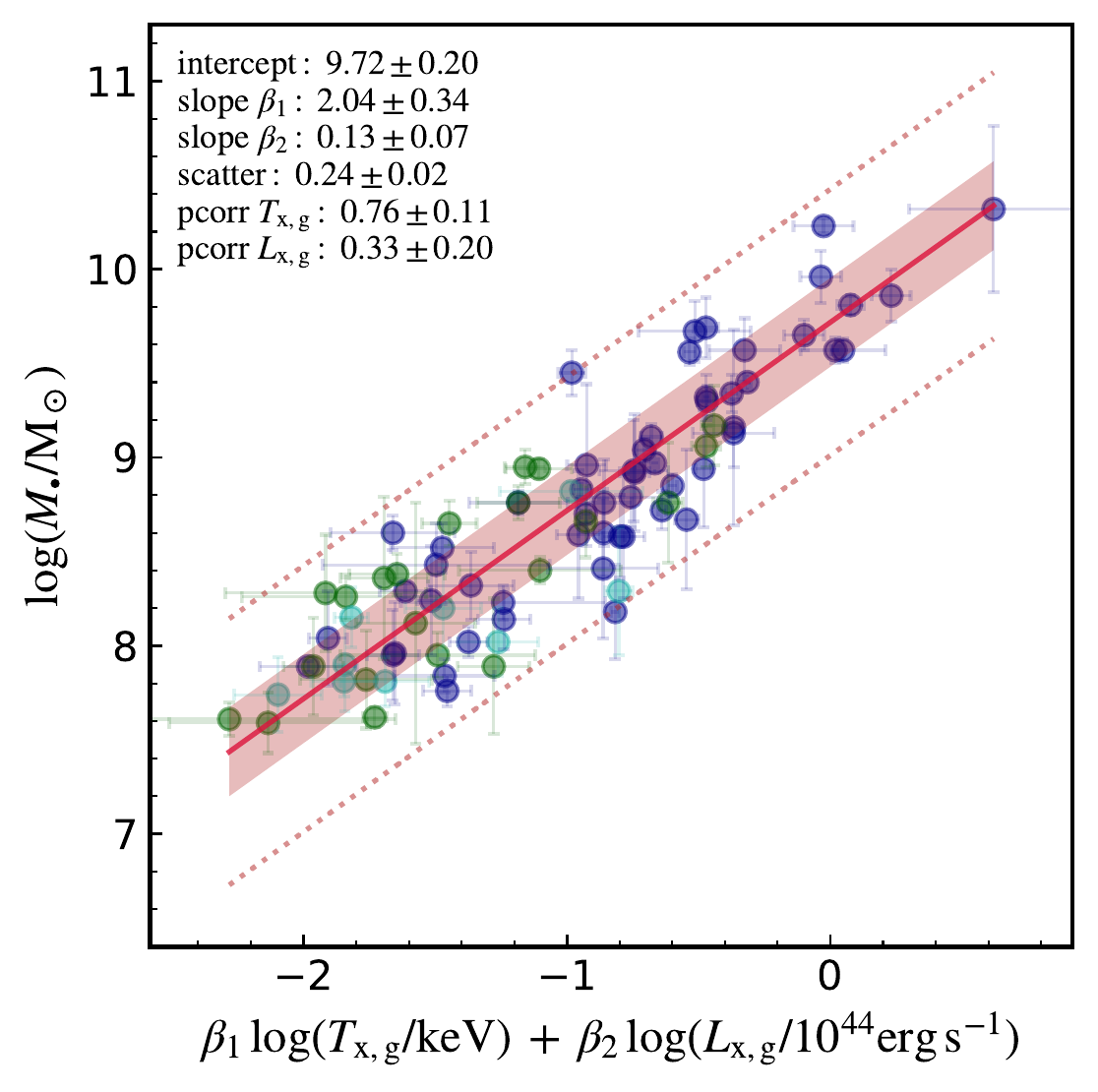}}
\hskip -0.21cm
\subfigure{\includegraphics[width=0.348\columnwidth]{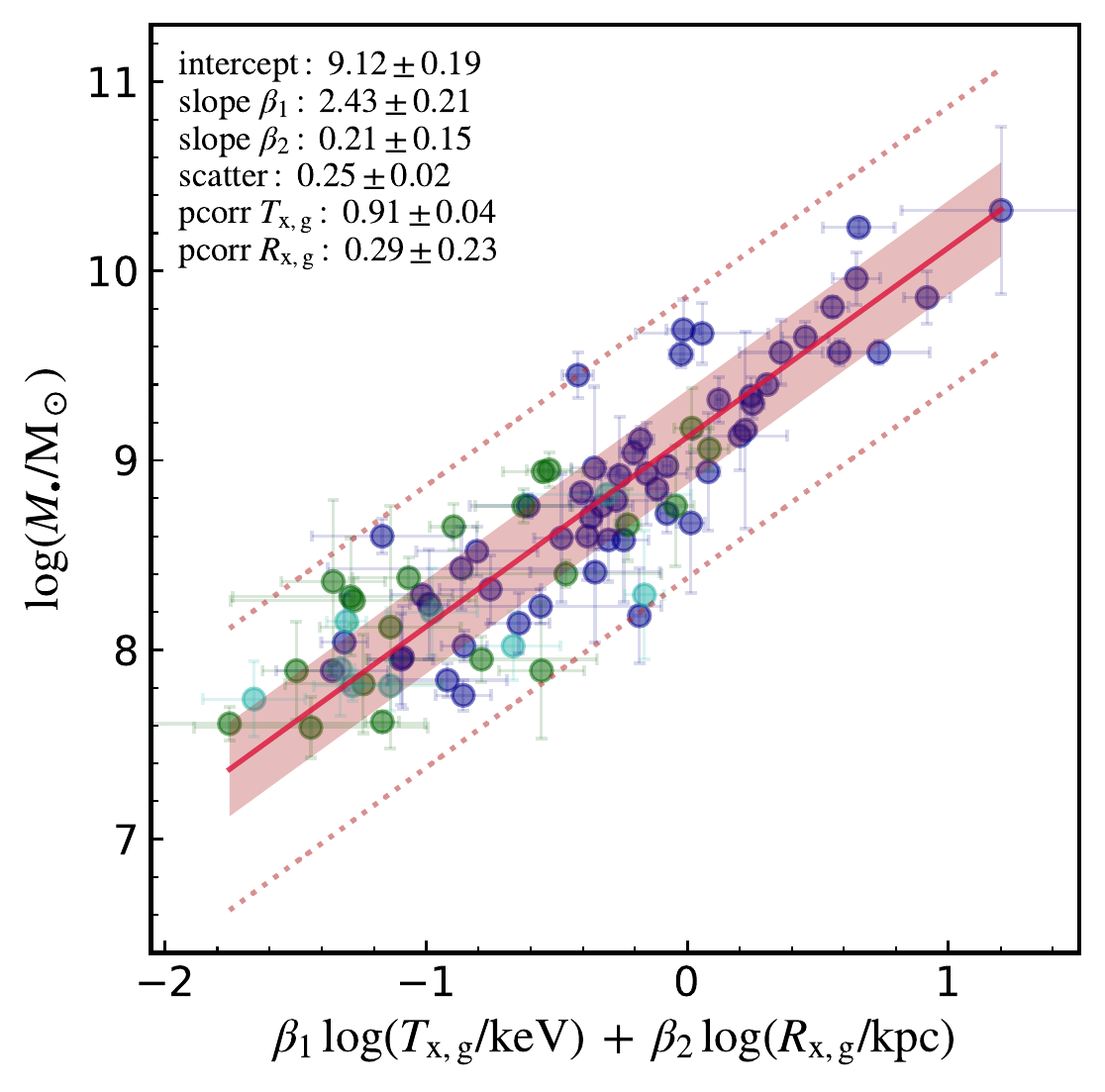}}
\vskip -0.4cm
\caption{BH mass vs.~dual properties of the xFP (edge-on view), permutating the X-ray luminosity, size, and temperature for the galactic/CGM region. Analog of Fig.~\ref{MbuoFP}. As for the other radial regions (\S\ref{s:MbhxFP}), using a multivariate fitting for the X-ray properties does not significantly improve the intrinsic scatter.
} 
\label{MbhxFPg}
\end{figure*}

\clearpage
\section{B. Optical and X-ray Data Sample Tables} \label{a:tables}
\LongTables 
\capstartfalse
\begin{center}
\begin{deluxetable*}{lccccccccccl} 
\tablecaption{Optical properties of our sample of host galaxies with direct dynamical SMBH measurements (and available X-ray hot halo data shown in Table~\ref{tabx}).} 
\tablehead{
Galaxy & {\ PGC\ } & {\ Type\ } & {$\ N_{\rm m}\ $} 
& {$D$} & {$M_\bullet$}  & {$\sigma_{\rm e}$}  & {$L_K$} & {$M_{\rm bulge}$} & {$R_{\rm e}$} & {$\ B/T\ $} & Refs.\,\&\\
    Name     &  \#   &    &    & {(Mpc)} &  {$\log(M_\odot)$} & {$\log(\kms)$}  & {$\log(L_\odot)$} & {$\ \ \ \log(M_\odot)\ \ \ $} & {$\log({\rm kpc})$} & & Notes    \\
 \ \ (i)   &   {(ii)} &  {(iii)} &  {(iv)}  & {(v)}  &  {(vi)}  &  {(vii)}  & {(viii)} & {(ix)} & {(x)} & {(xi)} & {(xii)}  
}

\startdata   
N0821 & 8160 & S0/E6 & 1 & 23.4${\scriptstyle\,\pm\,1.8}$ & 7.59${\scriptstyle\,\pm\,0.16}$ & 2.32${\scriptstyle\,\pm\,0.02}$ & 11.14${\scriptstyle\,\pm\,0.08}$ & 10.98${\scriptstyle\,\pm\,0.09}$ & 0.82${\scriptstyle\,\pm\,0.09}$ & 0.95 & (d,a,b);(1) \\
N2787 & 26341 & SB0 & 2 & 7.4${\scriptstyle\,\pm\,1.2}$ & 7.61${\scriptstyle\,\pm\,0.09}$ & 2.28${\scriptstyle\,\pm\,0.02}$ & 10.21${\scriptstyle\,\pm\,0.13}$ & 9.78${\scriptstyle\,\pm\,0.09}$ & -0.02${\scriptstyle\,\pm\,0.07}$ & 0.26 & (a,b);(2) \\
N1023 & 10123 & SB0 & 4 & 10.8${\scriptstyle\,\pm\,0.8}$ & 7.62${\scriptstyle\,\pm\,0.05}$ & 2.22${\scriptstyle\,\pm\,0.02}$ & 10.94${\scriptstyle\,\pm\,0.07}$ & 10.53${\scriptstyle\,\pm\,0.09}$ & 0.49${\scriptstyle\,\pm\,0.10}$ & 0.39 & (a,b);(3) \\
N7582 & 71001 & SBab & 8 & 22.3${\scriptstyle\,\pm\,9.8}$ & 7.74${\scriptstyle\,\pm\,0.20}$ & 2.19${\scriptstyle\,\pm\,0.05}$ & 11.21${\scriptstyle\,\pm\,0.32}$ & 10.02${\scriptstyle\,\pm\,0.10}$ & 0.89${\scriptstyle\,\pm\,0.19}$ & 0.1 & (a,b);(4) \\
N5128 & 46957 & E/S0 & 6 & 3.6${\scriptstyle\,\pm\,0.2}$ & 7.76${\scriptstyle\,\pm\,0.08}$ & 2.18${\scriptstyle\,\pm\,0.02}$ & 11.04${\scriptstyle\,\pm\,0.06}$ & 11.05${\scriptstyle\,\pm\,0.09}$ & 0.41${\scriptstyle\,\pm\,0.06}$ & 1.0 & (a,b);(5) \\
N3031 & 28630 & SAab & 3 & 3.6${\scriptstyle\,\pm\,0.1}$ & 7.81${\scriptstyle\,\pm\,0.13}$ & 2.15${\scriptstyle\,\pm\,0.02}$ & 10.93${\scriptstyle\,\pm\,0.08}$ & 10.42${\scriptstyle\,\pm\,0.09}$ & 0.44${\scriptstyle\,\pm\,0.11}$ & 0.34 & (a,c);(6) \\
N4151 & 38739 & Sab & 65 & 20.0${\scriptstyle\,\pm\,2.8}$ & 7.81${\scriptstyle\,\pm\,0.08}$ & 2.06${\scriptstyle\,\pm\,0.01}$ & 10.96${\scriptstyle\,\pm\,0.13}$ & 9.99${\scriptstyle\,\pm\,0.08}$ & 0.60${\scriptstyle\,\pm\,0.12}$ & 0.33 & (a,c,o,p);(7) \\
N4203 & 39158 & SAB0 & 1 & 14.1${\scriptstyle\,\pm\,1.4}$ & 7.82${\scriptstyle\,\pm\,0.26}$ & 2.11${\scriptstyle\,\pm\,0.02}$ & 10.76${\scriptstyle\,\pm\,0.11}$ & 10.30${\scriptstyle\,\pm\,0.14}$ & 0.42${\scriptstyle\,\pm\,0.13}$ & 0.37 & (a,d);(8) \\
N4459 & 41104 & E2 & 197 & 16.0${\scriptstyle\,\pm\,0.5}$ & 7.84${\scriptstyle\,\pm\,0.09}$ & 2.20${\scriptstyle\,\pm\,0.02}$ & 10.94${\scriptstyle\,\pm\,0.04}$ & 10.88${\scriptstyle\,\pm\,0.09}$ & 0.42${\scriptstyle\,\pm\,0.06}$ & 1.0 & (a,b);(9) \\
N4596 & 42401 & SB0 & 197 & 16.5${\scriptstyle\,\pm\,6.2}$ & 7.89${\scriptstyle\,\pm\,0.26}$ & 2.10${\scriptstyle\,\pm\,0.02}$ & 10.97${\scriptstyle\,\pm\,0.26}$ & 10.20${\scriptstyle\,\pm\,0.09}$ & 0.70${\scriptstyle\,\pm\,0.16}$ & 0.78 & (a,d);(10) \\
N3377 & 32249 & E5 & 11 & 11.0${\scriptstyle\,\pm\,0.5}$ & 7.89${\scriptstyle\,\pm\,0.03}$ & 2.11${\scriptstyle\,\pm\,0.02}$ & 10.52${\scriptstyle\,\pm\,0.06}$ & 10.50${\scriptstyle\,\pm\,0.09}$ & 0.36${\scriptstyle\,\pm\,0.11}$ & 1.0 & (e,a,b);(11) \\
N4036 & 7930 & S0 & 3 & 19.0${\scriptstyle\,\pm\,1.9}$ & 7.89${\scriptstyle\,\pm\,0.36}$ & 2.26${\scriptstyle\,\pm\,0.02}$ & 10.90${\scriptstyle\,\pm\,0.09}$ & 10.14${\scriptstyle\,\pm\,0.10}$ & 0.43${\scriptstyle\,\pm\,0.06}$ & 0.17 & (a,d);(12) \\
N4258 & 39600 & SABbc & 5 & 7.3${\scriptstyle\,\pm\,0.5}$ & 7.90${\scriptstyle\,\pm\,0.25}$ & 2.06${\scriptstyle\,\pm\,0.04}$ & 10.90${\scriptstyle\,\pm\,0.08}$ & 9.86${\scriptstyle\,\pm\,0.09}$ & 0.64${\scriptstyle\,\pm\,0.08}$ & 0.12 & (s,a,b);(13) \\
N4564 & 42051 & S0 & 197 & 15.9${\scriptstyle\,\pm\,0.5}$ & 7.95${\scriptstyle\,\pm\,0.12}$ & 2.19${\scriptstyle\,\pm\,0.02}$ & 10.67${\scriptstyle\,\pm\,0.04}$ & 10.38${\scriptstyle\,\pm\,0.09}$ & 0.42${\scriptstyle\,\pm\,0.06}$ & 0.67 & (a,b);(14) \\
N4473 & 41228 & E5 & 197 & 15.2${\scriptstyle\,\pm\,0.5}$ & 7.95${\scriptstyle\,\pm\,0.24}$ & 2.27${\scriptstyle\,\pm\,0.02}$ & 10.89${\scriptstyle\,\pm\,0.03}$ & 10.85${\scriptstyle\,\pm\,0.09}$ & 0.38${\scriptstyle\,\pm\,0.05}$ & 1.0 & (a,b);(15) \\
N4278 & 39764 & E1-2 & 15 & 15.0${\scriptstyle\,\pm\,1.5}$ & 7.96${\scriptstyle\,\pm\,0.27}$ & 2.33${\scriptstyle\,\pm\,0.02}$ & 10.87${\scriptstyle\,\pm\,0.09}$ & 10.90${\scriptstyle\,\pm\,0.09}$ & 0.24${\scriptstyle\,\pm\,0.09}$ & 1.0 & (a,d);(16) \\
N5018 & 45908 & E3 & 5 & 40.5${\scriptstyle\,\pm\,4.9}$ & 8.02${\scriptstyle\,\pm\,0.08}$ & 2.32${\scriptstyle\,\pm\,0.01}$ & 11.54${\scriptstyle\,\pm\,0.09}$ & 11.12${\scriptstyle\,\pm\,0.06}$ & 0.62${\scriptstyle\,\pm\,0.06}$ & 0.51 & (a,c);(17) \\
N7331 & 69327 & SAb & 2 & 12.2${\scriptstyle\,\pm\,1.2}$ & 8.02${\scriptstyle\,\pm\,0.18}$ & 2.06${\scriptstyle\,\pm\,0.02}$ & 11.14${\scriptstyle\,\pm\,0.10}$ & 10.77${\scriptstyle\,\pm\,0.13}$ & 0.57${\scriptstyle\,\pm\,0.08}$ & 0.47 & (a,n);(18) \\
N3379 & 32256 & E0 & 11 & 10.7${\scriptstyle\,\pm\,0.5}$ & 8.04${\scriptstyle\,\pm\,0.25}$ & 2.27${\scriptstyle\,\pm\,0.02}$ & 11.01${\scriptstyle\,\pm\,0.06}$ & 10.91${\scriptstyle\,\pm\,0.09}$ & 0.48${\scriptstyle\,\pm\,0.09}$ & 1.0 & (d,a,b);(19) \\
N2110 & 18030 & SAB0 & 2 & 29.1${\scriptstyle\,\pm\,2.9}$ & 8.12${\scriptstyle\,\pm\,0.64}$ & 2.30${\scriptstyle\,\pm\,0.05}$ & 11.04${\scriptstyle\,\pm\,0.08}$ & 10.65${\scriptstyle\,\pm\,0.09}$ & 0.40${\scriptstyle\,\pm\,0.06}$ & 0.39 & (a,p);(20) \\
N3607 & 34426 & E1 & 16 & 22.6${\scriptstyle\,\pm\,1.8}$ & 8.14${\scriptstyle\,\pm\,0.16}$ & 2.32${\scriptstyle\,\pm\,0.02}$ & 11.35${\scriptstyle\,\pm\,0.07}$ & 11.26${\scriptstyle\,\pm\,0.09}$ & 0.60${\scriptstyle\,\pm\,0.09}$ & 1.0 & (a,b);(21) \\
N0224 & 2557 & Sb & $>\,$4 & 0.77${\scriptstyle\,\pm\,0.03}$ & 8.15${\scriptstyle\,\pm\,0.16}$ & 2.20${\scriptstyle\,\pm\,0.02}$ & 10.69${\scriptstyle\,\pm\,0.05}$ & 10.35${\scriptstyle\,\pm\,0.09}$ & 0.29${\scriptstyle\,\pm\,0.02}$ & 0.57 & (a,r);(22) \\
N1316 & 12651 & E4 & 49 & 18.6${\scriptstyle\,\pm\,0.6}$ & 8.18${\scriptstyle\,\pm\,0.25}$ & 2.35${\scriptstyle\,\pm\,0.02}$ & 11.91${\scriptstyle\,\pm\,0.06}$ & 11.74${\scriptstyle\,\pm\,0.09}$ & 1.20${\scriptstyle\,\pm\,0.13}$ & 1.0 & (a,b);(23) \\
N1667 & 16062 & SABc & 4 & 56.1${\scriptstyle\,\pm\,5.6}$ & 8.20${\scriptstyle\,\pm\,0.23}$ & 2.24${\scriptstyle\,\pm\,0.07}$ & 11.32${\scriptstyle\,\pm\,0.10}$ & 9.99${\scriptstyle\,\pm\,0.09}$ & 0.69${\scriptstyle\,\pm\,0.08}$ & 0.16 & (a,j);(24) \\
N5576 & 51275 & E3 & 5 & 25.7${\scriptstyle\,\pm\,1.7}$ & 8.23${\scriptstyle\,\pm\,0.09}$ & 2.19${\scriptstyle\,\pm\,0.02}$ & 11.18${\scriptstyle\,\pm\,0.06}$ & 11.00${\scriptstyle\,\pm\,0.09}$ & 0.69${\scriptstyle\,\pm\,0.07}$ & 1.0 & (d,a);(25) \\
N1052 & 10175 & E4 & 8 & 18.1${\scriptstyle\,\pm\,1.8}$ & 8.24${\scriptstyle\,\pm\,0.29}$ & 2.28${\scriptstyle\,\pm\,0.01}$ & 10.92${\scriptstyle\,\pm\,0.08}$ & 10.54${\scriptstyle\,\pm\,0.09}$ & 0.34${\scriptstyle\,\pm\,0.05}$ & 1.0 & (a,h);(26) \\
N4026 & 37760 & S0 & 65 & 13.4${\scriptstyle\,\pm\,1.7}$ & 8.26${\scriptstyle\,\pm\,0.12}$ & 2.19${\scriptstyle\,\pm\,0.02}$ & 10.55${\scriptstyle\,\pm\,0.11}$ & 10.33${\scriptstyle\,\pm\,0.09}$ & 0.22${\scriptstyle\,\pm\,0.07}$ & 0.61 & (a,b);(27) \\
N3801 & 36200 & S0 & 7 & 46.3${\scriptstyle\,\pm\,4.6}$ & 8.28${\scriptstyle\,\pm\,0.31}$ & 2.32${\scriptstyle\,\pm\,0.04}$ & 11.22${\scriptstyle\,\pm\,0.09}$ & 10.82${\scriptstyle\,\pm\,0.09}$ & 0.78${\scriptstyle\,\pm\,0.09}$ & 0.38 & (a,h);(28) \\
N4697 & 43276 & E5 & 15 & 12.5${\scriptstyle\,\pm\,0.4}$ & 8.29${\scriptstyle\,\pm\,0.04}$ & 2.23${\scriptstyle\,\pm\,0.02}$ & 11.08${\scriptstyle\,\pm\,0.05}$ & 10.97${\scriptstyle\,\pm\,0.09}$ & 0.64${\scriptstyle\,\pm\,0.07}$ & 1.0 & (e,a,b);(29) \\
N1961 & 17625 & SABb & 9 & 48.6${\scriptstyle\,\pm\,4.9}$ & 8.29${\scriptstyle\,\pm\,0.34}$ & 2.34${\scriptstyle\,\pm\,0.08}$ & 11.55${\scriptstyle\,\pm\,0.05}$ & 10.71${\scriptstyle\,\pm\,0.10}$ & 1.20${\scriptstyle\,\pm\,0.07}$ & 0.15 & (a,j);(30) \\
N3608 & 34433 & E1 & 16 & 22.8${\scriptstyle\,\pm\,1.5}$ & 8.32${\scriptstyle\,\pm\,0.18}$ & 2.23${\scriptstyle\,\pm\,0.02}$ & 10.92${\scriptstyle\,\pm\,0.07}$ & 11.01${\scriptstyle\,\pm\,0.09}$ & 0.47${\scriptstyle\,\pm\,0.06}$ & 1.0 & (d,a);(31) \\
N3998 & 37642 & S0 & 65 & 14.3${\scriptstyle\,\pm\,1.3}$ & 8.36${\scriptstyle\,\pm\,0.43}$ & 2.35${\scriptstyle\,\pm\,0.02}$ & 10.71${\scriptstyle\,\pm\,0.07}$ & 10.67${\scriptstyle\,\pm\,0.09}$ & 0.11${\scriptstyle\,\pm\,0.06}$ & 0.85 & (d,a,b);(32) \\
N3245 & 30744 & S0 & 4 & 21.4${\scriptstyle\,\pm\,2.0}$ & 8.38${\scriptstyle\,\pm\,0.11}$ & 2.25${\scriptstyle\,\pm\,0.02}$ & 10.86${\scriptstyle\,\pm\,0.09}$ & 10.69${\scriptstyle\,\pm\,0.09}$ & 0.39${\scriptstyle\,\pm\,0.05}$ & 0.7 & (a,b);(33) \\
N3414 & 32533 & S0 & 5 & 25.2${\scriptstyle\,\pm\,2.7}$ & 8.40${\scriptstyle\,\pm\,0.07}$ & 2.28${\scriptstyle\,\pm\,0.02}$ & 10.99${\scriptstyle\,\pm\,0.11}$ & 11.10${\scriptstyle\,\pm\,0.08}$ & 0.46${\scriptstyle\,\pm\,0.08}$ & 0.79 & (a,c);(34) \\
N3862 & 36606 & E & 61 & 84.6${\scriptstyle\,\pm\,8.5}$ & 8.41${\scriptstyle\,\pm\,0.37}$ & 2.32${\scriptstyle\,\pm\,0.03}$ & 11.59${\scriptstyle\,\pm\,0.09}$ & 11.62${\scriptstyle\,\pm\,0.09}$ & 1.08${\scriptstyle\,\pm\,0.10}$ & 1.0 & (a,d);(35) \\
N5845 & 53901 & E3 & 13 & 25.9${\scriptstyle\,\pm\,4.1}$ & 8.43${\scriptstyle\,\pm\,0.22}$ & 2.36${\scriptstyle\,\pm\,0.02}$ & 10.53${\scriptstyle\,\pm\,0.13}$ & 10.57${\scriptstyle\,\pm\,0.09}$ & -0.31${\scriptstyle\,\pm\,0.10}$ & 1.0 & (e,a,b);(36) \\
N3585 & 34160 & E6/S0 & 3 & 20.5${\scriptstyle\,\pm\,1.7}$ & 8.52${\scriptstyle\,\pm\,0.13}$ & 2.33${\scriptstyle\,\pm\,0.02}$ & 11.42${\scriptstyle\,\pm\,0.07}$ & 11.26${\scriptstyle\,\pm\,0.09}$ & 0.80${\scriptstyle\,\pm\,0.07}$ & 0.93 & (a,b);(37) \\
N7626 & 71140 & E & 16 & 38.1${\scriptstyle\,\pm\,3.8}$ & 8.58${\scriptstyle\,\pm\,0.33}$ & 2.37${\scriptstyle\,\pm\,0.02}$ & 11.44${\scriptstyle\,\pm\,0.09}$ & 11.48${\scriptstyle\,\pm\,0.09}$ & 0.87${\scriptstyle\,\pm\,0.08}$ & 1.0 & (a,d);(38) \\
N4636 & 42734 & E0-1 & 11 & 13.7${\scriptstyle\,\pm\,1.4}$ & 8.58${\scriptstyle\,\pm\,0.22}$ & 2.26${\scriptstyle\,\pm\,0.02}$ & 11.25${\scriptstyle\,\pm\,0.09}$ & 10.71${\scriptstyle\,\pm\,0.09}$ & 0.96${\scriptstyle\,\pm\,0.08}$ & 0.28 & (a,d);(39) \\
N0541 & 5305 & E & 43 & 63.7${\scriptstyle\,\pm\,6.4}$ & 8.59${\scriptstyle\,\pm\,0.34}$ & 2.28${\scriptstyle\,\pm\,0.01}$ & 11.36${\scriptstyle\,\pm\,0.10}$ & 11.38${\scriptstyle\,\pm\,0.13}$ & 0.92${\scriptstyle\,\pm\,0.09}$ & 1.0 & (a,m);(40) \\
N7052 & 66537 & E3 & 1 & 70.4${\scriptstyle\,\pm\,8.4}$ & 8.60${\scriptstyle\,\pm\,0.23}$ & 2.42${\scriptstyle\,\pm\,0.02}$ & 11.68${\scriptstyle\,\pm\,0.09}$ & 11.61${\scriptstyle\,\pm\,0.10}$ & 1.00${\scriptstyle\,\pm\,0.05}$ & 1.0 & (a,b);(41) \\
N4621 & 42628 & E5 & 197 & 18.3${\scriptstyle\,\pm\,3.0}$ & 8.60${\scriptstyle\,\pm\,0.09}$ & 2.30${\scriptstyle\,\pm\,0.02}$ & 11.19${\scriptstyle\,\pm\,0.12}$ & 11.12${\scriptstyle\,\pm\,0.12}$ & 0.53${\scriptstyle\,\pm\,0.06}$ & 1.0 & (a,c);(42) \\
N4526 & 41772 & SAB0 & 197 & 16.4${\scriptstyle\,\pm\,1.8}$ & 8.65${\scriptstyle\,\pm\,0.12}$ & 2.32${\scriptstyle\,\pm\,0.02}$ & 11.22${\scriptstyle\,\pm\,0.09}$ & 11.02${\scriptstyle\,\pm\,0.09}$ & 0.54${\scriptstyle\,\pm\,0.10}$ & 0.65 & (a,b);(43) \\
N4342 & 40252 & S0 & 197 & 22.9${\scriptstyle\,\pm\,1.4}$ & 8.66${\scriptstyle\,\pm\,0.19}$ & 2.38${\scriptstyle\,\pm\,0.02}$ & 10.44${\scriptstyle\,\pm\,0.05}$ & 10.31${\scriptstyle\,\pm\,0.09}$ & -0.24${\scriptstyle\,\pm\,0.06}$ & 0.7 & (a,b);(44) \\
N0741 & 7252 & E0 & 8 & 65.7${\scriptstyle\,\pm\,6.6}$ & 8.67${\scriptstyle\,\pm\,0.37}$ & 2.37${\scriptstyle\,\pm\,0.02}$ & 11.82${\scriptstyle\,\pm\,0.09}$ & 11.86${\scriptstyle\,\pm\,0.09}$ & 1.18${\scriptstyle\,\pm\,0.07}$ & 1.0 & (a,b);(45) \\
N4552 & 41968 & E0-1 & 197 & 15.3${\scriptstyle\,\pm\,1.0}$ & 8.70${\scriptstyle\,\pm\,0.05}$ & 2.35${\scriptstyle\,\pm\,0.02}$ & 11.11${\scriptstyle\,\pm\,0.06}$ & 11.42${\scriptstyle\,\pm\,0.11}$ & 0.45${\scriptstyle\,\pm\,0.07}$ & 1.0 & (a,c);(46) \\
N4261 & 39659 & E2 & 39 & 32.4${\scriptstyle\,\pm\,2.8}$ & 8.72${\scriptstyle\,\pm\,0.10}$ & 2.42${\scriptstyle\,\pm\,0.02}$ & 11.60${\scriptstyle\,\pm\,0.09}$ & 11.65${\scriptstyle\,\pm\,0.09}$ & 0.86${\scriptstyle\,\pm\,0.12}$ & 1.0 & (a,b);(47) \\
N4291 & 39791 & E2 & 7 & 26.6${\scriptstyle\,\pm\,3.9}$ & 8.76${\scriptstyle\,\pm\,0.23}$ & 2.38${\scriptstyle\,\pm\,0.02}$ & 10.95${\scriptstyle\,\pm\,0.13}$ & 10.85${\scriptstyle\,\pm\,0.09}$ & 0.44${\scriptstyle\,\pm\,0.11}$ & 1.0 & (a+e,b);(48) \\
N3665 & 35064 & SA0 & 4 & 34.7${\scriptstyle\,\pm\,6.7}$ & 8.76${\scriptstyle\,\pm\,0.09}$ & 2.34${\scriptstyle\,\pm\,0.02}$ & 11.50${\scriptstyle\,\pm\,0.15}$ & 11.29${\scriptstyle\,\pm\,0.17}$ & 0.80${\scriptstyle\,\pm\,0.09}$ & 0.58 & (a,g);(49) \\
N1374 & 13267 & E0 & 49 & 19.2${\scriptstyle\,\pm\,0.7}$ & 8.76${\scriptstyle\,\pm\,0.06}$ & 2.31${\scriptstyle\,\pm\,0.02}$ & 10.67${\scriptstyle\,\pm\,0.03}$ & 10.63${\scriptstyle\,\pm\,0.09}$ & 0.24${\scriptstyle\,\pm\,0.03}$ & 1.0 & (a,b);(50) \\
N0383 & 3982 & S0 & 48 & 59.2${\scriptstyle\,\pm\,5.9}$ & 8.76${\scriptstyle\,\pm\,0.32}$ & 2.38${\scriptstyle\,\pm\,0.03}$ & 11.69${\scriptstyle\,\pm\,0.11}$ & 11.74${\scriptstyle\,\pm\,0.14}$ & 1.09${\scriptstyle\,\pm\,0.13}$ & 1.0 & (a,f);(51) \\
N6251 & 58472 & E1 & 17 & 108.4${\scriptstyle\,\pm\,9.0}$ & 8.79${\scriptstyle\,\pm\,0.16}$ & 2.46${\scriptstyle\,\pm\,0.02}$ & 11.95${\scriptstyle\,\pm\,0.07}$ & 11.88${\scriptstyle\,\pm\,0.09}$ & 1.13${\scriptstyle\,\pm\,0.07}$ & 1.0 & (a,b,q);(52) \\
N4594 & 42407 & Sa & 15 & 9.9${\scriptstyle\,\pm\,0.8}$ & 8.82${\scriptstyle\,\pm\,0.05}$ & 2.38${\scriptstyle\,\pm\,0.02}$ & 11.39${\scriptstyle\,\pm\,0.08}$ & 11.47${\scriptstyle\,\pm\,0.09}$ & 0.74${\scriptstyle\,\pm\,0.08}$ & 0.93 & (a,b);(53) \\
N1332 & 12838 & E6/S0 & 10 & 22.3${\scriptstyle\,\pm\,1.9}$ & 8.83${\scriptstyle\,\pm\,0.04}$ & 2.52${\scriptstyle\,\pm\,0.02}$ & 11.31${\scriptstyle\,\pm\,0.08}$ & 11.26${\scriptstyle\,\pm\,0.09}$ & 0.68${\scriptstyle\,\pm\,0.12}$ & 1.0 & (a,b);(54) \\
N5813 & 53643 & E1-2 & 13 & 32.2${\scriptstyle\,\pm\,2.7}$ & 8.85${\scriptstyle\,\pm\,0.06}$ & 2.32${\scriptstyle\,\pm\,0.02}$ & 11.50${\scriptstyle\,\pm\,0.08}$ & 11.77${\scriptstyle\,\pm\,0.09}$ & 0.96${\scriptstyle\,\pm\,0.07}$ & 1.0 & (a,c);(55) \\
N0315 & 3455 & E/cD & 14 & 57.7${\scriptstyle\,\pm\,2.8}$ & 8.92${\scriptstyle\,\pm\,0.31}$ & 2.49${\scriptstyle\,\pm\,0.04}$ & 11.79${\scriptstyle\,\pm\,0.08}$ & 11.87${\scriptstyle\,\pm\,0.09}$ & 1.01${\scriptstyle\,\pm\,0.05}$ & 1.0 & (a,b);(56) \\
N5077 & 46456 & E3 & 5 & 38.7${\scriptstyle\,\pm\,8.4}$ & 8.93${\scriptstyle\,\pm\,0.27}$ & 2.35${\scriptstyle\,\pm\,0.02}$ & 11.26${\scriptstyle\,\pm\,0.16}$ & 11.28${\scriptstyle\,\pm\,0.09}$ & 0.52${\scriptstyle\,\pm\,0.08}$ & 1.0 & (a,b);(57) \\
N0524 & 5222 & S0 & 9 & 24.2${\scriptstyle\,\pm\,2.2}$ & 8.94${\scriptstyle\,\pm\,0.05}$ & 2.37${\scriptstyle\,\pm\,0.02}$ & 11.29${\scriptstyle\,\pm\,0.08}$ & 11.26${\scriptstyle\,\pm\,0.09}$ & 0.56${\scriptstyle\,\pm\,0.07}$ & 0.92 & (a,b);(58) \\
N1399 & 13418 & E1 & 49 & 20.9${\scriptstyle\,\pm\,0.7}$ & 8.94${\scriptstyle\,\pm\,0.31}$ & 2.53${\scriptstyle\,\pm\,0.02}$ & 11.54${\scriptstyle\,\pm\,0.05}$ & 11.50${\scriptstyle\,\pm\,0.09}$ & 0.74${\scriptstyle\,\pm\,0.09}$ & 1.0 & (a,b);(59) \\
N3115 & 29265 & S0 & 2 & 9.5${\scriptstyle\,\pm\,0.4}$ & 8.95${\scriptstyle\,\pm\,0.09}$ & 2.36${\scriptstyle\,\pm\,0.02}$ & 10.98${\scriptstyle\,\pm\,0.04}$ & 10.92${\scriptstyle\,\pm\,0.09}$ & 0.42${\scriptstyle\,\pm\,0.06}$ & 0.9 & (a,b);(60) \\
IC1459 & 70090 & E4 & 6 & 28.9${\scriptstyle\,\pm\,3.7}$ & 8.96${\scriptstyle\,\pm\,0.43}$ & 2.53${\scriptstyle\,\pm\,0.02}$ & 11.64${\scriptstyle\,\pm\,0.11}$ & 11.60${\scriptstyle\,\pm\,0.09}$ & 0.80${\scriptstyle\,\pm\,0.08}$ & 1.0 & (a+i,b);(61) \\
N4374 & 40455 & E1 & 197 & 18.5${\scriptstyle\,\pm\,0.6}$ & 8.97${\scriptstyle\,\pm\,0.05}$ & 2.41${\scriptstyle\,\pm\,0.02}$ & 11.40${\scriptstyle\,\pm\,0.03}$ & 11.62${\scriptstyle\,\pm\,0.09}$ & 0.57${\scriptstyle\,\pm\,0.03}$ & 1.0 & (a,b);(62) \\
N5846 & 53932 & E0-1 & 13 & 24.9${\scriptstyle\,\pm\,2.3}$ & 9.04${\scriptstyle\,\pm\,0.06}$ & 2.35${\scriptstyle\,\pm\,0.02}$ & 11.46${\scriptstyle\,\pm\,0.09}$ & 11.62${\scriptstyle\,\pm\,0.16}$ & 0.80${\scriptstyle\,\pm\,0.10}$ & 1.0 & (a,c);(63) \\
N1277 & 12434 & S0 & 180 & 71.0${\scriptstyle\,\pm\,7.1}$ & 9.06${\scriptstyle\,\pm\,0.11}$ & 2.50${\scriptstyle\,\pm\,0.01}$ & 11.07${\scriptstyle\,\pm\,0.08}$ & 10.98${\scriptstyle\,\pm\,0.09}$ & 0.15${\scriptstyle\,\pm\,0.07}$ & 0.55 & (t,a,b);(64) \\
IC4296 & 48040 & E0 & 64 & 49.2${\scriptstyle\,\pm\,3.6}$ & 9.11${\scriptstyle\,\pm\,0.07}$ & 2.51${\scriptstyle\,\pm\,0.02}$ & 11.83${\scriptstyle\,\pm\,0.07}$ & 11.78${\scriptstyle\,\pm\,0.09}$ & 0.98${\scriptstyle\,\pm\,0.06}$ & 1.0 & (a,b);(65) \\
N7768 & 72605 & E4 & 9 & 116.0${\scriptstyle\,\pm\,27.5}$ & 9.13${\scriptstyle\,\pm\,0.18}$ & 2.42${\scriptstyle\,\pm\,0.02}$ & 11.91${\scriptstyle\,\pm\,0.19}$ & 11.75${\scriptstyle\,\pm\,0.09}$ & 1.15${\scriptstyle\,\pm\,0.12}$ & 1.0 & (a,b);(66) \\
U12064 & 69055 & E/S0 & 7 & 72.5${\scriptstyle\,\pm\,6.7}$ & 9.16${\scriptstyle\,\pm\,0.52}$ & 2.41${\scriptstyle\,\pm\,0.03}$ & 11.25${\scriptstyle\,\pm\,0.17}$ & 11.30${\scriptstyle\,\pm\,0.19}$ & 0.70${\scriptstyle\,\pm\,0.13}$ & 1.0 & (a,f);(67) \\
N6240S & 59186 & S0/I0 & 2 & 105.0${\scriptstyle\,\pm\,10.5}$ & 9.17${\scriptstyle\,\pm\,0.21}$ & 2.44${\scriptstyle\,\pm\,0.07}$ & 11.84${\scriptstyle\,\pm\,0.09}$ & 11.35${\scriptstyle\,\pm\,0.13}$ & 0.93${\scriptstyle\,\pm\,0.08}$ & 0.28 & (a,k);(68) \\
N6861 & 64136 & E4/SA0 & 12 & 27.3${\scriptstyle\,\pm\,4.5}$ & 9.30${\scriptstyle\,\pm\,0.08}$ & 2.59${\scriptstyle\,\pm\,0.02}$ & 11.14${\scriptstyle\,\pm\,0.13}$ & 11.21${\scriptstyle\,\pm\,0.09}$ & 0.32${\scriptstyle\,\pm\,0.08}$ & 1.0 & (a,b);(69) \\
N4649 & 42831 & E2 & 197 & 16.5${\scriptstyle\,\pm\,0.6}$ & 9.32${\scriptstyle\,\pm\,0.12}$ & 2.43${\scriptstyle\,\pm\,0.02}$ & 11.66${\scriptstyle\,\pm\,0.06}$ & 11.64${\scriptstyle\,\pm\,0.09}$ & 0.90${\scriptstyle\,\pm\,0.10}$ & 1.0 & (d,a);(70) \\
N7619 & 71121 & E3 & 16 & 51.5${\scriptstyle\,\pm\,7.4}$ & 9.34${\scriptstyle\,\pm\,0.10}$ & 2.51${\scriptstyle\,\pm\,0.02}$ & 11.61${\scriptstyle\,\pm\,0.12}$ & 11.65${\scriptstyle\,\pm\,0.09}$ & 0.87${\scriptstyle\,\pm\,0.09}$ & 1.0 & (b,a);(71) \\
N4472 & 41220 & E2 & 197 & 17.1${\scriptstyle\,\pm\,0.6}$ & 9.40${\scriptstyle\,\pm\,0.04}$ & 2.40${\scriptstyle\,\pm\,0.02}$ & 11.75${\scriptstyle\,\pm\,0.07}$ & 11.86${\scriptstyle\,\pm\,0.09}$ & 0.89${\scriptstyle\,\pm\,0.11}$ & 1.0 & (a,b);(72) \\
N3923 & 37061 & E4 & 10 & 20.9${\scriptstyle\,\pm\,2.7}$ & 9.45${\scriptstyle\,\pm\,0.12}$ & 2.35${\scriptstyle\,\pm\,0.02}$ & 11.50${\scriptstyle\,\pm\,0.11}$ & 11.56${\scriptstyle\,\pm\,0.09}$ & 0.89${\scriptstyle\,\pm\,0.10}$ & 1.0 & (a,c);(73) \\
N3091 & 28927 & E3 & 10 & 51.2${\scriptstyle\,\pm\,8.3}$ & 9.56${\scriptstyle\,\pm\,0.07}$ & 2.49${\scriptstyle\,\pm\,0.02}$ & 11.62${\scriptstyle\,\pm\,0.12}$ & 11.58${\scriptstyle\,\pm\,0.09}$ & 0.89${\scriptstyle\,\pm\,0.09}$ & 1.0 & (a,b);(74) \\
N1550 & 14880 & E1/cD & 15 & 51.6${\scriptstyle\,\pm\,5.6}$ & 9.57${\scriptstyle\,\pm\,0.07}$ & 2.48${\scriptstyle\,\pm\,0.02}$ & 11.32${\scriptstyle\,\pm\,0.10}$ & 11.31${\scriptstyle\,\pm\,0.09}$ & 0.66${\scriptstyle\,\pm\,0.08}$ & 1.0 & (a,b);(75) \\
N6086 & 57482 & E/cD & 37 & 138.0${\scriptstyle\,\pm\,11.5}$ & 9.57${\scriptstyle\,\pm\,0.17}$ & 2.50${\scriptstyle\,\pm\,0.02}$ & 11.87${\scriptstyle\,\pm\,0.08}$ & 11.69${\scriptstyle\,\pm\,0.09}$ & 1.20${\scriptstyle\,\pm\,0.08}$ & 1.0 & (a,b);(76) \\
A1836$_\textrm{B}$ & 49940 & E & 41 & 152.4${\scriptstyle\,\pm\,8.4}$ & 9.57${\scriptstyle\,\pm\,0.06}$ & 2.46${\scriptstyle\,\pm\,0.02}$ & 11.75${\scriptstyle\,\pm\,0.06}$ & 11.81${\scriptstyle\,\pm\,0.10}$ & 0.89${\scriptstyle\,\pm\,0.06}$ & 1.0 & (a,b);(77) \\
N1407 & 13505 & E0 & 15 & 28.0${\scriptstyle\,\pm\,3.4}$ & 9.65${\scriptstyle\,\pm\,0.08}$ & 2.45${\scriptstyle\,\pm\,0.02}$ & 11.72${\scriptstyle\,\pm\,0.12}$ & 11.71${\scriptstyle\,\pm\,0.09}$ & 0.97${\scriptstyle\,\pm\,0.11}$ & 1.0 & (a,b);(78) \\
N5328 & 49307 & E1 & 11 & 64.1${\scriptstyle\,\pm\,7.0}$ & 9.67${\scriptstyle\,\pm\,0.16}$ & 2.52${\scriptstyle\,\pm\,0.02}$ & 11.71${\scriptstyle\,\pm\,0.09}$ & 11.75${\scriptstyle\,\pm\,0.19}$ & 0.94${\scriptstyle\,\pm\,0.06}$ & 1.0 & (a,c);(79) \\
M1216 & 23789 & E & 2 & 94.0${\scriptstyle\,\pm\,9.4}$ & 9.69${\scriptstyle\,\pm\,0.16}$ & 2.49${\scriptstyle\,\pm\,0.01}$ & 11.29${\scriptstyle\,\pm\,0.09}$ & 11.37${\scriptstyle\,\pm\,0.09}$ & 0.47${\scriptstyle\,\pm\,0.07}$ & 1.0 & (l,a);(80) \\
N4486 & 41361 & E1/cD & 197 & 16.7${\scriptstyle\,\pm\,0.6}$ & 9.81${\scriptstyle\,\pm\,0.05}$ & 2.42${\scriptstyle\,\pm\,0.02}$ & 11.61${\scriptstyle\,\pm\,0.05}$ & 11.72${\scriptstyle\,\pm\,0.09}$ & 0.82${\scriptstyle\,\pm\,0.07}$ & 1.0 & (v,a,b);(81) \\
N5419 & 50100 & E & 43 & 56.2${\scriptstyle\,\pm\,6.1}$ & 9.86${\scriptstyle\,\pm\,0.14}$ & 2.57${\scriptstyle\,\pm\,0.01}$ & 12.00${\scriptstyle\,\pm\,0.09}$ & 12.01${\scriptstyle\,\pm\,0.15}$ & 1.26${\scriptstyle\,\pm\,0.07}$ & 1.0 & (a,c);(82) \\
N3842 & 36487 & E1 & 117 & 92.2${\scriptstyle\,\pm\,10.6}$ & 9.96${\scriptstyle\,\pm\,0.14}$ & 2.44${\scriptstyle\,\pm\,0.02}$ & 11.81${\scriptstyle\,\pm\,0.11}$ & 11.77${\scriptstyle\,\pm\,0.09}$ & 1.11${\scriptstyle\,\pm\,0.07}$ & 1.0 & (a,b);(83) \\
N1600 & 15406 & E3 & 30 & 64.0${\scriptstyle\,\pm\,6.4}$ & 10.23${\scriptstyle\,\pm\,0.04}$ & 2.47${\scriptstyle\,\pm\,0.02}$ & 11.86${\scriptstyle\,\pm\,0.08}$ & 11.92${\scriptstyle\,\pm\,0.10}$ & 1.08${\scriptstyle\,\pm\,0.05}$ & 1.0 & (a,u);(84) \\
N4889 & 44715 & E4/cD & 583 & 102.0${\scriptstyle\,\pm\,5.2}$ & 10.32${\scriptstyle\,\pm\,0.44}$ & 2.56${\scriptstyle\,\pm\,0.02}$ & 12.13${\scriptstyle\,\pm\,0.05}$ & 12.30${\scriptstyle\,\pm\,0.10}$ & 1.34${\scriptstyle\,\pm\,0.05}$ & 1.0 & (a,c);(85) \\

\vspace{-0.3cm}
\enddata

\tablenotetext{}{\small \textbf{\textit{Columns.}} 
(i) Galaxy name, with prefixes defined as N$=$NGC (New General Catalogue); IC (Index Catalogue); A$=$Abell Catalogue; M$=$Mrk (Markarian Catalogue); U$=$UGC (Uppsala General Catalogue). 
Top to bottom: galaxies are in order of ascending $\mbh$.
(ii) Principal Galaxies Catalog (PGC) identification number (HyperLEDA). 
(iii) Hubble morphological type (from RC3, \citealt{deVaucouleurs:1991,Beifiori:2012,Kormendy:2013}; NED).
(iv) Number of group/cluster members (from 2MASS 11.75 catalog; \citealt{Tully:2015}).
(v) Distance (from surface brightness fluctuations or redshifts; \citealt{vBosch:2016,Saglia:2016}). 
(vi) Dynamical BH masses measurements via stellar/gas kinematics (mostly from \citealt{vBosch:2016} and refs.\,therein).
(vii) Stellar velocity dispersion within the optical half-light radius $\re$ (mostly from \citealt{vBosch:2016}).
(viii) Total $K_{\rm s}$-band luminosity (mostly from \citealt{vBosch:2016}).
(ix) Stellar bulge mass (mostly from \citealt{Kormendy:2013}). 
(x) Effective, optical half-light radius (mostly from \citealt{vBosch:2016}).
(xi) (Classical or pseudo) bulge-to-total luminosity ratio (mostly from \citealt{Kormendy:2013}).
(xii) References used for the listed optical properties and single-object notes (if any).\\ 
\indent
\textbf{\textit{     References.}} 
(a) \citet{vBosch:2016};
(b) \citet{Kormendy:2013};
(c) \citet{Saglia:2016};
(d) \citet{Beifiori:2012};
(e) \citet{Graham:2013};
(f) \citet{Donzelli:2007};
(g) \citet{Graham:2015};
(h) \citet{Beifiori:2010phd};
(i) \citet{Cappellari:2002};
(j) \citet{Dong:2006};
(k) \citet{Medling:2015};
(l) \citet{Walsh:2017};
(m) \citet{deSouza:2004};
(n) \citet{Bottema:1999};
(o) \citet{Onken:2014};
(p) \citet{Gadotti:2008};
(q) \citet{Chen:2011};
(r) \citet{Seigar:2008};
(s) \citet{Pastorini:2007};
(t) \citet{Graham:2016};
(u) \citet{Thomas:2016};
(v) \citet{EHT:2019I}.
\\
\indent
\textbf{\textit{     Notes.}} 
(6, 13, 19, 22, 42, 46, 53, 62, 65, 67, 70, 77, 81, 85) Other common names (respectively): M81, M106, M105, M31/Andromeda, M59, M89, M104, M84, A3565$_{\rm BCG}$, 3C449, M60, PKS\,B1358-113, M87, A1656$_{\rm BCG}$.
(8, 10, 12, 16, 18, 20, 22, 24, 26, 28, 35, 38, 39, 40, 49, 51, 67, 68) Bulge mass computed via $B/T$ in conjunction with Eq.~\ref{e:ML} (\citealt{Kormendy:2013}).
(13) $N_{\rm m}$ from \citet{deVaucouleurs:1975}.
(48) Geometric average between (a) and (e) dynamical BH masses. 
(52) $N_{\rm m}$ from \citet{Chen:2011}.
(61) Geometric average between the stellar (a) and gas (i) dynamical BH masses. 
(65, 76, 77, 83) $N_{\rm m}$ from \citet{Abell:1989} catalog.
(67) NED redshift/distance adopted. 
(68) BH in the major southern (S) structure (merging pair).
(77) `B' subscript stands for BCG.
(81) Direct SMBH horizon imaging via EHT.
(85) $N_{\rm m}$ from \citet{Beijersbergen:2003}.
}
\label{tabop}
\end{deluxetable*}
\end{center}
\capstarttrue

\vspace{-0.2cm}
\capstartfalse
\begin{center}
\begin{deluxetable*}{llccccccccl} 
\tablecaption{X-ray properties and environment of our sample of galaxies, groups, and clusters hosting the SMBHs listed in Table~\ref{tabop}.}
\tablehead{
Galaxy & {Central} & {$L_{\rm x,g}$} & {$T_{\rm x,g}$}  & {$\rxg$} & {$L_{\rm x,c}$} & {$T_{\rm x,c}$} & {$\rxc$} & {$L_{\rm x,500}$} & $R_{500}$ & Refs.\,\&\\
 Name        &  Galaxy   & {$\log({\rm erg\,s^{-1}})$} & {$\log({\rm keV})$} & {$\log({\rm kpc})$}  & {$\log({\rm erg\,s^{-1}}) $} & {$\log({\rm keV})$} & {$\log({\rm kpc})$} & {$\log({\rm erg\,s^{-1}})$} & {$\log({\rm kpc})$} &  Notes \\
 \ \ (i)   &   \ \ \,(ii) &  {(iii)} &  {(iv)}  & {(v)}  &  {(vi)}  &  {(vii)}  & {(viii)} & {(ix)} & {(x)} & {(xi)} 
}

\startdata
N0821 & isolated & 38.40${\scriptstyle\,\pm\,0.21}$ & -0.70${\scriptstyle\,\pm\,0.19}$ & 1.20 & 38.40${\scriptstyle\,\pm\,0.21}$ & -0.70${\scriptstyle\,\pm\,0.19}$ & 1.48 & 38.40${\scriptstyle\,\pm\,0.21}$ & 2.31 & (a,b);(1) \\
N2787 & isolated & 37.97${\scriptstyle\,\pm\,0.48}$ & -0.74${\scriptstyle\,\pm\,0.31}$ & 0.26 & 37.97${\scriptstyle\,\pm\,0.48}$ & -0.74${\scriptstyle\,\pm\,0.31}$ & 1.46 & 37.97${\scriptstyle\,\pm\,0.48}$ & 2.28 & (C);(2) \\
N1023 & BFG:N1023 & 38.76${\scriptstyle\,\pm\,0.07}$ & -0.52${\scriptstyle\,\pm\,0.05}$ & 0.48 & 38.76${\scriptstyle\,\pm\,0.07}$ & -0.52${\scriptstyle\,\pm\,0.05}$ & 1.58 & 38.76${\scriptstyle\,\pm\,0.07}$ & 2.40 & (a);(3) \\
N7582 & BFG:Grus & 39.43${\scriptstyle\,\pm\,0.42}$ & -0.74${\scriptstyle\,\pm\,0.09}$ & 0.70 & 39.43${\scriptstyle\,\pm\,0.42}$ & -0.74${\scriptstyle\,\pm\,0.09}$ & 1.46 & 39.43${\scriptstyle\,\pm\,0.42}$ & 2.28 & (w,C);(4) \\
N5128 & BFG:CenA & 39.86${\scriptstyle\,\pm\,0.22}$ & -0.46${\scriptstyle\,\pm\,0.06}$ & 1.18 & 39.86${\scriptstyle\,\pm\,0.22}$ & -0.46${\scriptstyle\,\pm\,0.06}$ & 1.61 & 39.86${\scriptstyle\,\pm\,0.22}$ & 2.44 & (v);(5) \\
N3031 & BFG:M81 & 38.62${\scriptstyle\,\pm\,0.03}$ & -0.49${\scriptstyle\,\pm\,0.09}$ & 0.30 & 38.62${\scriptstyle\,\pm\,0.03}$ & -0.49${\scriptstyle\,\pm\,0.09}$ & 1.59 & 38.62${\scriptstyle\,\pm\,0.03}$ & 2.42 & (Q);(6) \\
N4151 & N4258(M106) & 39.11${\scriptstyle\,\pm\,0.14}$ & -0.60${\scriptstyle\,\pm\,0.07}$ & 0.85 & 39.11${\scriptstyle\,\pm\,0.14}$ & -0.60${\scriptstyle\,\pm\,0.07}$ & 1.54 & 39.11${\scriptstyle\,\pm\,0.14}$ & 2.36 & (J);(7) \\
N4203 & isolated & 38.74${\scriptstyle\,\pm\,0.11}$ & -0.54${\scriptstyle\,\pm\,0.13}$ & 0.30 & 38.74${\scriptstyle\,\pm\,0.11}$ & -0.54${\scriptstyle\,\pm\,0.13}$ & 1.57 & 38.74${\scriptstyle\,\pm\,0.11}$ & 2.39 & (c);(8) \\
N4459 & N4486/N4472 & 39.20${\scriptstyle\,\pm\,0.04}$ & -0.42${\scriptstyle\,\pm\,0.10}$ & 0.48 & 39.20${\scriptstyle\,\pm\,0.04}$ & -0.42${\scriptstyle\,\pm\,0.10}$ & 1.63 & 39.20${\scriptstyle\,\pm\,0.04}$ & 2.46 & (c);(9) \\
N4596 & N4486/N4472 & 39.41${\scriptstyle\,\pm\,0.35}$ & -0.68${\scriptstyle\,\pm\,0.07}$ & 0.70 & 39.41${\scriptstyle\,\pm\,0.35}$ & -0.68${\scriptstyle\,\pm\,0.07}$ & 1.49 & 39.41${\scriptstyle\,\pm\,0.35}$ & 2.32 & (a);(10) \\
N3377 & N3379 & 38.04${\scriptstyle\,\pm\,0.33}$ & -0.60${\scriptstyle\,\pm\,0.10}$ & 0.48 & 38.04${\scriptstyle\,\pm\,0.33}$ & -0.60${\scriptstyle\,\pm\,0.10}$ & 1.53 & 38.04${\scriptstyle\,\pm\,0.33}$ & 2.36 & (j,n);(11) \\
N4036 & N3945 & 39.34${\scriptstyle\,\pm\,0.09}$ & -0.34${\scriptstyle\,\pm\,0.08}$ & 1.23 & 39.34${\scriptstyle\,\pm\,0.09}$ & -0.34${\scriptstyle\,\pm\,0.08}$ & 1.68 & 39.34${\scriptstyle\,\pm\,0.09}$ & 2.50 & (b);(12) \\
N4258 & BFG:CVnI & 39.72${\scriptstyle\,\pm\,0.07}$ & -0.64${\scriptstyle\,\pm\,0.05}$ & 1.04 & 39.72${\scriptstyle\,\pm\,0.07}$ & -0.64${\scriptstyle\,\pm\,0.05}$ & 1.52 & 39.72${\scriptstyle\,\pm\,0.07}$ & 2.34 & (R);(13) \\
N4564 & N4486/N4472 & 39.00${\scriptstyle\,\pm\,0.09}$ & -0.42${\scriptstyle\,\pm\,0.19}$ & 1.10 & 39.00${\scriptstyle\,\pm\,0.09}$ & -0.42${\scriptstyle\,\pm\,0.19}$ & 1.63 & 39.00${\scriptstyle\,\pm\,0.09}$ & 2.46 & (b);(14) \\
N4473 & N4486/N4472 & 39.08${\scriptstyle\,\pm\,0.08}$ & -0.51${\scriptstyle\,\pm\,0.06}$ & 0.65 & 39.08${\scriptstyle\,\pm\,0.08}$ & -0.51${\scriptstyle\,\pm\,0.06}$ & 1.59 & 39.08${\scriptstyle\,\pm\,0.08}$ & 2.41 & (a);(15) \\
N4278 & N4414 & 39.36${\scriptstyle\,\pm\,0.09}$ & -0.52${\scriptstyle\,\pm\,0.04}$ & 0.85 & 39.36${\scriptstyle\,\pm\,0.09}$ & -0.52${\scriptstyle\,\pm\,0.04}$ & 1.58 & 39.36${\scriptstyle\,\pm\,0.09}$ & 2.40 & (a);(16) \\
N5018 & BGG:N5018 & 39.92${\scriptstyle\,\pm\,0.13}$ & -0.42${\scriptstyle\,\pm\,0.05}$ & 0.78 & 39.92${\scriptstyle\,\pm\,0.13}$ & -0.42${\scriptstyle\,\pm\,0.05}$ & 1.63 & 39.92${\scriptstyle\,\pm\,0.13}$ & 2.46 & (l);(17) \\
N7331 & isolated & 38.89${\scriptstyle\,\pm\,0.10}$ & -0.30${\scriptstyle\,\pm\,0.09}$ & 0.30 & 38.89${\scriptstyle\,\pm\,0.10}$ & -0.30${\scriptstyle\,\pm\,0.09}$ & 1.70 & 38.89${\scriptstyle\,\pm\,0.10}$ & 2.52 & (F,G);(18) \\
N3379 & BFG:N3379 & 38.63${\scriptstyle\,\pm\,0.08}$ & -0.60${\scriptstyle\,\pm\,0.05}$ & 0.70 & 38.63${\scriptstyle\,\pm\,0.08}$ & -0.60${\scriptstyle\,\pm\,0.05}$ & 1.54 & 38.63${\scriptstyle\,\pm\,0.08}$ & 2.36 & (a);(19) \\
N2110 & isolated & 39.11${\scriptstyle\,\pm\,0.09}$ & -0.47${\scriptstyle\,\pm\,0.12}$ & 0.00 & 39.11${\scriptstyle\,\pm\,0.09}$ & -0.47${\scriptstyle\,\pm\,0.12}$ & 1.61 & 39.11${\scriptstyle\,\pm\,0.09}$ & 2.43 & (O);(20) \\
N3607 & BGG:LeoII & 39.80${\scriptstyle\,\pm\,0.07}$ & -0.35${\scriptstyle\,\pm\,0.06}$ & 0.93 & 40.40${\scriptstyle\,\pm\,0.36}$ & -0.35${\scriptstyle\,\pm\,0.07}$ & 1.69 & 41.36${\scriptstyle\,\pm\,0.09}$ & 2.50 & (f,v,d);(21) \\
N0224 & BFG:LG & 38.30${\scriptstyle\,\pm\,0.12}$ & -0.54${\scriptstyle\,\pm\,0.05}$ & 0.00 & 38.30${\scriptstyle\,\pm\,0.12}$ & -0.54${\scriptstyle\,\pm\,0.05}$ & 1.57 & 38.30${\scriptstyle\,\pm\,0.12}$ & 2.39 & (M,N);(22) \\
N1316 & BGG:FornaxA & 40.46${\scriptstyle\,\pm\,0.05}$ & -0.18${\scriptstyle\,\pm\,0.04}$ & 1.20 & 40.64${\scriptstyle\,\pm\,0.05}$ & -0.17${\scriptstyle\,\pm\,0.05}$ & 1.80 & 40.79${\scriptstyle\,\pm\,0.05}$ & 2.59 & (e);(23) \\
N1667 & BFG:N1667 & 40.15${\scriptstyle\,\pm\,0.18}$ & -0.48${\scriptstyle\,\pm\,0.08}$ & 0.90 & 40.15${\scriptstyle\,\pm\,0.18}$ & -0.48${\scriptstyle\,\pm\,0.08}$ & 1.60 & 40.15${\scriptstyle\,\pm\,0.18}$ & 2.42 & (w);(24) \\
N5576 & N5566 & 38.77${\scriptstyle\,\pm\,0.21}$ & -0.28${\scriptstyle\,\pm\,0.19}$ & 0.60 & 38.77${\scriptstyle\,\pm\,0.21}$ & -0.28${\scriptstyle\,\pm\,0.19}$ & 1.71 & 38.77${\scriptstyle\,\pm\,0.21}$ & 2.53 & (a);(25) \\
N1052 & N0988 & 39.58${\scriptstyle\,\pm\,0.09}$ & -0.47${\scriptstyle\,\pm\,0.05}$ & 0.70 & 39.58${\scriptstyle\,\pm\,0.09}$ & -0.47${\scriptstyle\,\pm\,0.05}$ & 1.61 & 39.58${\scriptstyle\,\pm\,0.09}$ & 2.43 & (j);(26) \\
N4026 & N4258(M106) & 38.38${\scriptstyle\,\pm\,0.27}$ & -0.55${\scriptstyle\,\pm\,0.20}$ & 0.30 & 38.38${\scriptstyle\,\pm\,0.27}$ & -0.55${\scriptstyle\,\pm\,0.20}$ & 1.56 & 38.38${\scriptstyle\,\pm\,0.27}$ & 2.39 & (a);(27) \\
N3801 & BFG:N3801 & 39.15${\scriptstyle\,\pm\,0.30}$ & -0.64${\scriptstyle\,\pm\,0.19}$ & 1.23 & 39.15${\scriptstyle\,\pm\,0.30}$ & -0.64${\scriptstyle\,\pm\,0.19}$ & 1.52 & 39.15${\scriptstyle\,\pm\,0.30}$ & 2.34 & (x);(28) \\
N4697 & N4594(M104) & 39.23${\scriptstyle\,\pm\,0.08}$ & -0.49${\scriptstyle\,\pm\,0.06}$ & 0.88 & 39.23${\scriptstyle\,\pm\,0.08}$ & -0.49${\scriptstyle\,\pm\,0.06}$ & 1.59 & 39.23${\scriptstyle\,\pm\,0.08}$ & 2.42 & (f);(29) \\
N1961 & BGG:N1961 & 40.36${\scriptstyle\,\pm\,0.09}$ & -0.17${\scriptstyle\,\pm\,0.05}$ & 1.15 & 40.62${\scriptstyle\,\pm\,0.09}$ & -0.21${\scriptstyle\,\pm\,0.05}$ & 1.62 & 41.08${\scriptstyle\,\pm\,0.17}$ & 2.57 & (y);(30) \\
N3608 & N3607(LeoII) & 39.63${\scriptstyle\,\pm\,0.09}$ & -0.40${\scriptstyle\,\pm\,0.09}$ & 1.00 & 39.63${\scriptstyle\,\pm\,0.09}$ & -0.40${\scriptstyle\,\pm\,0.09}$ & 1.64 & 39.63${\scriptstyle\,\pm\,0.09}$ & 2.47 & (a);(31) \\
N3998 & N4258(M106) & 40.04${\scriptstyle\,\pm\,0.12}$ & -0.59${\scriptstyle\,\pm\,0.09}$ & 0.30 & 40.04${\scriptstyle\,\pm\,0.12}$ & -0.59${\scriptstyle\,\pm\,0.09}$ & 1.55 & 40.04${\scriptstyle\,\pm\,0.12}$ & 2.37 & (a);(32) \\
N3245 & BFG:N3245 & 39.43${\scriptstyle\,\pm\,0.16}$ & -0.52${\scriptstyle\,\pm\,0.10}$ & 0.95 & 39.43${\scriptstyle\,\pm\,0.14}$ & -0.52${\scriptstyle\,\pm\,0.10}$ & 1.58 & 39.43${\scriptstyle\,\pm\,0.14}$ & 2.40 & (a);(33) \\
N3414 & BFG:N3414 & 39.23${\scriptstyle\,\pm\,0.16}$ & -0.24${\scriptstyle\,\pm\,0.16}$ & 0.60 & 39.23${\scriptstyle\,\pm\,0.16}$ & -0.24${\scriptstyle\,\pm\,0.16}$ & 1.73 & 39.23${\scriptstyle\,\pm\,0.16}$ & 2.55 & (a);(34) \\
N3862 & N3842 & 40.20${\scriptstyle\,\pm\,0.10}$ & -0.19${\scriptstyle\,\pm\,0.11}$ & 0.48 & 40.20${\scriptstyle\,\pm\,0.10}$ & -0.19${\scriptstyle\,\pm\,0.11}$ & 1.76 & 40.20${\scriptstyle\,\pm\,0.10}$ & 2.58 & (H);(35) \\
N5845 & N5846 & 38.77${\scriptstyle\,\pm\,0.25}$ & -0.41${\scriptstyle\,\pm\,0.21}$ & 0.60 & 38.77${\scriptstyle\,\pm\,0.25}$ & -0.41${\scriptstyle\,\pm\,0.21}$ & 1.64 & 38.77${\scriptstyle\,\pm\,0.25}$ & 2.46 & (a);(36) \\
N3585 & BFG:N3585 & 39.32${\scriptstyle\,\pm\,0.10}$ & -0.43${\scriptstyle\,\pm\,0.10}$ & 1.15 & 39.32${\scriptstyle\,\pm\,0.10}$ & -0.43${\scriptstyle\,\pm\,0.10}$ & 1.63 & 39.32${\scriptstyle\,\pm\,0.10}$ & 2.45 & (e);(37) \\
N7626 & N7619(Pegasus) & 40.40${\scriptstyle\,\pm\,0.10}$ & -0.16${\scriptstyle\,\pm\,0.06}$ & 0.70 & 40.40${\scriptstyle\,\pm\,0.10}$ & -0.16${\scriptstyle\,\pm\,0.06}$ & 1.77 & 40.40${\scriptstyle\,\pm\,0.10}$ & 2.59 & (c);(38) \\
N4636 & BGG:N4636 & 41.04${\scriptstyle\,\pm\,0.10}$ & -0.21${\scriptstyle\,\pm\,0.04}$ & 0.95 & 41.26${\scriptstyle\,\pm\,0.09}$ & -0.12${\scriptstyle\,\pm\,0.04}$ & 1.49 & 41.88${\scriptstyle\,\pm\,0.09}$ & 2.61 & (e,b,d);(39) \\
N0541 & N0547 & 40.63${\scriptstyle\,\pm\,0.14}$ & -0.26${\scriptstyle\,\pm\,0.06}$ & 0.70 & 40.63${\scriptstyle\,\pm\,0.14}$ & -0.26${\scriptstyle\,\pm\,0.06}$ & 1.72 & 40.63${\scriptstyle\,\pm\,0.14}$ & 2.54 & (k,B);(40) \\
N7052 & isolated & 41.26${\scriptstyle\,\pm\,0.11}$ & -0.25${\scriptstyle\,\pm\,0.06}$ & 1.08 & 41.26${\scriptstyle\,\pm\,0.11}$ & -0.25${\scriptstyle\,\pm\,0.06}$ & 1.72 & 41.26${\scriptstyle\,\pm\,0.11}$ & 2.55 & (c);(41) \\
N4621 & N4486/N4472 & 40.04${\scriptstyle\,\pm\,0.52}$ & -0.57${\scriptstyle\,\pm\,0.12}$ & 1.00 & 40.04${\scriptstyle\,\pm\,0.52}$ & -0.57${\scriptstyle\,\pm\,0.12}$ & 1.55 & 40.04${\scriptstyle\,\pm\,0.52}$ & 2.38 & (u,a);(42) \\
N4526 & N4486/N4472 & 39.92${\scriptstyle\,\pm\,0.56}$ & -0.46${\scriptstyle\,\pm\,0.06}$ & 1.00 & 39.92${\scriptstyle\,\pm\,0.56}$ & -0.46${\scriptstyle\,\pm\,0.06}$ & 1.60 & 39.92${\scriptstyle\,\pm\,0.56}$ & 2.44 & (u);(43) \\
N4342 & N4486/N4472 & 39.79${\scriptstyle\,\pm\,0.05}$ & -0.19${\scriptstyle\,\pm\,0.04}$ & 1.15 & 39.79${\scriptstyle\,\pm\,0.05}$ & -0.19${\scriptstyle\,\pm\,0.04}$ & 1.75 & 39.79${\scriptstyle\,\pm\,0.05}$ & 2.58 & (b);(44) \\
N0741 & BGG:N0741 & 41.26${\scriptstyle\,\pm\,0.09}$ & -0.10${\scriptstyle\,\pm\,0.05}$ & 1.18 & 41.49${\scriptstyle\,\pm\,0.09}$ & 0.01${\scriptstyle\,\pm\,0.04}$ & 1.93 & 42.23${\scriptstyle\,\pm\,0.09}$ & 2.68 & (c,b,d);(45) \\
N4552 & N4486/N4472 & 40.30${\scriptstyle\,\pm\,0.06}$ & -0.23${\scriptstyle\,\pm\,0.04}$ & 0.85 & 40.30${\scriptstyle\,\pm\,0.06}$ & -0.23${\scriptstyle\,\pm\,0.04}$ & 1.74 & 40.30${\scriptstyle\,\pm\,0.06}$ & 2.56 & (a);(46) \\
N4261 & N4303(M61) & 40.88${\scriptstyle\,\pm\,0.08}$ & -0.12${\scriptstyle\,\pm\,0.04}$ & 1.00 & 40.88${\scriptstyle\,\pm\,0.08}$ & -0.12${\scriptstyle\,\pm\,0.04}$ & 1.79 & 40.88${\scriptstyle\,\pm\,0.08}$ & 2.62 & (a);(47) \\
N4291 & BFG:N4291 & 40.91${\scriptstyle\,\pm\,0.25}$ & -0.23${\scriptstyle\,\pm\,0.06}$ & 1.08 & 40.91${\scriptstyle\,\pm\,0.25}$ & -0.23${\scriptstyle\,\pm\,0.06}$ & 1.73 & 40.91${\scriptstyle\,\pm\,0.25}$ & 2.56 & (v);(48) \\
N3665 & BFG:N3665 & 40.23${\scriptstyle\,\pm\,0.19}$ & -0.35${\scriptstyle\,\pm\,0.09}$ & 1.02 & 40.23${\scriptstyle\,\pm\,0.19}$ & -0.35${\scriptstyle\,\pm\,0.09}$ & 1.67 & 40.23${\scriptstyle\,\pm\,0.19}$ & 2.50 & (f);(49) \\
N1374 & N1316(ForA) & 39.78${\scriptstyle\,\pm\,0.05}$ & -0.32${\scriptstyle\,\pm\,0.10}$ & 0.78 & 39.78${\scriptstyle\,\pm\,0.05}$ & -0.32${\scriptstyle\,\pm\,0.10}$ & 1.69 & 39.78${\scriptstyle\,\pm\,0.05}$ & 2.51 & (k,c);(50) \\
N0383 & N0410 & 40.53${\scriptstyle\,\pm\,0.10}$ & -0.09${\scriptstyle\,\pm\,0.05}$ & 0.78 & 40.53${\scriptstyle\,\pm\,0.10}$ & -0.09${\scriptstyle\,\pm\,0.05}$ & 1.81 & 40.53${\scriptstyle\,\pm\,0.10}$ & 2.63 & (c);(51) \\
N6251 & BGG:N6251 & 41.59${\scriptstyle\,\pm\,0.13}$ & -0.22${\scriptstyle\,\pm\,0.08}$ & 1.26 & 41.75${\scriptstyle\,\pm\,0.07}$ & -0.08${\scriptstyle\,\pm\,0.04}$ & 1.83 & 42.11${\scriptstyle\,\pm\,0.31}$ & 2.64 & (K,b,L);(52) \\
N4594 & BGG:Sombrero & 39.93${\scriptstyle\,\pm\,0.08}$ & -0.23${\scriptstyle\,\pm\,0.14}$ & 1.18 & 40.51${\scriptstyle\,\pm\,0.25}$ & -0.20${\scriptstyle\,\pm\,0.12}$ & 1.72 & 40.51${\scriptstyle\,\pm\,0.25}$ & 2.57 & (U,V,X);(53) \\
N1332 & BGG:N1332 & 40.00${\scriptstyle\,\pm\,0.09}$ & -0.21${\scriptstyle\,\pm\,0.04}$ & 0.54 & 40.32${\scriptstyle\,\pm\,0.08}$ & -0.27${\scriptstyle\,\pm\,0.04}$ & 1.45 & 40.80${\scriptstyle\,\pm\,0.09}$ & 2.54 & (e,d);(54) \\
N5813 & N5846 & 41.76${\scriptstyle\,\pm\,0.07}$ & -0.15${\scriptstyle\,\pm\,0.04}$ & 1.23 & 41.76${\scriptstyle\,\pm\,0.07}$ & -0.15${\scriptstyle\,\pm\,0.04}$ & 1.77 & 41.76${\scriptstyle\,\pm\,0.07}$ & 2.60 & (f);(55) \\
N0315 & BGG:N0315 & 41.04${\scriptstyle\,\pm\,0.13}$ & -0.18${\scriptstyle\,\pm\,0.05}$ & 0.85 & 41.30${\scriptstyle\,\pm\,0.05}$ & -0.12${\scriptstyle\,\pm\,0.04}$ & 1.86 & 41.60${\scriptstyle\,\pm\,0.15}$ & 2.62 & (c,b,h);(56) \\
N5077 & BFG:N5077 & 40.60${\scriptstyle\,\pm\,0.20}$ & -0.15${\scriptstyle\,\pm\,0.07}$ & 1.04 & 40.60${\scriptstyle\,\pm\,0.20}$ & -0.15${\scriptstyle\,\pm\,0.07}$ & 1.77 & 40.60${\scriptstyle\,\pm\,0.20}$ & 2.60 & (k,c);(57) \\
N0524 & BGG:N0524 & 40.11${\scriptstyle\,\pm\,0.10}$ & -0.30${\scriptstyle\,\pm\,0.07}$ & 0.85 & 40.62${\scriptstyle\,\pm\,0.09}$ & -0.19${\scriptstyle\,\pm\,0.06}$ & 1.75 & 40.95${\scriptstyle\,\pm\,0.09}$ & 2.58 & (a,d);(58) \\
N1399 & BCG:Fornax & 40.94${\scriptstyle\,\pm\,0.05}$ & -0.05${\scriptstyle\,\pm\,0.04}$ & 0.90 & 41.46${\scriptstyle\,\pm\,0.05}$ & 0.09${\scriptstyle\,\pm\,0.04}$ & 1.54 & 42.52${\scriptstyle\,\pm\,0.04}$ & 2.73 & (e,i);(59) \\
N3115 & isolated & 39.70${\scriptstyle\,\pm\,0.24}$ & -0.30${\scriptstyle\,\pm\,0.05}$ & 0.95 & 39.70${\scriptstyle\,\pm\,0.24}$ & -0.30${\scriptstyle\,\pm\,0.05}$ & 1.70 & 39.70${\scriptstyle\,\pm\,0.24}$ & 2.52 & (k,u);(60) \\
IC1459 & BGG:IC1459 & 40.04${\scriptstyle\,\pm\,0.14}$ & -0.21${\scriptstyle\,\pm\,0.05}$ & 0.70 & 40.52${\scriptstyle\,\pm\,0.12}$ & -0.22${\scriptstyle\,\pm\,0.06}$ & 1.59 & 41.45${\scriptstyle\,\pm\,0.12}$ & 2.56 & (e,d);(61) \\
N4374 & N4486/N4472 & 40.66${\scriptstyle\,\pm\,0.06}$ & -0.12${\scriptstyle\,\pm\,0.04}$ & 1.00 & 40.66${\scriptstyle\,\pm\,0.06}$ & -0.12${\scriptstyle\,\pm\,0.04}$ & 1.79 & 40.66${\scriptstyle\,\pm\,0.06}$ & 2.62 & (f);(62) \\
N5846 & BGG:N5846 & 41.43${\scriptstyle\,\pm\,0.09}$ & -0.19${\scriptstyle\,\pm\,0.04}$ & 1.18 & 41.64${\scriptstyle\,\pm\,0.09}$ & -0.12${\scriptstyle\,\pm\,0.05}$ & 1.78 & 42.08${\scriptstyle\,\pm\,0.09}$ & 2.62 & (e,d);(63) \\
N1277 & N1275(Perseus) & 40.15${\scriptstyle\,\pm\,0.11}$ & 0.01${\scriptstyle\,\pm\,0.05}$ & 0.30 & 40.15${\scriptstyle\,\pm\,0.11}$ & 0.01${\scriptstyle\,\pm\,0.05}$ & 1.86 & 40.15${\scriptstyle\,\pm\,0.11}$ & 2.68 & (T);(64) \\
IC4296 & BCG:A3565 & 41.23${\scriptstyle\,\pm\,0.07}$ & -0.16${\scriptstyle\,\pm\,0.05}$ & 1.00 & 41.83${\scriptstyle\,\pm\,0.11}$ & 0.06${\scriptstyle\,\pm\,0.06}$ & 2.00 & 42.00${\scriptstyle\,\pm\,0.08}$ & 2.71 & (l,m);(65) \\
N7768 & BFG:A2666 & 41.84${\scriptstyle\,\pm\,0.31}$ & -0.05${\scriptstyle\,\pm\,0.08}$ & 1.48 & 41.84${\scriptstyle\,\pm\,0.31}$ & -0.05${\scriptstyle\,\pm\,0.08}$ & 1.83 & 41.84${\scriptstyle\,\pm\,0.31}$ & 2.66 & (k,c);(66) \\
U12064 & BFG:3C449 & 41.04${\scriptstyle\,\pm\,0.09}$ & 0.00${\scriptstyle\,\pm\,0.04}$ & 1.00 & 41.04${\scriptstyle\,\pm\,0.09}$ & 0.00${\scriptstyle\,\pm\,0.04}$ & 1.86 & 41.04${\scriptstyle\,\pm\,0.09}$ & 2.68 & (g);(67) \\
N6240S & isolated & 41.49${\scriptstyle\,\pm\,0.10}$ & -0.06${\scriptstyle\,\pm\,0.05}$ & 0.78 & 41.49${\scriptstyle\,\pm\,0.10}$ & -0.06${\scriptstyle\,\pm\,0.05}$ & 1.82 & 41.49${\scriptstyle\,\pm\,0.10}$ & 2.65 & (z);(68) \\
N6861 & N6868(Tel.) & 40.74${\scriptstyle\,\pm\,0.15}$ & -0.03${\scriptstyle\,\pm\,0.04}$ & 1.49 & 40.74${\scriptstyle\,\pm\,0.15}$ & -0.03${\scriptstyle\,\pm\,0.04}$ & 1.84 & 40.74${\scriptstyle\,\pm\,0.15}$ & 2.67 & (n);(69) \\
N4649 & N4486/N4472 & 40.94${\scriptstyle\,\pm\,0.05}$ & -0.04${\scriptstyle\,\pm\,0.04}$ & 1.04 & 40.94${\scriptstyle\,\pm\,0.05}$ & -0.04${\scriptstyle\,\pm\,0.04}$ & 1.83 & 40.94${\scriptstyle\,\pm\,0.05}$ & 2.66 & (f);(70) \\
N7619 & BGG:Pegasus & 41.48${\scriptstyle\,\pm\,0.13}$ & -0.03${\scriptstyle\,\pm\,0.04}$ & 1.46 & 42.41${\scriptstyle\,\pm\,0.13}$ & 0.00${\scriptstyle\,\pm\,0.04}$ & 2.20 & 42.65${\scriptstyle\,\pm\,0.13}$ & 2.68 & (n,o);(71) \\
N4472 & BCG$_1$:Virgo & 41.18${\scriptstyle\,\pm\,0.07}$ & 0.02${\scriptstyle\,\pm\,0.04}$ & 1.20 & 43.08${\scriptstyle\,\pm\,0.05}$ & 0.30${\scriptstyle\,\pm\,0.05}$ & 1.86 & 43.38${\scriptstyle\,\pm\,0.05}$ & 2.84 & (f,p,q);(72) \\
N3923 & BGG:N3923 & 40.43${\scriptstyle\,\pm\,0.12}$ & -0.26${\scriptstyle\,\pm\,0.05}$ & 1.00 & 40.60${\scriptstyle\,\pm\,0.11}$ & -0.24${\scriptstyle\,\pm\,0.04}$ & 1.62 & 40.92${\scriptstyle\,\pm\,0.25}$ & 2.55 & (e,b,d);(73) \\
N3091 & BGG:HCG42 & 41.34${\scriptstyle\,\pm\,0.15}$ & -0.10${\scriptstyle\,\pm\,0.04}$ & 1.00 & 41.68${\scriptstyle\,\pm\,0.15}$ & -0.06${\scriptstyle\,\pm\,0.04}$ & 1.90 & 42.08${\scriptstyle\,\pm\,0.15}$ & 2.65 & (g,d,b);(74) \\
N1550 & BGG:N1550 & 42.32${\scriptstyle\,\pm\,0.10}$ & 0.11${\scriptstyle\,\pm\,0.04}$ & 1.45 & 43.04${\scriptstyle\,\pm\,0.10}$ & 0.14${\scriptstyle\,\pm\,0.04}$ & 2.28 & 43.20${\scriptstyle\,\pm\,0.11}$ & 2.75 & (n,r,i);(75) \\
N6086 & BGG:A2162 & 41.43${\scriptstyle\,\pm\,0.17}$ & 0.00${\scriptstyle\,\pm\,0.08}$ & 1.70 & 42.48${\scriptstyle\,\pm\,0.17}$ & 0.18${\scriptstyle\,\pm\,0.07}$ & 2.45 & 42.59${\scriptstyle\,\pm\,0.18}$ & 2.77 & (S);(76) \\
A1836$_\textrm{B}$ & BCG:A1836 & 41.51${\scriptstyle\,\pm\,0.08}$ & 0.18${\scriptstyle\,\pm\,0.09}$ & 1.41 & 42.54${\scriptstyle\,\pm\,0.05}$ & 0.22${\scriptstyle\,\pm\,0.05}$ & 2.11 & 42.70${\scriptstyle\,\pm\,0.10}$ & 2.80 & (P);(77) \\
N1407 & BGG:Eridanus & 41.94${\scriptstyle\,\pm\,0.11}$ & 0.08${\scriptstyle\,\pm\,0.06}$ & 1.23 & 42.00${\scriptstyle\,\pm\,0.11}$ & 0.08${\scriptstyle\,\pm\,0.06}$ & 2.03 & 42.08${\scriptstyle\,\pm\,0.11}$ & 2.72 & (s);(78) \\
N5328 & BGG:N5328 & 41.08${\scriptstyle\,\pm\,0.12}$ & -0.07${\scriptstyle\,\pm\,0.11}$ & 1.08 & 41.67${\scriptstyle\,\pm\,0.11}$ & -0.02${\scriptstyle\,\pm\,0.20}$ & 2.03 & 41.80${\scriptstyle\,\pm\,0.11}$ & 2.67 & (I,k);(79) \\
M1216 & isolated & 41.83${\scriptstyle\,\pm\,0.12}$ & -0.10${\scriptstyle\,\pm\,0.05}$ & 1.04 & 41.83${\scriptstyle\,\pm\,0.12}$ & -0.10${\scriptstyle\,\pm\,0.05}$ & 1.80 & 41.83${\scriptstyle\,\pm\,0.12}$ & 2.63 & (A);(80) \\
N4486 & BCG$_2$:Virgo & 42.26${\scriptstyle\,\pm\,0.06}$ & 0.15${\scriptstyle\,\pm\,0.05}$ & 0.95 & 43.08${\scriptstyle\,\pm\,0.05}$ & 0.30${\scriptstyle\,\pm\,0.05}$ & 1.86 & 43.38${\scriptstyle\,\pm\,0.05}$ & 2.84 & (l,p,q);(81) \\
N5419 & BCG:AS753 & 41.53${\scriptstyle\,\pm\,0.12}$ & 0.27${\scriptstyle\,\pm\,0.05}$ & 1.28 & 42.26${\scriptstyle\,\pm\,0.14}$ & 0.32${\scriptstyle\,\pm\,0.04}$ & 2.32 & 42.52${\scriptstyle\,\pm\,0.12}$ & 2.85 & (t);(82) \\
N3842 & BCG:A1367 & 41.78${\scriptstyle\,\pm\,0.18}$ & 0.12${\scriptstyle\,\pm\,0.06}$ & 1.68 & 43.93${\scriptstyle\,\pm\,0.11}$ & 0.51${\scriptstyle\,\pm\,0.04}$ & 2.58 & 44.06${\scriptstyle\,\pm\,0.12}$ & 2.95 & (k,c,D,i,E);(83) \\
N1600 & BGG:N1600 & 41.30${\scriptstyle\,\pm\,0.11}$ & 0.16${\scriptstyle\,\pm\,0.07}$ & 1.32 & 41.52${\scriptstyle\,\pm\,0.11}$ & 0.21${\scriptstyle\,\pm\,0.09}$ & 1.75 & 41.52${\scriptstyle\,\pm\,0.11}$ & 2.79 & (\"e);(84) \\
N4889 & BCG:Coma & 42.78${\scriptstyle\,\pm\,0.07}$ & 0.38${\scriptstyle\,\pm\,0.16}$ & 1.32 & 44.51${\scriptstyle\,\pm\,0.07}$ & 0.86${\scriptstyle\,\pm\,0.06}$ & 2.56 & 44.88${\scriptstyle\,\pm\,0.06}$ & 3.14 & (k,Y,Z,\"a);(85) \\
  
\vspace{-0.3cm}        
\enddata

\tablenotetext{}{\small \textbf{\textit{Columns.}} 
(i) Galaxy name, with prefixes defined as N$=$NGC (New General Catalogue); IC (Index Catalogue); A$=$Abell Catalogue; M$=$Mrk (Markarian Catalogue); U$=$UGC (Uppsala General Catalogue). 
Top to bottom: galaxies are in order of ascending $\mbh$.
(ii) Central galaxy of the macro-scale cluster, group, or field halo (from \citealt{Tully:2015} PGC1 catalog and (xi) references); whenever matching the considered galaxy, we label it `brightest cluster/group/field galaxy' (BCG/BGG/BFG) of the related macro-scale halo (e.g., `BGG:Pegasus'). Clusters typically have $N_{\rm m}\sim50$\,-\,1000; groups $N_{\rm m}\sim8$\,-\,50; fields $N_{\rm m}\sim 2$\,-\,8; galaxies with $N_{\rm m} \le 2$ are labeled as `isolated' (which includes pairs). 
(iii) X-ray luminosity (0.3\,-\,7\,keV) of the hot halo within the galactic/CGM radius.
(iv) X-ray temperature within the galactic/CGM radius.
(v) Galactic/CGM radius ($\sim1$\,-\,3\;$\re$). 
(vi) X-ray luminosity (0.3\,-\,7\,keV) of the hot halo within the core radius (different from (iii) only for BCGs/BGGs). 
(vii) X-ray temperature within the core radius (analog of $T_{\rm x, 500}$ and group/cluster virial temperature; different from (iv) only for BCGs/BGGs).
(viii) Group/cluster core radius ($\sim 0.15\,R_{500}$).
(ix) X-ray luminosity (0.3\,-\,7\,keV) of the hot halo within the macro-scale cluster/group halo ($\sim$$\,R_{500}$; different from (iii) only for BCGs/BGGs).
(x) Group/cluster $R_{500}$ ($\simeq R_{\rm vir}/1.7$).
(xi) References used for the listed X-ray properties and single-object notes (if any).\\
\indent
\textbf{\textit{    References.}} 
(a) \citet{Kim:2015};
(b) \citet{Babyk:2018};
(c) \citet{Goulding:2016};
(d) \citet{Osmond:2004};
(e) \citet{Nagino:2009};
(f) \citet{Su:2015};
(g) \citet{Lakhchaura:2018};
(h) \citet{Helsdon:2000};
(i) \citet{Reiprich:2002};
(j) \citet{Boroson:2011};
(k) \citet{OSullivan:2001};
(l) \citet{Athey:2007};
(m) \citet{Horner:2001};
(n) \citet{Fukazawa:2006};
(o) \citet{OSullivan:2017};
(p) \citet{Peres:1998};
(q) \citet{Urban:2011};
(r) \citet{Sun:2003};
(s) \citet{Su:2014};
(t) \citet{Eckmiller:2011};
(u) \citet{Diehl:2008b};
(v) \citet{OSullivan:2003};
(w) \citet{LaMassa:2012};
(x) \citet{Croston:2007};
(y) \citet{Anderson:2016}; 
(z) \citet{Grimes:2005};
(A) \citet{Buote:2019}; 
(B) \citet{Bogdan:2011};
(C) \citet{Li:2013_halo};
(D) \citet{Hudson:2010};
(E) \citet{Ebeling:1996};
(F) \citet{Tyler:2004};
(G) \citet{Gallo:2006};
(H) \citet{Sun:2007};
(I) \citet{Trinchieri:2012};
(J) \citet{Wang:2010};
(K) \citet{Gliozzi:2004};
(L) \citet{Mulchaey:2003};
(M) \citet{Supper:2001};
(N) \citet{Liu:2010};
(O) \citet{Evans:2006};
(P) \citet{Stawarz:2014};
(Q) \citet{Swartz:2003};
(R) \citet{Konami:2009};
(S) \citet{Burns:1994};
(T) \citet{Fabian:2013};
(U) \citet{Pellegrini:2002};
(V) \citet{Li:2007};
(X) \citet{Benson:2000};
(Y) \citet{Sanders:2014};
(Z) \citet{White:1997};
(\"a) \citet{Mittal:2011};
(\"e) \citet{Sivakoff:2004}.
\\
\indent
\textbf{\textit{    Notes.}} 
(4, 13, 20) Cooler component of the 2-$T$ thermal plasma fit (the hotter one driven by a starburst or jet-driven shock).
(23, 30, 37, 39, 54, 59, 61, 63, 73, 77, 78, 80, 82, 84) \Lx\ and/or \Tx\ retrieved via the integration of the plasma density (emissivity) profile and/or LW temperature profile within the given \Rx\ (avoiding extrapolations). 
(28) Only the diffuse ISM component is considered (excluding the shock-heated lobes).
(50, 57, 66, 76) Temperature available only via the $\lx-\tx$ relation (\citealt{Goulding:2016}; full aperture). 
(65, 76, 77) $L_{\rm x,500}$ newly retrieved via archival ROSAT data by using \citet{Eckert:2012} procedure. 
(71) BGG of LGG\,473 (Lyon Galaxy Group catalog).
(72, 81) N4472 and N4486 are equally massive and central to Virgo cluster.
}

\label{tabx}
\end{deluxetable*}
\end{center}
\capstarttrue


\end{appendix}

\label{lastpage}
\end{document}